\pdfoutput=1
\documentclass[12pt]{caltech_thesis}
\usepackage[hyphens]{url}
\usepackage{lipsum}
\usepackage{graphicx}

\usepackage{todonotes}

\usepackage[utf8]{inputenc}
\usepackage[T1]{fontenc}
\usepackage{newtxtext,newtxmath}

\usepackage[numbers,sort&compress]{natbib}
\usepackage[sectionbib]{bibunits}
\usepackage{braket}
\usepackage{caption}
\usepackage{subcaption}
\usepackage{mathrsfs}
\usepackage{hyperref}

\defaultbibliographystyle{plainnat}
\usepackage{enumitem}

\begin{document}

\title{A Deep Dive into the Connections Between the Renormalization Group and Deep Learning in the Ising Model}
\author{Kelsie Taylor}

\degreeaward{Bachelor of Science in Physics}    
\university{California Institute of Technology} 
\address{Pasadena, California}                  
\unilogo{caltech.pdf}                           
\copyyear{2023}
\defenddate{June 5th}
\orcid{0009-0001-7510-2306}

\rightsstatement{All rights reserved}

\maketitle[logo]

\begin{acknowledgements}

  I would like to acknowledge my thesis advisors Dr. Joseph Lykken and Dr. Maria Spiropulu for making this project possible. I'd also like to acknowledge the assistance of Jonathan Booker for his programming of the Wolff algorithm in C and Damian Musk for his assistance in moving the Restricted Boltzmann Machines into TensorFlow. I'd also like to thank my academic advisor Dr. Mark Wise for being a useful resource for me all four years. Furthermore, I'd like to thank my former Caltech research mentors: Dr. Alan Weinstein, Dr. Derek Davis, Dr. Michele Papucci, Dr. Nikolai Lauk, and Dr. Neil Sinclair. Without them, I would not have developed the research skills necessary to complete this project. In addition, I'd like to thank the members of Dabney House and Venerable House for supporting me through the many long nights I had during both my thesis and my entire degree. 

  This thesis work was supported by the Department of Energy Office of High Energy Physics QuantISED program grant SC0019219 on QCCFP-QMLQCF; the Fermi Research Alliance, LLC under Contract No. DE-AC02-07CH11359 with the U.S. Department of Energy, Office of Science, Office of High Energy Physics; the Caltech SURF program and the Brinson Foundation. 
\end{acknowledgements}

\begin{abstract}

  The renormalization group (RG) is an essential technique in statistical physics and quantum field theory, which considers scale-invariant properties of physical theories and how these theories' parameters change with scaling. Deep learning is a powerful computational technique that uses multi-layered neural networks to solve a myriad of complicated problems. Previous research suggests the possibility that unsupervised deep learning may be a form of RG flow, by being a layer-by-layer coarse graining of the original data. We examined this connection on a more rigorous basis for the simple example of Kadanoff block renormalization of the 2D nearest-neighbor Ising model, with our deep learning accomplished via Restricted Boltzmann Machines (RBMs). We developed extensive renormalization techniques for the 1D and 2D Ising model to provide a baseline for comparison.  For the 1D Ising model, we successfully used Adam optimization on a correlation length loss function to learn the group flow, yielding results consistent with the analytical model for infinite N. For the 2D Ising model, we successfully generated Ising model samples using the Wolff algorithm, and performed the group flow using a quasi-deterministic method, validating these results by calculating the critical exponent $\nu$. We then examined RBM learning of the Ising model layer by layer, finding a blocking structure in the learning that is qualitatively similar to RG. Lastly, we directly compared the weights of each layer from the learning to Ising spin renormalization, but found quantitative inconsistencies for the simple case of nearest-neighbor Ising models.
\end{abstract}

\tableofcontents
\listoffigures
\listoftables
\printnomenclature

\mainmatter
\mainmatter

\chapter{Introduction}

Within quantum field theory and statistical physics, it is often necessary to change parameters within a theory in order to treat infinities and correct for the effects of self-interactions. This technique is called renormalization. In general, renormalization arises when there are problems of scale, where parameters describing a process at short-distance scales may disagree with those describing a process at long-distance scales. This idea can be generalized, forming the concept of the renormalization group (RG): the apparatus which allows us to deal with multi-scale features in physics.

Most phenomena have features that are characterized by multiple scales, and near certain critical points, the dynamics of these phenomena become scale-invariant. This implies that we can take a theory with some parameters when examined at small scales near a critical point, and instead examine it at coarser scale with new, "renormalized" parameters. Usually, this results in fewer relevant parameters being needed to describe the dynamics. This process is called an "RG flow". RG flows show that even systems that are microscopically different can have similar macroscopic behaviors. Due to this, we can define a class of systems consisting of all systems with the same renormalization parameters: the "universality class" of a model. Models in the same class can apply to any number of fields, including statistical mechanics, quantum behavior, social dynamics, or even the stock market \cite{slides}.

Deep learning is another useful technique used in all areas of physics. It uses multiple layers of representations to learn features directly from training data, allowing for massive improvements in image processing \cite{NIPS2012_c399862d}, language modeling \cite{liu2023summary}, and more \cite{Mapping}. In particular, we focus on algorithms known as Deep Neural Networks (DNNs), graphical statistical models where each layer receives inputs from the layer before them. DNNs have shown massive success, but it is still not completely theoretically understood why they work so well.

In a 2014 paper \cite{Mapping}, Mehta and Schwab argued that one reason DNNs work so well is because they perform an effective coarse-graining of the training data, in the same way that an RG flow does. In particular, they claimed that an exact mapping exists between the RG flow of the one and two dimensional Ising models and a DNN comprised of stacks of "Restricted Boltzmann Machines" (RBMs) (one of the simplest possible DNNs). This claim was controversial \cite{DeepAndCheap, DeepCheepComment}, warranting further discussion.

The Ising model was chosen because it is a simple model to which renormalization group flows can be applied. It is a classical model that is directly solvable in the one-dimensional case, with a simple Hamiltonian that only depends on spins, interaction strength, and magnetic field. Because the model is simple, we are able to more easily prove results for the Ising model, results which can be generalized to the entire Ising universality class. This includes models not just in physics, but also models in medicine, sociology, and economics \cite{PhysRevLett.114.108101, 2013}. 

The one dimensional Ising Model RG flow can be solved exactly using easy formulas, although the model does not contain any non-trivial critical behavior. One can use this exact solution to show mathematically that RG flow has a one-to-one mapping with stacks of RBMs. One may also apply RBM learning to the 2D Ising model, yielding a weight plot that gives results similar to the two dimensional Ising RG flow \cite{Mapping} . 

The existence of a connection between RG flow and deep learning could be a paradigm shift in not just the field of physics, but also in other data-driven fields that rely on similar renormalization or learning techniques. The work showing that there is a mapping between RBMs and RG flow in the one dimensional Ising model is promising, and the similarities in deep learning in the 2D Ising model to renormalization group flow is compelling. Other studies in group theory, renormalization, and deep learning also suggest a compelling connection \cite{PCA, GroupTheoryMachineLearning}. 

We thus re-examine the claims of \cite{Mapping} in detail, providing a "deep dive" into the qualitative and quantitative nature of this question. To do so, for the remainder of this chapter, we discuss in greater detail the specifics of RG flows and RBM learning to provide the necessary background for the rest of the work. Then, we develop techniques for renormalization group flows in the 1D and 2D Ising models to provide a baseline to which we compare our RBM models, with the 1D Ising model discussed in Chapter \ref{chap:1d} and the 2D Ising model discussed in Chapter \ref{chap:renorm}. Following this, we use RBM learning to reconstruct the Ising model, and perform a qualitative analysis of the weight structure in Chapter \ref{chap:dl}.  Lastly, we combine the RG structure and RBM structures to perform a quantitative analysis in Chapter \ref{chap:connection} and reach our general conclusions in Chapter \ref{chap:conclusions}. In addition, we include additional figures in Appendices \ref{chap:2d_layer_plots} and \ref{chap:connection_plots}.

\section{Renormalization Group Techniques}

For a general renormalization group flow, we have some set of couplings $\{ K \}$ that fully characterize the system for one scale. Then, in applying the renormalization group transformation, we receive a new set of couplings $\{ K' \}$ which preserve some properties of the system. In order to perform a renormalization group flow, we must be able to find all of the couplings $\{ K' \}$ as functions of the old couplings $\{ K \}$. These new couplings will define a new, renormalized model.

Within the universality class of the $d$-dimensional Ising model, we apply renormalization group flows via a block spin procedure: grouping spins together and assigning a new spin to them based on the old ones. These groups of spins each have a respective size $b$. Within the group flow, we expect the free energy of the system to follow the inhomogeneous transformation law 
\begin{equation}
    \label{translaw}
    f(\{ K \}) = g(\{K \}) + b^{-d} f(\{K' \})
\end{equation}

and the correlation length of the system to follow the transformation law
\begin{equation}
\label{eq:xi_varying_gen}
\xi(\{K' \}) = b^{-1} \xi(\{K \}),
\end{equation}
either of which can be used to calculate the new coupling constants $\{ K' \}$ \cite{cardy_1996}. 

We attempt to use these equations to perform accurate RG flows, in order to compare the group flow results we get to RBM learning results. We apply some basic learning techniques to perform accurate RG flows, first learning the 1D Ising model to check the method's validity, and then using it to calculate the RG flow of the 2D Ising model. We run the learning in TensorFlow, a Python package for machine learning \cite{45381}, attempting to calculate the new coupling constants $\{ K'\}$ in terms of the old ones, $\textbf\{ K\}$. To do so, we provide a loss function using either the free energy or the correlation length, then provide the original free energy or correlation length of the system, along with the original coupling constants and guesses for the new coupling constants. Then, the machine learns how the coupling constant runs through Adam optimization, a standard machine learning optimization algorithm \cite{adam_opt, ruder2017overview}.

We also can run our method in reverse: inputting into the machine the running of the coupling constant, and having it determine the Hamiltonian from this coupling constant. This will allow us to see if the basic learning algorithm can learn locality properties we consider essential to nature. We only did a simple exploration of this for the 1D Ising model, however.

\section{The Restricted Boltzmann Machine} \label{sec:RBMS}

We use simple energy-based models called Restricted Boltzmann Machines (RBMs) to analyze the connection between the RG and deep learning. Our following discussion follows closely to that in \cite{Mapping, Hinton2012}. 

We consider RBMs that act on simple binary data drawn from a probability distribution $P(\{v_i)\})$ with $\{v_i\}$ a set of $N$ binary "visible" spins indexed by $i=1, ... ,N.$ For example, the data could consists of a black and white picture, with black pixels being denoted as 1 and white pixels being denoted as 0. Similarly, the data could be Ising spins.

RBMs model the data by introducing a new set of "hidden" spins, $\{h_j\},$ a set of $M$ binary spins indexed by $j=1, ... ,M.$ These spins couple to the visible spins through the following energy function:

\begin{equation}
    E(\{v_i\}, \{h_j\}) = \sum_j b_j h_j + \sum_{ij} v_i w_{ij} h_j + \sum_{i} v_i a_i, 
\end{equation}

where $\lambda = \{ b_j, w_{ij}, a_i \}$ forms the variational parameters of the model, with $w_{ij}$ falling between $-\infty$ and $\infty$. The joint distribution for a given set of $\{v_i\}$ and $\{h_j\}$ is then given in the expected form from statistical mechanics:
\begin{equation}
    p_\lambda(\{v_i\}, \{h_j\}) = \frac{ \exp(-E(\{v_i\}, \{h_j\}))}{Z},
\end{equation}
where $Z$ denotes the relevant partition factor. 
We can thus define marginal distributions for both visible spins, 
\begin{equation}
    p_\lambda(\{v_i\}) = \text{Tr}_{{h_j}} \left\{\frac{ \exp(-E(\{v_i\}, \{h_j\}))}{Z}\right\},
\end{equation}
and hidden spins
\begin{equation}
    p_\lambda(\{h_j\}) = \text{Tr}_{v_i} \left\{\frac{ \exp(-E(\{v_i\}, \{h_j\}))}{Z}\right\}.
\end{equation}

From this, we can define new variational Hamiltonians as follows for the visible spins, 
\begin{equation}
    p_\lambda(\{v_i\}) = \frac{ \exp(-H_{RBM}(\{v_i\}))}{Z},
\end{equation}
and the hidden spins
\begin{equation}
    p_\lambda(\{v_i\}) = \frac{ \exp(-H_{RBM}(\{h_j\}))}{Z}.
\end{equation}

For unsupervised learning, RBM parameters are generally chosen to minimize the Kullback-Leibler divergence between the actual distribution $P(\{v_i\})$ and the estimated one $p_\lambda(\{v_i\})$:

\begin{equation}
    D_{KL}(P(\{v_i\})||p_\lambda(\{v_i\})) = \text{Tr}_{v_i} \left\{ P(\{v_i\}) \log\left(\frac{ P(\{v_i\})}{p_\lambda(\{v_i\})}\right)\right\}.
\end{equation}
Of course, when the RBM reproduces the visible data exactly, 
\begin{equation}
    D_{KL}(P(\{v_i\})||p_\lambda(\{v_i\})) = 0.
\end{equation}

However, minimizing  $D_{KL}(P(\{v_i\})||p_\lambda(\{v_i\}))$ cannot be done analytically in most cases. Instead, we minimize  the Contrastive Divergence, the difference between two Kullback-Leibler functions, given by
\begin{equation}
CD_1 = D_{KL}(P(\{v_i\})||p_\lambda(\{v_i\})) - D_{KL}(P(\{v_i\})||p_{\lambda,1}(\{v_i\})), \ \ \ \text{\cite{min_contrastive_div, on_contrastive_div}} 
\end{equation}
where $p_{\lambda,1}(\{v_i\})$ is a sample of the distribution $p_\lambda(\{v_i\})$. We minimize this equation because it is easier to approximate the gradient of this function than just the Kullback-Leibler function. Then, we can minimize the function using stochastic gradient descent with momentum, another standard machine learning optimization algorithm \cite{ruder2017overview}. 

To do so, we convert our weight tensors to to binary spins. It suffices to generate 64 numbers $n_{ij}$ between 0 and 1, then convert $n_{ij} > .5$ to 1 and the rest to 0. In particular, we convert the weight tensor values to the range $ [0, 1] $ via the logistic function:
\begin{equation}
\sigma(w) = \frac{1}{1 + e^{-w}}
\label{eq:logistic}
\end{equation}
 
Thus, a large positive weight is close to 1, a large negative weight is close to 0, and a zero weight is .5. 

We can thus construct a probability distribution from the hidden variables using
\begin{equation}
    \text{pHid}_j = \sigma(h_j),
    \label{eq:pHid}
\end{equation}
and take a random sample from this distribution, denoted $\text{sampHid}_j$. We can then define a new reconstructed lattice of 
\begin{equation}
    \text{vReco}_i =  \sum_j w_{ij}^T \cdot \text{sampHid}_j
\end{equation}
and reconstructed hidden weights of 
\begin{equation}
    \text{pHidReco}_j = \sum_i \sigma(w_{ij} \cdot \text{vReco}_i).
\end{equation}

From these distributions, the approximate 
gradient used in learning is given by 
\begin{equation}
    \nabla w_{ij} \approx \langle v_i \otimes \text{pHid}_{j} - \text{vReco}_i \otimes \text{pHidReco}_j \rangle.
\end{equation}

To make this learning "deep", these RBMs are stacked upon one another, such that the hidden layer of the first RBM is the visible layer of the second RBM, and so on. To do so, one maps a visible spin configuration to a hidden configuration using $\text{pHid}_j$. Then, the hidden spins' response to the visible spins can be treated as a new layer, and the cycle continues.

\chapter{Preliminary Connections in The One Dimensional Ising Model}
\label{chap:1d}

\section{Methods}
\label{sec:1d_meth}

In the one dimensional Ising model, the renormalization group flow is exactly solvable. In particular, this model contains $N_0$ classical spins, with each spin taking a value of either 1 or -1. The Hamiltonian is a nearest-neighbor Hamiltonian, given by 
\begin{equation}
\label{1d-Hamil0}
H_0 = -K \sum_{k=0}^{N_0-1} s_0^k s_0^{k+1}.
\end{equation}
Here, we take $K=1$ without loss of generality, absorbing it into the inverse temperature $\beta_0$. We assume periodic boundary conditions, in which $s_0^0 = s_0^N$. \cite{cardy_1996}

From the Hamiltonian, we may calculate the partition function \begin{equation}
\label{1d-Z}
Z(\beta_0) = \text{Tr}_{s_0} e^{- \beta_0 H_0}.\end{equation}

We may then calculate the free energy 
\begin{equation}
\label{f_0-express}
f_0(\beta_0) = - \frac{1}{N_0} \log (Z(\beta_0)), 
\end{equation}
which has an analytical value of 
\begin{equation}
\label{f_0-analy}
f_0(\beta_0) = - \frac{1}{N_0} \log \left((2 \cosh \beta_0)^{N_0} + (2 \sinh \beta_0)^{N_0}\right).
\end{equation}

We can also derive an analytical expression for the correlation length of the model, given by \begin{equation}
\label{xi_0-analy}
\xi_0(\beta_0) = \frac{c}{\log \tanh \beta_0}, 
\end{equation}
where $c = -1$ for simplicity (this constant does not affect the group flow). 

In applying the renormalization flow using blocks of $b$ spins, we yield a new system with $N_1 = N_0/b$ sites and a new inverse temperature $\beta_1(\beta_0)$. Similarly to Equations \ref{1d-Hamil0}-\ref{f_0-analy}, we may define for these new parameters a new Hamiltonian
\begin{equation}
\label{1d-Hamil1}
H_1 = -K \sum_{k=0}^{N_1-1} s_1^k s_1^{k+1},
\end{equation}
a new free energy
\begin{equation}
\label{f_1-analy}
f_1(\beta_1) = - \frac{1}{N_1} \log ((2 \cosh (\beta_1)^{N_1} + (2 \sinh \beta_1)^{N_1}),
\end{equation}
and a new correlation length
\begin{equation}
\label{xi_1-analy}
\xi_1(\beta_1) = \frac{c}{\log \tanh \beta_1},
\end{equation}
where $c = -1$ for consistency.

We thus have a way to calculate free energies and correlation lengths directly from the temperatures, allowing for them to be calculated in loss functions. To create a free energy loss function, we note that equation \ref{translaw} becomes 
\begin{equation}
    \label{1dtranslaw}
     f_0(\beta_0) = g_1(\beta_0) + f_1(\beta_1)/b,
\end{equation}
and we can minimize the loss function 
\begin{equation}
    \label{1dtransloss}
     L(\beta_1) = f_0(\beta_0) - g_1(\beta_0) - f_1(\beta_1)/b.
\end{equation}

Alternatively, we also know that
\begin{equation}
\label{eq:xi_varying}
\xi_1(\beta_1) = b^{-1} \xi_0(\beta_0),
\end{equation}
and we can thus minimize a loss function defined by
\begin{equation}
\label{eq:xi_loss}
L(\beta_1) = b \cdot \frac{-1}{\log \tanh \beta_1} + \frac{1}{\log \tanh \beta_0},
\end{equation}

This running of the coupling constant is analytically solvable, so we can validate our results. We have that:
 \begin{equation}
    \label{1db0}
     \beta_1(\beta_0) = \tanh^{-1}(\tanh (\beta_0)^b)
\end{equation}
and
\begin{equation}
    \label{1dg1}
     g_1(\beta_0) = \frac{\log(\cosh(\beta_1))}{b} -\log(\cosh(\beta_0)) - \frac{b-1}{b} \log(2) \ \ \text{ \cite{cardy_1996}}.
\end{equation}

The main loss function we use for the 1D Ising model is the correlation length loss function, equation \ref{eq:xi_loss}. We use equations \ref{xi_0-analy} and \ref{xi_1-analy} to attempt to learn equation \ref{1db0} through this loss function. Alternatively, we may also use equation \ref{1dtransloss} as a free energy loss function. Then, we'd use equations \ref{f_0-analy}, \ref{f_1-analy} to attempt to derive equations \ref{1db0} and \ref{1dg1}. In addition, we also use the analytical forms of equation \ref{f_0-analy}, \ref{f_1-analy}, \ref{1db0} and \ref{1dg1}, along with an L1 regularization loss function, to attempt to learn the Hamiltonian couplings in equation \ref{1d-Hamil1}. Due to the fact that all of these cases have analytical solutions, we can then check our results against the analytical solutions to see how well or how poorly the learning has done. 

\section{Results}
\begin{figure}
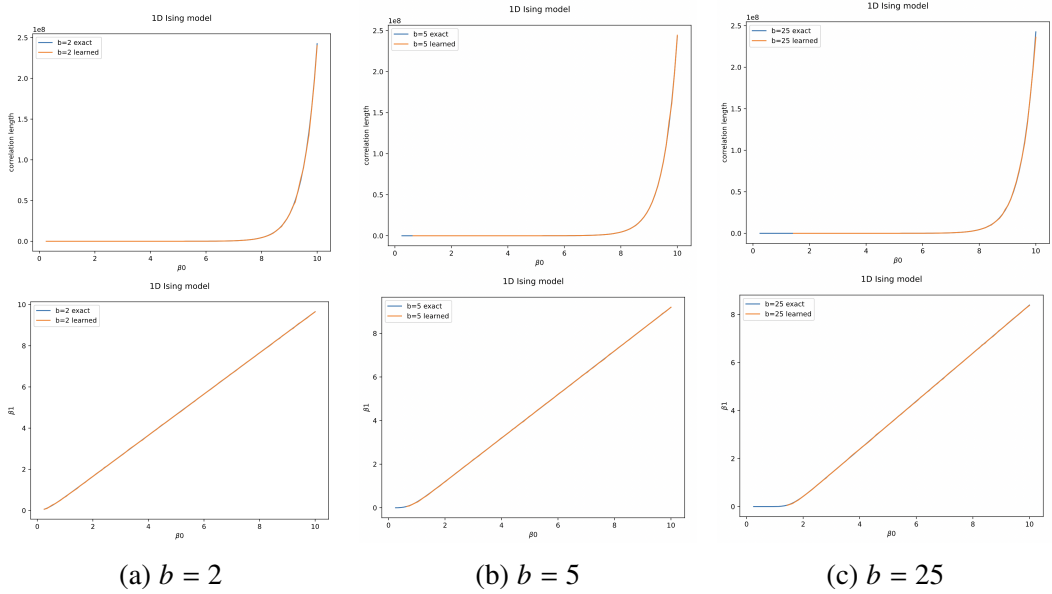

     \centering
     \begin{subfigure}{0.32\textwidth}
         \centering
         \includegraphics[width=\textwidth]{ScreenShot2022-08-13at45619AM.pdf}
         \caption{$b=2$}
         \label{fig:b=2}
     \end{subfigure}
     \hfill
     \begin{subfigure}{0.32\textwidth}
         \centering
         \includegraphics[width=\textwidth]{ScreenShot2022-08-13at52341AM.pdf}
         \caption{$b=5$}
         \label{fig:b=5}
     \end{subfigure}
     \hfill
     \begin{subfigure}{0.32\textwidth}
         \centering
         \includegraphics[width=\textwidth]{ScreenShot2022-08-13at52410AM.pdf}
         \caption{$b=25$}
         \label{fig:b=25}
     \end{subfigure}
        \caption{Comparison of the analytical results and learned results from RG flow for different $b$ values, using the correlation length loss function given in Equation \ref{eq:xi_loss}. Graphs on the top indicate the difference between the real correlation lengths and the learned ones: a measure of how well the loss function is minimized. The bottom graph compares the actual running of the coupling constant $\beta_1$ to the learned value. We successfully learn the running of the coupling constant for each $b$ value, while running into a few problems at low $\beta_0$.}
        \label{fig:1d_ising}
\end{figure}

The results using the correlation length loss function given in Equation \ref{eq:xi_loss} are shown in Figure \ref{fig:1d_ising}. For the most part, our learning model works perfectly, with no way to tell the difference between the learned function and the analytical function. The running of the coupling constant is successfully learned and this works for $b$ values up to 100. 

There are a couple of minor problems in this model, however. For whatever reason, the loss function has trouble with learning on the lower values of $\beta$, with the function returning undefined values instead of actual numbers. This is likely a computing issue. Additionally, the correlation length model only applies to infinite systems, as the analytical correlation length value is defined for a infinite system. Given that all our 2D Ising model results deal with finite systems, it would be useful to also have a model using a finite system.

\begin{figure}
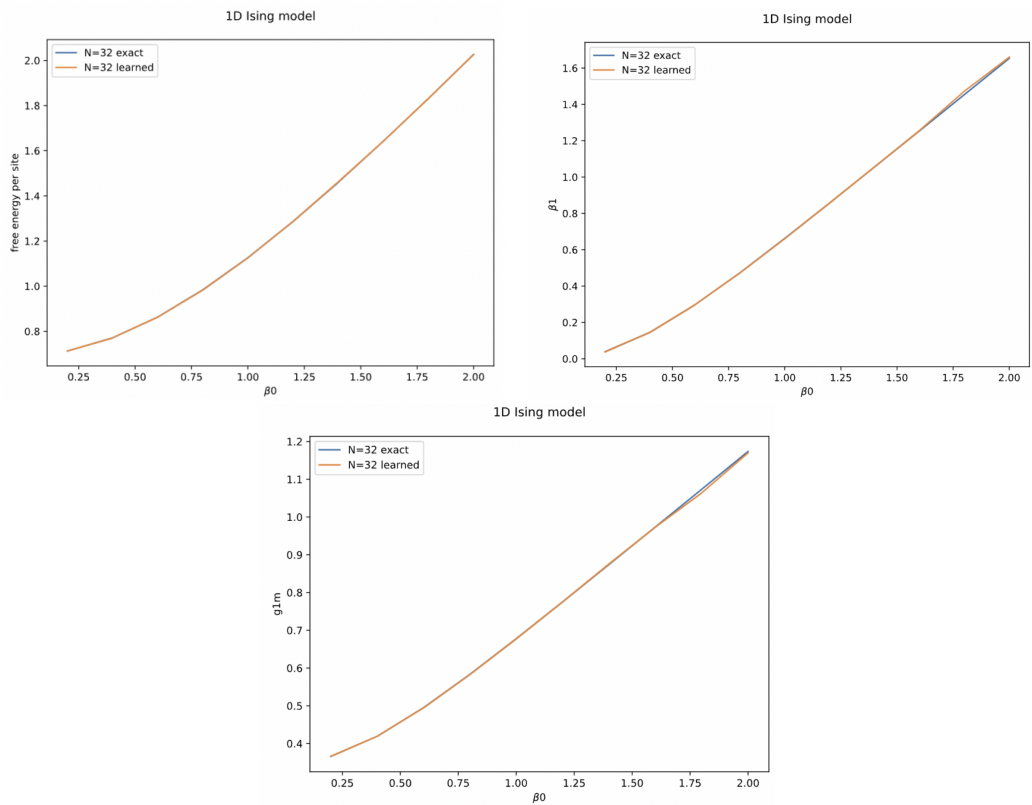

     \centering
     \begin{subfigure}{0.49\textwidth}
         \centering
         \includegraphics[width=\textwidth]{N32_1.pdf}
     \end{subfigure}
     \hfill
     \begin{subfigure}{0.49\textwidth}
         \centering
         \includegraphics[width=\textwidth]{N32_2.pdf}
     \end{subfigure}
     \hfill
     \begin{subfigure}{0.49\textwidth}
         \centering
         \includegraphics[width=\textwidth]{N32_3.pdf}
     \end{subfigure}
    \caption{Comparison of the analytical results and learned results from RG flow for 1D Ising using the free energy loss function Equation \ref{1dtransloss} at $N_0=32$. We let $N_0=32$, $b=2$, evaluated the loss function over four iterations of the renormalization group, guessed $\beta_1 = .2, g_1 = 0$, and ran over 1000 epochs at a learning rate of .005. The results for both $\beta_1$ and $g_1$ seem to match the analytical results very well, and so the learning is working successfully in this case. }
    \label{fig:good}
\end{figure}

\begin{figure}
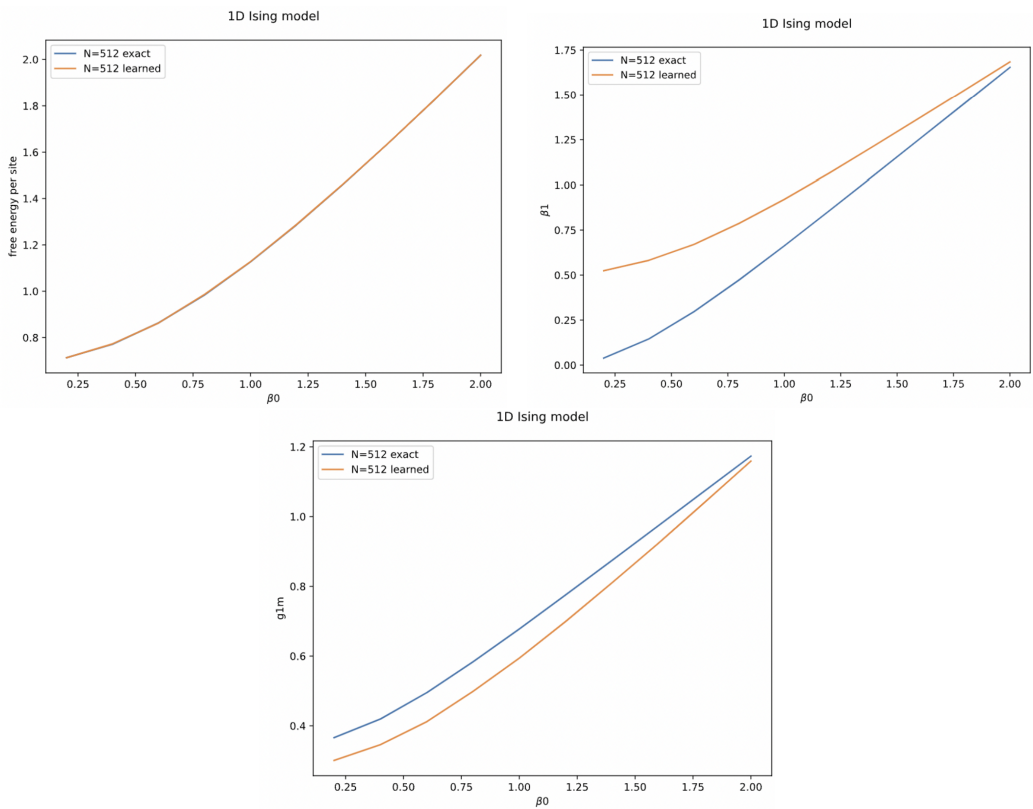

     \centering
     \begin{subfigure}{0.49\textwidth}
         \centering
         \includegraphics[width=\textwidth]{N512_bad.pdf}
     \end{subfigure}
     \hfill
     \begin{subfigure}{0.49\textwidth}
         \centering
         \includegraphics[width=\textwidth]{N512_bad2.pdf}
     \end{subfigure}
     \hfill
     \begin{subfigure}{0.49\textwidth}
         \centering
         \includegraphics[width=\textwidth]{N512_bad3.pdf}
     \end{subfigure}
    \caption{Comparison of the analytical results and learned results from RG flow for 1D Ising using the free energy loss function Equation \ref{1dtransloss} at $N_0=512$, a poor fit.  The case where we let $N_0 = 512$, $b=2$, evaluated the loss function over five iterations of the renormalization group, guessed $\beta_1 = .2, g_1 = 0$, and ran over 1000 epochs at a learning rate of .005. These results are quite poor, with the model over-predicting $\beta_1$ and under-predicting $g_0$, even though the free energy is minimized.}
    \label{fig:bad}
\end{figure}

\begin{figure}
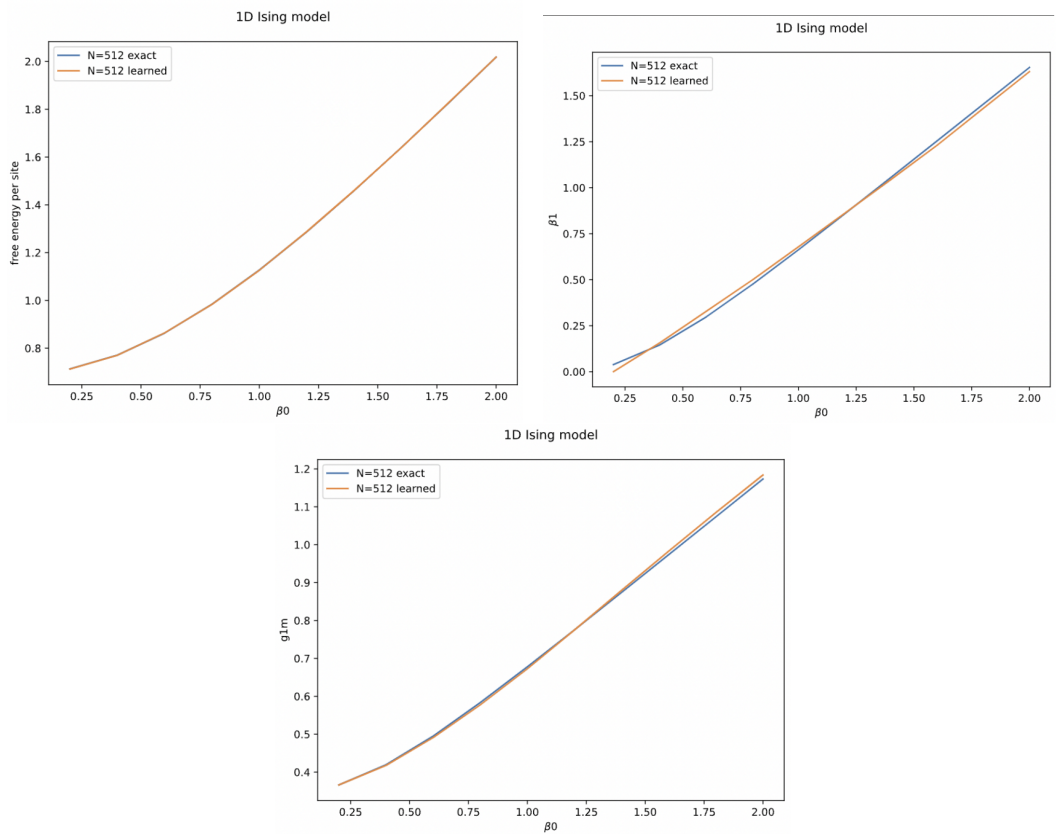

     \centering
     \begin{subfigure}{0.49\textwidth}
         \centering
         \includegraphics[width=\textwidth]{N512_good.pdf}
     \end{subfigure}
     \hfill
     \begin{subfigure}{0.49\textwidth}
         \centering
         \includegraphics[width=\textwidth]{N512_good2.pdf}
     \end{subfigure}
     \hfill
     \begin{subfigure}{0.49\textwidth}
         \centering
         \includegraphics[width=\textwidth]{N512_good3.pdf}
     \end{subfigure}
    \caption{Comparison of the analytical results and learned results from RG flow for 1D Ising using the free energy loss function Equation \ref{1dtransloss} at $N_0=512$, a good fit. The case where we let $N_0 = 512$, $b=2$, evaluated the loss function over five iterations of the renormalization group, guessed $\beta_1 = .833 ( \beta_0 - .2) + .1, g_1 = .444(\beta_0-.2) + .4$, and ran over 1000 epochs at a learning rate of .005. These results are much better than those in Figure \ref{fig:bad}, showing the importance of good starting guesses to resolve under-constraining problems.}
    \label{fig:better}
\end{figure}

We attempted to do this using the free energy loss function Equation \ref{1dtransloss}, finding estimations for equations \ref{1db0} and \ref{1dg1}. We are able to do it for $\beta_0$ values between .2 and 2, for up to $N_0 = 1024$. Higher than that, we run into overflow problems in the analytical solutions. 

For the case of low $N_0$, we were able to yield learning results that were good. In Figure \ref{fig:good},  we let $N_0=32$, $b=2$, evaluated the loss function over four iterations of the renormalization group, guessed $\beta_1 = .2, g_1 = 0$, and ran over 1000 epochs at a learning rate of .005. As can be seen here, the results for both $\beta_1$ and $g_1$ seem to match the analytical results very well, and so the learning is working successfully in this case. 

However, for most other cases, these results are less good, and are instead highly dependent on our starting values for $\beta_1, g_1.$ In Figure \ref{fig:bad}, we let $N_0 = 512$, $b=2$, evaluating the loss function over five iterations of the renormalization group, guessing $\beta_1 = .2, g_1 = 0$, and running over 1000 epochs at a learning rate of .005. These results are quite poor, with the model over-predicting $\beta_1$ and under-predicting $g_0$, even though the free energy is minimized. Other results in other parameter spaces show similar patterns, with one of the parameters overestimated and the other underestimated. This usually means that very precise guesses are needed to properly learn the functions. For example, in the $N_0 = 512$ case, we can change our initial guesses to  $\beta_1 = .833 ( \beta_0 - .2) + .1$ and $g_1 = .444(\beta_0-.2) + .4$, yielding the results in \ref{fig:better}. 

These results suggest a general under-constraining of our learning that seems to be resultant from trying to calculate two values, $\beta_1, g_1$ from our singular loss function. This is, of course to be expected when using a single inhomogeneous equation to determine two functions. It is yet unclear as to how to resolve such an issue. Perhaps this method could be combined with using the correlation length method in order to learn both functions successfully. 

\begin{figure}
    \centering
     \includegraphics[scale=.6]{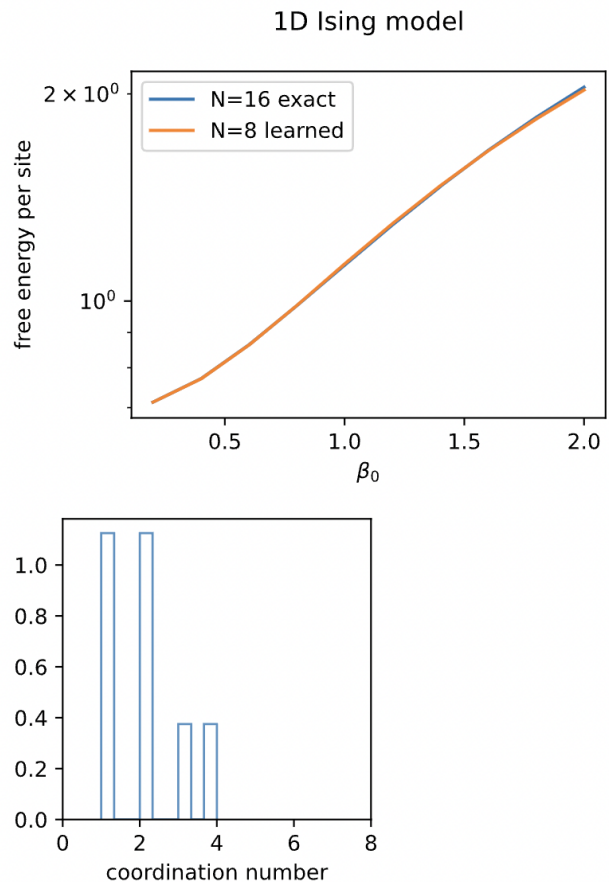}
    \caption{Learning the renormalized Hamiltonian. Here, we let $\beta_0$ go from .2 to 2, and initialize with a random coupling matrix with standard deviation of .2. Although the couplings aren't fully correct (they are given in equation \ref{eq:hamilt}), the resultant Hamiltonian does have an average non-zero correlation number of 2 and 8 total couplings, just like the 1D Ising model should.}
    \label{fig:hamilton}
\end{figure}

In addition, we also performed a method where we learned the renormalized Hamiltonian from the running of the coupling constant, letting the Hamiltonian be any possible coupling of the spin pairs and using the same free energy loss function as in equation \ref{1dtransloss}. This was only run for the $N_0 = 16$ case, as higher spin values lead to computational problems that have not yet been resolved, due to the high amount of coupling constants. Here, we let $\beta_0$ go from .2 to 2, and initialize with a random coupling matrix with standard deviation of .2. We run for 2000 epochs with a learning rate of .001 and a L1 value of 5. The results of the learning are shown in Figure \ref{fig:hamilton}. The recovered coupling matrix is given by
\begin{equation}
\label{eq:hamilt}
\begin{pmatrix}
 0. &  0. &  0. &  0. &  0. &  0. &  0. &  1.07 \\
 0. &  0. &  0.84 & 0. &  0. &  0.84 & 0. &  0. \\
 0.&   0.&   0. &  0. &  0. &  0.85 &1.23 & 0. \\
 0. &  0.  & 0. &  0.  & 0.97& 0. &  0.  &  0. \\
 0. &  0.  & 0.  & 0. &  0.  & 0. &  0.  &  0.  \\
 0. &  0.  & 0.  & 0. &  0.  & 0. &  0.63 & 0.9 \\
 0. &  0. &  0.  & 0. &  0.  & 0. &  0.  & 0.  \\
 0. &  0. &  0.  & 0.&   0.  & 0. &  0.   & 0.  \\
 \end{pmatrix}
\end{equation}
Obviously, this result is not the same as the correct coupling matrix, but it does have an average non-zero correlation number of 2 and 8 total couplings, just like the 1D Ising model. Thus, results seem promising to at least derive some dynamics from renormalization group flows, but more work should be done in this topic. Once again, perhaps this method could be combined with using the correlation length method in order to learn the coupling successfully.

Additionally, current progress in using the free energy loss function is hampered by overflow errors in the analytical solutions, making it impossible to reasonably calculate systems with $N_0>1024$ when we learn the group flow. Additionally, when we attempt to learn the Hamiltonian from the group flow, we run into larger scaling errors, as the number of parameters needed gets absurdly high for $N_0 > 16$. Further efforts may be done to expand the learning to as large of finite models as possible.

\chapter{Renormalization Techniques in the Two Dimensional Ising Model}
\label{chap:renorm}

\section{Methods}

For the 2-dimensional Ising model, we consider a model with $N^2$ classical spins on an $N$ by $N$ lattice, where each spin $s[i, j]$ takes on a value $+1$ or $-1$.  Using nearest neighbor couplings, we get a Hamiltonian of the form 
\begin{equation}
\label{eq:nn_hamil}
    H_0 = -K_1 \sum_{i=0}^{N-1} \sum_{j=0}^{N-1} \sum_{\langle nn \rangle } s[i, j] s[nn],
\end{equation}
where $\langle nn \rangle$ means to sum over all nearest neighbors such that each possible spin-spin coupling only occurs in the sum once. The overall coupling constant $K_1$ we take to be $1$, as we absorb it into the inverse temperature $\beta = 1/T.$ We also assume periodic boundary conditions.

Unfortunately, for the two dimensional Ising model, there are not analytical solutions for the free energy, correlation length, or the general transformation laws in equation \ref{translaw}. This presents us with two main questions we must answer to perform the group flow: namely, how to create and validate accurate computational samples of the nearest-neighbor 2D Ising model and how to actually accomplish the renormalization. To generate samples, we use an algorithm known as the Wolff algorithm \cite{kyimba_3AD} and we use the critical exponent $\nu$ to validate it. We then perform the renormalization directly by changing the spins of each $b$ by $b$ block according to an algorithm, and then using either an analytical or machine learning method to yield $\beta_1(\beta_0).$ These two techniques are discussed below. 

\subsection{Wolff Algorithm}
\label{sec:wolff_meth}
There are multiple Monte Carlo methods that are used in the generation of representative Ising model samples. Away from the critical temperature, the typical Metropolis algorithm works well for the Ising model. Here, a random spin is flipped, and then the algorithm either accepts or rejects the new spin with probability 
\begin{equation}
    \label{Wolff_prob}
    p = \min(1, -e^{-\beta \Delta E}).
\end{equation}
Over many epochs, the algorithm gives representative results at temperatures away from the critical temperature. 

However, near the critical temperature, this algorithm is extremely slow and inefficient. Instead, we use a cluster algorithm called the Wolff algorithm, which is specifically designed for dealing with Ising model critical behaviors. This algorithm begins by choosing a random spin within the lattice, and then establishing bonds between nearest neighbors with the same spin with a probability of \begin{equation}
    \label{Wolff_prob2}
    p = 1-e^{-2 \beta J}.
\end{equation}
It then proceeds to do this for each nearest neighbor spins, adding them to a stack and flipping their signs to avoid repeats. This proceeds until there are no more nearest neighbors or a stack maximum is hit. Over many epochs, this can quickly produce effective samples near the critical temperature for the 2D Ising model. \cite{kyimba_3AD}

In order to confirm that our models are representative of the 2D Ising model, we calculate the model's critical exponent $\nu$ from the data's correlation lengths. The correlation length is defined by computing the 2-point spin-spin correlator $\langle s(0)s(r) \rangle$, where $s(0)$ is defined as $s(0) = s[0, 0]$ and $s(r)$ is a distance $r$ away, where $r$ is the shortest distance between the two spins. To find the correlation length, we then fit to the form \begin{equation}
\label{eq:correl-length}
\langle s(0)s(r) \rangle = \left(\frac{a}{a+r}\right)^{1/4} e^{-r/\xi},
\end{equation}
where $a$ is a parameter on the order of the lattice spacing, included to ensure $\langle s(0)s(r) \rangle = 1.$ We choose $a = .2$ for a $64$ by $64$ grid, as we determined this value to fit the best empirically. 

For the infinite nearest neighbor Ising model, we expect this correlation length to diverge at a critical temperature $T_c = 2.269$ or $\beta_c = .44.$ As we approach this critical temperature, $\xi \to \infty, $ and for $r >> a$ (as expected), we thus have 
 \begin{equation}
\lim_{T\to T_c} \langle s(0)s(r) \rangle =\frac{1}{r^{1/4}},
\end{equation}
which describes the universal behavior of the universality class of the 2D Ising model. 

In order to then calculate the critical exponent $\nu$, we can then fit the temperature dependence of the correlation length to the form \begin{equation}
\label{eq:nu-behav}
    \xi(T) = c(T-T_c)^{-\nu},
\end{equation} where simulated results show that $c \approx 1$. The critical exponent $\nu$ here should be 1 in the critical limit: where $N \to \infty $ and $T \to T_c$. \cite{cardy_1996}

By fitting to these critical exponent functions, we are able to see if the Wolff algorithm produces samples as expected; while noting that the effective critical temperature $T_c$ may be slightly different due to the finiteness of our model. 
 
\subsection{Renormalization Group Flow}

For the renormalization group flow, we take Ising models at temperatures slightly higher than $T_c$, such that we are still in the critical limit. We then proceed using block spinning of our $N$ by $N$ lattice, where we replace blocks of $b$ by $b$ spins with a single spin. The model thus becomes a lattice of $N/b$ by $N/b$ spins.

There are two methods we use for this block spinning algorithm. The first is a quasi-deterministic method, determined by the average of the spins in the $b$ by $b$ block. If the average is positive, we set the spin to 1. If it is negative, we set the spin to -1. If it is 0, we set it to 1 or -1 with equal probability. 

The second method is called a "Hinton-like" method, as it is structured to be analogous to Hinton's RBM networks.  In this method, for a given block, we define $s_{av}$ as the average spin, and $w$ as some nonnegative parameter. We then set the block spin to 1 with probability 
\begin{equation}
    P = \frac{1}{1+e^{-s_{av} \cdot w}},
\end{equation}
and to -1 otherwise. If we let $w \to \infty$, this "Hinton-like" method then becomes the quasi-deterministic one. Due to this, we actually only implement the Hinton-method, but take $w$ large to test out this quasi-deterministic method.

With this step complete, we must also calculate the running of the coupling constant $\beta$. However, this presents us with a problem. In the 2D Ising model, the nearest-neighbor Hamiltonian \ref{eq:nn_hamil} does not, in general, block spin into a nearest-neighbor Hamiltonian. Instead, it block-spins into a model with next-to-nearest neighbor couplings. This new Hamiltonian is given by 
\begin{equation}
    H_0 = -K_1 \sum_{i=0}^{N-1} \sum_{j=0}^{N-1} \sum_{\langle nn \rangle } s[i, j] s[nn] - K_2 \sum_{i=0}^{N-1} \sum_{j=0}^{N-1} \sum_{\langle nnn \rangle } s[i, j] s[nnn] ,
\label{eq:nnn_hamil}
\end{equation}
where $\langle nnn \rangle$ means to sum over all next-to-nearest neighbors such that each possible spin-spin coupling only occurs in the sum once. This means we're actually in a 2D-space, with the running of $\beta \cdot K_1$ and $\beta \cdot K_2$. We choose to assume $\beta \cdot K_2 << \beta \cdot K_1$ and pay attention only to the running of $K_1 \cdot \beta$, but this does not include the full picture.

In order to calculate the running of the coupling constant, we consider two possible methods. The first method is using an analytical function developed by Maris and Kadanoff. For $b =2$, we have that 
\begin{equation}
    \beta_1 = \frac{\alpha}{4} \log \cosh (4 \beta_0), \text{\cite{https://doi.org/10.48550/arxiv.2107.06898}}
    \label{eq:mk}
\end{equation}
where $\alpha = 1.604521$.

A second method to calculate $\beta_1$ is to use a method akin to that in Section \ref{sec:1d_meth}, in which we minimize a loss function based on the correlation lengths. Using critical behavior described in equation \ref{eq:nu-behav}, along with noting that equation \ref{eq:xi_varying} also holds in 2D, we can define a loss function given by 
\begin{equation}
L(\beta_1) = b \cdot \frac{1}{1/\beta_1 - T_c} - \frac{1}{1/\beta_0 - T_c},
\end{equation}
and minimize it using Adam optimization to produce the running of the coupling constant.

From here, we can generate new Ising models that have gone through RG flow, and we can validate them through calculating the critical exponent $\nu$ as described in Section \ref{sec:wolff_meth}. We can use this to compare the different methods of block spinning and the different methods of calculating $\beta$, to find which is the most accurate. 

\section{Results}
\label{sec:2d_results}

\subsection{Wolff Algorithm}
\label{sec:wolff_res}
\begin{figure}
    \centering
    \includegraphics[width=\textwidth]{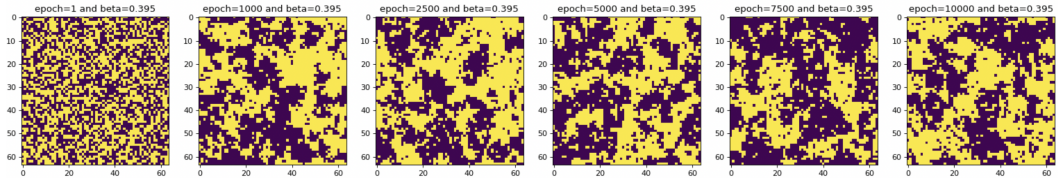}
    \caption{Generated example of the 2D Ising model from the Wolff algorithm for $\beta =.395$. The Ising model keeps its macroscopic properties after these 10000 epochs. Running the algorithm longer will not change the behavior, suggesting that 10000 epochs is enough to successfully train this value near the critical point. This is consistent for all $\beta$, except $\beta > .42$ must run for 20000 epochs. }
    \label{fig:wolff}
\end{figure}

For the Wolff algorithm, we started by using Python code from Github that successfully generates 2D Ising model examples using the Wolff algorithm \cite{Zhao}, later switching to C code that does the same thing but faster. One example of output provided by such code is given in Figure \ref{fig:wolff}, which has $\beta = .395$ and is run over 10000 epochs. As seen in the figure, the Ising model keeps its macroscopic properties after 10000 epochs. Running the algorithm longer will not change the behavior, suggesting that 10000 epochs is enough to successfully train this value near the critical point. This is true for all examined $\beta$, except $\beta > .42$, which must run over 20000 epochs. Running over this many epochs is essential as it allows us to train many representative samples of the Ising model successfully. 

\begin{figure}
    \centering
    \includegraphics[width=.5\textwidth]{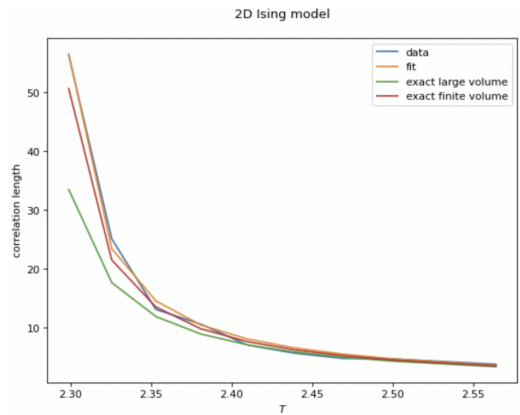}
    \caption{Fitting the temperature dependence of the correlation length, allowing for the calculation of $\nu$. The fit here is good, yielding a $\nu$ value of 1.028 and a $\beta_c$ value of .4388. The exact finite volume is calculated using the fitted value of $\beta_c$ while the exact large volume uses the true value of $\beta_c=.44$. Results suggest a lower critical temperature due to finiteness, but seem to validate the model.}
    \label{fig:nu}
\end{figure}

For the validation of our model, we use a 64 by 64 Ising model. We first check that the results are consistent with the Ising model, calculating the critical exponent $\nu$. The results are given in Figure \ref{fig:nu}. The results are good, yielding a $\nu$ value of 1.028 and a $\beta_c$ value of .4388. This seems reasonably consistent with the true $\nu$ value of $1$ and $\beta_c$ value of .44.

However, models near criticality took much longer to run than expected for a critical temperature of $.44$. This suggests to us that the effective critical inverse temperature of the finite model is lower than $.44$, and is instead closer to $.435$. This is supported further by the fact that values of the correlation length in the data are larger than 32 for $\beta = .435$, and that our fit yields a value smaller than .44. This could suggest an effective critical temperature closer to $.435$, which we will find essential to note for Section \ref{2d_RG_flow_results}.

\subsection{Renormalization Group Flow}
\label{2d_RG_flow_results}

\begin{figure}
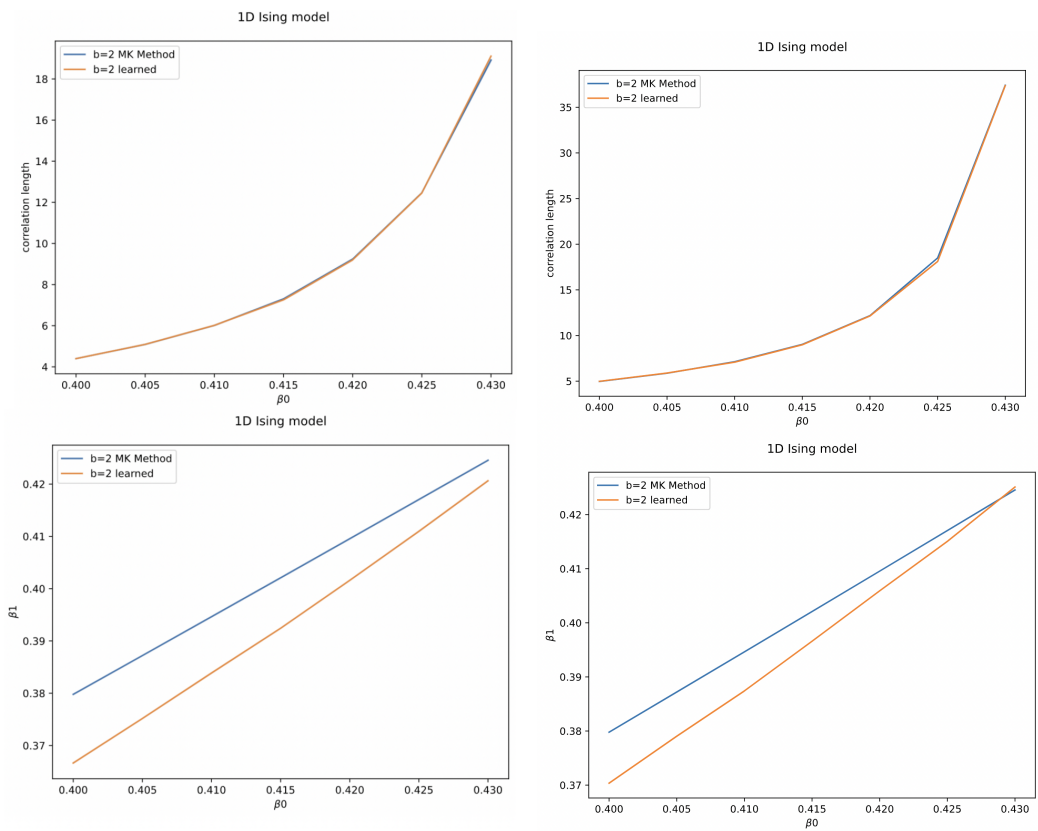

     \centering
     \begin{subfigure}{0.49\textwidth}
         \centering
         \includegraphics[width=\textwidth]{ScreenShot2022-08-13at52219AM.pdf}
         \caption{ Optimization Method with $\beta_c = .44$}
         \label{fig:opt_comp_.44}
     \end{subfigure}
     \hfill
     \begin{subfigure}{0.49\textwidth}
         \centering
         \includegraphics[width=\textwidth]{ScreenShot2022-08-13at52242AM.pdf}
          \caption{Optimization Method with $\beta_c = .435$}
         \label{fig:opt_comp_.435}
     \end{subfigure}
        \caption{Results of using Adam optimization for $\beta_1$ compared to the Maris-Kadanoff method. The optimization seems to fit to the loss function well in both cases, as we would expect from Adam optimization. However, it is closer to Maris-Kadanoff when we take a lower inverse critical temperature of $\beta_c = .435$ than $\beta_c = .44$, yet is still very different.}
        \label{fig:opt_method}
\end{figure}

\begin{figure}
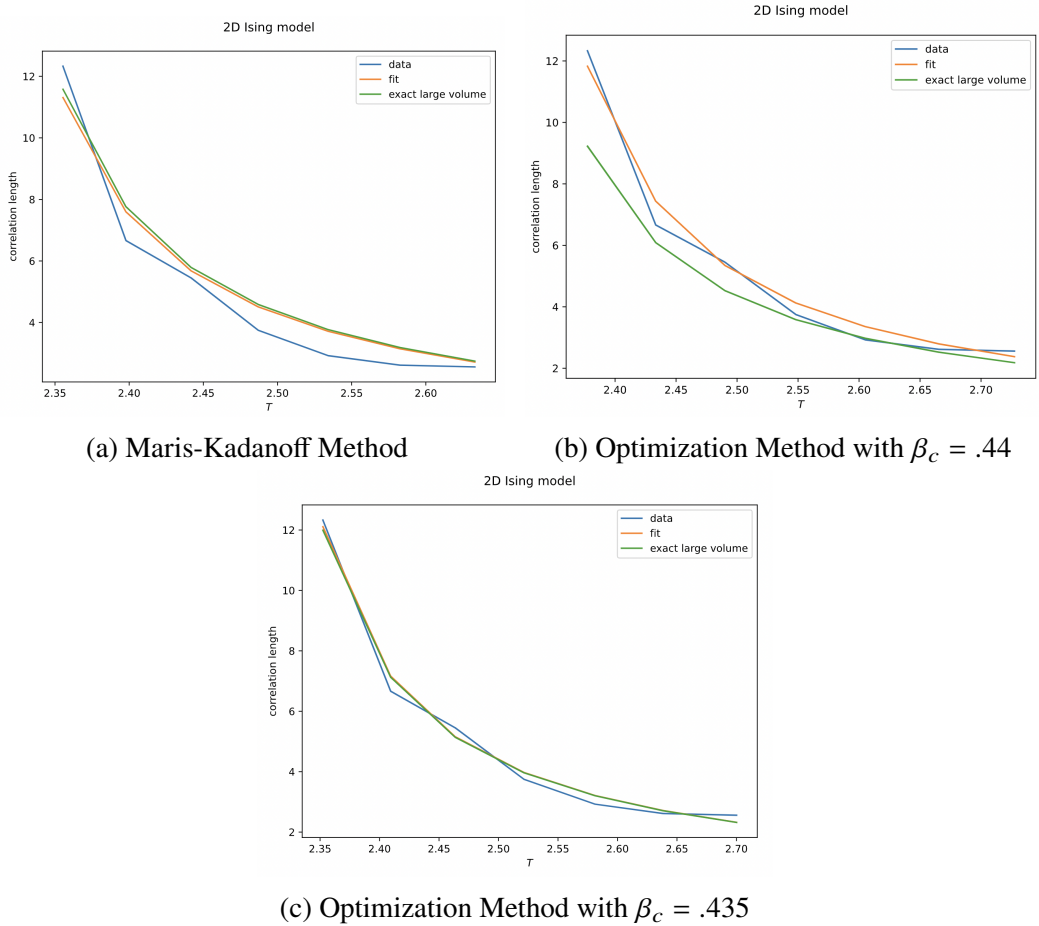

     \centering
     \begin{subfigure}{0.49\textwidth}
         \centering
         \includegraphics[width=\textwidth]{ScreenShot2022-08-13at50423AM.pdf}
         \caption{Maris-Kadanoff Method}
         \label{fig:mk}
     \end{subfigure}
     \hfill
     \begin{subfigure}{0.49\textwidth}
         \centering
         \includegraphics[width=\textwidth]{ScreenShot2022-08-13at50320AM.pdf}
         \caption{Optimization Method with $\beta_c = .44$}
         \label{fig:opt_.44}
     \end{subfigure}
     \hfill
     \begin{subfigure}{0.49\textwidth}
         \centering
         \includegraphics[width=\textwidth]{ScreenShot2022-08-13at50253AM.pdf}
         \caption{Optimization Method with $\beta_c = .435$}
         \label{fig:opt_.435}
     \end{subfigure}
        \caption{Fitting the temperature dependence of the correlation length of renormalized data, allowing for the calculation of $\nu$, by method of calculating $\beta_1$ values. All figures are calculated for $N=64$ and $w=20$. The $\nu$ value for Figure \ref{fig:mk} is $\nu = .986$, for  Figure \ref{fig:opt_.44} is $\nu = 1.107$, and for Figure \ref{fig:opt_.435} is $\nu = 0.999$. The fits seem to be best when we use learning methods with a lower critical inverse temperature of $\beta_c = .435$, and worst when we use it with $\beta_c = .44$. Maris-Kadanoff is in the middle of the two, suggesting the importance of accounting for finiteness lowering the  inverse effective critical temperature. }
        \label{fig:meth_compar}
\end{figure}

We proceeded using the validated model from Section \ref{sec:wolff_res} and performed four steps (64 to 4) of a $b=2$ RG flow on it for values of $w=20$, $w=5$ and $w=1$. There are several conclusions we can draw from the results of our flow, which are summarized below. 

The first conclusion we reach is that the best method for calculating the running of the coupling constant $\beta$ is given via optimization using the effective critical temperature \textit{of the finite case}. When we went through the first step of the group flow, we compared fit results using three different values of $\beta_1$: those calculated via the Maris-Kadanoff equation, Equation \ref{eq:mk}, those using Adam optimization assuming $\beta_c = .44$, and those using Adam optimization assuming $\beta_c = .435$ (as suggested by results from Section \ref{sec:wolff_res}). Comparisons of the $\beta_1$ values are shown in Figure \ref{fig:opt_method} and the actual fits of these values are shown in Figure \ref{fig:meth_compar}, where we assume $w=20$. The $\nu$ value for Maris-Kadanoff is $\nu = .986$, for $\beta_c = .44$ optimization is $\nu = 1.107$, and for $\beta_c = .435$ optimization is $\nu = 0.999$. The fits seem to be best when we use learning methods with a lower critical inverse temperature of $\beta_c = .435$, and worst when we use it with $\beta_c = .44$. Maris-Kadanoff is in the middle of the two. This suggests that the finite nature of our Ising model pushes the effective critical temperature closer to $\beta_c = .435$ than $\beta_c = .44$ 

\begin{figure}
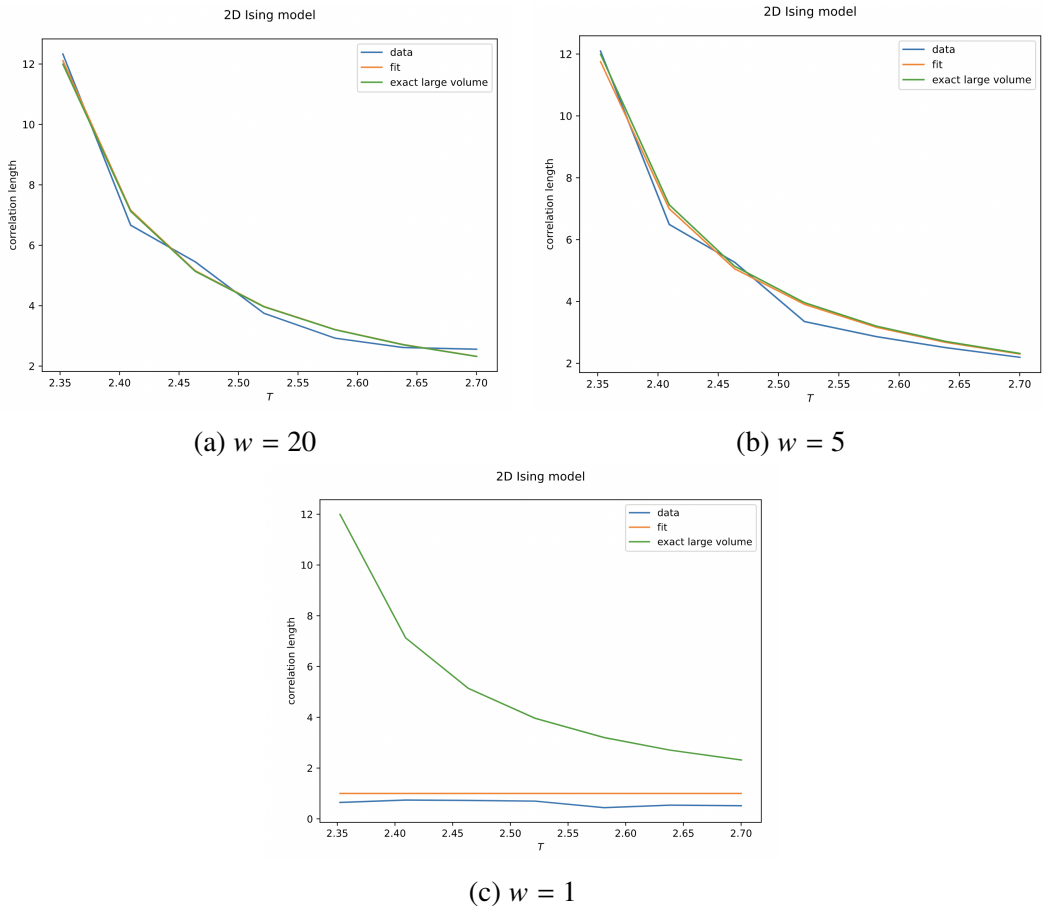

     \centering
     \begin{subfigure}{0.49\textwidth}
         \centering
         \includegraphics[width=\textwidth]{ScreenShot2022-08-13at50253AM.pdf}
         \caption{$w=20$}
         \label{fig:w=20}
     \end{subfigure}
     \hfill
     \begin{subfigure}{0.49\textwidth}
         \centering
         \includegraphics[width=\textwidth]{ScreenShot2022-08-13at50505AM.pdf}
         \caption{$w=5$}
         \label{fig:w=5}
     \end{subfigure}
     \hfill
     \begin{subfigure}{0.49\textwidth}
         \centering
         \includegraphics[width=\textwidth]{ScreenShot2022-08-13at50525AM.pdf}
         \caption{$w=1$}
         \label{fig:w=1}
     \end{subfigure}
        \caption{Fitting the temperature dependence of the correlation length of renormalized data, allowing for the calculation of $\nu$, by value of $w$. All figures are calculated for $N=64$ and using optimization with $\beta_c = .435$. The $\nu$ value for Figure \ref{fig:w=20} is $\nu = .999$, for  Figure \ref{fig:w=5} is $\nu = .987$, and Figure \ref{fig:w=1} is $\nu = 0$. The fits seem to be best when we use methods in the pseudo-deterministic limit, and that they get worse as we get more and more Hinton-like. This can lead to dramatic problems in the fit for low $w$; $w=1$ is not a valid renormalization group flow for the Ising model. }
        \label{fig:w_compar}
\end{figure}

The second conclusion we reach is that block spinning in the pseudo-deterministic limit is more accurate than Hinton-like block spinning. When we went through the first step of the group flow, we compared fit results using three different values of $w$: $w=20$, $w=5$, $w=1$. These results for using $\beta_c = .435$ optimization are shown in Figure \ref{fig:w_compar}. The $\nu$ value for Figure \ref{fig:w=20} is $\nu = .999$, for  Figure \ref{fig:w=5} is $\nu = .987$, and Figure \ref{fig:w=1} is $\nu = 0$. The fits seem to be best when we use methods in the pseudo-deterministic limit, and that they get worse as we get more and more Hinton-like. This can lead to dramatic problems in the fit for low $w$; $w=1$ is not a valid renormalization group flow for the Ising model. This  suggests problems with RBM learning, if it is more similar to low $w$ Hinton learning than high $w$ Hinton learning.

\begin{figure}
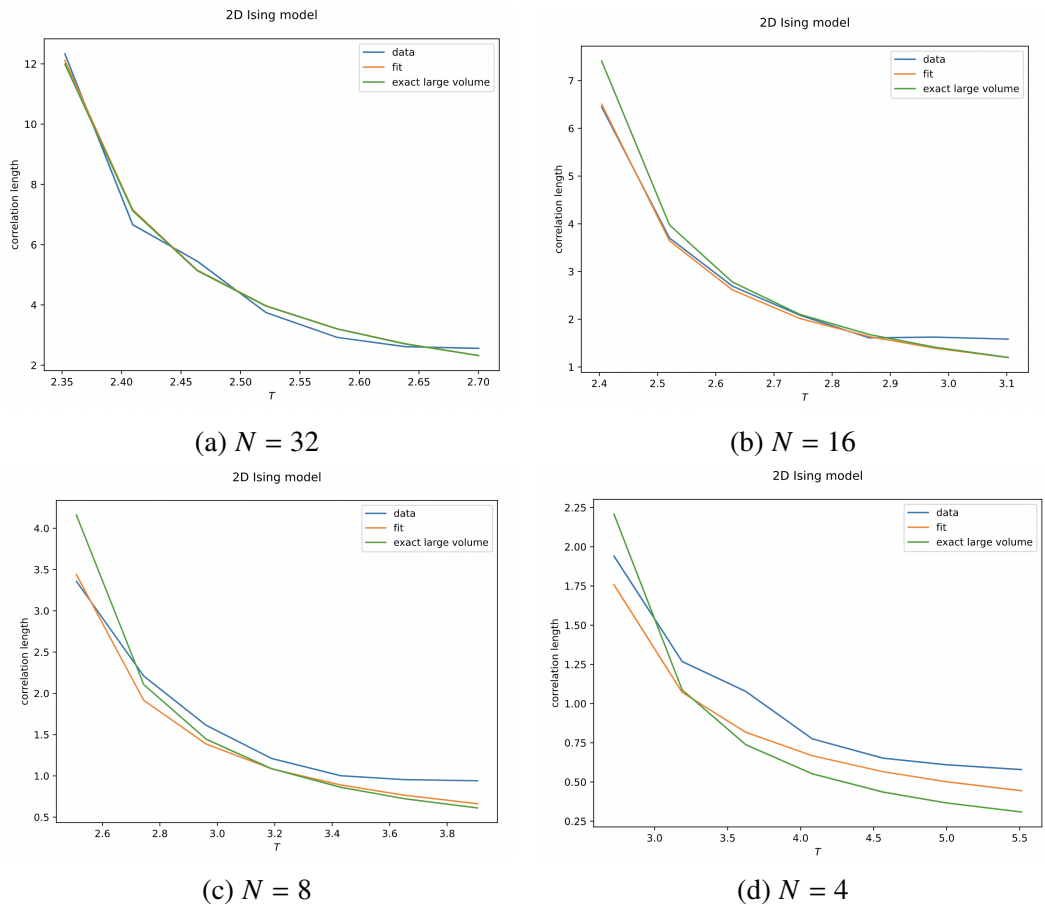

     \centering
     \begin{subfigure}{0.49\textwidth}
         \centering
         \includegraphics[width=\textwidth]{ScreenShot2022-08-13at50253AM.pdf}
         \caption{$N=32$}
         \label{fig:N=32}
     \end{subfigure}
     \hfill
     \begin{subfigure}{0.49\textwidth}
         \centering
         \includegraphics[width=\textwidth]{ScreenShot2022-08-13at50656AM.pdf}
         \caption{$N=16$}
         \label{fig:N=16}
     \end{subfigure}
     \hfill
     \begin{subfigure}{0.49\textwidth}
         \centering
         \includegraphics[width=\textwidth]{ScreenShot2022-08-13at50745AM.pdf}
         \caption{$N=8$}
         \label{fig:N=8}
     \end{subfigure}
     \hfill
     \begin{subfigure}{0.49\textwidth}
         \centering
         \includegraphics[width=\textwidth]{ScreenShot2022-08-13at50846AM.pdf}
         \caption{$N=4$}
         \label{fig:N=4}
     \end{subfigure}
        \caption{Fitting the temperature dependence of the correlation length of renormalized data after multiple renormalization group flow steps, allowing for the calculation of $\nu$, by value of $N$. All figures are calculated for $w=20$ and using optimization with $\beta_c = .435$. The $\nu$ value for Figure \ref{fig:N=32} is $\nu = .999$, for Figure \ref{fig:N=16} is $\nu = 0.93$, for Figure \ref{fig:N=8} is $\nu = .857$, and for Figure \ref{fig:N=4} is $\nu = .698$. As the renormalization group flow continues, the fits get worse and worse, but this is expected as we are restricting the RG flow to one dimension. We have a reasonable RG flow for 2 iterations, with it getting a bit worse after that. Allowing for next-to-nearest neighbors, or double-checking critical temperatures may fix this.}
        \label{fig:N_compar}
\end{figure}

Lastly, we can conclude that the renormalization group flow is pretty accurate for the first two steps, assuming $w=20$ and we use $\beta_c =.435$ optimization to calculate the $\beta_1$ values. After that, the renormalization group flow seems to get less reasonable. This can be seen in Figure \ref{fig:N_compar} both visually and numerically, as the $\nu$ value for Figure \ref{fig:N=32} is $\nu = .999$, for Figure \ref{fig:N=16} is $\nu = 0.93$, for Figure \ref{fig:N=8} is $\nu = .857$, and for Figure \ref{fig:N=4} is $\nu = .698$. As the renormalization group flow continues, the fits get worse and worse,  with a reasonable renormalization group flow for 2 iterations, but not for the last two.

This of course, is expected, as we are working in a one-dimensional parameter space when we should be working with a higher dimensional one. Allowing for next-to-nearest neighbors, or changing critical temperatures in the optimization for each layer may fix this.

\chapter{Analysis of the Deep Learning of the Two Dimensional Ising Model}
\label{chap:dl}

\section{Introduction}

We now leave our discussion of the renormalization group to discuss the coarse-graining of the Ising model through RBM learning. To do so, we implemented  a three-layer stack of Restricted Boltzmann Machines as described in Section \ref{sec:RBMS}, the same way as in \cite{Mapping}. For simplicity and speed, we made these RBMs in TensorFlow. 

We made 20,000 instantiations of a 64x64 = 4096 spin Ising model for varying $\beta$ values, produced by the Wolff algorithm discussed in \ref{sec:wolff_meth}. The first layer of the RBM had 32x32 = 1024 nodes, the second had 16x16=256 nodes, and the third had 8x8=64 nodes. In this way, we mirrored the renormalization procedure as discussed in Chapter \ref{chap:renorm}. 

\begin{table}
    \centering
    \begin{tabular}{|p{5cm}|p{5cm}|}
    \hline
    Parameter & Value(s)\\
    \hline
    $\beta$ Values & \{.395, .4, 45, .41, .415, .42, .425, .43\}\\
    \hline
    Number of Lattices & 20000 \\
    \hline
    Epochs in Wolff Algorithm & 10000 if $\beta \leq .42$, 20000 if $\beta \geq .425$\\
    \hline
    Epochs in Learning & 1500 \\
    \hline
    Learning Rate & .004 \\
    \hline
    Learning Rate Reduction & .998 \\
    \hline
    L1 Regularization Weight & .0008 \\
    \hline
    Batch Size & 200 \\
    \hline
    Momentum & .9 \\
    \hline
    
\end{tabular}
    \caption{Deep learning parameters for the RBM stacks. Parameters were chosen to best reproduce the learning in \cite{Mapping}. In particular, we apply an L1 regulator to penalize over-fitting, as is done in \cite{Mapping}.  }
    \label{tab:DL_params}
\end{table}

The parameters of the learning are given in Table \ref{tab:DL_params}, chosen to match standard deep learning procedures and the procedures in \cite{Mapping} to effectively learn the coarse-graining of the Ising model. In particular, we use a non-zero L1 penalty on the weights to sparsify the results and penalize over-fitting. Some authors consider this to bias the results towards learning locality \cite{DeepAndCheap}, but our goal is to more rigorously examine the claims made in \cite{Mapping} with the regularization, so we continue to use the regulator. 

For the remainder of this chapter, we will discuss the results of this coarse-graining on its own, in detail. Then, in Chapter \ref{chap:connection}, we discuss the results via numerical comparisons to the previously discussed renormalization results. 

In this chapter, we first discuss the reproduction of plots from \cite{Mapping} in detail in \ref{sec:receptive}. We then discuss the structure of the trained weights and how they can form blocks in \ref{sec:weight_structures}, followed by a deeper analysis of distances between spins in these blocks in \ref{sec:blocks}. We conclude by using these results to analyze the use of the RBM network as an autoencoder in \ref{sec:reconstruction}. In the following analysis, we only present plots from our $\beta = .41$ results, as this is the closest to the $\beta = .408$ used in \cite{Mapping}. However, we will discuss how these results vary across $\beta$ values. We also present the full set of plots in Appendix \ref{chap:2d_layer_plots}. 

\section{Receptive Field Analysis} \label{sec:receptive}

\begin{figure}[h!]
    \centering
    \includegraphics[scale=.3]{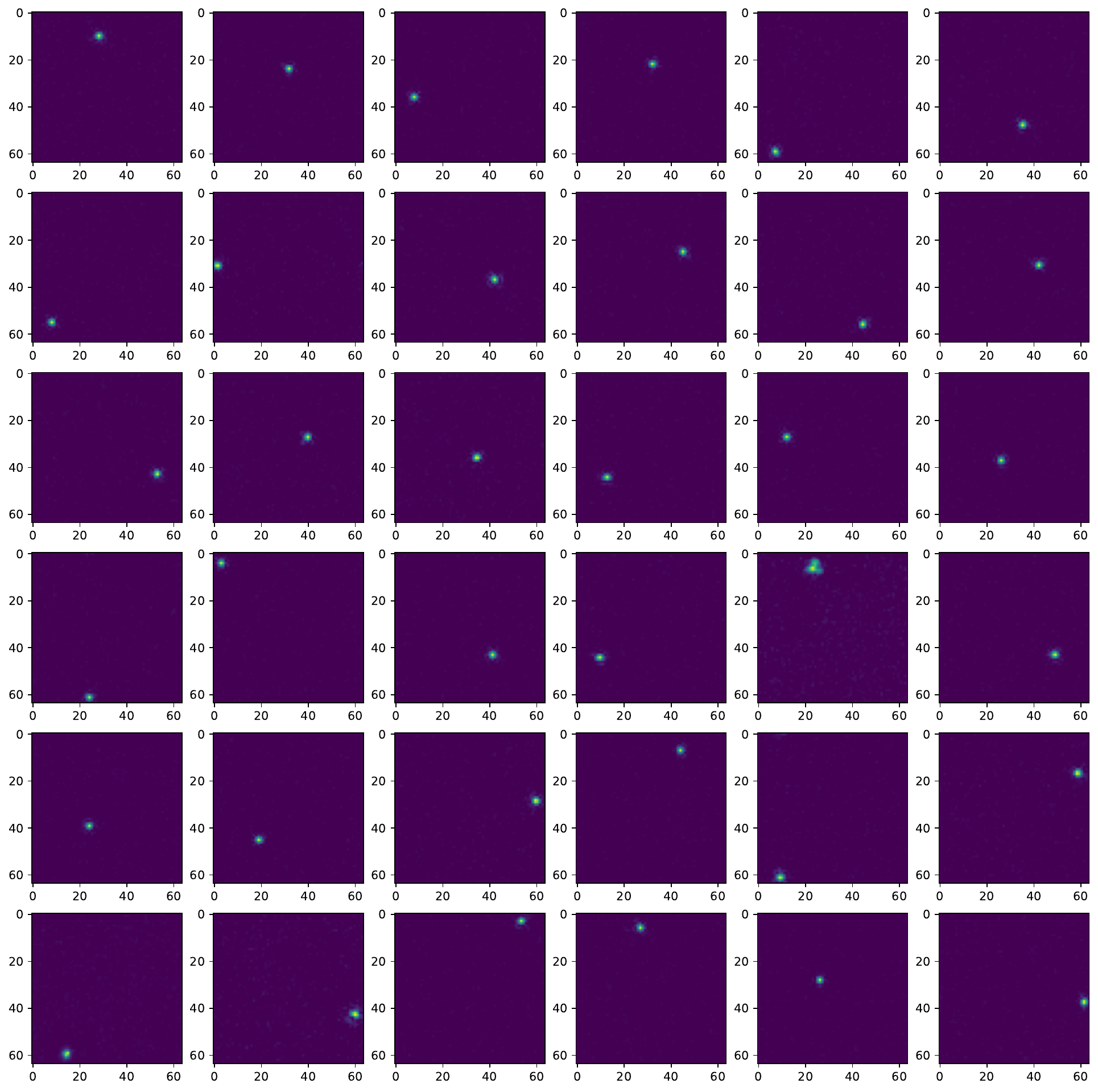}
    \caption{Layer 1 receptive field plot for $\beta = .41$. Each individual plot represents a node in Layer 1 and the values on the plot represents the full set of absolute weights corresponding to that node. The clustering of non-zero weights next to each other suggests the model is qualitatively learning some form of locality. }
    \label{fig:rec_1_.41_analyzed}
\end{figure}

\begin{figure}[h!]
    \centering
    \includegraphics[scale=.25]{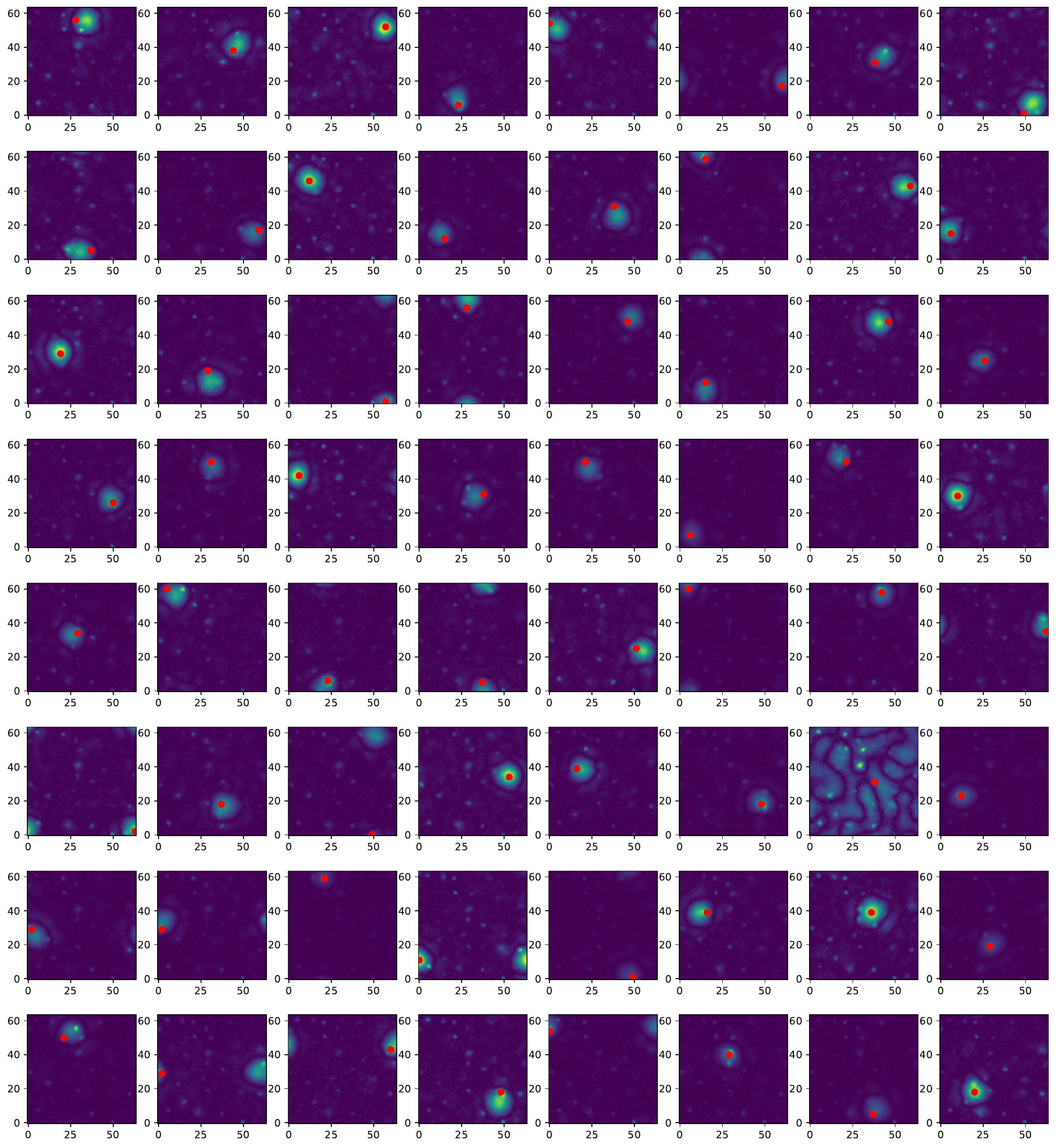}
    \caption{Layer 3 receptive field plot for $\beta = .41$. Each individual plot represents a node in Layer 3 and the values on the plot are those of a tensor  representing the original Ising spins that fed information to the node. The clustering of non-zero values next to each other suggests the model is qualitatively learning some form of locality. }
    \label{fig:rec_3_.41_analyzed}
\end{figure}

One of the main qualitative results from \cite{Mapping} were the "receptive field plots", which we have reproduced in Figures \ref{fig:rec_1_.41_analyzed} and \ref{fig:rec_3_.41_analyzed} for $\beta = .41$. To understand what these plots represent, consider Layer 1. Each of the 32x32 = 1024 hidden nodes has an associated tensor of 64x64= 4096 weights, one for each Ising spin input, with each weight falling between  $-\infty$ and $\infty$. Figure \ref{fig:rec_1_.41_analyzed} is a color map of the absolute values of these weights for the first 36 nodes. Most of the 4096 weights are close to zero, with a few O(1) weights that are clustered together in terms of the original lattice.

To yield a similar map for other layers, we define a tensor similar to that from Layer 1: one which contains information on which of the original Ising spins fed information to each Layer 3 hidden spin. To do so, we denote the Layer 3 weights by $w^{L3}_{i_3j_3;i_2j_2}$ where indices $i_3, j_3$ run $1, ...8$ and $i_2, j_2$ run $1, ...16$. In a similar vein, we denote the Layer 2 weights by $w^{L2}_{i_2j_2;i_1j_1}$ and the Layer 1 weights by $w^{L2}_{i_1j_1;ij}$, where $i_1, j_1$ run $1, ...32$ and $i, j$ run $1, ...64$. We can thus define the Layer 3 "receptive field tensor" as

\begin{equation}
    \text{rec}^{L3}_{i_3j_3;ij} = \sum_{i_2, j_2, i_1, j_1} w^{L3}_{i_3j_3;i_2j_2} \cdot w^{L2}_{i_2j_2;i_1j_1} \cdot w^{L1}_{i_1j_1;ij}.
\end{equation}

Similarly, we can define the Layer 2 and Layer 1 receptive field tensors as

\begin{equation}
    \text{rec}^{L2}_{i_2j_2;ij} = \sum_{i_1, j_1} w^{L2}_{i_2j_2;i_1j_1} \cdot w^{L1}_{i_1j_1;ij},
\end{equation}
and 
\begin{equation}
    \text{rec}^{L1}_{i_1j_1;ij} = w^{L1}_{i_1j_1;ij}.
\end{equation}

Thus, we can define receptive field maps for every layer. In particular, we examine the Layer 3 receptive field, which is given for all Layer 3 spins, in Figure \ref{fig:rec_3_.41_analyzed}. The Layer 3 receptive field plot is similar in structure to the Layer 1 receptive field plot, except with larger weight clusters. This structure is similarly found for receptive field plots of all the tested $\beta$ values, as shown in Appendix \ref{sec:recep_plots}. 

These figures form the basis of the argument in \cite{Mapping} that the trained network is accomplishing a variation of Kadanoff block spinning layer by layer. In this interpretation, each cluster of spins in the receptive field maps corresponds to a block in the Ising lattice, and each RBM node can be converted into a form equivalent to an Ising block-spinning.

\begin{figure}
    \centering
    \includegraphics[scale=.35]{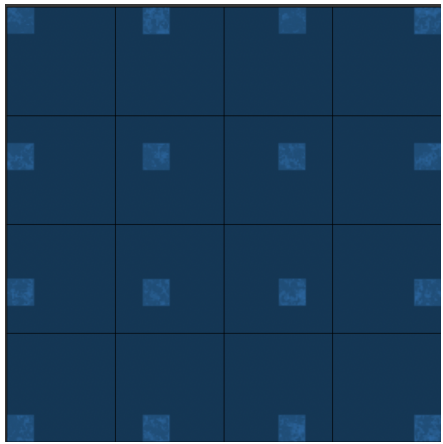}
    \caption{Plot for RG analogous to an RBM receptive field plot. We let $\beta= .41$ averaged over 1000 samples with $w=20$ and using optimization techniques. We determined how the initial 64 by 64 spins affected a given spin in the final 4 by 4 model as follows: if an initial spin is both part of the block that determines the final spin and the same as the final spin, it gets a value of 1, otherwise it gets a value of 0; then, we average over all samples.}
    \label{fig:big_blue}
\end{figure}

We further clarify this argument by considering what an analogous plot would look like using RG techniques instead of the RBM.
To do so, we take our renormalization procedure from Chapter \ref{chap:renorm} and block spin the results from an initial 64 by 64 lattice to a 4 by 4 lattice. We then determine how the initial 64 by 64 spins affect a given spin in the final 4 by 4 model as follows: if an initial spin is both part of the block that determines the final spin and the same as the final spin, it gets a value of 1, otherwise it gets a value of 0; then, we average over all samples. These results for $\beta= .41$ averaged over 1000 samples with $w=20$ and using optimization techniques are shown in Figure \ref{fig:big_blue}. 

In this plot, we see a situation similar to that in Figures \ref{fig:rec_1_.41_analyzed} and \ref{fig:rec_3_.41_analyzed}. Each node is only determined by a small block of the original 64 by 64 lattice, as expected from block spinning. In this way, we find that qualitatively, RBM learning looks like block spinning. Additionally, this qualitative connection holds for all the $\beta$ values we tested, outside of just the $\beta = .408$ used in \cite{Mapping}.

In some ways, this result is already surprising, given that the RBM network only uses 1D vector representations of the tensors, but seems to derive locality in the receptive fields. We further examine this locality qualitatively in the next sections. 

\section{Structure of Weights}
\label{sec:weight_structures}

To better understand the structure of this derived locality, we look at the weight tensor structure directly. We first note that the bias weights $a_i, b_j$, as described in Section \ref{sec:RBMS}, are 0 in our learning. This means that we only need to examine the properties of the weight tensors $w$ for each layer. For layers 2 and 3, we consider the properties of both the weight tensor and the receptive field tensor separately, as we should consider the weight properties both alone and in the context of the layers above them.

\begin{figure}
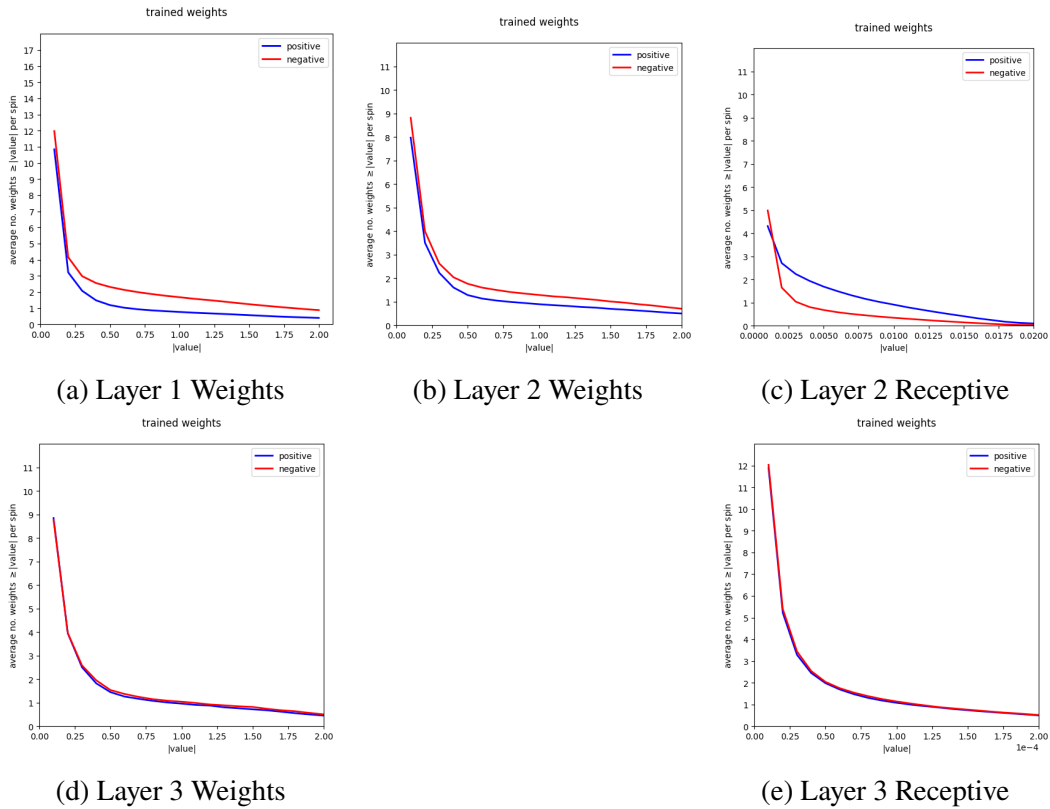

     \centering
     \begin{subfigure}{0.32\textwidth}
         \centering
         \includegraphics[width=\textwidth]{av_weights_1__41.pdf}
         \caption{Layer 1 Weights}
     \end{subfigure}
     \hfill
     \begin{subfigure}{0.32\textwidth}
         \centering
         \includegraphics[width=\textwidth]{av_weights_2w__41.pdf}
         \caption{Layer 2 Weights}
     \end{subfigure}
     \hfill
          \begin{subfigure}{0.32\textwidth}
         \centering
         \includegraphics[width=\textwidth]{av_weights_2r__41.pdf}
         \caption{Layer 2 Receptive}
     \end{subfigure}
     \hfill
    \begin{subfigure}{0.32\textwidth}
         \centering
         \includegraphics[width=\textwidth]{av_weights_3w__41.pdf}
         \caption{Layer 3 Weights}
        \end{subfigure}
     \hfill
         \begin{subfigure}{0.32\textwidth}
         \centering
         \includegraphics[width=\textwidth]{av_weights_3r__41.pdf}
         \caption{Layer 3 Receptive}
         \end{subfigure}
        \caption{Average number of trained positive/negative weights with magnitude above a given value connected to a given spin location in input lattices. Results are given for both weight tensors and receptive field tensors for $\beta = .41.$ The results suggest that we can think of each node in the RBM architecture being associated with two spins: a positive and negative. }
        \label{fig:weight_analysis_.41_analyzed}
\end{figure}

For layer 1, we take each of the 64x64 = 4096 spin locations, examine the 32x32=1024 layer 1 weights connected to it, and calculate how many of these nodes have a positive spin or a negative spin greater than a given magnitude. We do the same thing for the layer 2 and 3 receptive field tensors so that for each of the 64x64 = 4096 weights, we examine the 16x16=256 layer 2 receptive field tensors  and 8x8=64 layer 3 receptive field tensors. In addition, we can also consider the layer 2 and 3 weights instead of the receptive field tensors, so that for each of the 32x32=1024 spins fed into layer 2, we examine the 16x16=256 layer 2 weights and for each of the 16x16=256 spins fed into layer 3, we examine the 8x8=64 layer 3 weights. In each case, we then average over all the lattices to reach the plots in Figure \ref{fig:weight_analysis_.41_analyzed}.

From these results, we find that the structure of the weights and receptive tensors for all three layers are similar. (The receptive tensors have lower magnitude, but that's because there are more weights multiplied together). We find that we may approximate the weight tensor and receptive field plots as assigning only a single large positive weight and a single large negative weight to each spin, significantly lowering 
the number of relevant weight tensor components. 

The results match for all $\beta$, as shown in Appendix \ref{sec:weight_plots}. This seems to suggest two things. First, it suggests that that models can be reconstructed using only the largest positive and negative weight, instead of the full weight tensor, which we examine in \ref{sec:reconstruction}. Secondly, it further qualitatively connects the RBM's coarse-graining algorithm to that of 2D Ising model RG flow. Each node in the machine learning architecture can be thought qualitatively as corresponding to two spins: positive and negative, just as the renormalized Ising lattice "nodes" correspond to one spin. Thus, the structure of the two systems are qualitatively much more similar than at first glance, with 2 weights per node instead of hundreds to thousands.

\begin{figure}
     \centering
     \begin{subfigure}{0.32\textwidth}
         \centering
         \includegraphics[width=\textwidth]{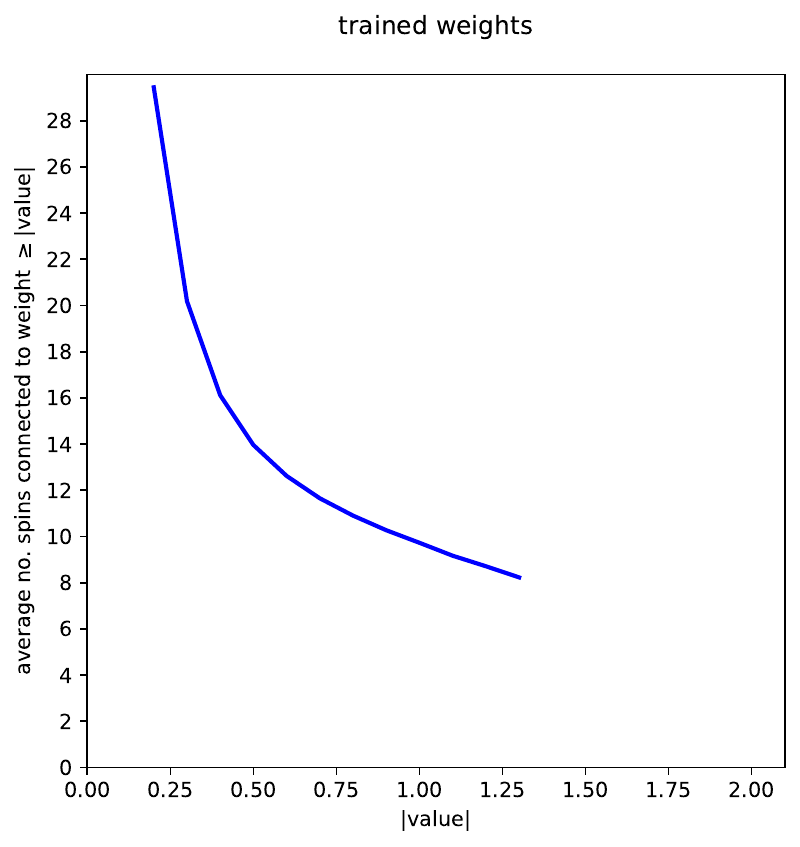}
         \caption{Layer 1 Weights}
     \end{subfigure}
     \hfill
     \begin{subfigure}{0.32\textwidth}
         \centering
         \includegraphics[width=\textwidth]{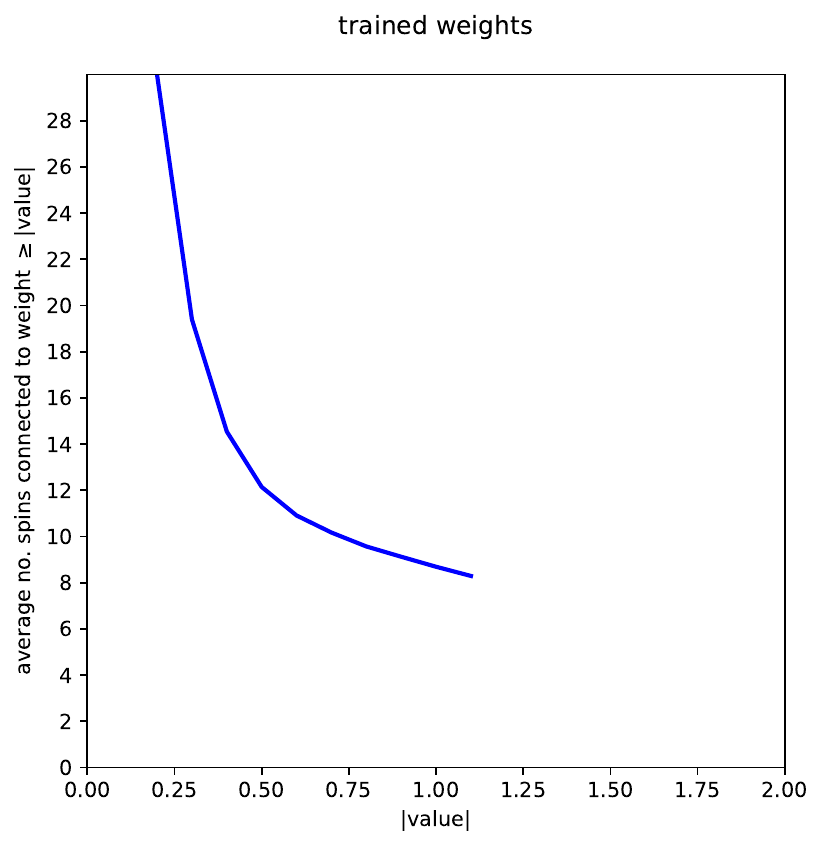}
         \caption{Layer 2 Weights}
     \end{subfigure}
     \hfill
     \begin{subfigure}{0.32\textwidth}
         \centering
         \includegraphics[width=\textwidth]{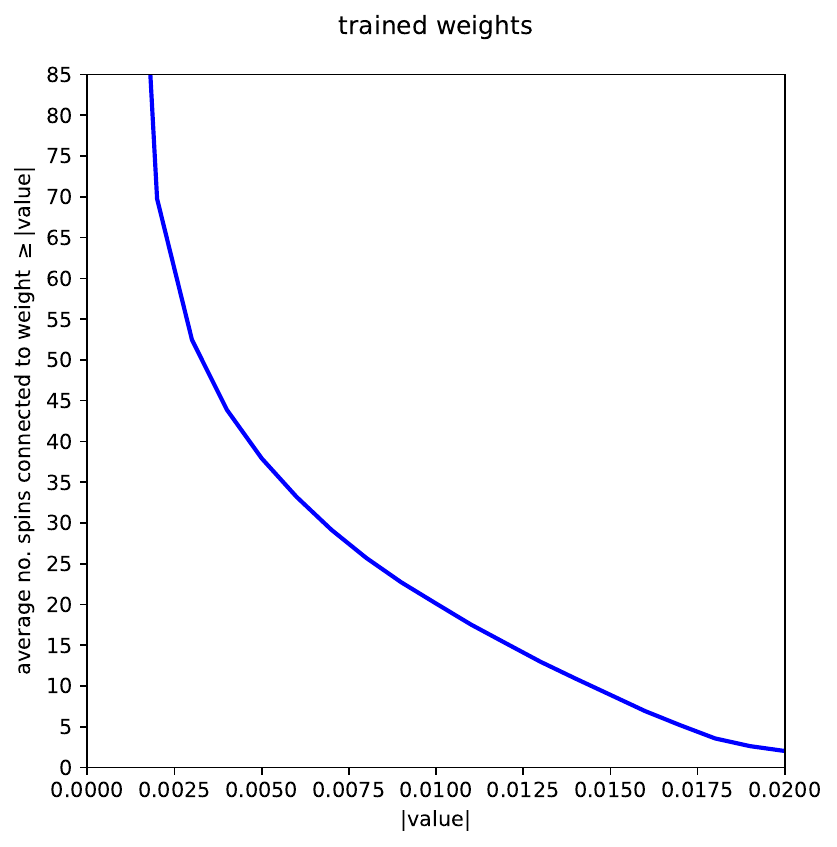}
         \caption{Layer 2 Receptive}
     \end{subfigure}
           \begin{subfigure}{0.32\textwidth}
         \centering
         \includegraphics[width=\textwidth]{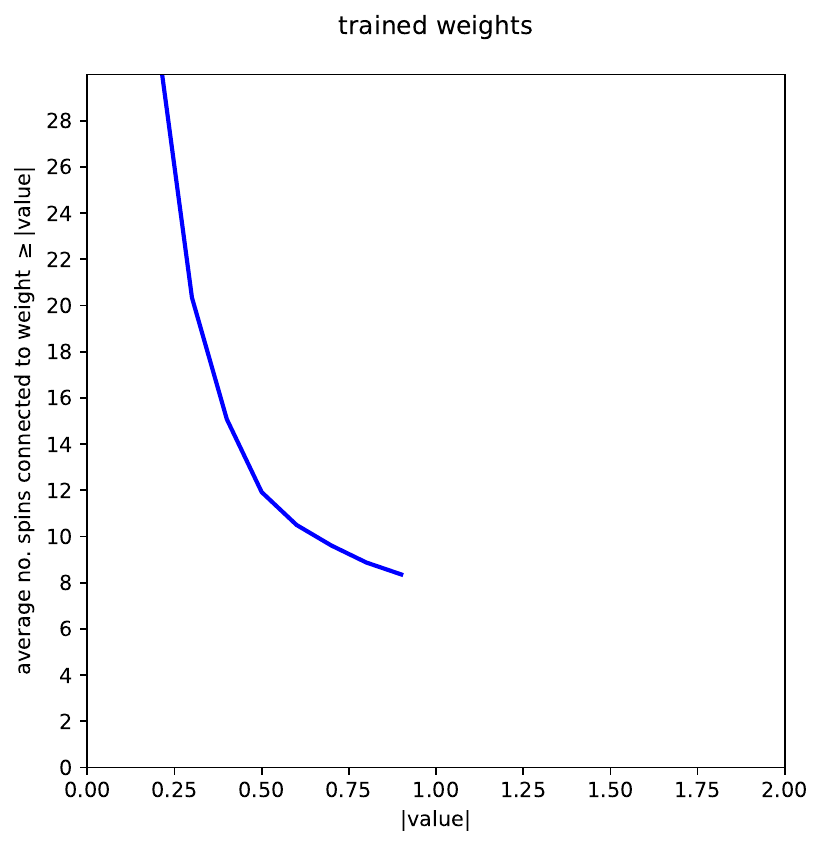}
         \caption{Layer 3 Weights}
     \end{subfigure}
     \hfill
           \begin{subfigure}{0.32\textwidth}
         \centering
         \includegraphics[width=\textwidth]{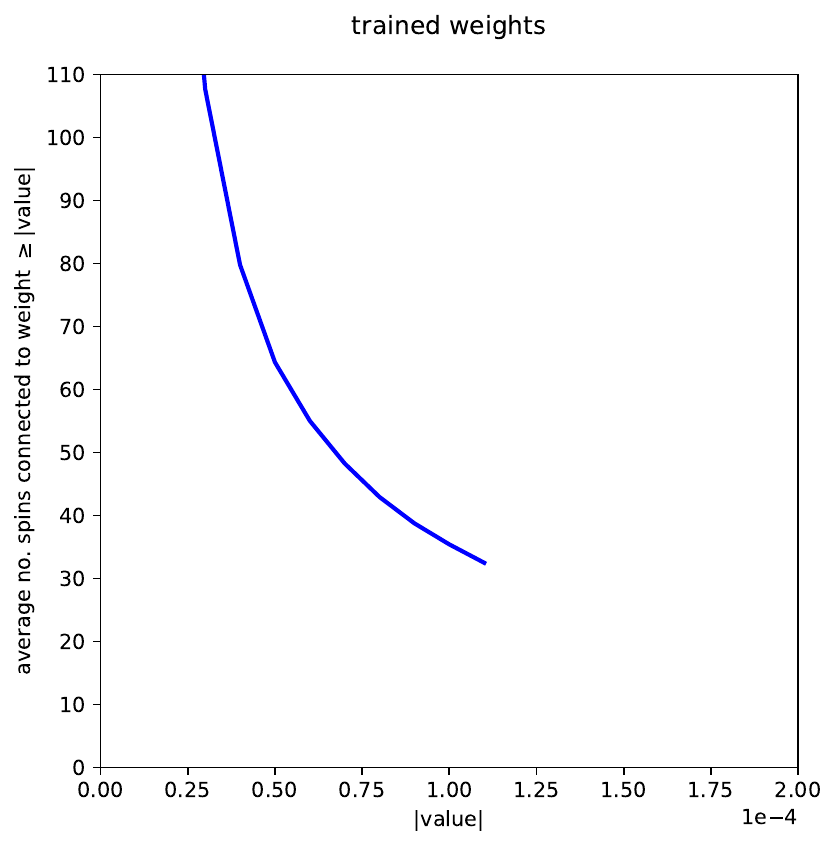}
         \caption{Layer 3 Receptive}
     \end{subfigure}
        \caption{Average number of spin locations connected to each weight with magnitudes above a given value, such that all locations connect to at least two weights. Results are given for both weight tensors and receptive field tensors for $\beta = .41.$ They show that we can approximate the RBM coarse-graining as a block spinning technique.}
        \label{fig:spin_analysis_.41}
\end{figure}

This viewpoint of the weights as spins allows us to do another analysis on the weight tensors. Now, instead of taking each Ising spin, and calculating how many weights are connected to each spin, we do the opposite: we take a magnitude and calculate how many spins are connected to a weight larger than that magnitude. These results are shown in Figure \ref{fig:spin_analysis_.41}. 

We find for both the weight tensors and the receptive field tensors that as we increase the magnitude of the weights, the number of spins connected to the weight drops off quickly. The higher magnitude weights are connected to only a few spins, forming "blocks". As discussed, these few high magnitude weights consist of the majority of the weights coupled to the RBM nodes. Thus, to some approximation, the RBM coarse-grains by creating blocks of nodes and connecting each block to two weights. These results are consistent for all $\beta$, as shown in Appendix \ref{sec:weight_plots}, 

\section{Block Analysis}
\label{sec:blocks}
From the analysis in Section \ref{sec:weight_structures}, we have shown that large weights in the RBMs only have a small number of nodes connected to them, forming blocks similar to those found in Kadanoff block spinning. In this section, we further examine the properties of these blocks to characterize the qualitative behavior of the model. We once again consider the properties of both the weight tensors and the receptive field tensors for all three layers. 

\begin{figure}
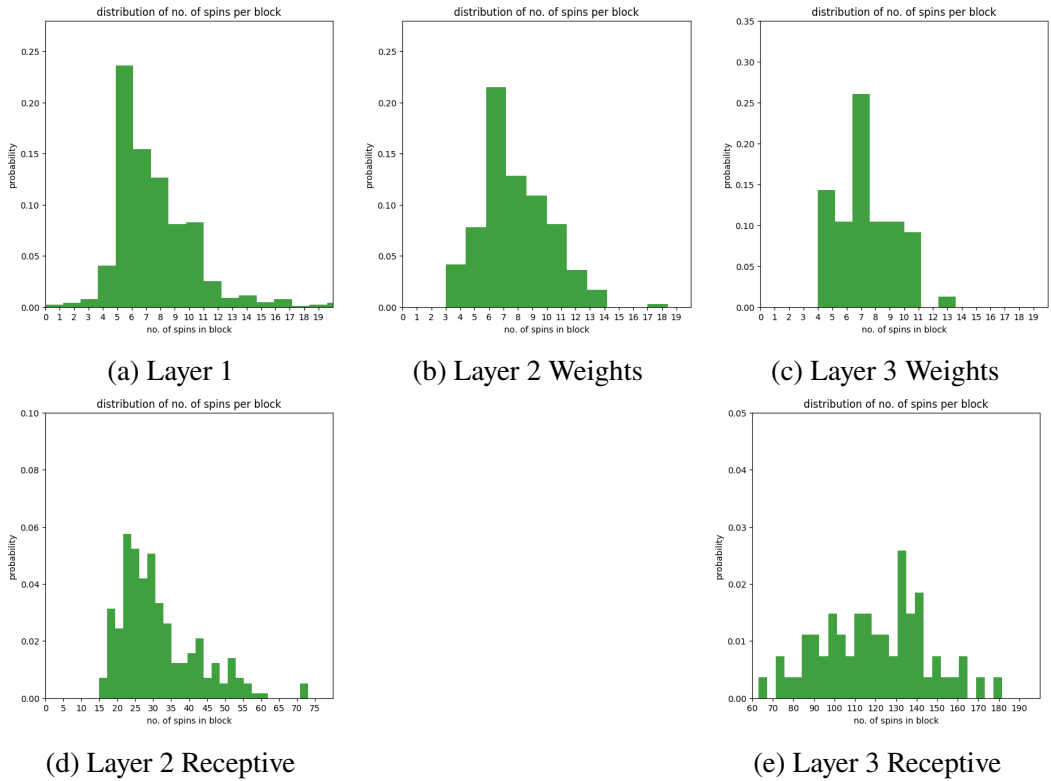

     \centering
     \begin{subfigure}{0.32\textwidth}
         \centering
         \includegraphics[width=\textwidth]{dist_spins_1__41.pdf}
         \caption{Layer 1}
     \end{subfigure}
     \hfill
     \begin{subfigure}{0.32\textwidth}
         \centering
         \includegraphics[width=\textwidth]{dist_spins_2w__41.pdf}
         \caption{Layer 2 Weights}
     \end{subfigure}
     \hfill
     \begin{subfigure}{0.32\textwidth}
         \centering
         \includegraphics[width=\textwidth]{dist_spins_3w__41.pdf}
         \caption{Layer 3 Weights}
     \end{subfigure}
     \hfill
          \begin{subfigure}{0.32\textwidth}
         \centering
         \includegraphics[width=\textwidth]{dist_spins_2r__41.pdf}
         \caption{Layer 2 Receptive}
     \end{subfigure}
     \hfill
     \begin{subfigure}{0.32\textwidth}
         \centering
         \includegraphics[width=\textwidth]{dist_spins_3r__41.pdf}
         \caption{Layer 3 Receptive}
     \end{subfigure}
     \hfill
        \caption{Histograms for the number of spins per block for all three layers' weight and receptive field tensors for $\beta = .41$. Results are highly consistent with Kadanoff $b=2$ block spinning, as the blocks are about the size we'd expect for block spinning with two weights.  }
        \label{fig:dist_spins_analysis_.41}
\end{figure}

To start, we plot the number of spins per block averaged over in each layer in \ref{fig:dist_spins_analysis_.41}. 
The results we find are quite similar to what we would expect from $b=2$ block spinning. For the weight tensors of all three layers, we find that the block size peaks at slightly less than 8 spins per block. This is in line with what we expect for Kadanoff block spinning, in which there are 4 spins per block, and we multiply by 2 to account for both the positive and negative weights. The same result is seen in the receptive field tensors, with the number of spins per block peaking slightly less than 32 for Layer 2 and around 128 for Layer 3. This is consistent with the blocks of 16 and 64 expected from block renormalization, once again multiplied by 2 for both signs of spin. These results are consistent for all $\beta$, as shown in Appendix \ref{sec:block_plots}.

\begin{figure}
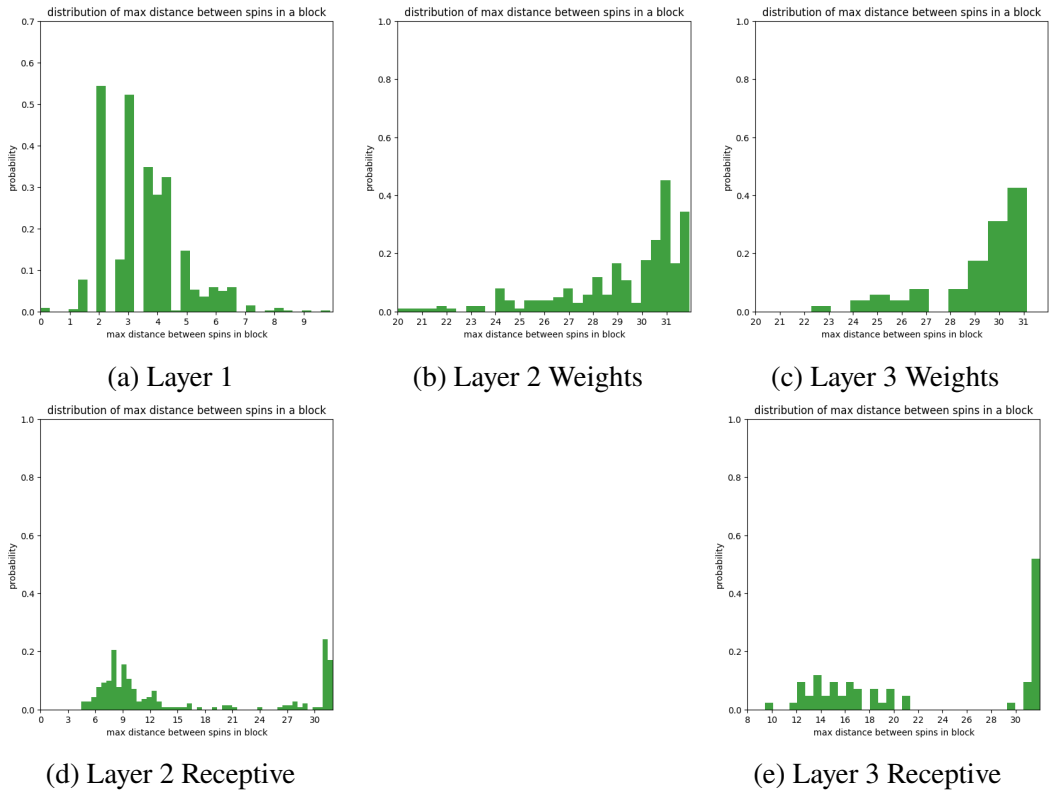

     \centering
     \begin{subfigure}{0.32\textwidth}
         \centering
         \includegraphics[width=\textwidth]{dist_max_1__41.pdf}
         \caption{Layer 1}
     \end{subfigure}
     \hfill
     \begin{subfigure}{0.32\textwidth}
         \centering
         \includegraphics[width=\textwidth]{dist_max_2w__41.pdf}
         \caption{Layer 2 Weights}
     \end{subfigure}
     \hfill
     \begin{subfigure}{0.32\textwidth}
         \centering
         \includegraphics[width=\textwidth]{dist_max_3w__41.pdf}
         \caption{Layer 3 Weights}
     \end{subfigure}
     \hfill
     \begin{subfigure}{0.32\textwidth}
         \centering
         \includegraphics[width=\textwidth]{dist_max_2r__41.pdf}
         \caption{Layer 2 Receptive}
     \end{subfigure}
     \hfill
     \begin{subfigure}{0.32\textwidth}
         \centering
         \includegraphics[width=\textwidth]{dist_max_3r__41.pdf}
         \caption{Layer 3 Receptive}
     \end{subfigure}
     \hfill
        \caption{Histogram of average maximum distance per block for $\beta = .41.$ Results are shown for both weight tensors and receptive field tensors. For layer 1, it seems that the system successfully learns locality. However, in layers 2 and 3, the individual weights contain nearly no information about locality, most blocks having a max 32 spins apart. The locality information is instead shown in the receptive field tensors, with about half of the layer 2 tensors and about a quarter of the layer 3 tensors learning locality. }
        \label{fig:dist_max_analysis_.41}
\end{figure}

From here, we examine the distances between spins in the blocks. We start by plotting the maximum distance between spins in a block in Figure \ref{fig:dist_max_analysis_.41}. We find that for the weight tensor in layer 1, the distribution peaks around a maximum distance of 3 spins between spins in a block. These short range distances imply that the blocks in layer 1 are formed from short range correlations, as expected in the Ising model. We find that this pattern breaks for the weight tensors in layers 2 and 3, with the maximum distance between spins in blocks almost always reaching its maximum of 32 spins. This implies that the weight tensors themselves do not contain information about Ising locality, as blocks hold over long distances. The receptive field tensors, however, do contain some information about locality, with about half the layer 2 tensors learning some form of locality (with a distance of about 7 to 8 between spins), and a quarter of the layer 3 tensors learning locality (with a distance of about 12 to 16 between spins). 

\begin{figure}
     \centering
     \begin{subfigure}{0.32\textwidth}
         \centering
         \includegraphics[width=\textwidth]{dist_1__41.pdf}
         \caption{Layer 1}
     \end{subfigure}
     \hfill
     \begin{subfigure}{0.32\textwidth}
         \centering
         \includegraphics[width=\textwidth]{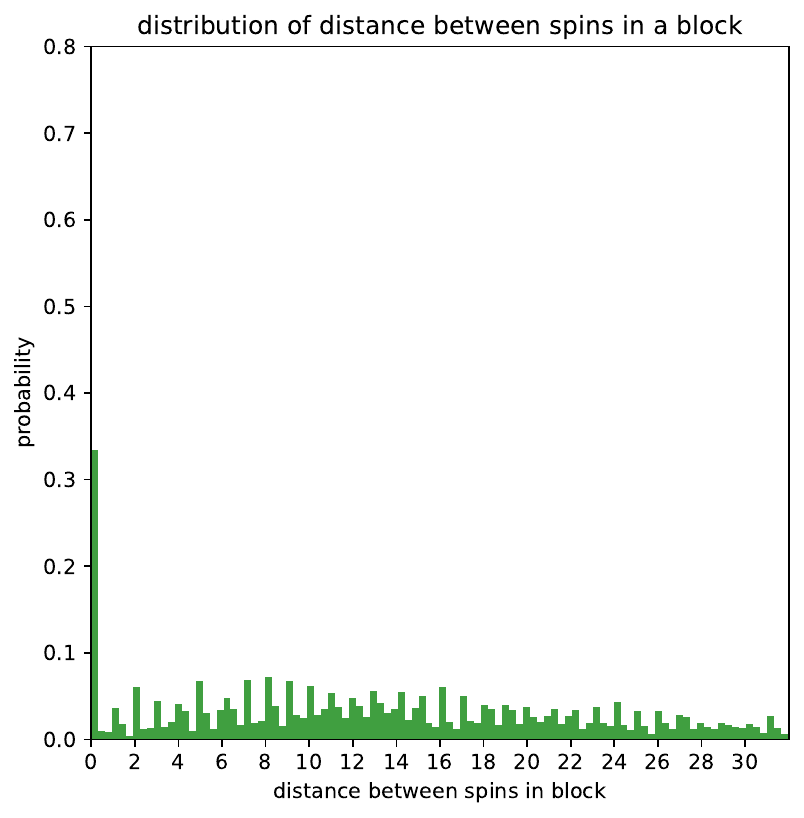}
         \caption{Layer 2 Weights}
     \end{subfigure}
          \hfill
          \begin{subfigure}{0.32\textwidth}
         \centering
         \includegraphics[width=\textwidth]{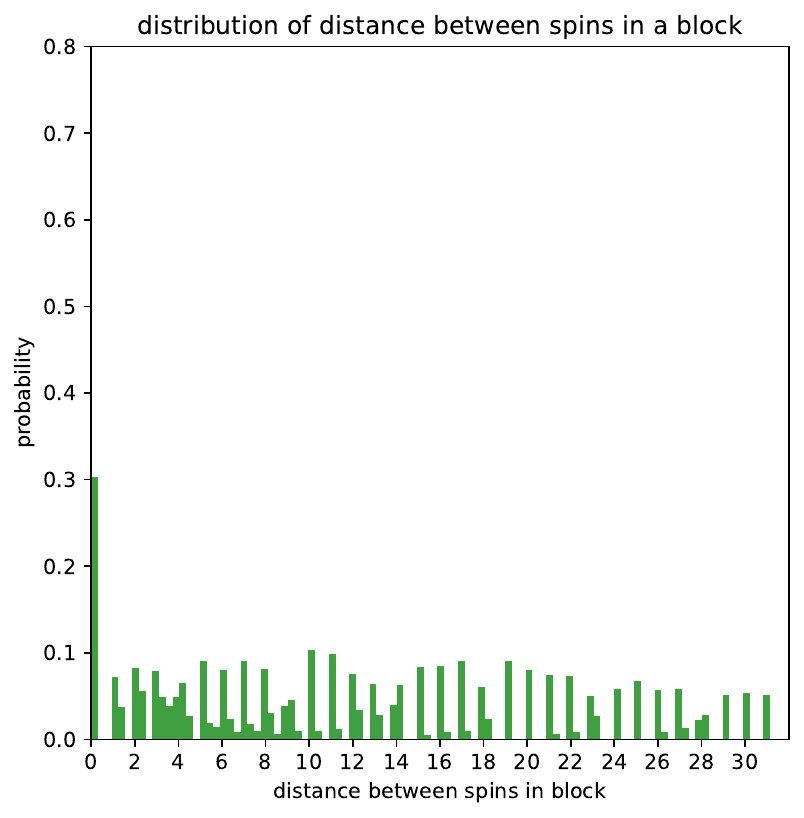}
         \caption{Layer 3 Weights}
     \end{subfigure}
     \hfill
     \begin{subfigure}{0.32\textwidth}
         \centering
         \includegraphics[width=\textwidth]{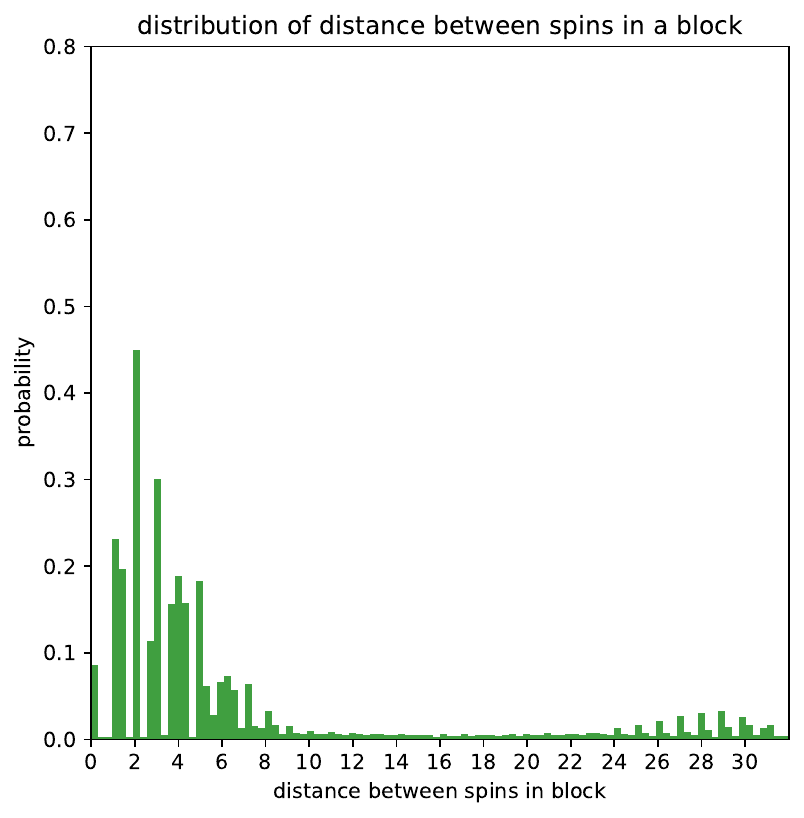}
         \caption{Layer 2 Receptive}
     \end{subfigure}
          \hfill
          \begin{subfigure}{0.32\textwidth}
         \centering
         \includegraphics[width=\textwidth]{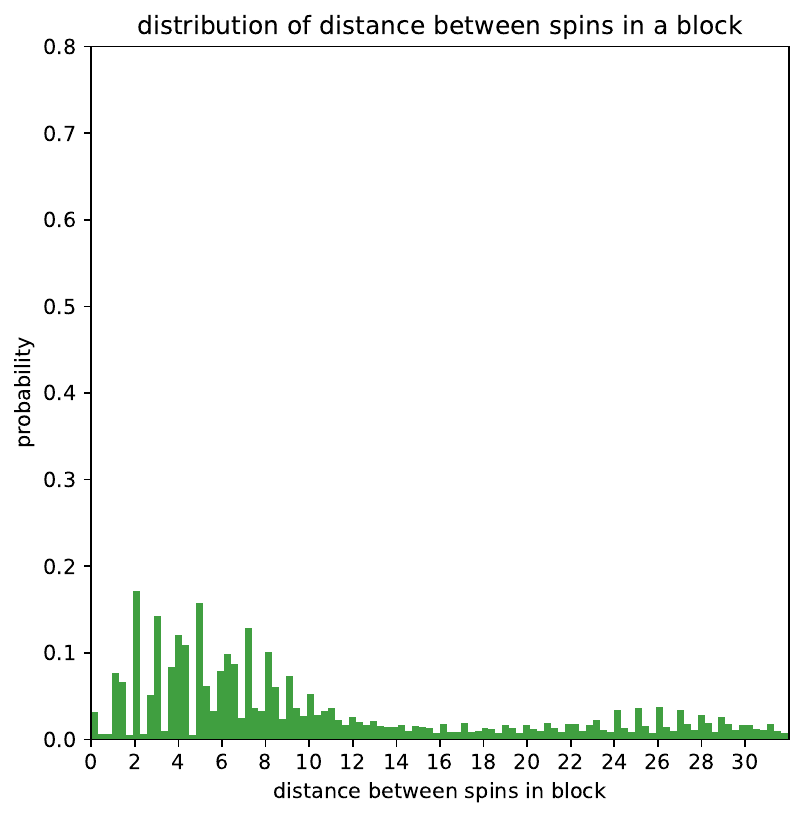}
         \caption{Layer 3 Receptive}
     \end{subfigure}
     \hfill
        \caption{Histogram of distance between all spins, averaged over all lattices. We analyze both the weight tensors and receptive field tensors. We find that locality is mostly learned for all three tensors, with most of the receptive tensors being close together, and only a couple far away. This is not true for the individual weights, which do not learn locality for layers 2 and 3.}
        \label{fig:dists_analysis_.41}
\end{figure}

We can further examine the structure of our block spinning via examining all distances between spins, instead of just the maximums. These results are shown in Figure \ref{fig:dists_analysis_.41}. We find similar results to our previous analysis. In particular, we find that the weight tensor for layer 1 learns locality that peaks with an average distance of 2 between nearby spins. We also find that the layer 2 and 3 weight tensors contain no indication of such locality, almost uniformly distributed throughout the distances. The receptive field tensors contain some information about locality, with the vast majority of the distances between spins in layer 2 and layer 3 being lower than 10 distances apart. 

These results suggest that while the spin size of the blocks from the RBM's are highly consistent with $b=2$ renormalization group flow, they also slightly struggle with learning locality. The results also imply that weight tensors from layers 2 and 3 do not characterize the system well, but the receptive field tensors characterize the system better, as expected from our receptive field plots. We find from our results that the receptive field tensor for layer 1 contains blocks that are entirely local (only 6 spins wide). On the other hand, layers 2 and 3 contain blocks in which most spins are close to each other with a few spins up to 32 spins away. 

We further note from plots in Appendix \ref{sec:block_plots} that the average distance between spins typically does not change as $\beta$ changes, though as $\beta$ reaches the critical point less and less locality is learned (likely due to diverging correlation lengths). This lack of peak change is different than what we'd expect from direct comparison of the weights to Ising model spins, due to correlation length changing with temperature. Even though block spinning is a useful model for the RBM, this suggests a deeper numerical analysis is needed to continue to connect the two concepts past this qualitative level. This numerical analysis is discussed in Chapter \ref{chap:connection}.

\section{Model Reconstructions}

\label{sec:reconstruction}

\begin{figure}
    \centering
    \includegraphics[width=\textwidth]{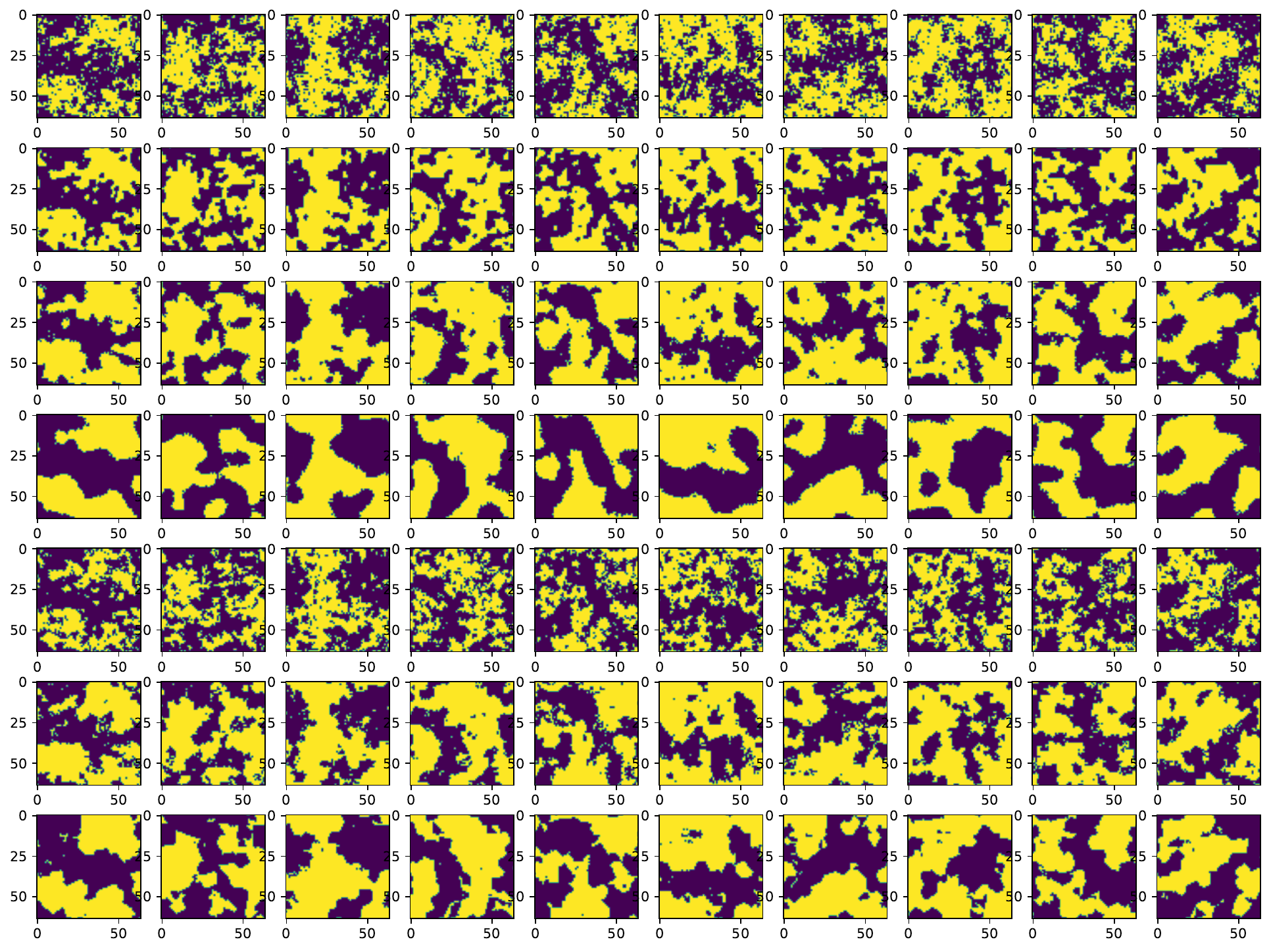}
    \caption{Reconstructed model plot for $\beta = .41$. The first row consisting of the original lattices, the next three rows consisting of the reconstructions from each layer in numerical order, and the last three rows consisting of the reconstructions with just using the large weights. The reconstructions are accurate, with the ones using only the large weights more accurate.}
    \label{fig:reco_analysis_.41}
\end{figure}

In addition to analyzing the weights of the model, we also consider how it works as an auto-encoder to reconstruct the training data. To do so, we consider the hidden weight probabilities $\text{pHid}_{i_nj_n}$ as defined in \ref{eq:pHid}, where $n$ is the layer number, to create a reconstructed lattice
\begin{equation}
    \text{vReco}_{ij}^{Ln} = \sum_{i_n, j_n} \sigma(\text{rec}_{ij;i_nj_n}^{Ln} \cdot \text{pHid}_{i_nj_n}),
    \label{eq:reco}
\end{equation}
where $\sigma$ is Equation \ref{eq:logistic}. We also consider the results of replacing $\text{rec}_{ij;i_nj}$ in Equation \ref{eq:reco} with just the two large weights. These results are shown in Figure \ref{fig:reco_analysis_.41} with the first row consisting of the original lattices, the next three rows consisting of the reconstructions from each layer in numerical order, and the last three rows consisting of the reconstructions with just using the large weights.

We find that the reconstructions are quite good for $\beta = .41$, with just the finer detail disappearing as we go down the layers, reproducing a result from \cite{Mapping}. We also find that by only including the large positive weights, the reconstructions are slightly better than the ones without them, keeping together some of the finer detail lost in the other reconstructions. These results are qualitatively consistent with the idea that the majority of the RBM's information in stored in only two weights. In addition, it is consistent with the idea that the RBM coarse-grains in a similar manner to RG flow, keeping the data's macroscopic structure, but not its the microscopic structure. 

However, these results are not consistent for all values of $\beta$. We find from plots in Appendix \ref{sec:reco_plots} that the reconstructions get worse as we approach criticality, suggesting that the network fails as an autoencoder in these cases, likely due to the excess of one spin over another. However, the large weight reconstructions are more consistent across all the $\beta$ values (though they still get worse near criticality). This implies that \cite{Mapping}'s argument that the RBM's make a good autoencoder does not apply as one gets closer to criticality, meaning the RG flow connection may fail at these temperatures. However, it also implies that most of the important information in the RBM is encoded within the two large weights, giving us an interesting view on the structure of the RBM learning.

\begin{figure}
    \centering
    \includegraphics[width=\textwidth]{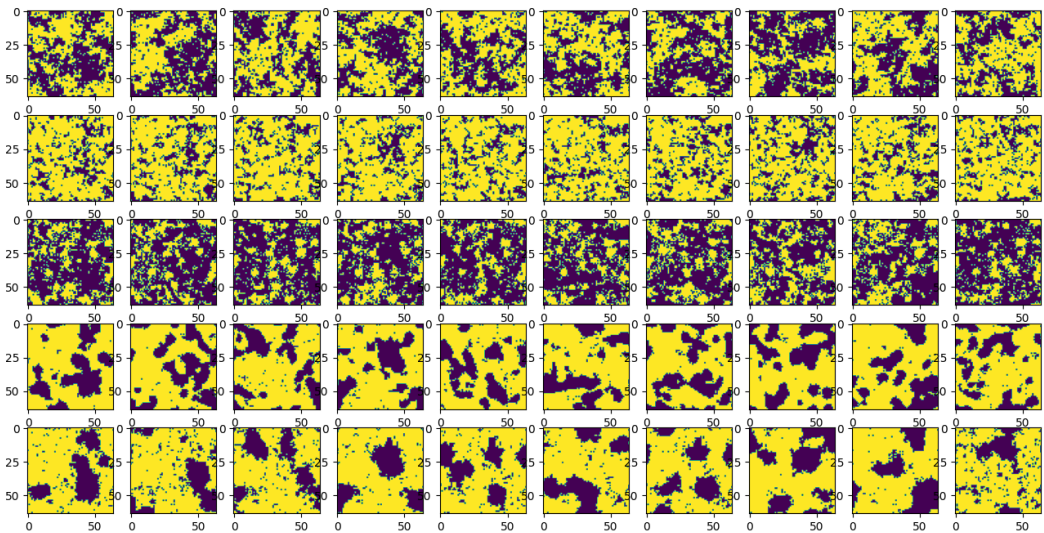}
    \caption{Another reconstructed model plot for $\beta = .41$. We consider the case in which we only use large weights to calculate the hidden spins and reconstructed spins, shown in row 2. Row 1 consists of original lattices while rows 3-5 consist of the reconstructed lattices using full weights for all three layers. The reconstruction that changes the hidden weights failed, suggesting that the full weight tensors are needed in the calculation of hidden weights, instead of just two large weights.}
    \label{fig:reco_analysis_fail}
\end{figure}

We also attempted to reconstruct the model by using only the two large weights to create the hidden weights used in calculating $pHid_{i_nj_n}$, along with using the two large weights in the receptive field tensor. We did this only for the first layer at $\beta = .41$. These results are shown in Figure \ref{fig:reco_analysis_fail}, where the first row is the original lattice, the second row is the reconstruction with only the large weights, and the last three row are the reconstructions from all three layers for comparison. Obviously, from the figure, the reconstruction where we changed the hidden weights to only consider large weights does not work well, implying that the full weight tensor is needed when creating the hidden weights. In this sense, the smaller weights have an effect on the outcome of the model: they are needed to get correct hidden weights. We thus conclude that the two weight model is a good approximation of how the RBM learns, but is nowhere near the full extent of the RBM learning. We did not run this test for other layers or other values of $\beta$ because of the drastic failure of the reconstruction.

\chapter{Connecting Renormalization Techniques and Deep Learning} \label{chap:connection}

In the previous sections, we discussed the renormalization group and machine learning separately, focusing on the validation of our renormalization techniques in Chapters \ref{chap:1d} and \ref{chap:renorm}, and a qualitative analysis of the deep learning of RBMs in Chapter \ref{chap:dl}. It is now time to put the two concepts together by creating direct Ising representations from the RBMs based on our qualitative analysis, feeding them into the renormalization group validation, and comparing results. We do this in two different ways: running the learning in reverse and generating 64 by 64 lattices from renormalized 8 by 8 lattices, and creating a direct Ising spin representation from our RBM weights.

\section{Generative Models}
\label{sec:gen_models}

In examining the quantitative success of the deep learning infrastructure as a renormalization group flow, we first used the learning weights as a way to generate new 64 by 64 lattice models from 8 by 8 lattice models. We then compared these generated models to those derived from the Wolff algorithm. 

\begin{table}
    \centering
    \begin{tabular}{|p{5cm}|p{5cm}|}
    \hline
    64x64 $\beta$ Value & 8x8 $\beta$ Value \\
    \hline
    .395 & .241 \\
    \hline
    .4 &  .256 \\
    \hline
    .405 & .274 \\
    \hline
    .41 & .291\\
    \hline
    .415 & .315\\
    \hline
    .42  & .338\\
    \hline
    .425 & .367\\
    \hline
    .43 & .397\\
    \hline
\end{tabular}
    \caption{$\beta$ values for the 64 by 64 Ising modes and the corresponding $\beta$ values for the 8 by 8 Ising models the original values flow to. The 8 by 8 Ising models are used to generate new 64 by 64 models by running the learning backwards. }
    \label{tab:beta_vals}
\end{table}
To do so, for each value of $\beta$ in the 64 by 64 lattice, we find the 8 by 8 $\beta$ value it flows to in the RG flow in Chapter \ref{chap:renorm}. These results are shown in Table \ref{tab:beta_vals}. We then generate 8 by 8 lattices for each of these 8 by 8 $\beta$ values using the Wolff algorithm, which we denote here as $x_{i_3j_3}^{L3}$. From here, we generate the model by taking:
\begin{equation}
    x_{i_2j_2}^{L2} = 2\sigma\left(\sum_{i_3, j_3} w_{i_2j_2;i_3j_3}^{L3} \cdot x_{i_3j_3}\right) -1,
\end{equation}
then
\begin{equation}
    x_{i_1j_1}^{L1} = 2\sigma\left(\sum_{i_2, j_2} w_{i_1j_1;i_2j_2}^{L2} \cdot x_{i_2j_2}\right) -1,
\end{equation}

and lastly, 

\begin{equation}
    v_{ij}^{gen} = 2\sigma\left(\sum_{i_1, j_1} w_{ij;i_1j_1}^{L1} \cdot x_{i_1j_1}\right) -1.
\end{equation}

\begin{figure}
    \centering
    \includegraphics[width=\textwidth]{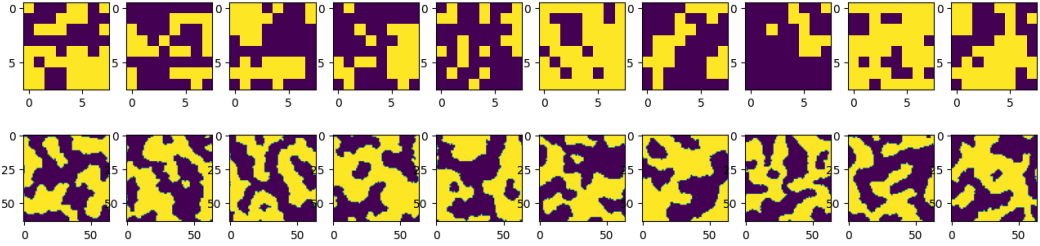}
    \caption{Generative model for $\beta = .41$. The first row contains the original 8 by 8 lattice models and the second row contains the 64 by 64 generated model.}
    \label{fig:gen_example}
\end{figure}
We do this for 20,000 lattices for all $\beta$. The result of this procedure for $\beta=.41$ is shown in Figure \ref{fig:gen_example}. The first row contains the original 8 by 8 lattice models and the second row contains the 64 by 64 generated models. At first glance, there seems to be little relation between the original lattices and the generated model, seeming to throw out the idea of the connection. However, this is not a problem. The RBM takes in our lattices as one dimensional and uses this information to rederive the spin connections. Due to this, the spin blocks get "mixed-up" and moved to different locations, and sometimes all the blocks get flipped. We deal with this more effectively in Section \ref{sec:weights_and_spins}. However, the generative model results are still useful, as the quantitative information, including the correlations, stays the same. This means that these generative models should accurately reflect the Ising model quantitatively.  These general results hold for all $\beta$, as shown in Appendix \ref{sec:gen_plots}.

\begin{figure}
     \centering
    \begin{subfigure}{0.49\textwidth}
         \centering
         \includegraphics[width=\textwidth]{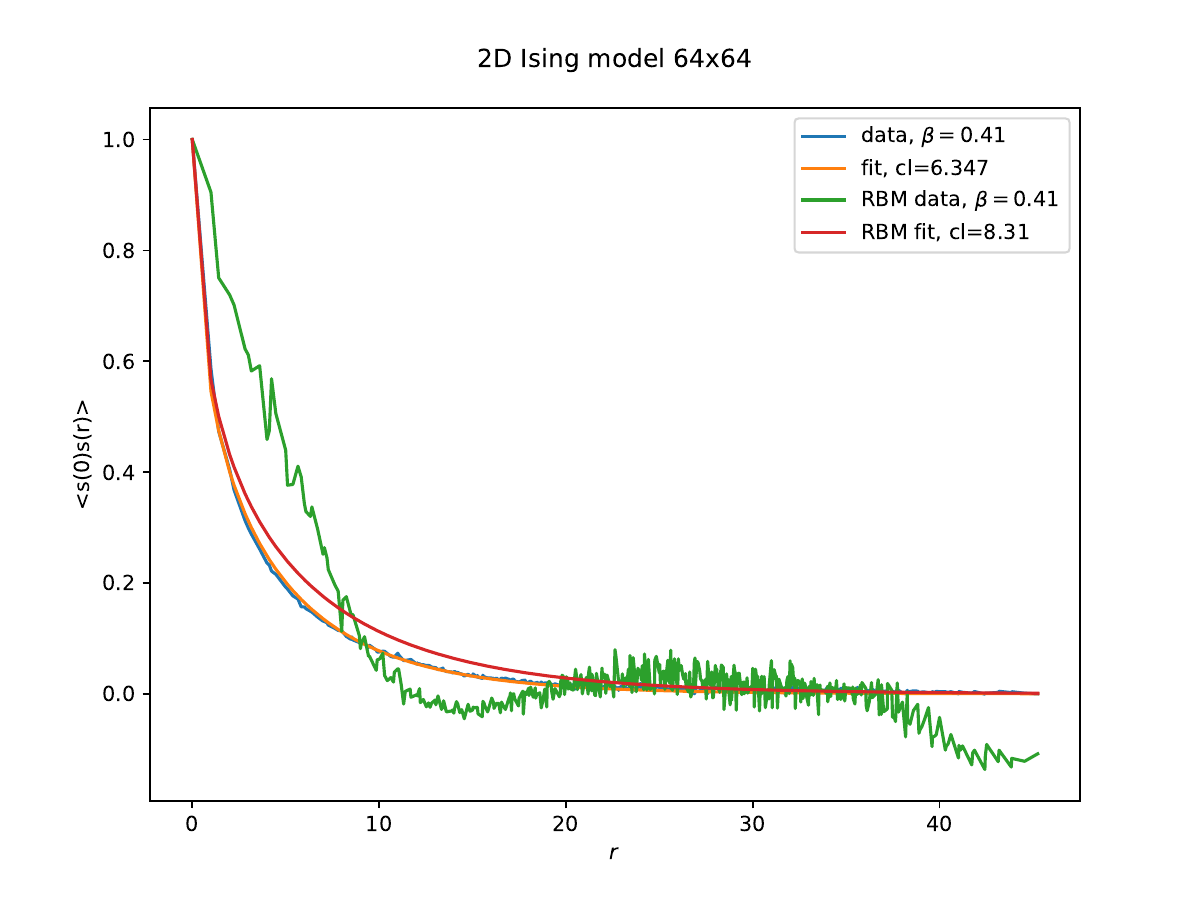}
         \caption{Correlation length for $\beta = .41$.}
     \end{subfigure}
      \hfill
    \begin{subfigure}{0.49\textwidth}
         \centering
         \includegraphics[width=\textwidth]{generative_models.pdf}
         \caption{Generative Model}
     \end{subfigure}
     \hfill
     \begin{subfigure}{0.49\textwidth}
         \centering
         \includegraphics[width=\textwidth]{generative_data_comparison.pdf}
         \caption{Generative Model and Wolff Algorithm Comparison}
     \end{subfigure}
     \hfill
     \begin{subfigure}{0.49\textwidth}
         \centering
         \includegraphics[width=\textwidth]{generative_fit_comparison.pdf}
         \caption{Generative Model and Wolff Algorithm Comparison}
     \end{subfigure}
          \hfill
        \caption{Comparing the Wolff algorithm with the generative model. The generative model tends to have a higher correlation function than expected for $0 < r < 10$ and a lower correlation function for $10<r< 20$, for all $\beta$. For the correlation length temperature dependence, we find that $\nu = 0.763$ for the generative model, as opposed to the Wolff algorithm value of $\nu = 0.973$. The generative model only somewhat matches the fit at $T > 2.35$, and does not for $T < 2.35$.  }
        \label{fig:gen_model_analysis}
\end{figure}

After doing this for all of our values of $\beta$, we proceed with numerically analyzing our results, in the same manner that we analyzed the Wolff algorithm in Chapter \ref{chap:renorm}. Some results of this analysis are shown in  Figure \ref{fig:gen_model_analysis}, including the correlation length fit for $\beta=.41$ for both the Wolff algorithm and generative models, the generative model correlation length data and fits, and comparisons between the generative model and Wolff model. The rest of the correlation length fits are shown in Appendix $\ref{sec:gen_plots}.$ 

We find that the correlation length fits to the generative models tend to have higher correlation functions than expected for $0 < r < 10$ and a lower correlation function for $10<r< 20$. This suggests that the reproductions do not perfectly match that of the Ising model, and on some level, the RBM fails to reproduce the Ising model due to these offsets. However, we still proceed with the analysis assuming that these over-estimations and under-estimations cancel out. 

When we calculate the temperature dependence of these correlation lengths, we find that the data only somewhat matches the fit at $T > 2.35$, and does not for the data point at $T < 2.35$. The fit suggests that $\nu = 0.763$ for the generative model, as opposed to the Wolff algorithm value of $\nu = 0.973$. These results imply that the generative models forms a poor approximation of the nearest-neighbor Ising model, with a few qualitative connections (such as the fit going in the correct direction), but little quantitative connections. 

While this data weakens the idea of a naive connection between the learning and the renormalization group, it does not rule out the connection completely. The models here were generating assuming that the group flow could be approximated using only the nearest-neighbor Ising models and only the coupling $\beta$. To further rule out a quantitative connection between RBM generative models and RG flow, this extra parameter space must be explored. 

\section{From Weights to Ising Spins}
\label{sec:weights_and_spins}

Using generative models, however, is not the only way to connect the RBM learning to the Ising spins. Instead, we can sample from our probability distributions pHid from Equation \ref{eq:pHid} for each layer, and then convert the binary spins of those layers to spins of $\pm 1.$ 

\begin{figure}
    \centering
    \includegraphics[width=\textwidth]{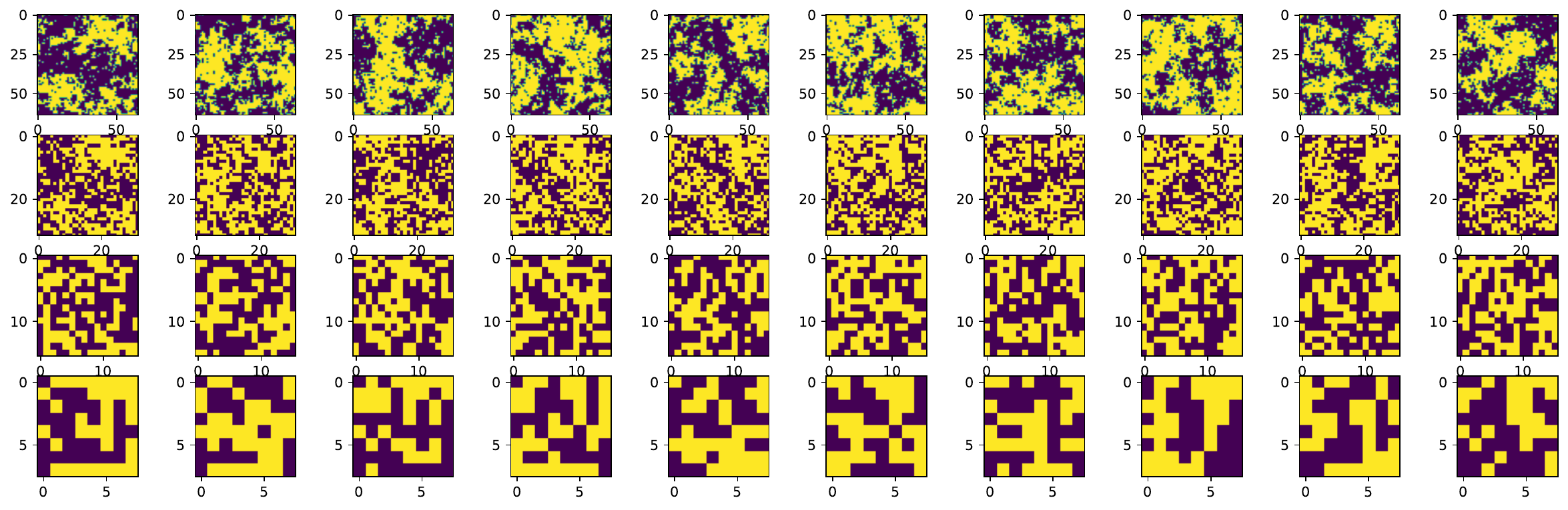}
    \caption{RBM coarse graining for $\beta = .41$. The first row consists of the original Wolff-generated Ising lattices and the next three rows consist of Ising spin representations of the RBM learning layers. We find that the first two layers are qualitatively similar to the renormalization group.}
    \label{fig:.41_grain_analysis}
\end{figure}

As stated in Section \ref{sec:gen_models}, the locations of these new lattices do not have a simple relationship to the locations of the original lattices. We fix this by taking the receptive field tensors $\text{rec}_{ij;i_nj_n}^{Ln}$ and finding which of the entries $i,j$ have the largest absolute value for each of the $i_n,j_n$. This is the location of the original spin that the weight is most connected to. We then coarse grain and define this as the location of the corresponding spin.

The results of this coarse graining procedure for $\beta = .41$ are given in Figure \ref{fig:.41_grain_analysis}. The first row consists of the original Wolff lattices, while the next 3 rows consist of the 3 layers of coarse graining in order. We do this for all $\beta$ as shown in Appendix \ref{sec:spin_plots}. 

The results here qualitatively show a coarse graining that is similar to Kadanoff-block spinning in some ways, with the large scale structure of the system being most strongly preserved in the second layer. The first layer of the coarse graining still preserves large scale structure, but not as well as the second layer, having extra noise. The last layer of the coarse graining, however, loses most of the macroscopic structure. This result is similar throughout all $\beta$ values. This strengthens the idea of a qualitative connection between the renormalization group and deep learning in the first two layers, while the third layer loses this qualitative connection. 

\begin{figure}
     \centering
    \begin{subfigure}{0.49\textwidth}
         \centering
\includegraphics[width=\textwidth]{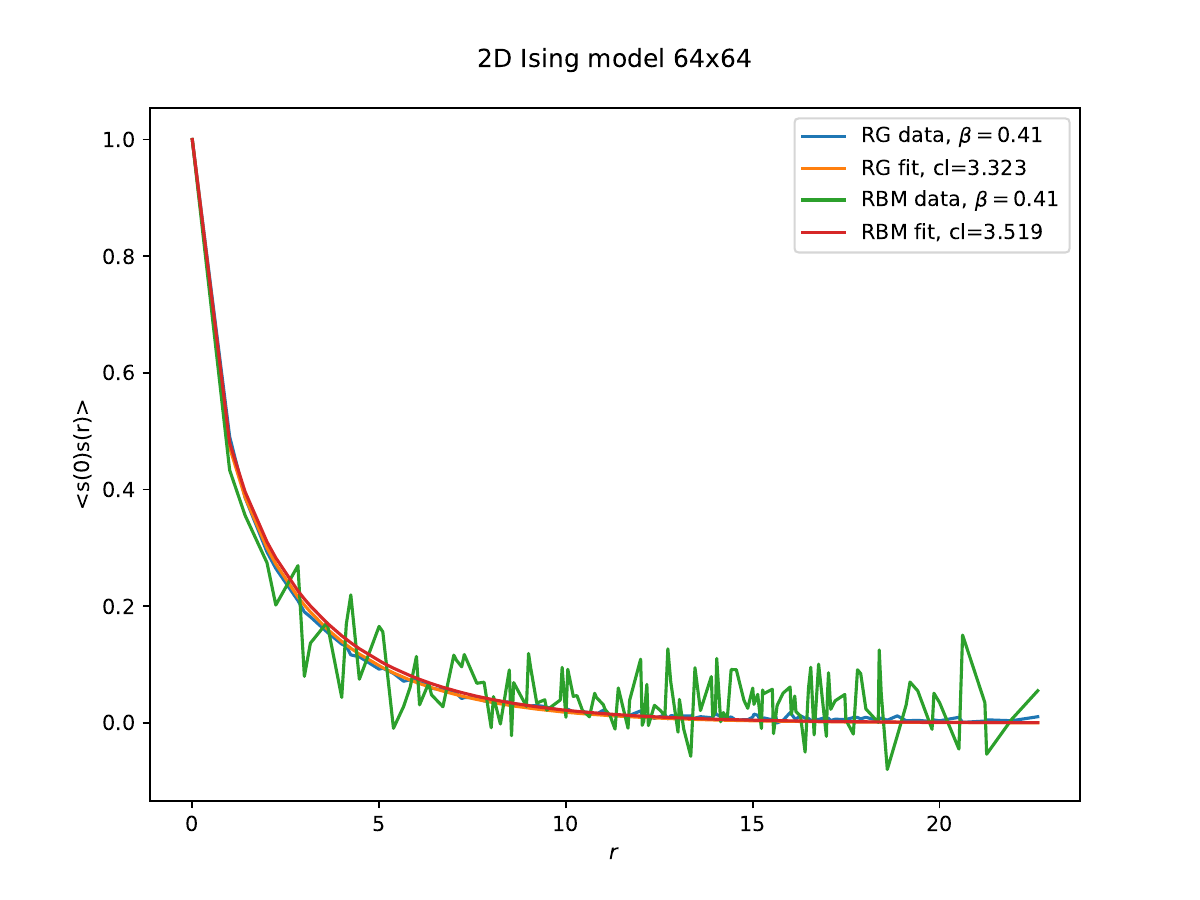}
         \caption{Correlation length comparison for $\beta = .41$.}
     \end{subfigure}
      \hfill
    \begin{subfigure}{0.49\textwidth}
         \centering
         \includegraphics[width=\textwidth]{Layer1-RBMreg.pdf}
         \caption{RBM Results}
     \end{subfigure}
     \hfill
     \begin{subfigure}{0.49\textwidth}
         \centering
         \includegraphics[width=\textwidth]{Layer1-hinton.pdf}
         \caption{RG Blocks}
     \end{subfigure}
     \hfill
     \begin{subfigure}{0.49\textwidth}
         \centering
         \includegraphics[width=\textwidth]{Layer1-wolff.pdf}
         \caption{Wolff Algorithm}
     \end{subfigure}
          \hfill
        \caption{Analysis of layer 1 RBM Ising representation. The fits for the RBM correlation function are consistent with the data for most $\beta$. For the correlation length temperature dependence, we find that $\nu =1.178$ for the Wolff algorithm, $\nu = 0.983$ for the RG flow and $\nu = 1.174$ for the RBM training. The RBM data, however,  does not match this fit well.}
        \label{fig:layer_1_analysis}
\end{figure}
We further examine this quantitatively by running the RBM coarse-grained models through the validation procedure from Chapter \ref{chap:renorm}, and comparing these results with both our traditional renormalization results and the Wolff algorithm. In our plots, we also include a correlation length fit for $\beta = .41$ for reference. The remaining correlation length plots are given in Appendix \ref{sec:spin_plots}. 

For layer 1, our results are shown in Figure \ref{fig:layer_1_analysis}. We have that most of our correlation function data matches the fits decently well, with the exception of a couple of points, like $\beta = .415$ and $\beta = .43$. In particular, when we fit the data, we get $\nu =1.178$ for the Wolff algorithm, $\nu = 0.983$ for the RG flow and $\nu = 1.174$ for the RBM training. Looking only at the fits, the three cases seem to match quantitatively! 

However, the RBM data does not match up with its fit very well. The overestimates and underestimates cancel to make the fit work, but the data and fit mismatch prevent us from concluding the first layer as qualitatively equivalent to the renormalization group results. 

\begin{figure}
     \centering
    \begin{subfigure}{0.49\textwidth}
         \centering
         \includegraphics[width=\textwidth]{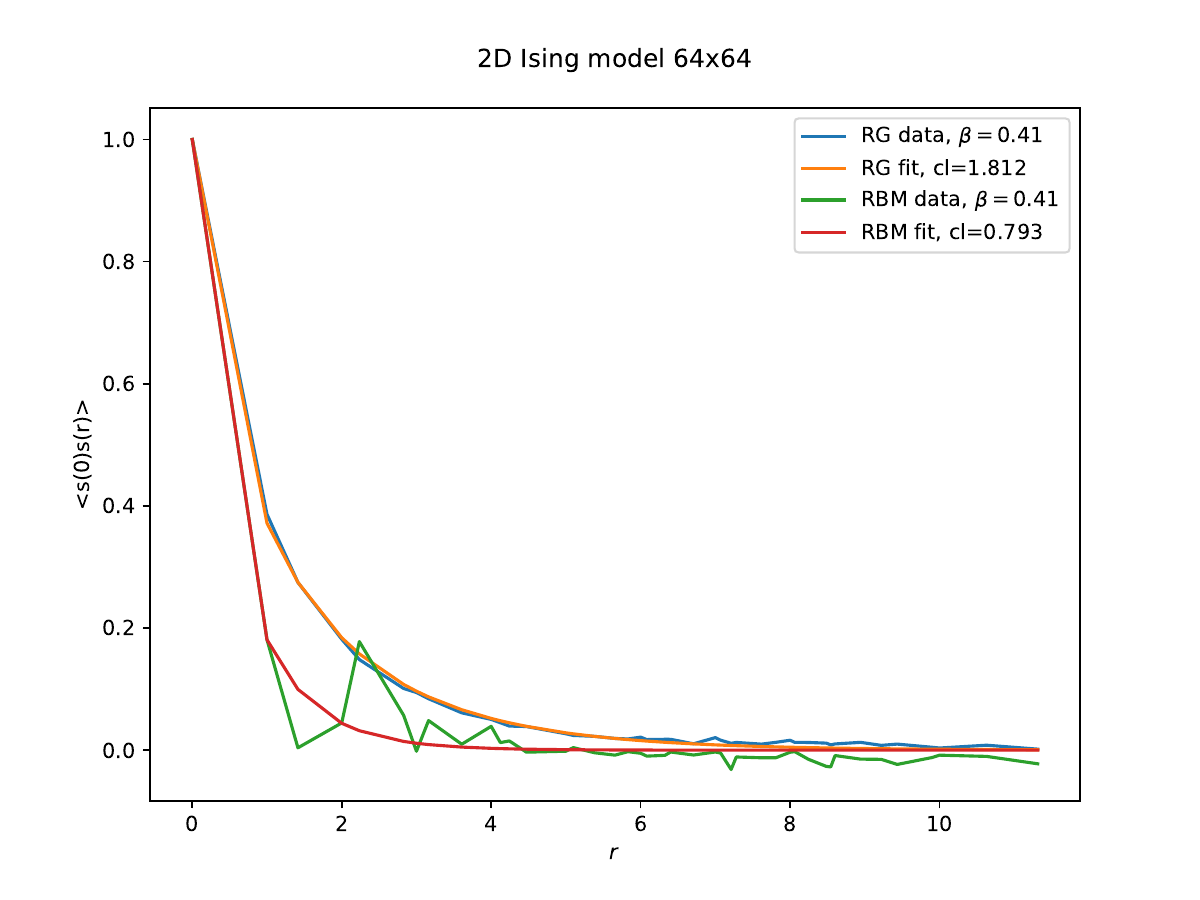}
         \caption{Correlation length for $\beta = .41$.}
     \end{subfigure}
      \hfill
    \begin{subfigure}{0.49\textwidth}
         \centering
         \includegraphics[width=\textwidth]{Layer2-RBMreg.pdf}
         \caption{RBM Results}
     \end{subfigure}
     \hfill
     \begin{subfigure}{0.49\textwidth}
         \centering
         \includegraphics[width=\textwidth]{Layer2-hinton.pdf}
         \caption{RG Blocks}
     \end{subfigure}
     \hfill
     \begin{subfigure}{0.49\textwidth}
         \centering
         \includegraphics[width=\textwidth]{Layer2-wolff.pdf}
         \caption{Wolff Algorithm}
     \end{subfigure}
          \hfill
        \caption{Analysis of layer 2 RBM Ising representation. The fits for the RBM correlation function are consistent with the data for most $\beta$. For the correlation length temperature dependence, we find that $\nu = 1.452$ for the Wolff algorithm, $\nu  = 0.969$ for block spinning, and $\nu = 0.729$ for the RBM training. The RBM data does not match this fit well.}
        \label{fig:layer_2_analysis}
\end{figure}

For layer 2, our results are shown in Figure \ref{fig:layer_2_analysis}. As in layer 1, the correlation lengths tend to match their fitting function, with the exception of $\beta = .415$. When we fit the correlation length temperature dependence, we found the value of $\nu = 1.452$ for the Wolff algorithm, $\nu  = 0.969$ for block spinning, and $\nu = 0.729$ for the RBM training. All three fits are close together above criticality, but diverge from one another as criticality is approached. This causes some quantitative mismatching.

Additionally, the same issue of a mismatch between data and fit appears here. The fit we use is rather restrictive, and our data undershoots it. The renormalization does appear to be going in the right direction, qualitatively, however.
\begin{figure}
     \centering
    \begin{subfigure}{0.49\textwidth}
         \centering
         \includegraphics[width=\textwidth]{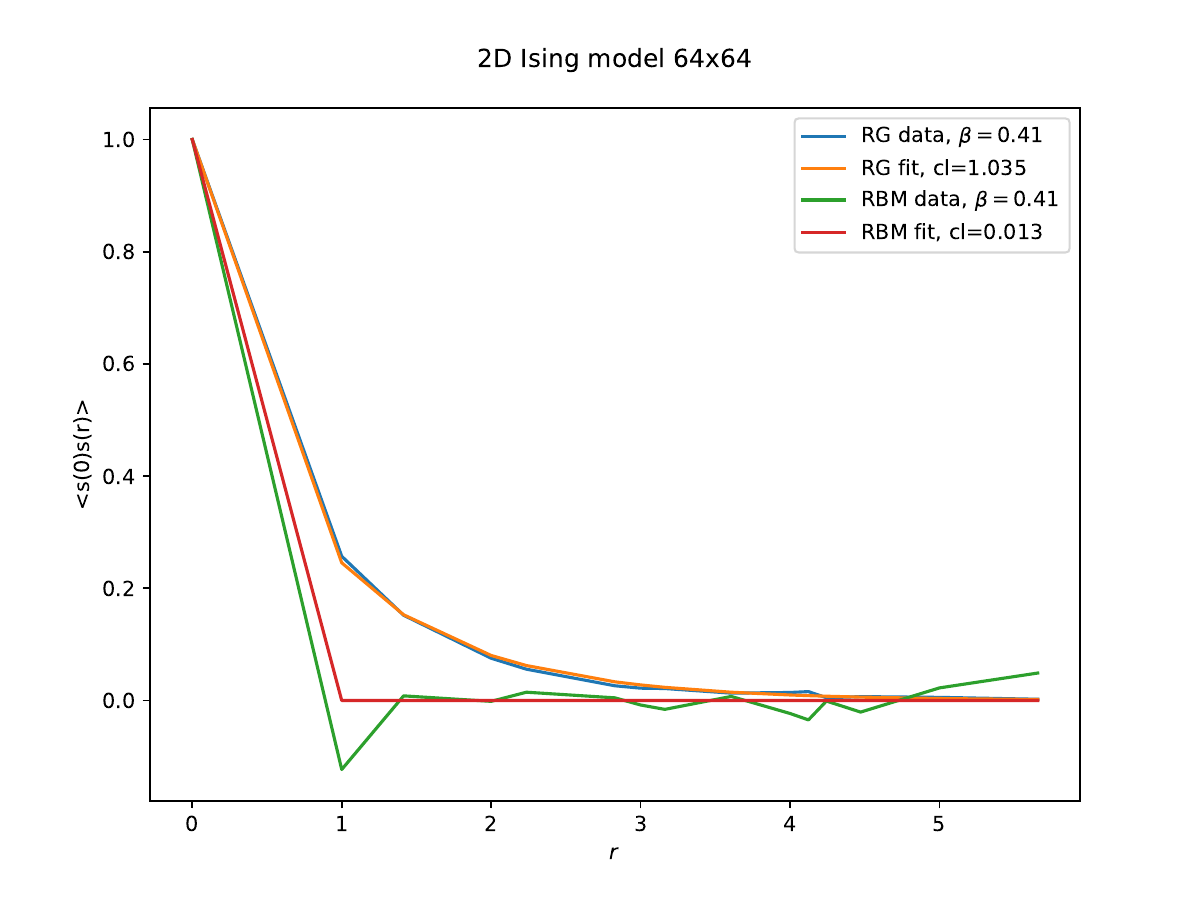}
         \caption{Correlation length for $\beta = .41$.}
     \end{subfigure}
      \hfill
    \begin{subfigure}{0.49\textwidth}
         \centering
         \includegraphics[width=\textwidth]{Layer3-RBMreg.pdf}
         \caption{RBM Results}
     \end{subfigure}
     \hfill
     \begin{subfigure}{0.49\textwidth}
         \centering
         \includegraphics[width=\textwidth]{Layer3-hinton.pdf}
         \caption{RG Blocks}
     \end{subfigure}
     \hfill
     \begin{subfigure}{0.49\textwidth}
         \centering
         \includegraphics[width=\textwidth]{Layer3-wolff.pdf}
         \caption{Wolff Algorithm}
     \end{subfigure}
          \hfill
        \caption{Analysis of layer 3 RBM Ising representation. The fits for the RBM correlation function are often inconsistent with the data and correlation lengths are close to zero. For the correlation length temperature dependence, we find that $\nu = 0.972$ for the RG flow, $\nu =  1.933$ for the Wolff algorithm and $\nu = 0.0$ for the RBM. The RBM data is not numerically consistent with RG flow at all for this layer.}
        \label{fig:layer_3_analysis}
\end{figure}

For layer 3, our results are shown in Figure \ref{fig:layer_3_analysis}. Here, even the qualitative results seem to change. Fits of the correlation function start being more inconsistent with the fits, and correlation lengths are all close to zero. For temperature dependence, we get $\nu = 0.972$ for the RG flow, $\nu =  1.933$ for the Wolff algorithm and $\nu = 0.0$ for the RBM. Here, even the qualitative results suggesting that the RBM is connected to RG completely fall apart, reinforcing the earlier qualitative conclusion we had from looking at the coarse-graining directly. We also have discrepancies near criticality for the Woolf algorithm and RG flow, suggesting that the 8 by 8 model may be too small to effectively do work with.

\chapter{Conclusions and Further Work}
\label{chap:conclusions}

\section{General Conclusions}

Throughout this work, we have considered applications of basic machine learning techniques to renormalization in the one dimensional and two dimensional Ising models (in Chapters \ref{chap:1d} and \ref{chap:renorm}), discussed the qualitative results of deep learning using RBMs of the the two dimensional Ising model (in Chapter \ref{chap:dl}), and attempted to connect the two paradigms quantitatively  (in Chapter \ref{chap:connection}). What, then, can we conclude?

Our first conclusion is that machine learning techniques through Adam optimization can be successfully applied in renormalizing both the 1D and 2D Ising models. These optimization techniques work best when using correlation lengths as loss functions, and when paying close attention to how finiteness affects criticality. We found better renormalization results using optimization techniques than we did using accepted analytical techniques, like the Maris-Kadanoff equation. 

Our second conclusion is that there are strong qualitative connections between deep learning through RBMs and RG flow, just as Mehta and Schwab argued in \cite{Mapping}. This is shown through the reproduction of layer 1 and 3 receptive field plots and model reconstruction for 8 $\beta$ values. In addition, we discovered the emergence of a two-weight blocking structure similar to the Ising model one-weight blocking structure, and found that most of these blocks contain some sense of locality. Furthermore, when converting these blocks to Ising spin representations, layer 1 and 2 had correlation function structures that match how they should look in the Ising model. In addition, the layer 1 fit $\nu$ value is very close to its expected value, and layer 2 correlation lengths tend to go in the correct direction for renormalization.

Our third conclusion, however, is that we do not have enough evidence to rule in favor of this connection existing on a quantitative level. In particular, the layer 3 model failed to produce results that looked like the Ising model, even qualitatively, with $\nu = 0$. Similarly, the generative model had trouble with deriving the correlation lengths due to fitting issues. Models for layers 1 and 2 had data that did not necessarily correspond to their fits, preventing us from establishing a qualitative connection there. Only 8 data points were used, some with poor reconstructions, and we have yet to examine the case of RBMs without L1 regularization. 

That being said, these quantitative discrepancies do not imply that the connection is non-existent either. As discussed in Chapter \ref{chap:renorm}, the nearest-neighbor renormalization is just a subspace of the total parameter space. It is possible that our results here tell only part of the story because we only accomplish the group flow in $\beta$. This is discussed more in the next section.

\section{Future Quantitative Work}

To make a more solid conclusion about the quantitative connection between the renormalization group and deep learning, further developments need to be made on both renormalization group and deep learning structures. 

For the renormalization group, extending the group flow to be a two parameter flow using the next-to-nearest neighbor Hamiltonian would provide more parameter space to examine the RBM flow in, getting a clearer picture on why the numerical connections are not present in the 1D space. In addition, criticality issues from using finite systems likely also played a role in the quantitative disconnect. Using larger systems should help prevent these criticality issues.

In addition, the layer 3 failure could have just been an issue of the 8 by 8 system being too small. Changing the entire system to start at 128 by 128 or 256 by 256 would allow us to ignore this as a possible problem. In addition, running the numerical analysis for more $\beta$ values would give us clearer data points on the connection between the two, and make our fits clearer. 

Lastly, our results here only considered what happens when we use an L1 penalty for learning. Removing this penalty would allow us to see how much the penalty affects how clearly and quickly the system is learning locality. 

\section{Further Extensions}

If these numerical issues are resolved and the connection made clearer, one could extend the work further using other renormalization models and more modern machine learning methods.

In particular, one could extend the work from the 2D Ising model to the Ising model in 3 or more dimensions. Similarly, one could look at Ising variations, such as the quantum Ising model or the tricritical Ising model. One could also extend it to other O(N) models, like the XY or Heisenberg models. 

Similarly, one could use more modern deep learning techniques, instead of the simple case of RBMs. One example in particular could be the use of variational autoencoders. 

All in all, though the examination of a possible connection between machine learning and renormalization has been greatly expanded on in our work here, there still remains plenty of work to be done to fully understand the extent of such a connection.

\bibliographystyle{plainnat}
\bibliography{bib}

\appendix

\chapter{Two Dimensional Ising Layer-by-Layer Analysis Plots}
\label{chap:2d_layer_plots}

\section{Receptive Field Plots}
\label{sec:recep_plots}

Each individual plot represents a node in a given layer and the values on the plot represent the full set of receptive field tensors corresponding to that node. The clustering of non-zero tensors next to each other suggests the model is qualitatively learning some form of locality. This is discussed in depth in Section \ref{sec:receptive}.

\begin{figure}[h!]
    \centering
    \includegraphics[scale=.3]{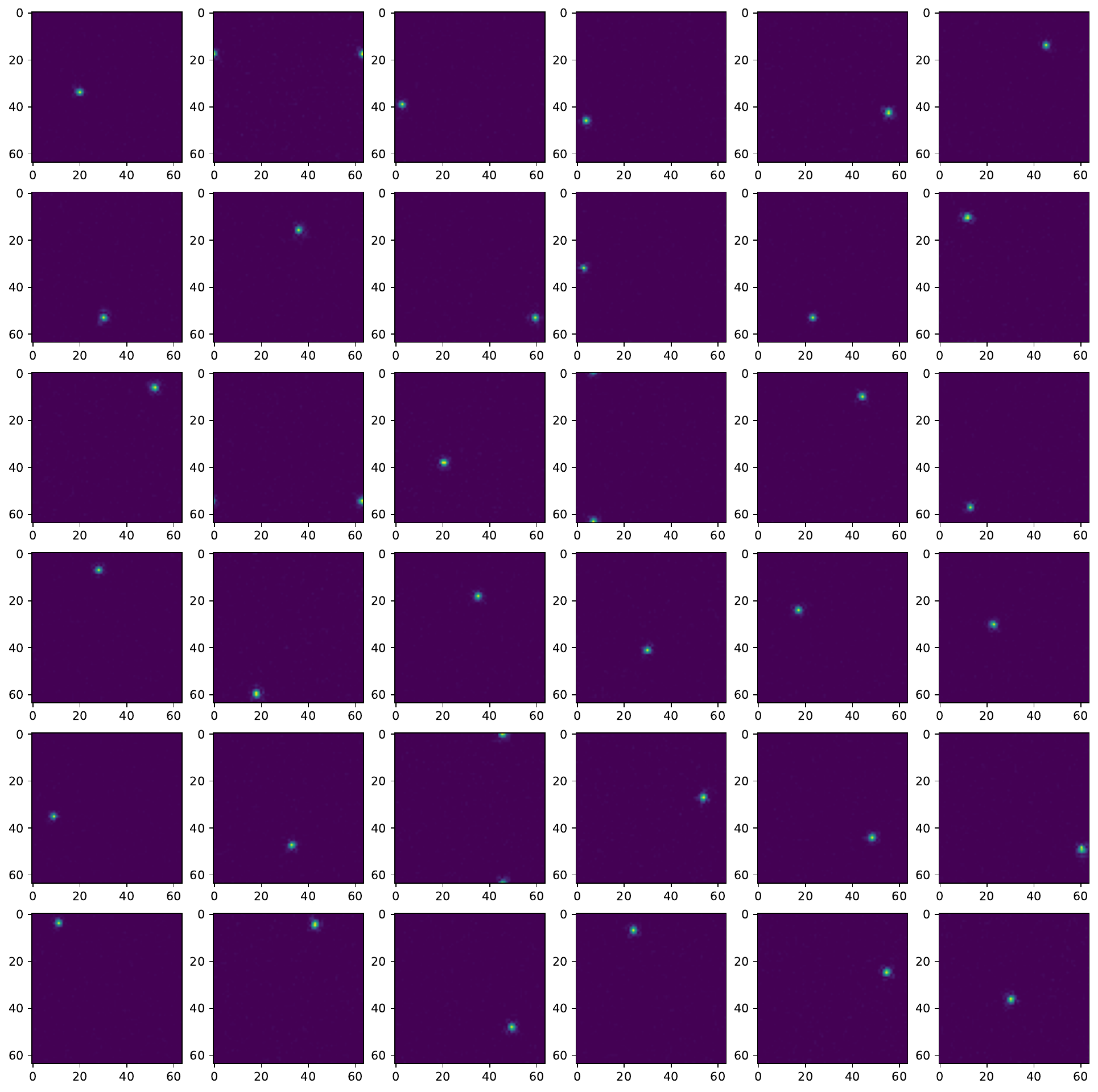}
    \caption{Layer 1 receptive field plot for $\beta = .395$.}
    \label{fig:rec_1_.395}
\end{figure}

\begin{figure}[h!]
    \centering
    \includegraphics[scale=.3]{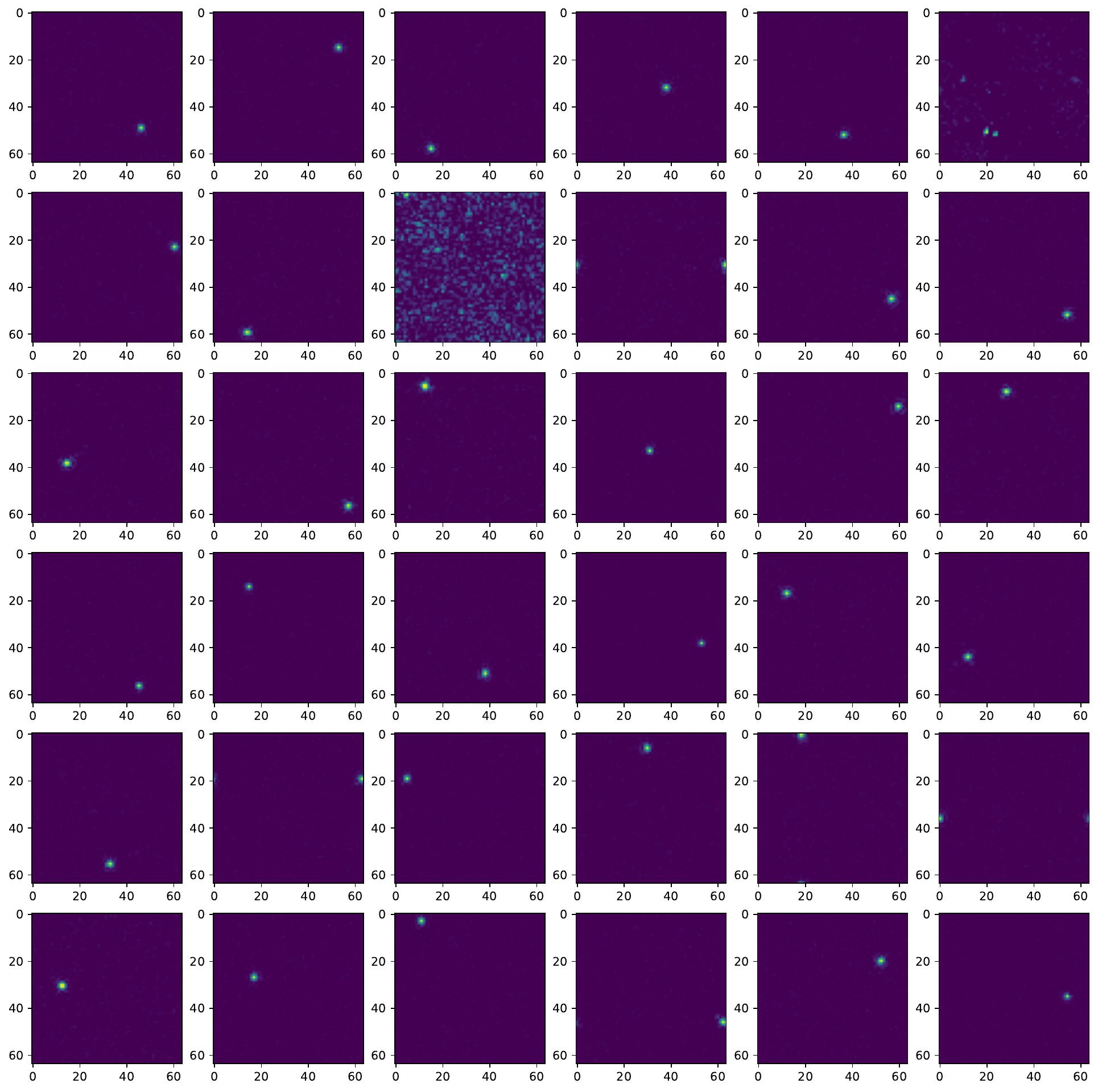}
    \caption{Layer 1 receptive field plot for $\beta = .4$.}
    \label{fig:rec_1_.4}
\end{figure}

\begin{figure}[h!]
    \centering
    \includegraphics[scale=.3]{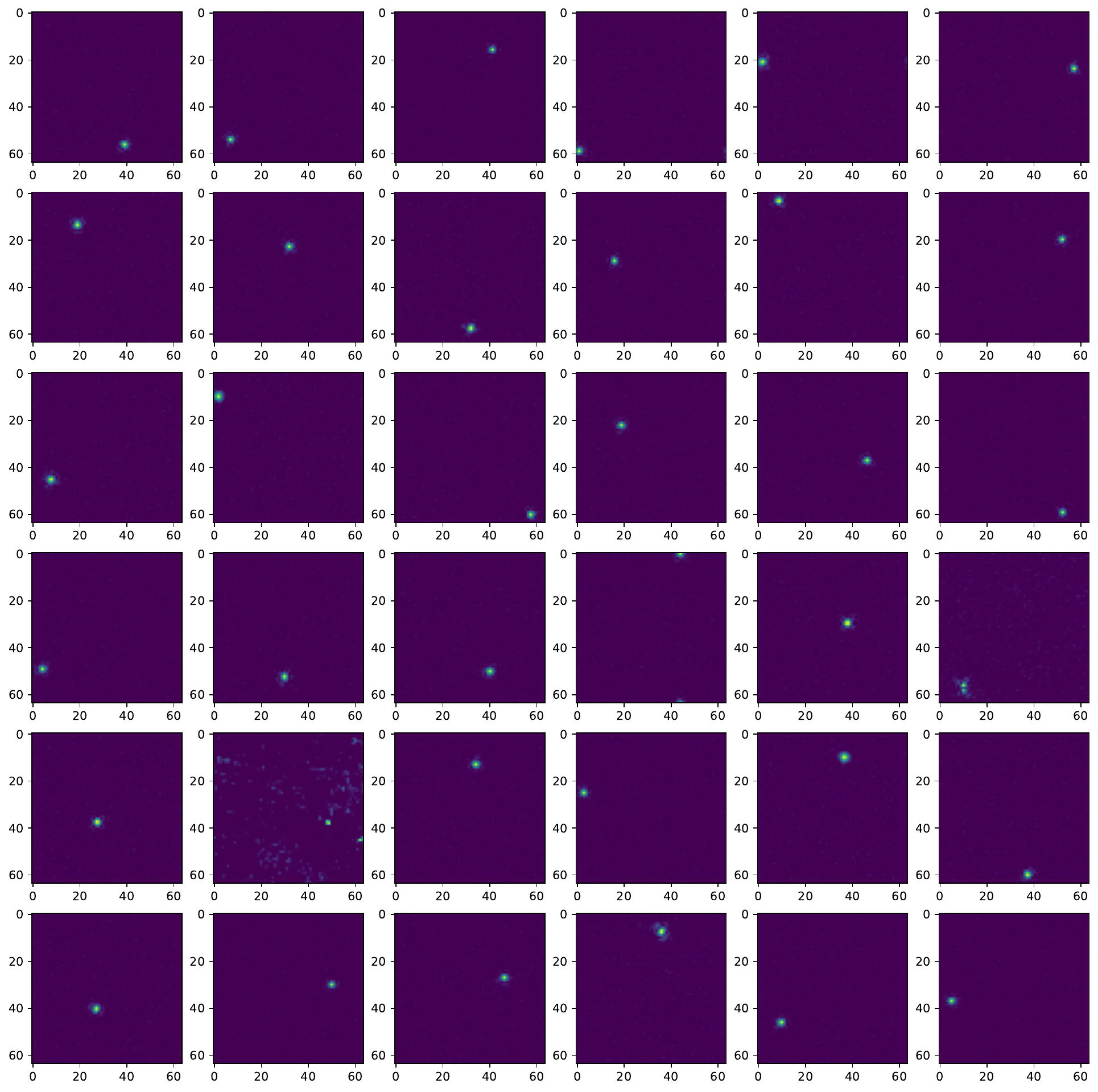}
    \caption{Layer 1 receptive field plot for $\beta = .405$.}
    \label{fig:rec_1_.405}
\end{figure}

\begin{figure}[h!]
    \centering
    \includegraphics[scale=.3]{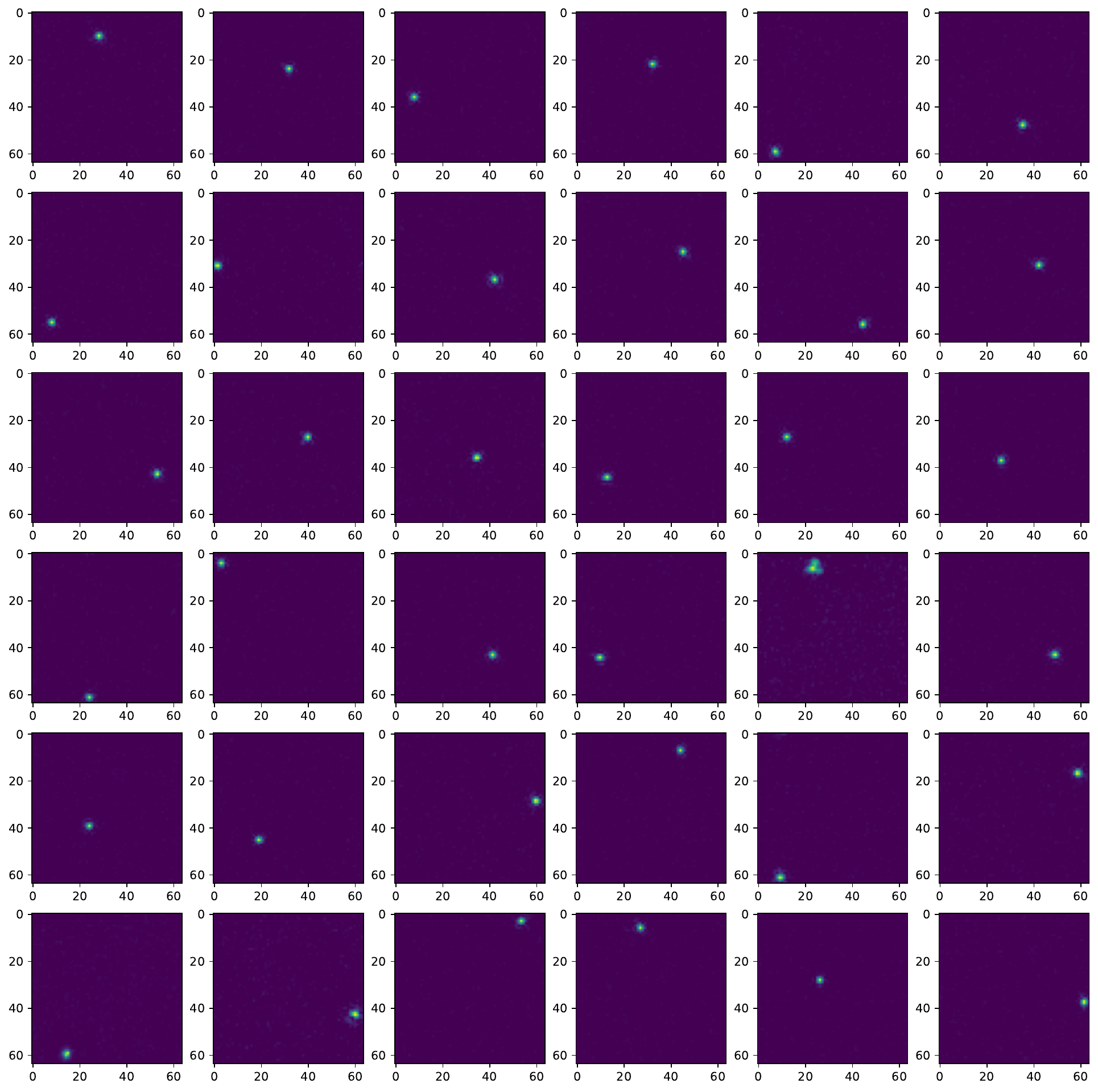}
    \caption{Layer 1 receptive field plot for $\beta = .41$.}
    \label{fig:rec_1_.41}
\end{figure}

\begin{figure}[h!]
    \centering
    \includegraphics[scale=.3]{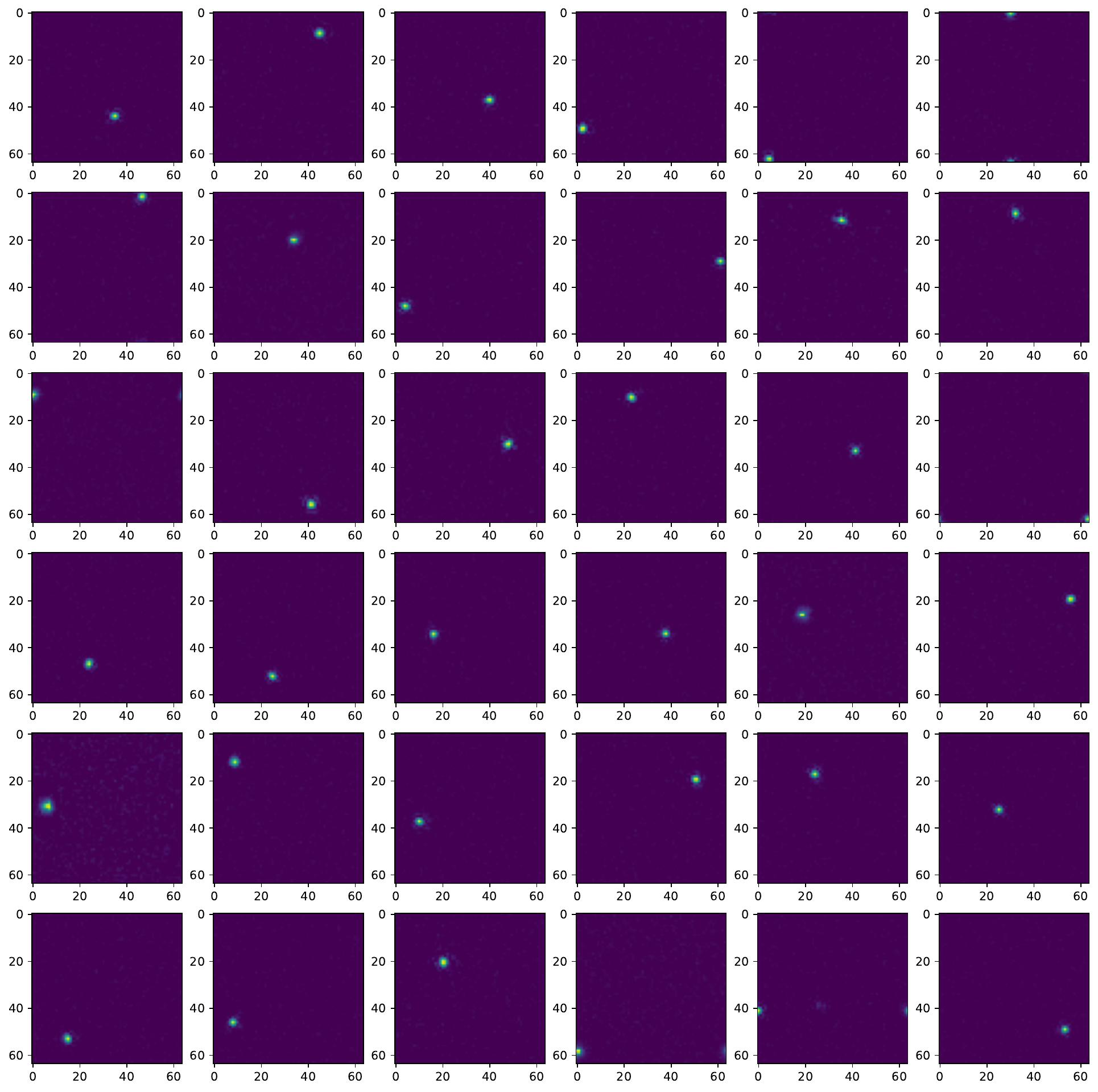}
    \caption{Layer 1 receptive field plot for $\beta = .415$.}
    \label{fig:rec_1_.415}
\end{figure}

\begin{figure}[h!]
    \centering
    \includegraphics[scale=.3]{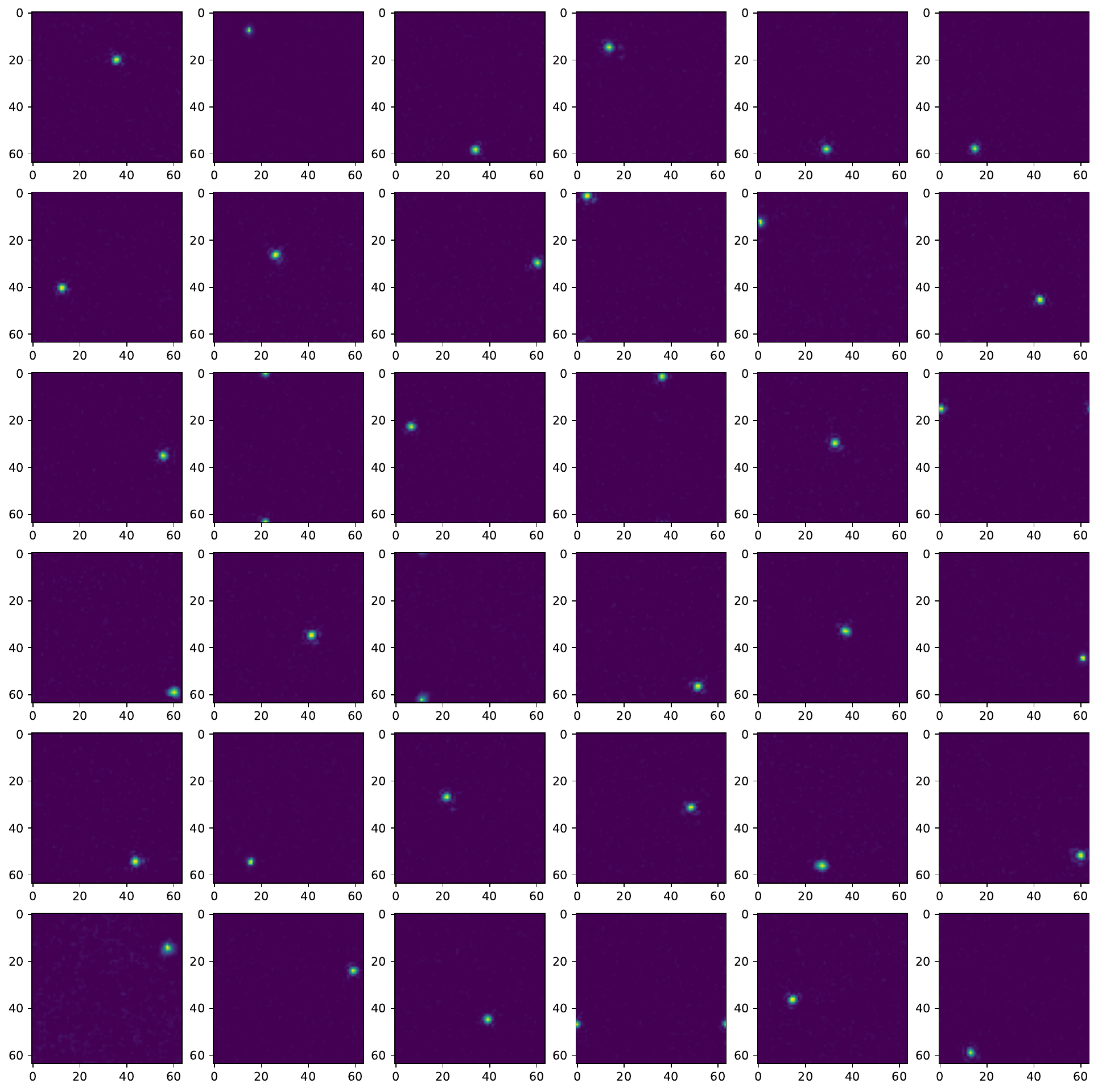}
    \caption{Layer 1 receptive field plot for $\beta = .42$.}
    \label{fig:rec_1_.42}
\end{figure}

\begin{figure}[h!]
    \centering
    \includegraphics[scale=.3]{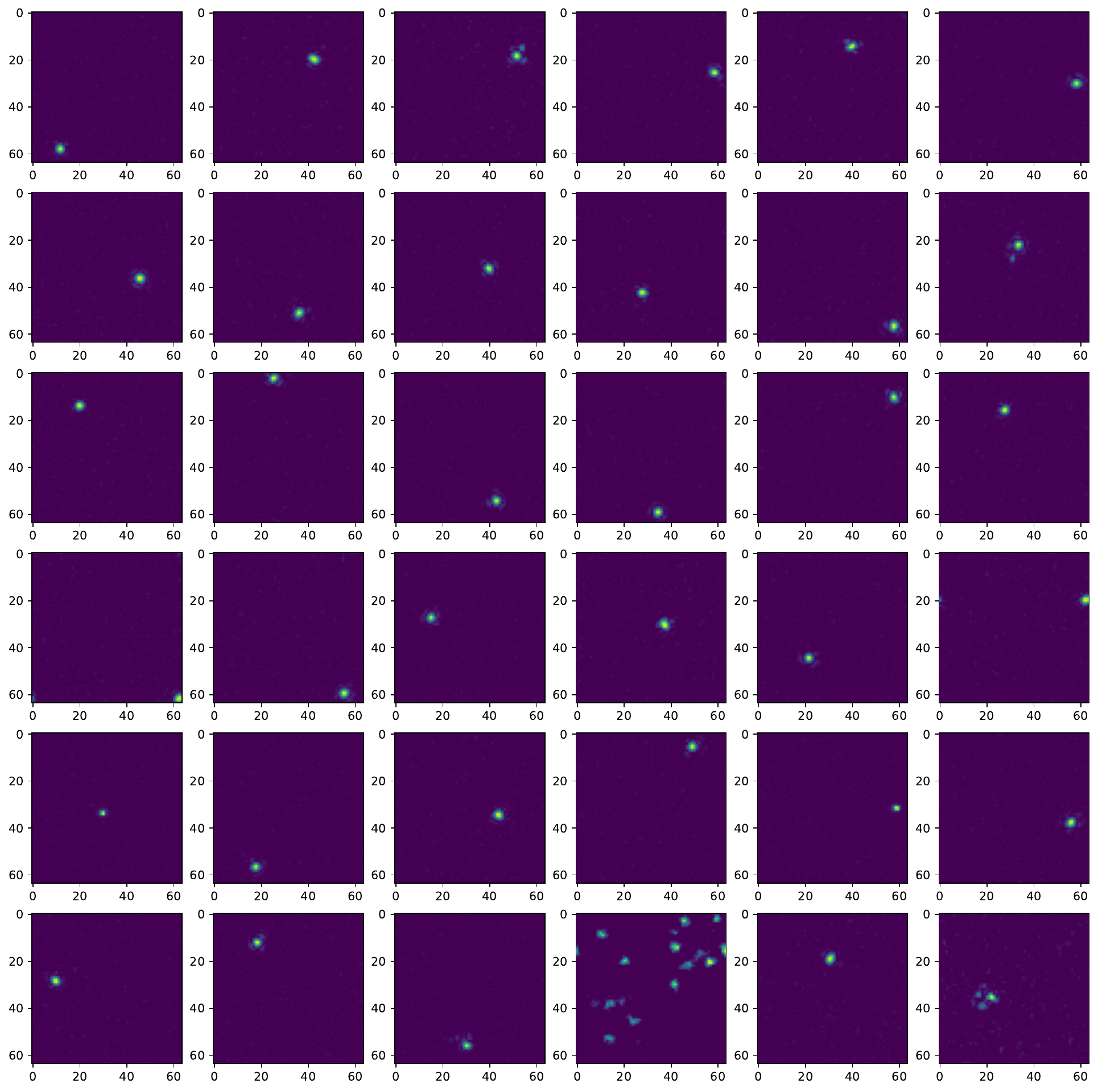}
    \caption{Layer 1 receptive field plot for $\beta = .425$.}
    \label{fig:rec_1_.425}
\end{figure}

\begin{figure}[h!]
    \centering
    \includegraphics[scale=.25]{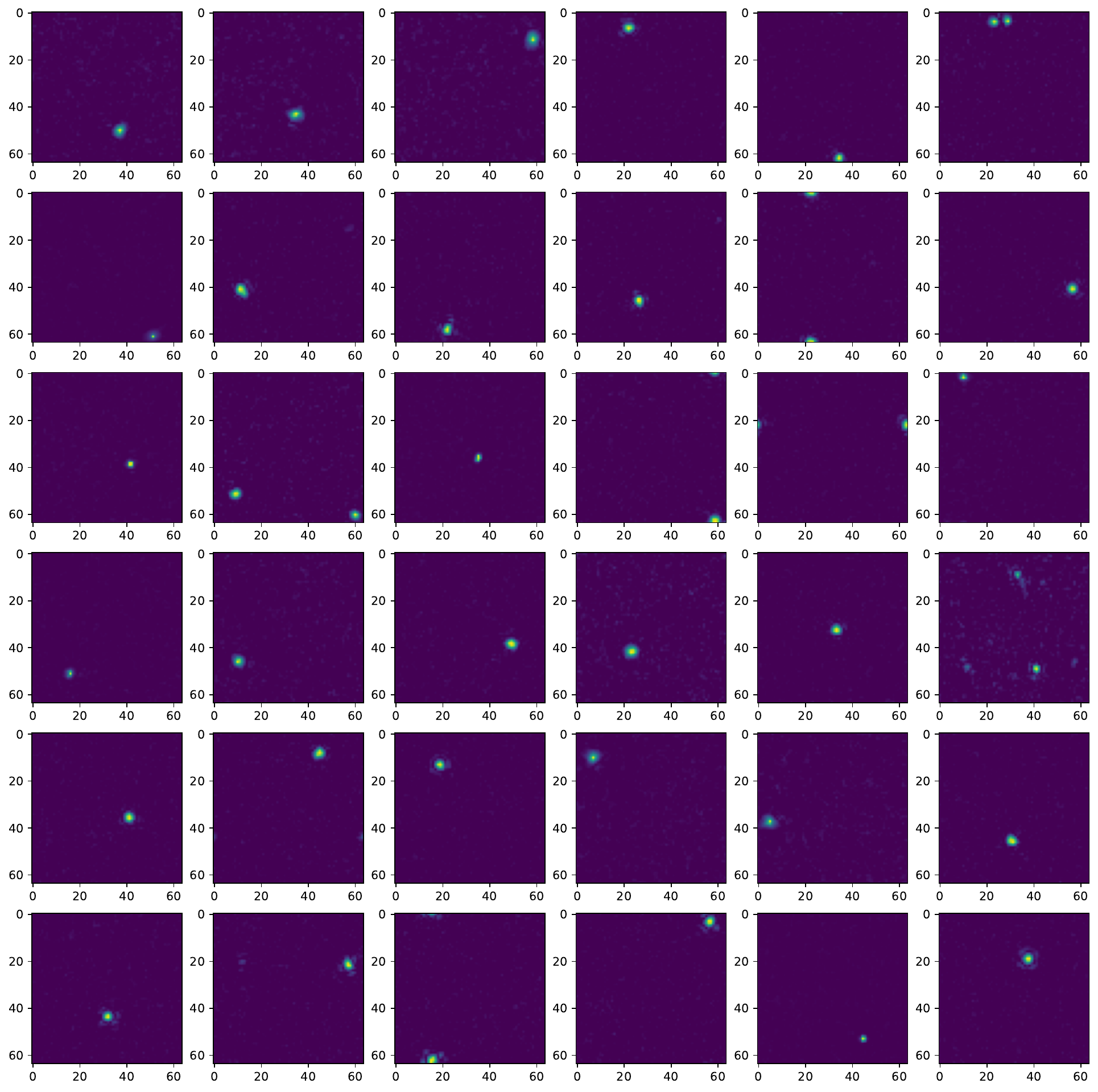}
    \caption{Layer 1 receptive field plot for $\beta = .43$.}
    \label{fig:rec_1_.43}
\end{figure}

\begin{figure}[h!]
    \centering
    \includegraphics[scale=.25]{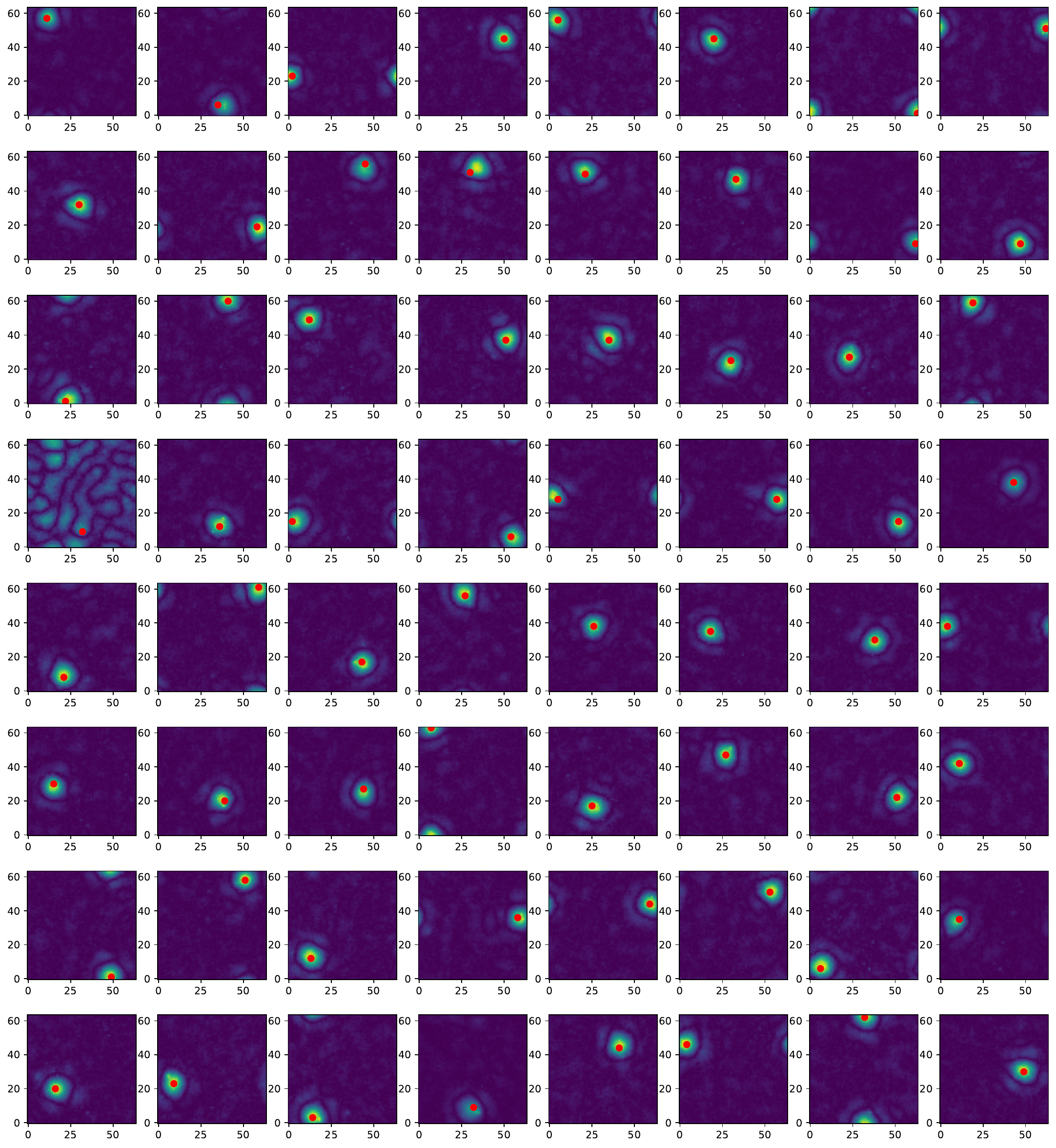}
    \caption{Layer 3 receptive field plot for $\beta = .395$.}
    \label{fig:rec_3_.395}
\end{figure}

\begin{figure}[h!]
    \centering
    \includegraphics[scale=.25]{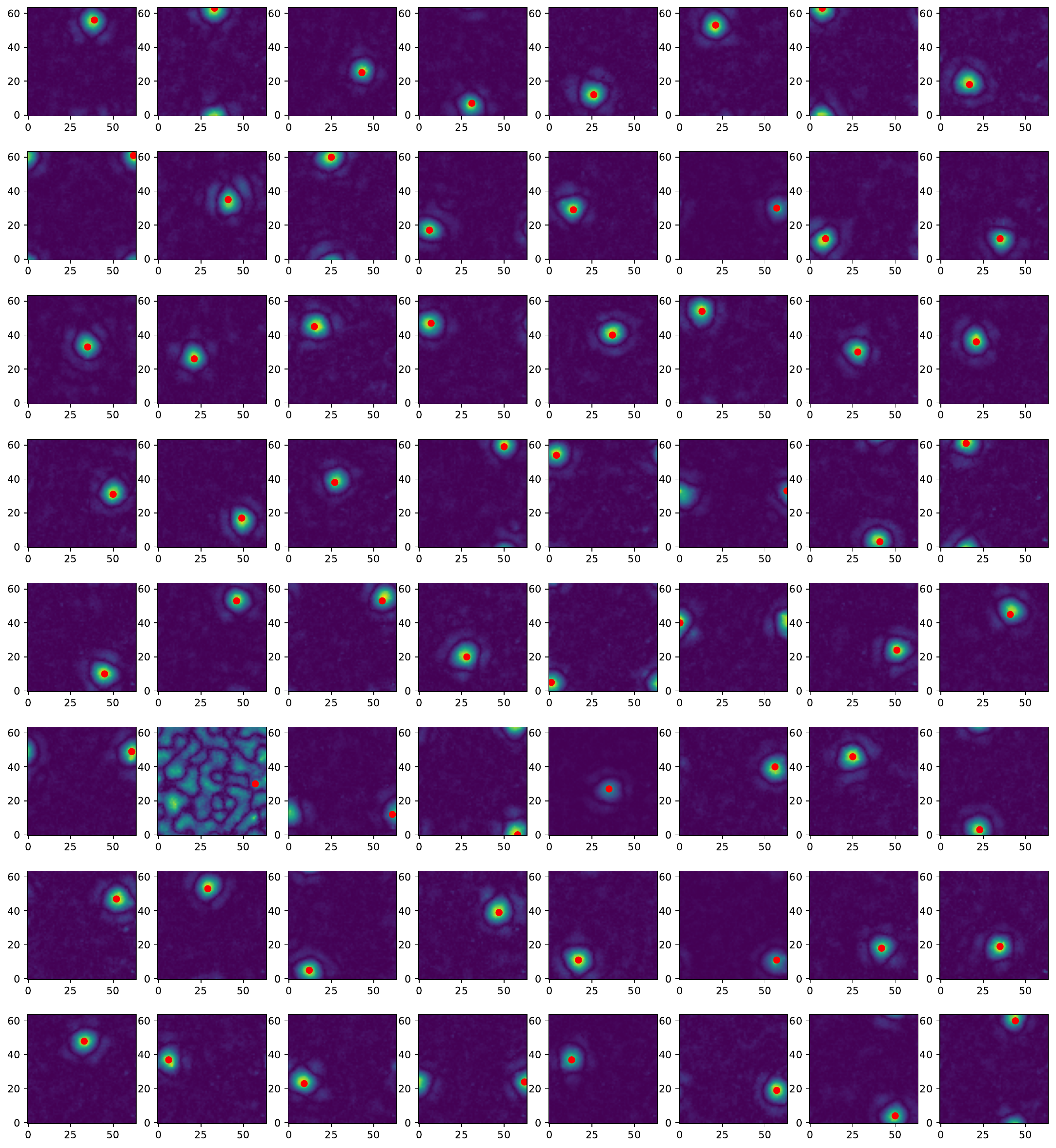}
    \caption{Layer 3 receptive field plot for $\beta = .4$.}
    \label{fig:rec_3_.4}
\end{figure}

\begin{figure}[h!]
    \centering
    \includegraphics[scale=.25]{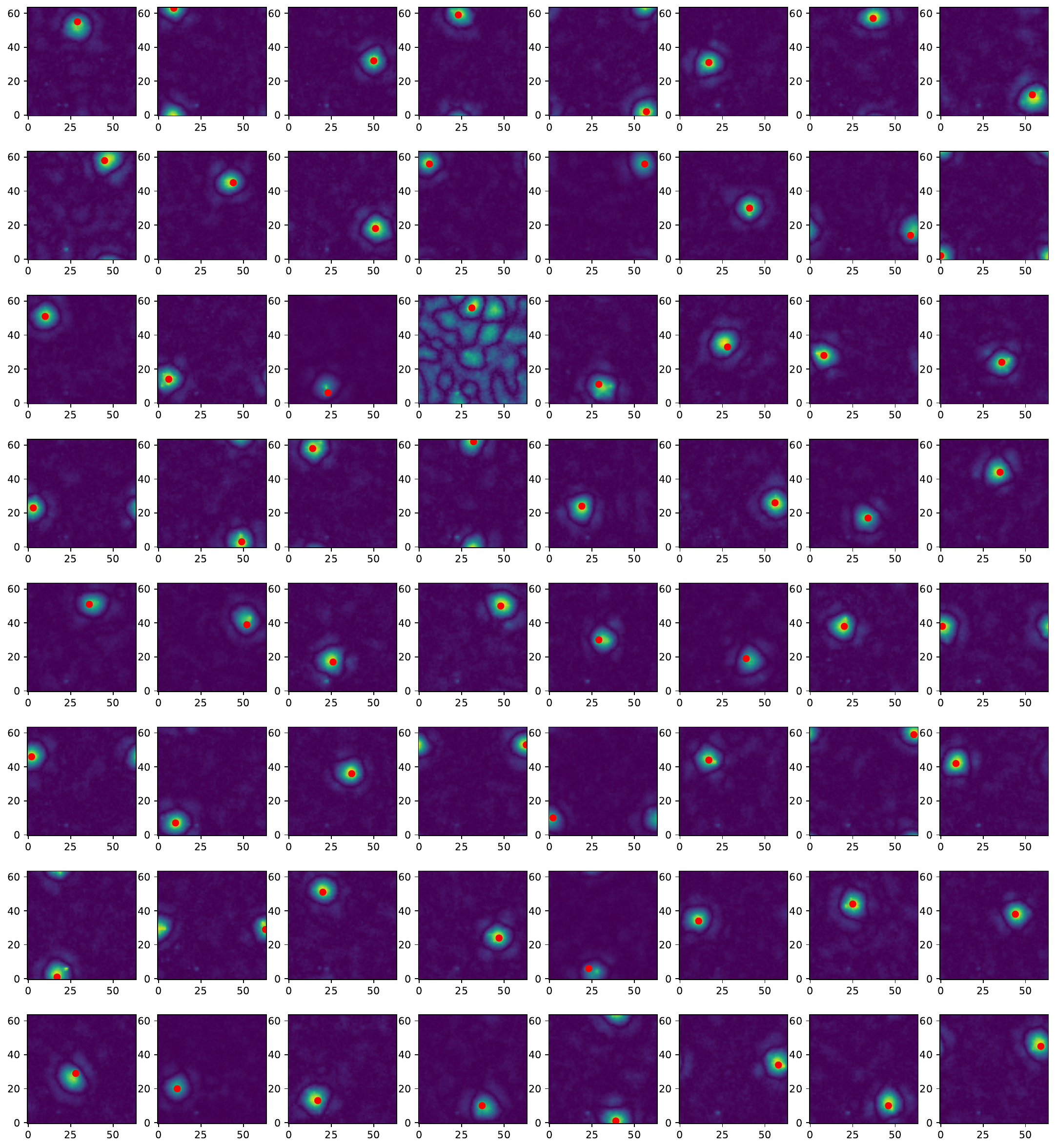}
    \caption{Layer 3 receptive field plot for $\beta = .405$.}
    \label{fig:rec_3_.405}
\end{figure}

\begin{figure}[h!]
    \centering
    \includegraphics[scale=.25]{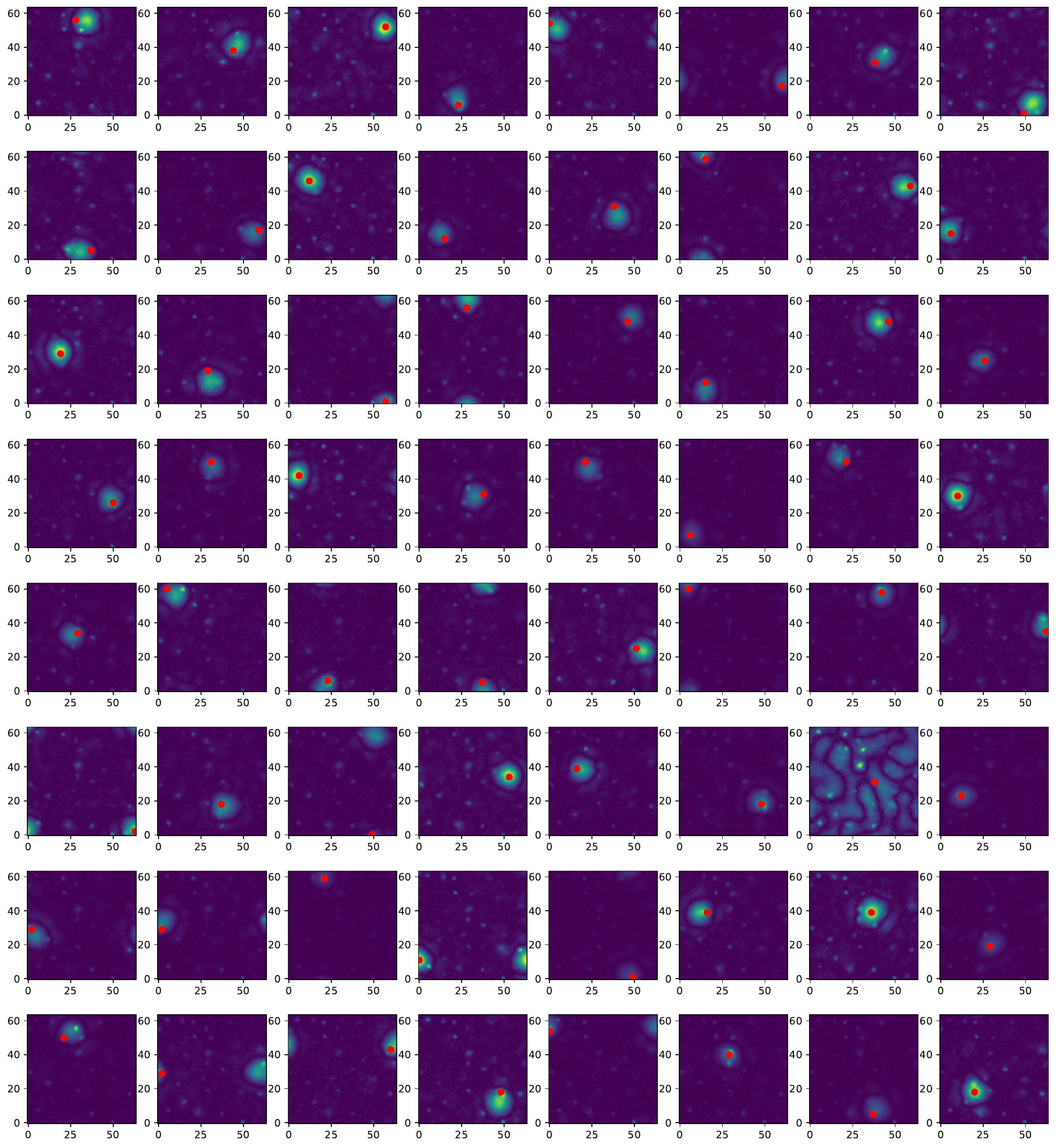}
    \caption{Layer 3 receptive field plot for $\beta = .41$.}
    \label{fig:rec_3_.41}
\end{figure}

\begin{figure}[h!]
    \centering
    \includegraphics[scale=.25]{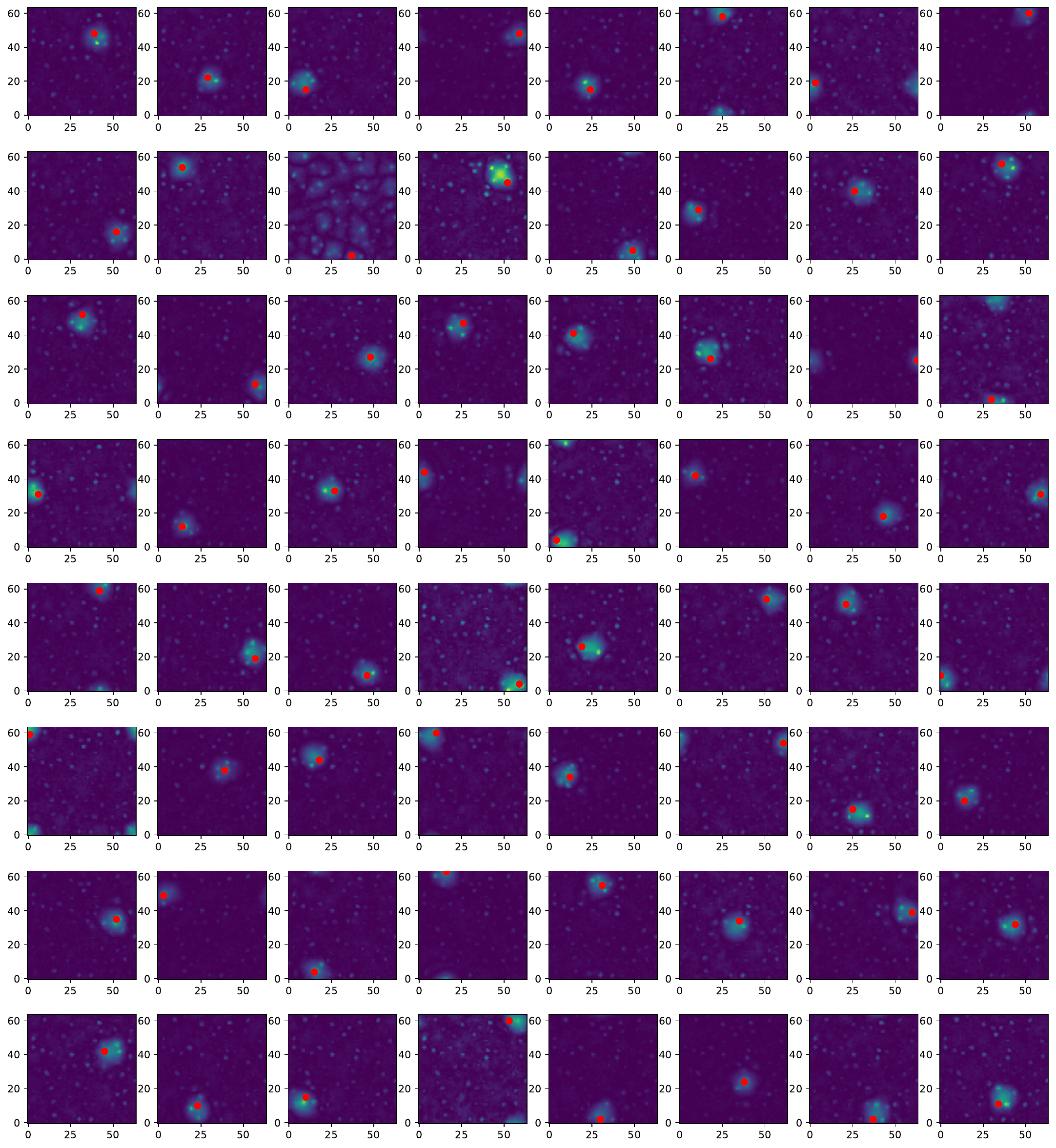}
    \caption{Layer 3 receptive field plot for $\beta = .415$.}
    \label{fig:rec_3_.415}
\end{figure}

\begin{figure}[h!]
    \centering
    \includegraphics[scale=.25]{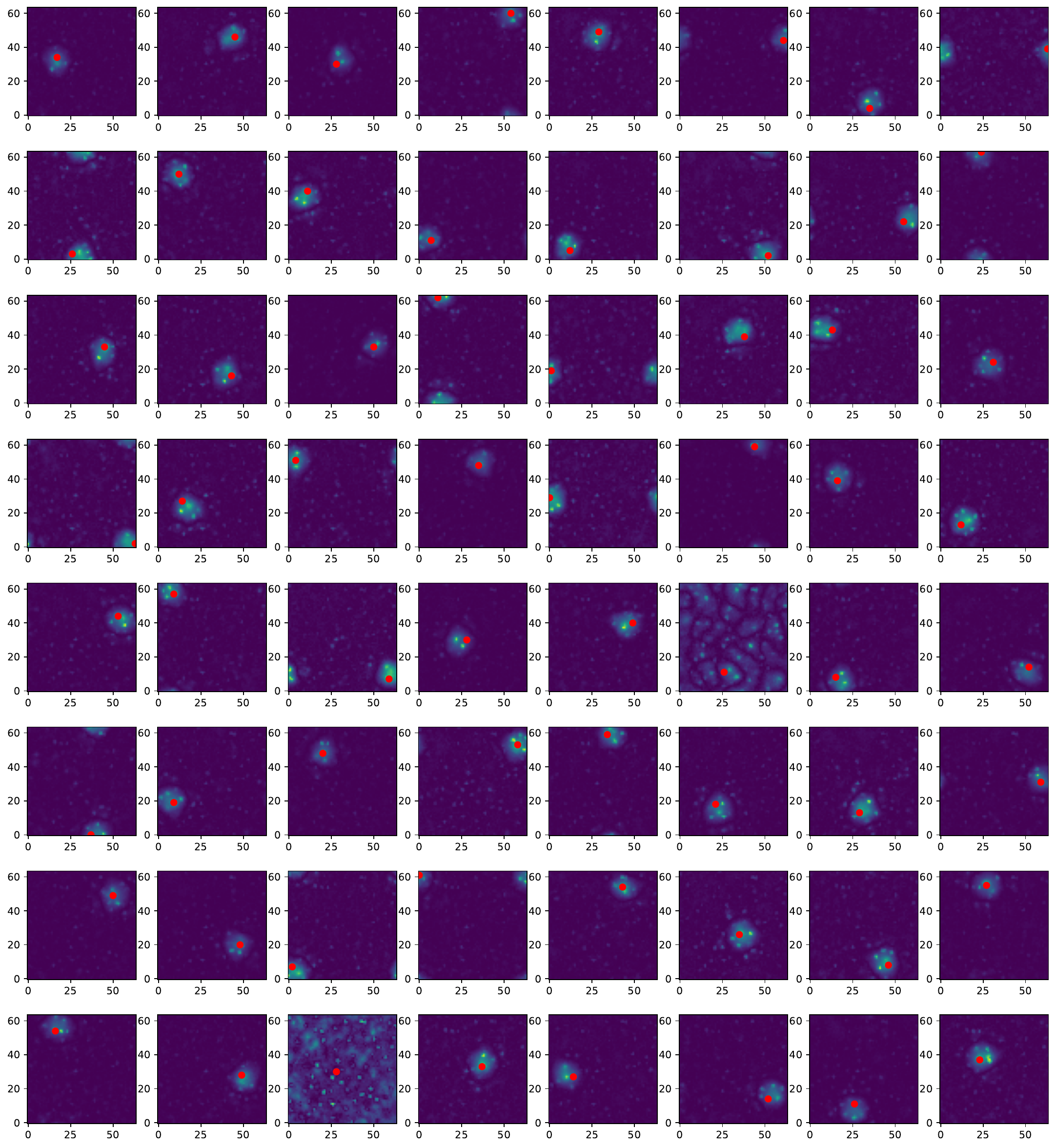}
    \caption{Layer 3 receptive field plot for $\beta = .42$.}
    \label{fig:rec_3_.42}
\end{figure}

\begin{figure}[h!]
    \centering
    \includegraphics[scale=.25]{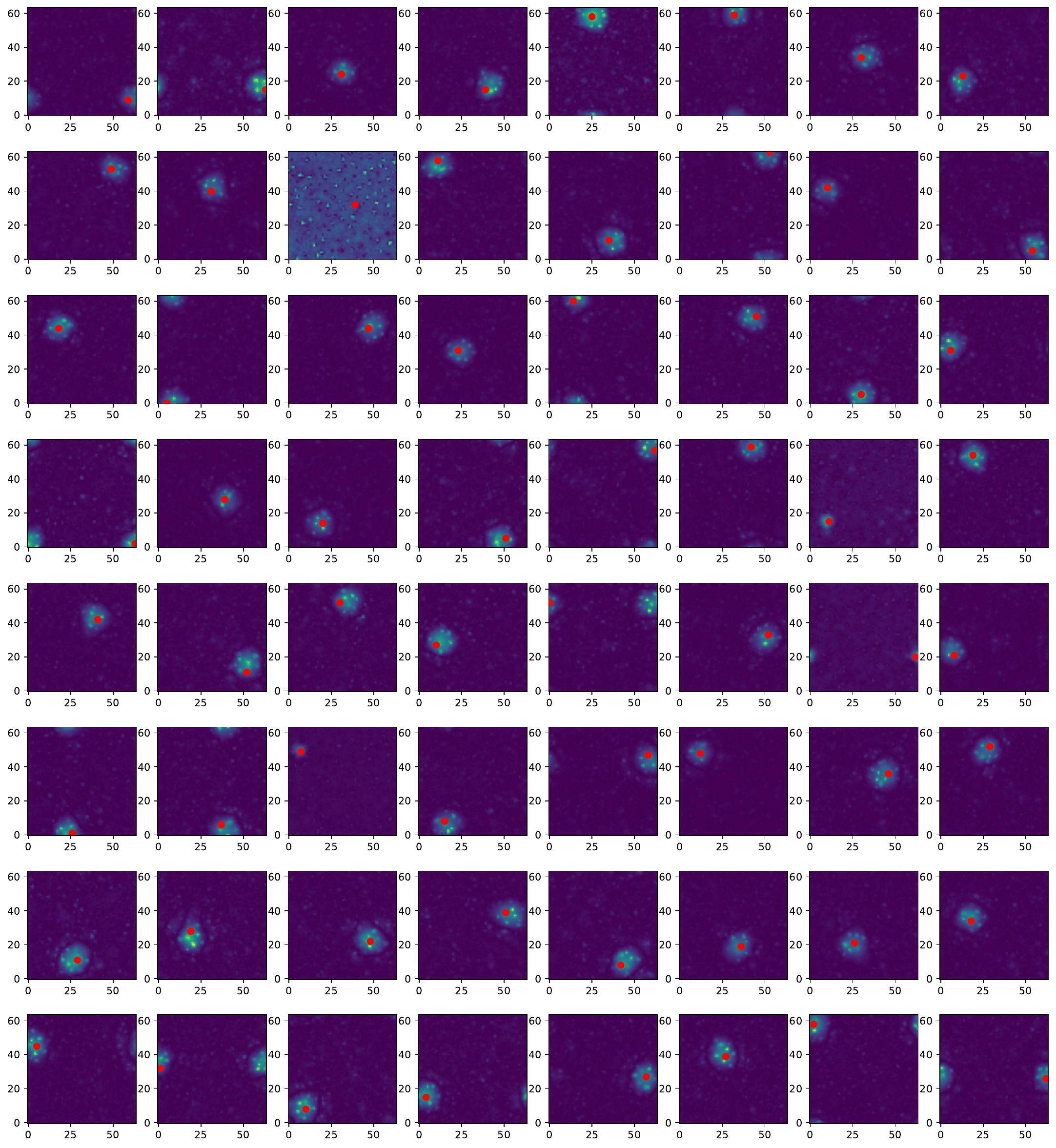}
    \caption{Layer 3 receptive field plot for $\beta = .425$.}
    \label{fig:rec_3_.425}
\end{figure}

\begin{figure}[h!]
    \centering
    \includegraphics[scale=.25]{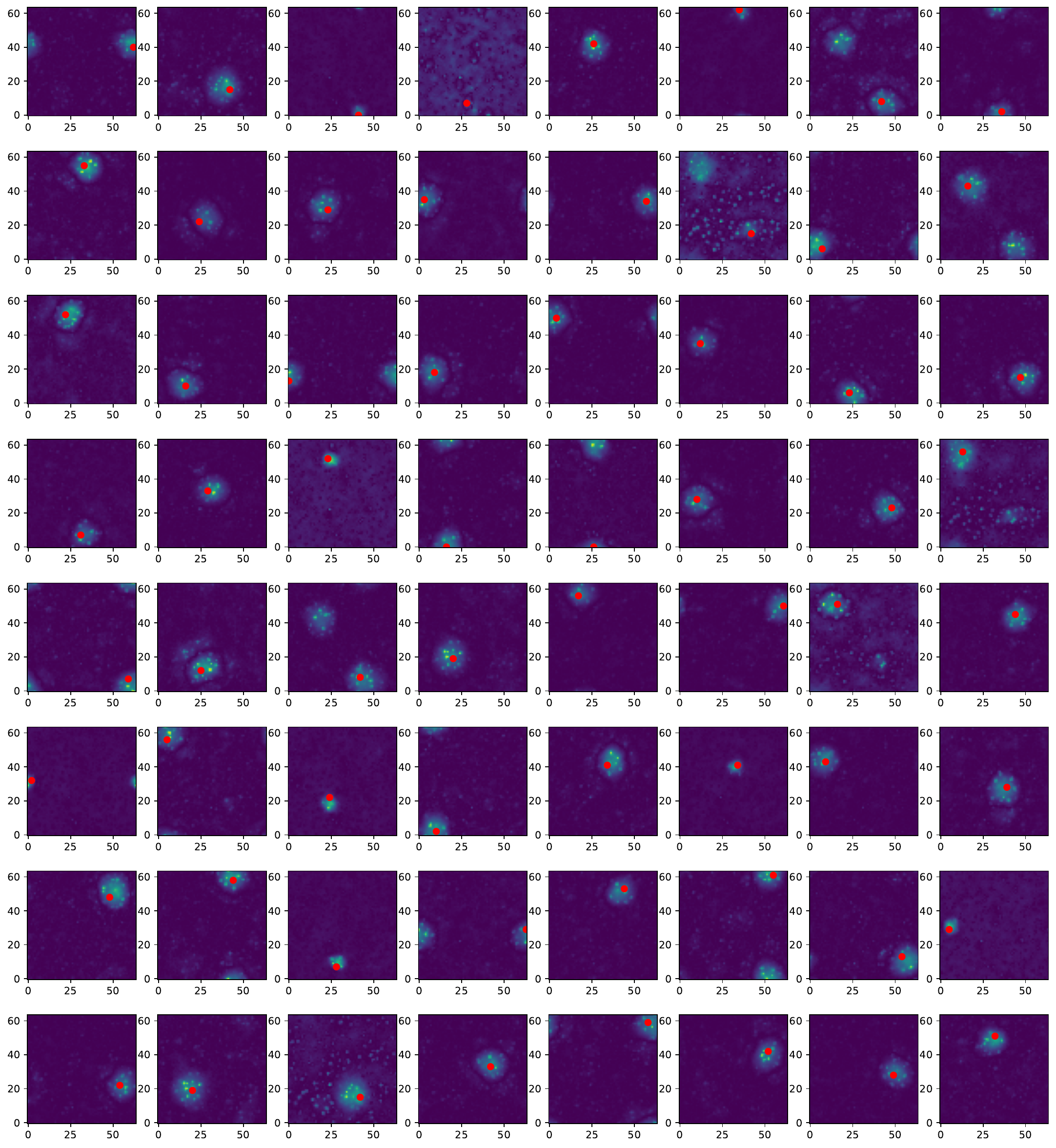}
    \caption{Layer 3 receptive field plot for $\beta = .43$.}
    \label{fig:rec_3_.43}
\end{figure}
\clearpage

\section{Weight Analysis Plots}
\label{sec:weight_plots}
The plots starting on the next page analyze the weight structure for each value of $\beta$, analyzing both the weight tensors and receptive field tensors. Each tensor has two plots. The first is the average number of trained positive/negative weights with a magnitude above a given value connected to a given spin location in an input lattice. The second is the average number of spin locations connected to each weight with magnitudes above a given value, such that all locations connect to at least two weights. More information about this analysis and its results are in Section \ref{sec:weight_structures}.

\clearpage

\begin{figure}[h!]
     \centering
     \begin{subfigure}{0.32\textwidth}
         \centering
         \includegraphics[width=\textwidth]{av_weights_1__395.pdf}
         \caption{Layer 1 Weights}
     \end{subfigure}
     \hfill
     \begin{subfigure}{0.32\textwidth}
         \centering
         \includegraphics[width=\textwidth]{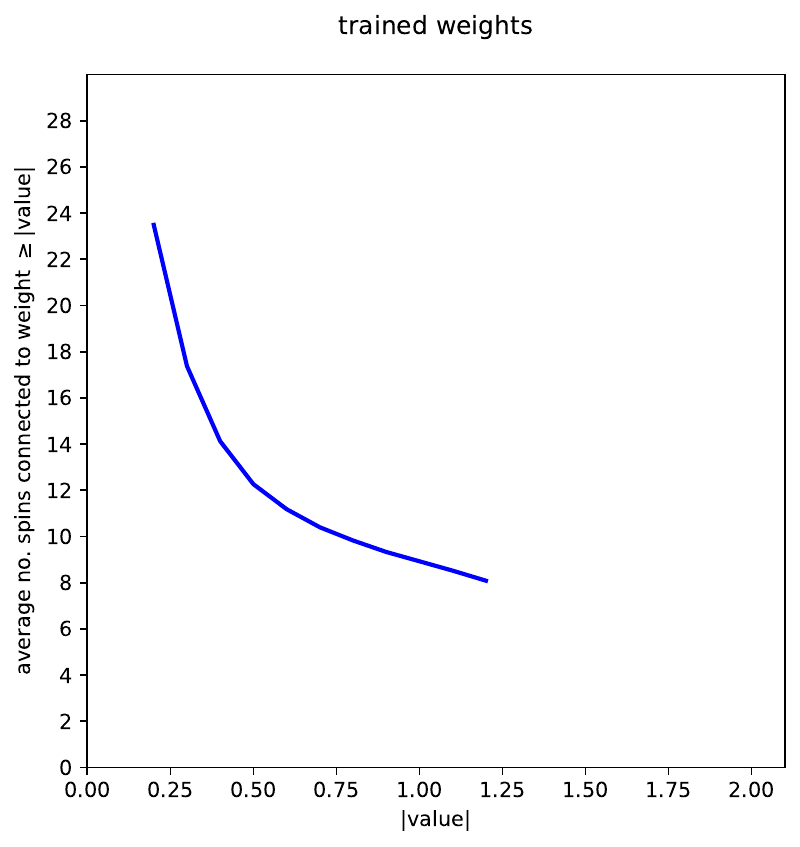}
         \caption{Layer 1 Weights}
     \end{subfigure}
     \hfill
     \begin{subfigure}{0.32\textwidth}
         \centering
         \includegraphics[width=\textwidth]{av_weights_2w__395.pdf}
         \caption{Layer 2 Weights}
     \end{subfigure}
     \hfill
     \begin{subfigure}{0.32\textwidth}
         \centering
         \includegraphics[width=\textwidth]{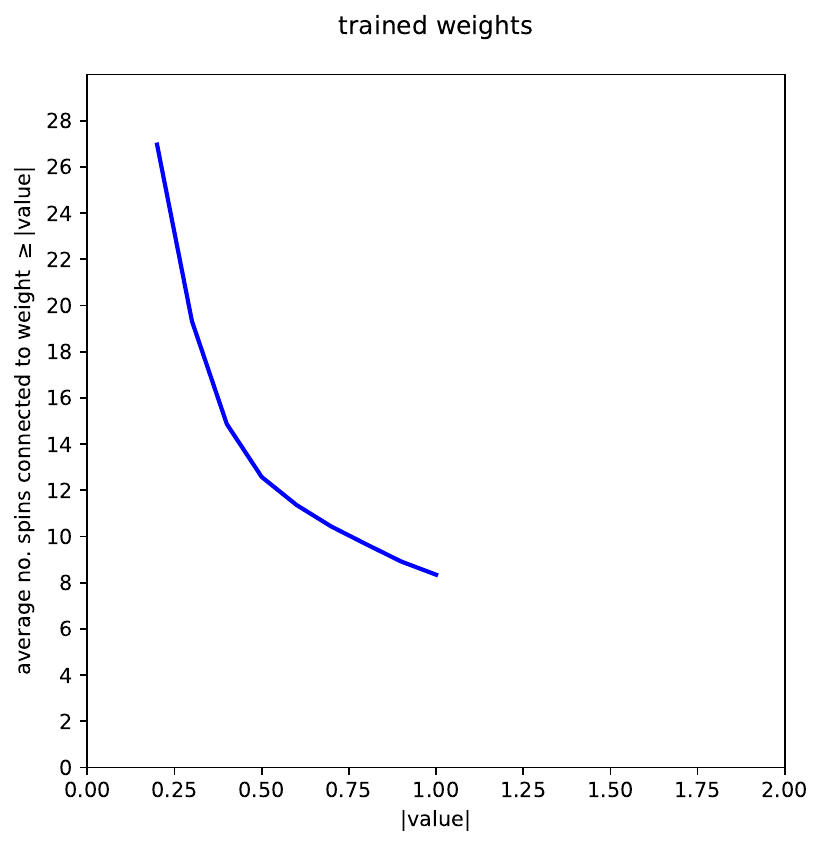}
         \caption{Layer 2 Weights}
     \end{subfigure}
          \begin{subfigure}{0.32\textwidth}
         \centering
         \includegraphics[width=\textwidth]{av_weights_2r__395.pdf}
         \caption{Layer 2 Receptive}
     \end{subfigure}
     \hfill
     \begin{subfigure}{0.32\textwidth}
         \centering
         \includegraphics[width=\textwidth]{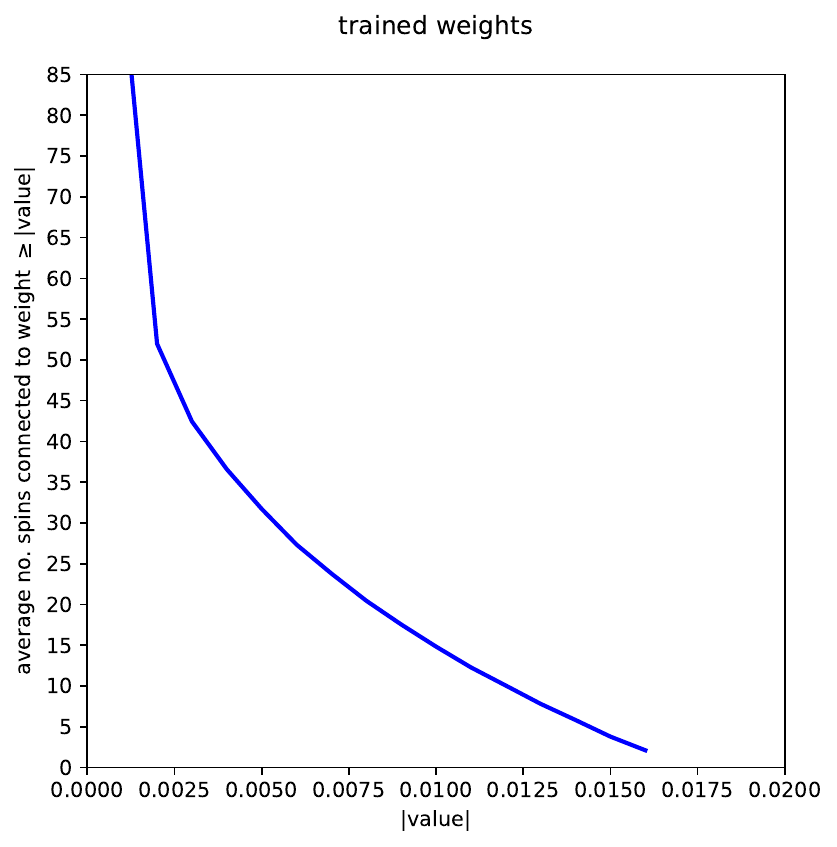}
         \caption{Layer 2 Receptive}
     \end{subfigure}
    \begin{subfigure}{0.32\textwidth}
         \centering
         \includegraphics[width=\textwidth]{av_weights_3w__395.pdf}
         \caption{Layer 3 Weights}
     \end{subfigure}
           \begin{subfigure}{0.32\textwidth}
         \centering
         \includegraphics[width=\textwidth]{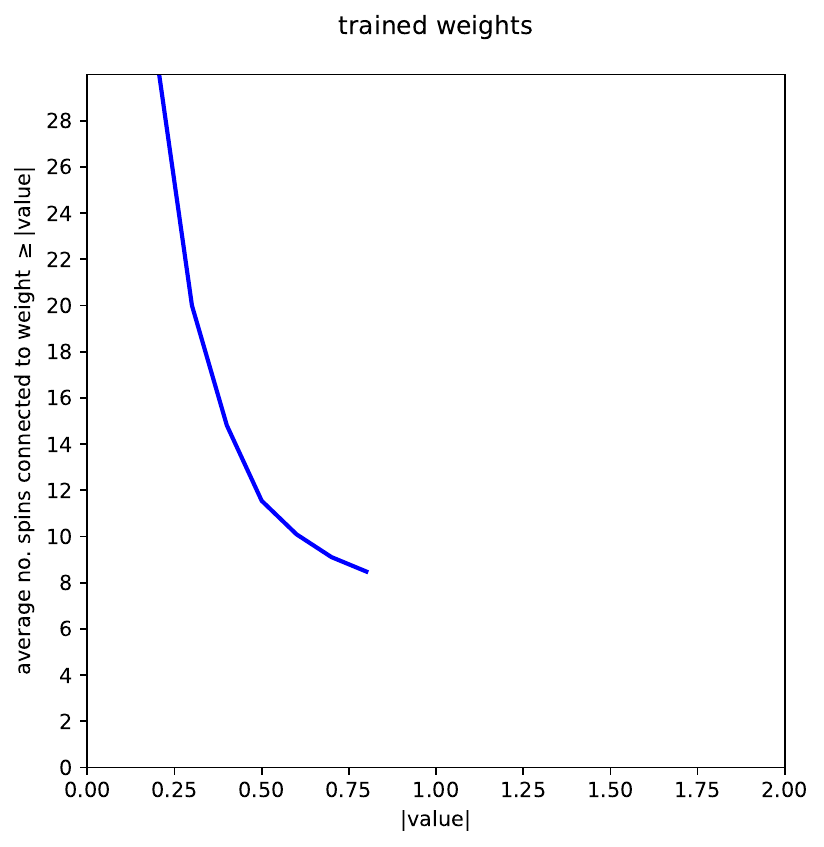}
         \caption{Layer 3 Weights}
     \end{subfigure}
     \hfill
         \begin{subfigure}{0.32\textwidth}
         \centering
         \includegraphics[width=\textwidth]{av_weights_3r__395.pdf}
         \caption{Layer 3 Receptive}
     \end{subfigure}
           \begin{subfigure}{0.32\textwidth}
         \centering
         \includegraphics[width=\textwidth]{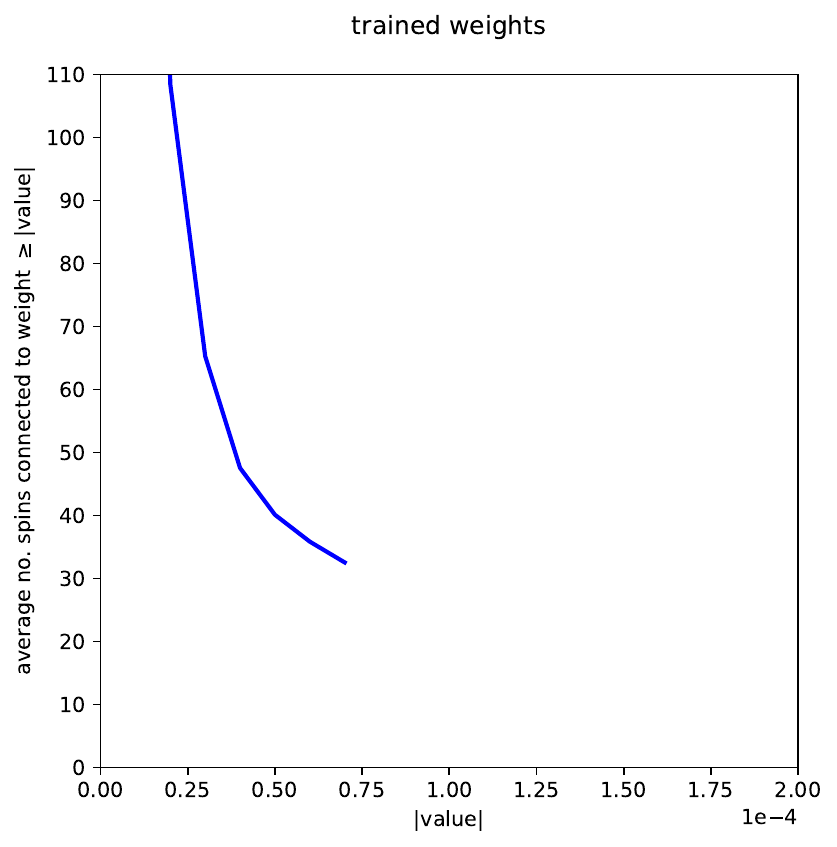}
         \caption{Layer 3 Receptive}
     \end{subfigure}
        \caption{Weight analysis plot for $\beta = .395.$}
        \label{fig:weight_analysis_.395}
\end{figure}

\begin{figure}[h!]
     \centering
     \begin{subfigure}{0.32\textwidth}
         \centering
         \includegraphics[width=\textwidth]{av_weights_1__4.pdf}
         \caption{Layer 1 Weights}
     \end{subfigure}
     \hfill
     \begin{subfigure}{0.32\textwidth}
         \centering
         \includegraphics[width=\textwidth]{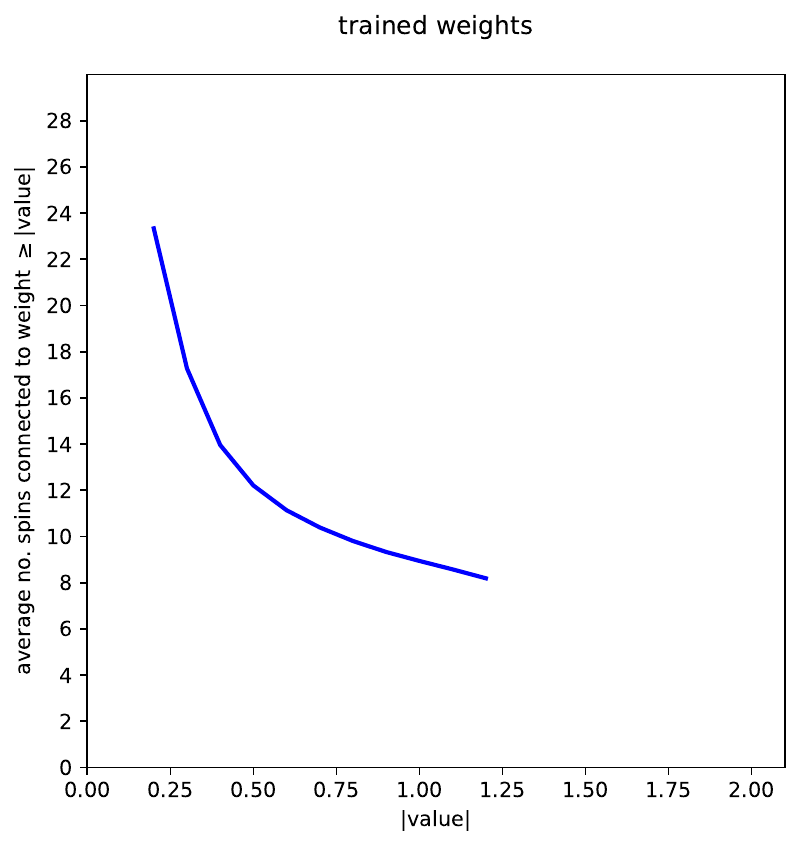}
         \caption{Layer 1 Weights}
     \end{subfigure}
     \hfill
     \begin{subfigure}{0.32\textwidth}
         \centering
         \includegraphics[width=\textwidth]{av_weights_2w__4.pdf}
         \caption{Layer 2 Weights}
     \end{subfigure}
     \hfill
     \begin{subfigure}{0.32\textwidth}
         \centering
         \includegraphics[width=\textwidth]{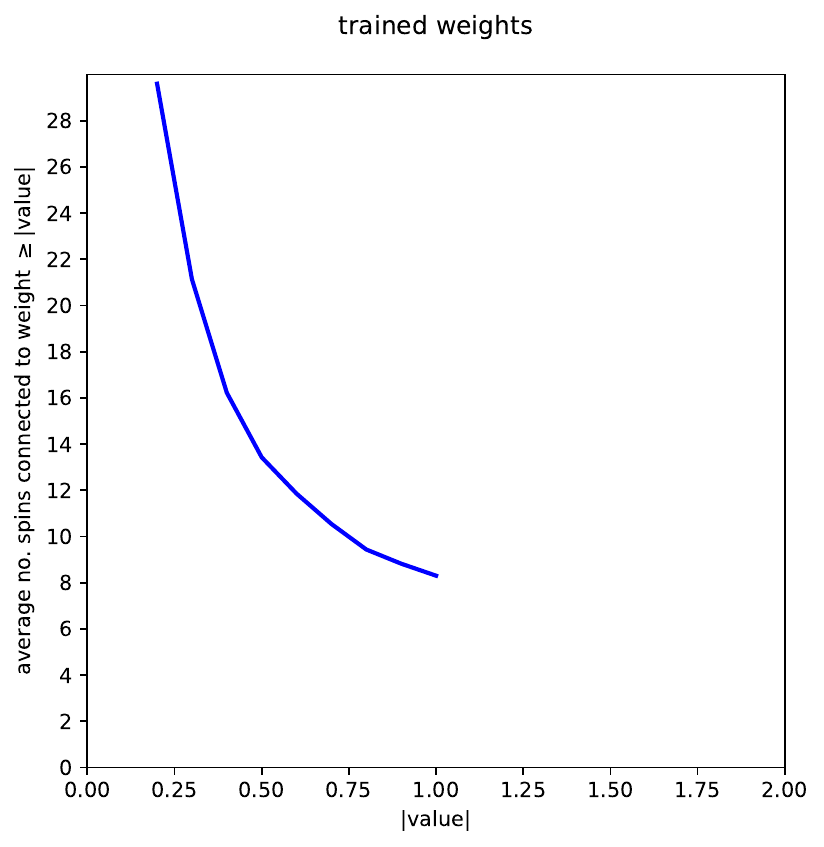}
         \caption{Layer 2 Weights}
     \end{subfigure}
          \begin{subfigure}{0.32\textwidth}
         \centering
         \includegraphics[width=\textwidth]{av_weights_2r__4.pdf}
         \caption{Layer 2 Receptive}
     \end{subfigure}
     \hfill
     \begin{subfigure}{0.32\textwidth}
         \centering
         \includegraphics[width=\textwidth]{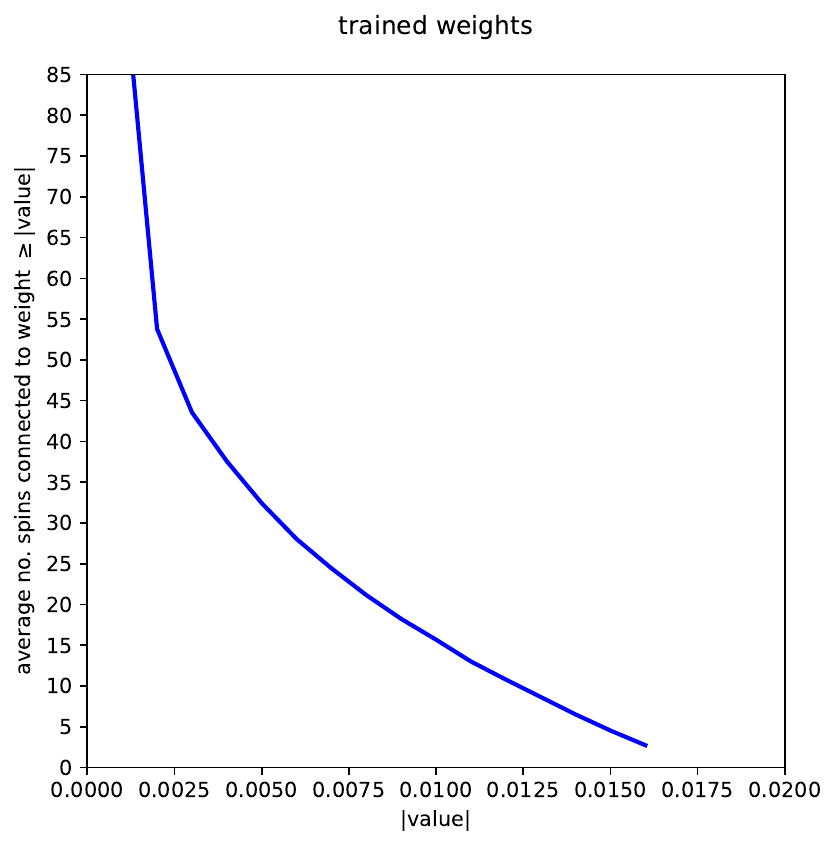}
         \caption{Layer 2 Receptive}
     \end{subfigure}
    \begin{subfigure}{0.32\textwidth}
         \centering
         \includegraphics[width=\textwidth]{av_weights_3w__4.pdf}
         \caption{Layer 3 Weights}
     \end{subfigure}
           \begin{subfigure}{0.32\textwidth}
         \centering
         \includegraphics[width=\textwidth]{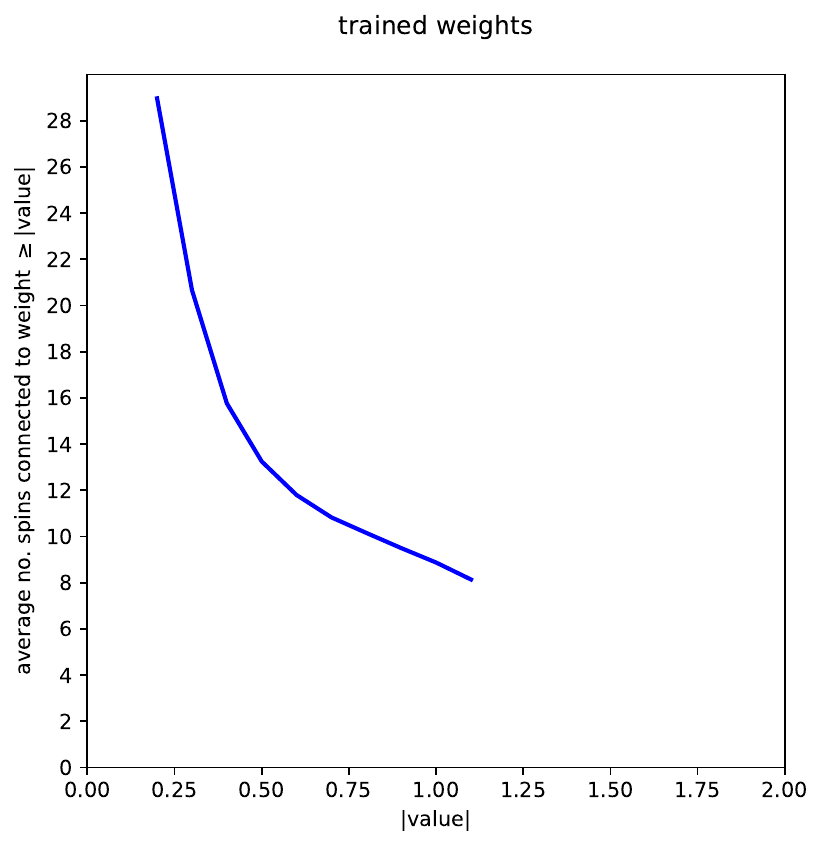}
         \caption{Layer 3 Weights}
     \end{subfigure}
     \hfill
         \begin{subfigure}{0.32\textwidth}
         \centering
         \includegraphics[width=\textwidth]{av_weights_3r__4.pdf}
         \caption{Layer 3 Receptive}
     \end{subfigure}
           \begin{subfigure}{0.32\textwidth}
         \centering
         \includegraphics[width=\textwidth]{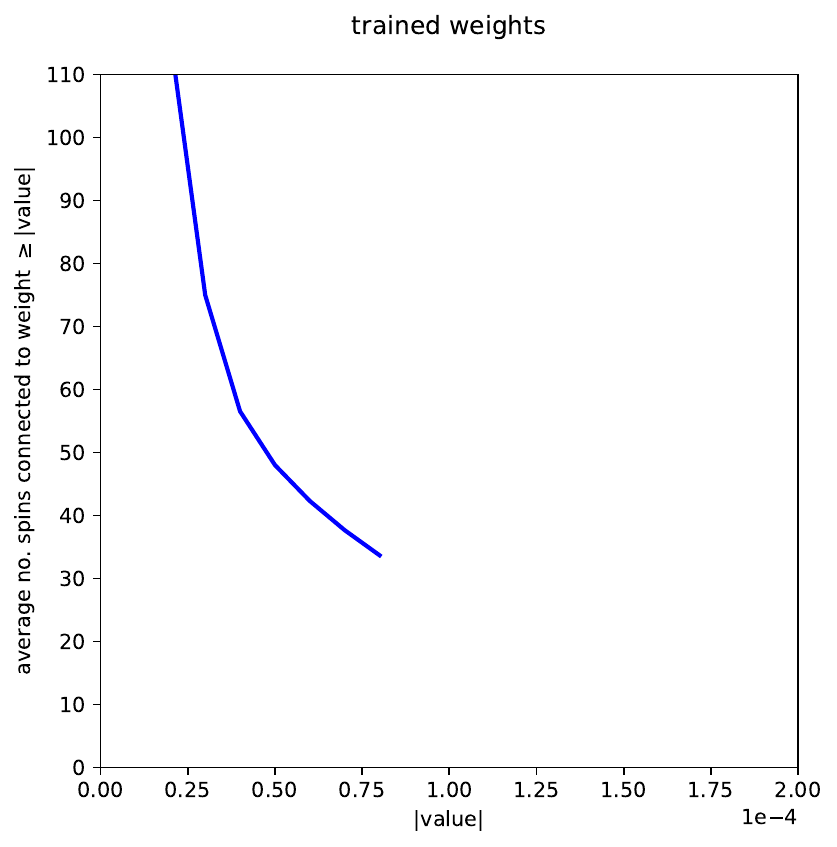}
         \caption{Layer 3 Receptive}
     \end{subfigure}
        \caption{Weight analysis plot for $\beta = .4.$}
        \label{fig:weight_analysis_.4}
\end{figure}

\begin{figure}[h!]
     \centering
     \begin{subfigure}{0.32\textwidth}
         \centering
         \includegraphics[width=\textwidth]{av_weights_1__405.pdf}
         \caption{Layer 1 Weights}
     \end{subfigure}
     \hfill
     \begin{subfigure}{0.32\textwidth}
         \centering
         \includegraphics[width=\textwidth]{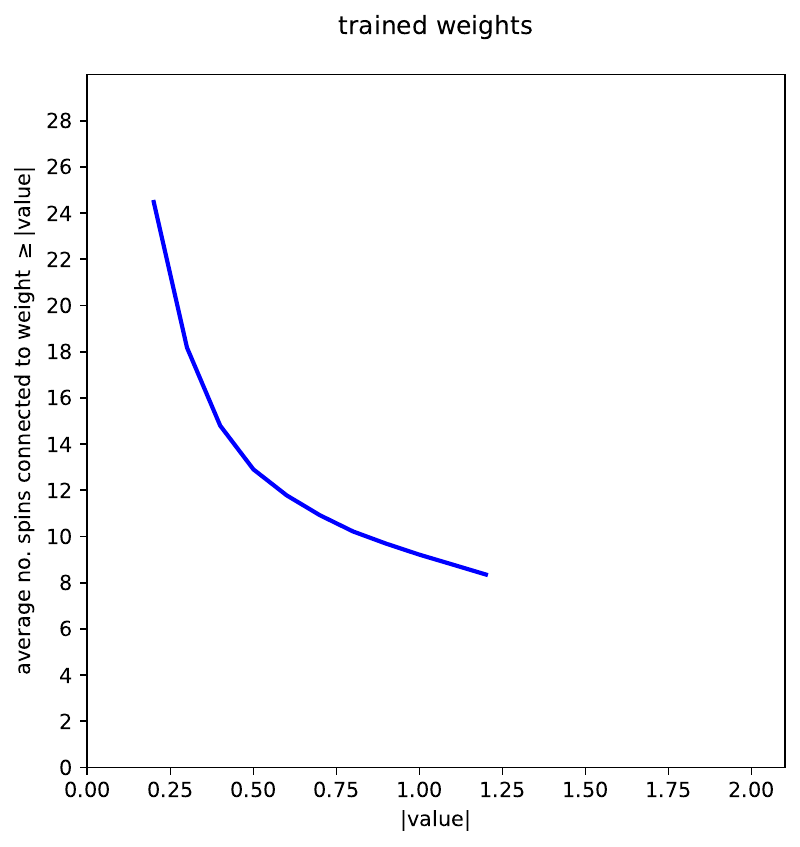}
         \caption{Layer 1 Weights}
     \end{subfigure}
     \hfill
     \begin{subfigure}{0.32\textwidth}
         \centering
         \includegraphics[width=\textwidth]{av_weights_2w__405.pdf}
         \caption{Layer 2 Weights}
     \end{subfigure}
     \hfill
     \begin{subfigure}{0.32\textwidth}
         \centering
         \includegraphics[width=\textwidth]{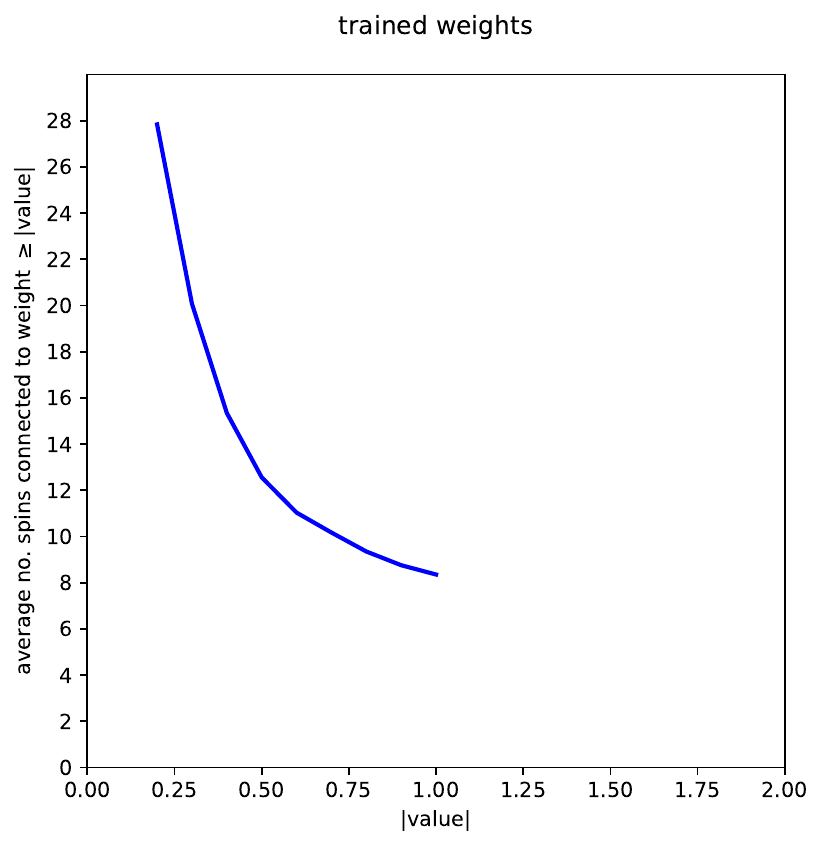}
         \caption{Layer 2 Weights}
     \end{subfigure}
          \begin{subfigure}{0.32\textwidth}
         \centering
         \includegraphics[width=\textwidth]{av_weights_2r__405.pdf}
         \caption{Layer 2 Receptive}
     \end{subfigure}
     \hfill
     \begin{subfigure}{0.32\textwidth}
         \centering
         \includegraphics[width=\textwidth]{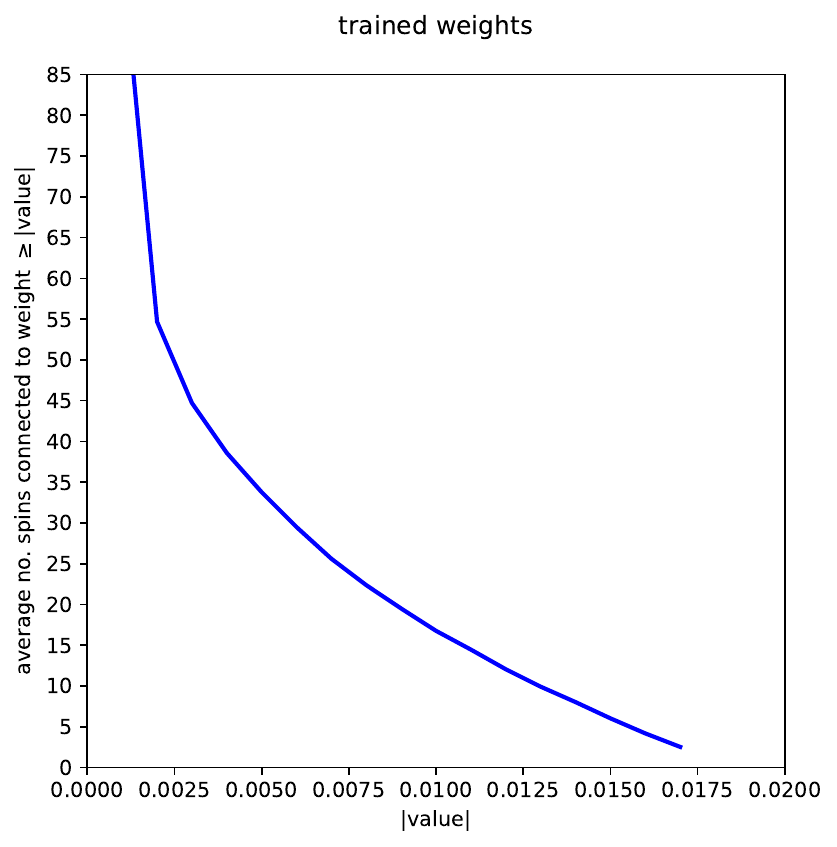}
         \caption{Layer 2 Receptive}
     \end{subfigure}
    \begin{subfigure}{0.32\textwidth}
         \centering
         \includegraphics[width=\textwidth]{av_weights_3w__405.pdf}
         \caption{Layer 3 Weights}
     \end{subfigure}
           \begin{subfigure}{0.32\textwidth}
         \centering
         \includegraphics[width=\textwidth]{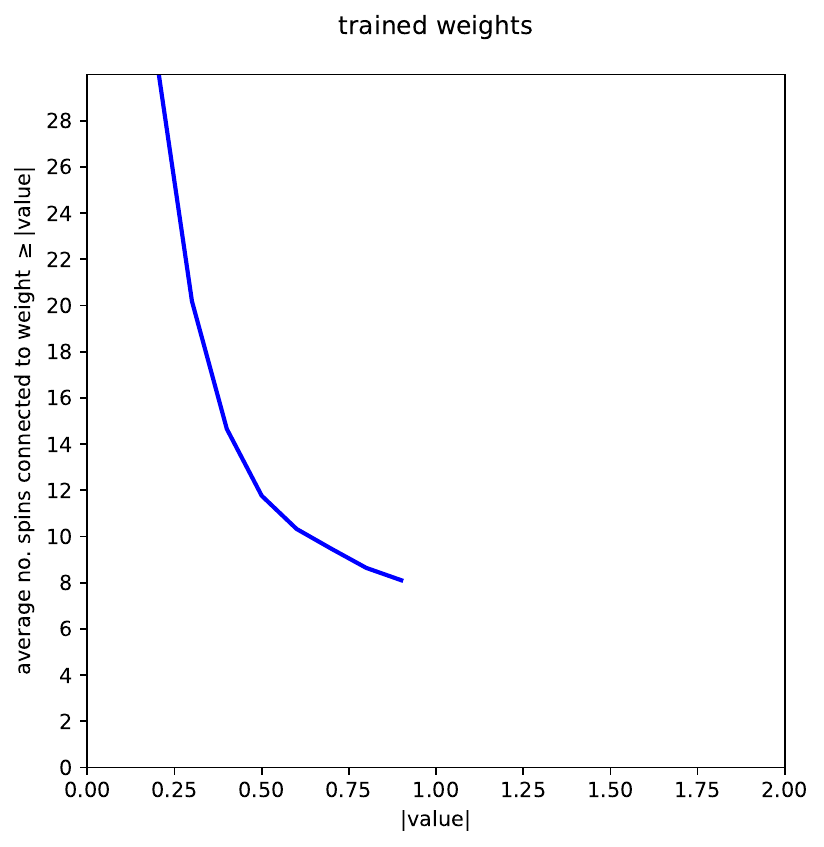}
         \caption{Layer 3 Weights}
     \end{subfigure}
     \hfill
         \begin{subfigure}{0.32\textwidth}
         \centering 
         \includegraphics[width=\textwidth]{av_weights_3r__405.pdf}
         \caption{Layer 3 Receptive}
     \end{subfigure}
           \begin{subfigure}{0.32\textwidth}
         \centering
         \includegraphics[width=\textwidth]{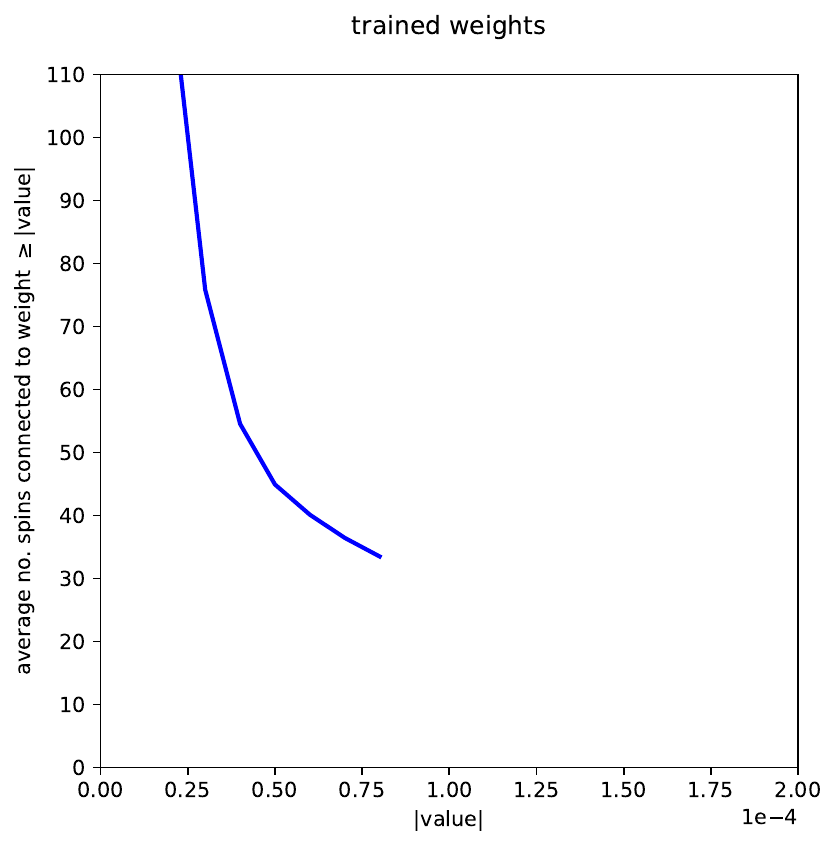}
         \caption{Layer 3 Receptive}
     \end{subfigure}
        \caption{Weight analysis plot for $\beta = .405.$}
        \label{fig:weight_analysis_.405}
\end{figure}

\begin{figure}[h!]
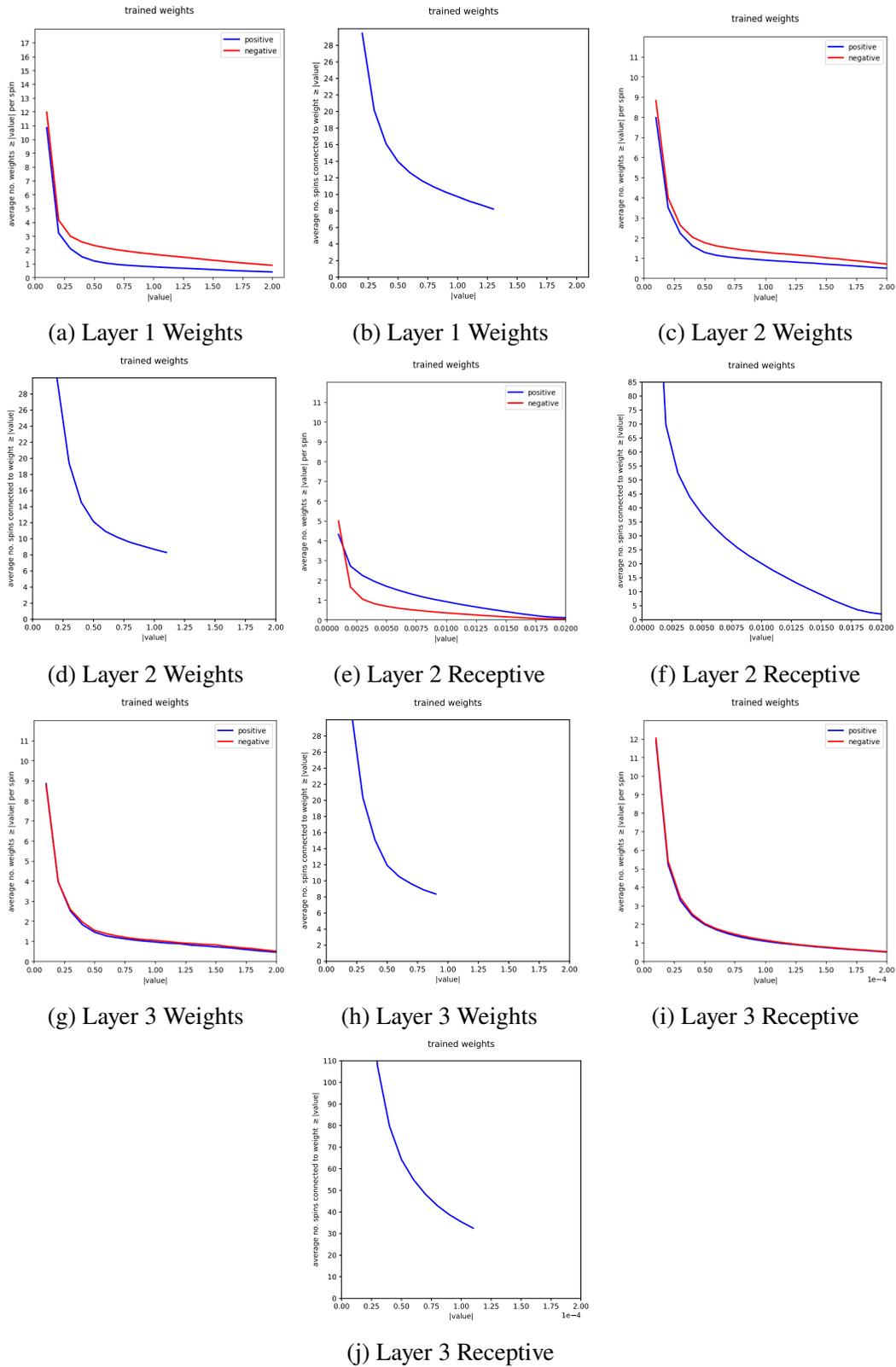

     \centering
     \begin{subfigure}{0.32\textwidth}
         \centering
         \includegraphics[width=\textwidth]{av_weights_1__41.pdf}
         \caption{Layer 1 Weights}
     \end{subfigure}
     \hfill
     \begin{subfigure}{0.32\textwidth}
         \centering
         \includegraphics[width=\textwidth]{aver_no_spins1_0_41_.pdf}
         \caption{Layer 1 Weights}
     \end{subfigure}
     \hfill
     \begin{subfigure}{0.32\textwidth}
         \centering
         \includegraphics[width=\textwidth]{av_weights_2w__41.pdf}
         \caption{Layer 2 Weights}
     \end{subfigure}
     \hfill
     \begin{subfigure}{0.32\textwidth}
         \centering
         \includegraphics[width=\textwidth]{aver_no_spins_2w_0_41_.pdf}
         \caption{Layer 2 Weights}
     \end{subfigure}
          \begin{subfigure}{0.32\textwidth}
         \centering
         \includegraphics[width=\textwidth]{av_weights_2r__41.pdf}
         \caption{Layer 2 Receptive}
     \end{subfigure}
     \hfill
     \begin{subfigure}{0.32\textwidth}
         \centering
         \includegraphics[width=\textwidth]{aver_no_spins_2r_0_41_.pdf}
         \caption{Layer 2 Receptive}
     \end{subfigure}
    \begin{subfigure}{0.32\textwidth}
         \centering
         \includegraphics[width=\textwidth]{av_weights_3w__41.pdf}
         \caption{Layer 3 Weights}
     \end{subfigure}
           \begin{subfigure}{0.32\textwidth}
         \centering
         \includegraphics[width=\textwidth]{aver_no_spins_3w_0_41_.pdf}
         \caption{Layer 3 Weights}
     \end{subfigure}
     \hfill
         \begin{subfigure}{0.32\textwidth}
         \centering
         \includegraphics[width=\textwidth]{av_weights_3r__41.pdf}
         \caption{Layer 3 Receptive}
     \end{subfigure}
           \begin{subfigure}{0.32\textwidth}
         \centering
         \includegraphics[width=\textwidth]{aver_no_spins_3r_0_41_.pdf}
         \caption{Layer 3 Receptive}
     \end{subfigure}
        \caption{Weight analysis plot for $\beta = .41.$}
        \label{fig:weight_analysis_.41}
\end{figure}

\begin{figure}[h!]
     \centering
     \begin{subfigure}{0.32\textwidth}
         \centering
         \includegraphics[width=\textwidth]{av_weights_1__415.pdf}
         \caption{Layer 1 Weights}
     \end{subfigure}
     \hfill
     \begin{subfigure}{0.32\textwidth}
         \centering
         \includegraphics[width=\textwidth]{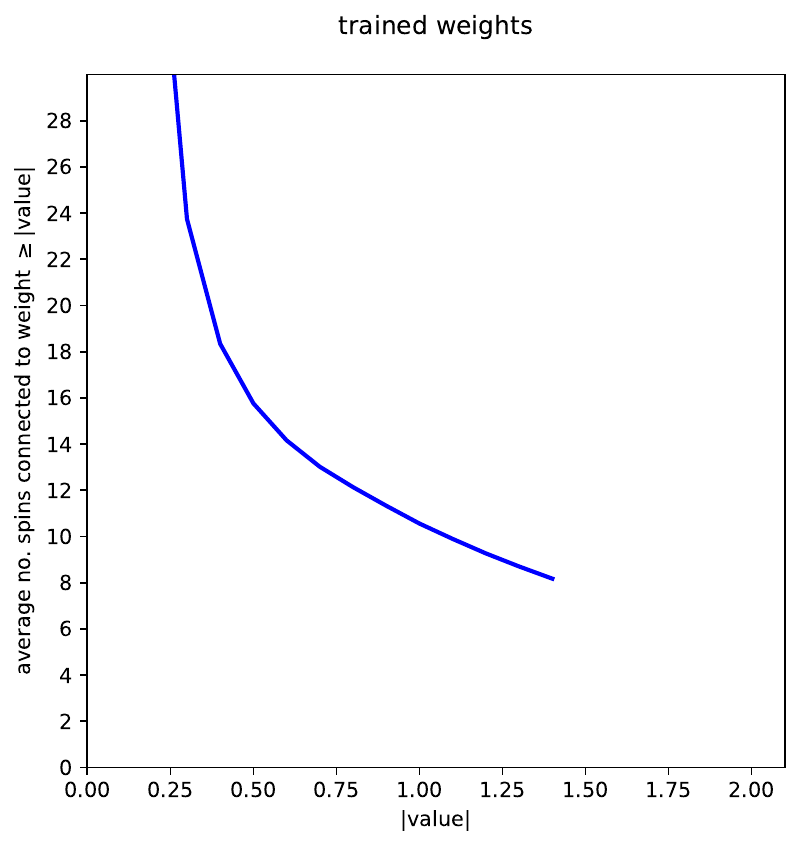}
         \caption{Layer 1 Weights}
     \end{subfigure}
     \hfill
     \begin{subfigure}{0.32\textwidth}
         \centering
         \includegraphics[width=\textwidth]{av_weights_2w__415.pdf}
         \caption{Layer 2 Weights}
     \end{subfigure}
     \hfill
     \begin{subfigure}{0.32\textwidth}
         \centering
         \includegraphics[width=\textwidth]{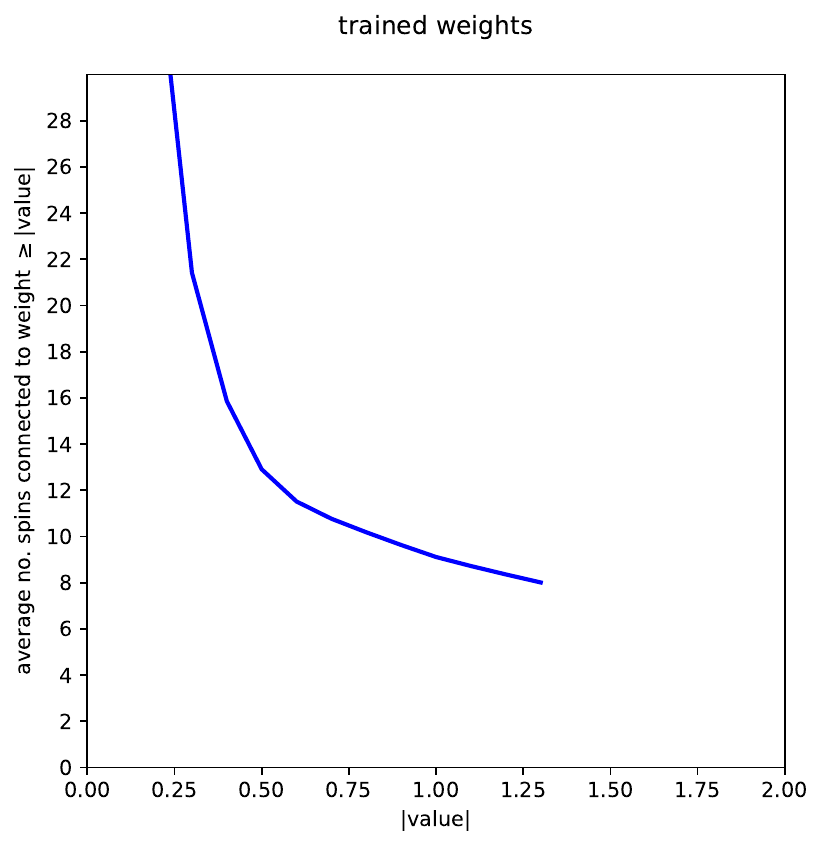}
         \caption{Layer 2 Weights}
     \end{subfigure}
          \begin{subfigure}{0.32\textwidth}
         \centering
         \includegraphics[width=\textwidth]{av_weights_2r__415.pdf}
         \caption{Layer 2 Receptive}
     \end{subfigure}
     \hfill
     \begin{subfigure}{0.32\textwidth}
         \centering
         \includegraphics[width=\textwidth]{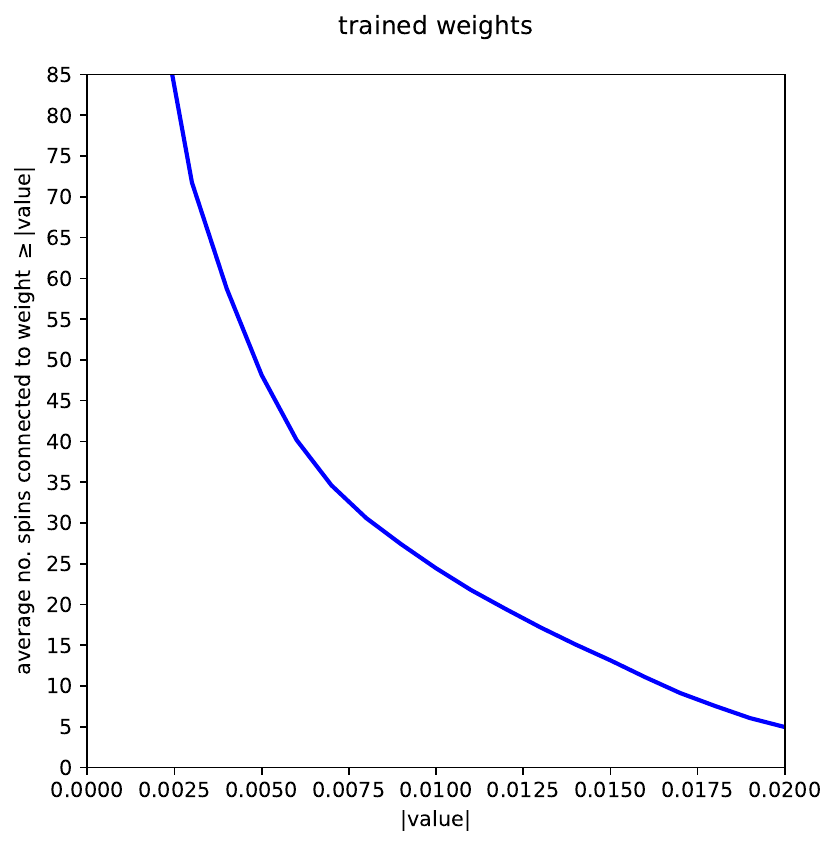}
         \caption{Layer 2 Receptive}
     \end{subfigure}
    \begin{subfigure}{0.32\textwidth}
         \centering
         \includegraphics[width=\textwidth]{av_weights_3w__415.pdf}
         \caption{Layer 3 Weights}
     \end{subfigure}
           \begin{subfigure}{0.32\textwidth}
         \centering
         \includegraphics[width=\textwidth]{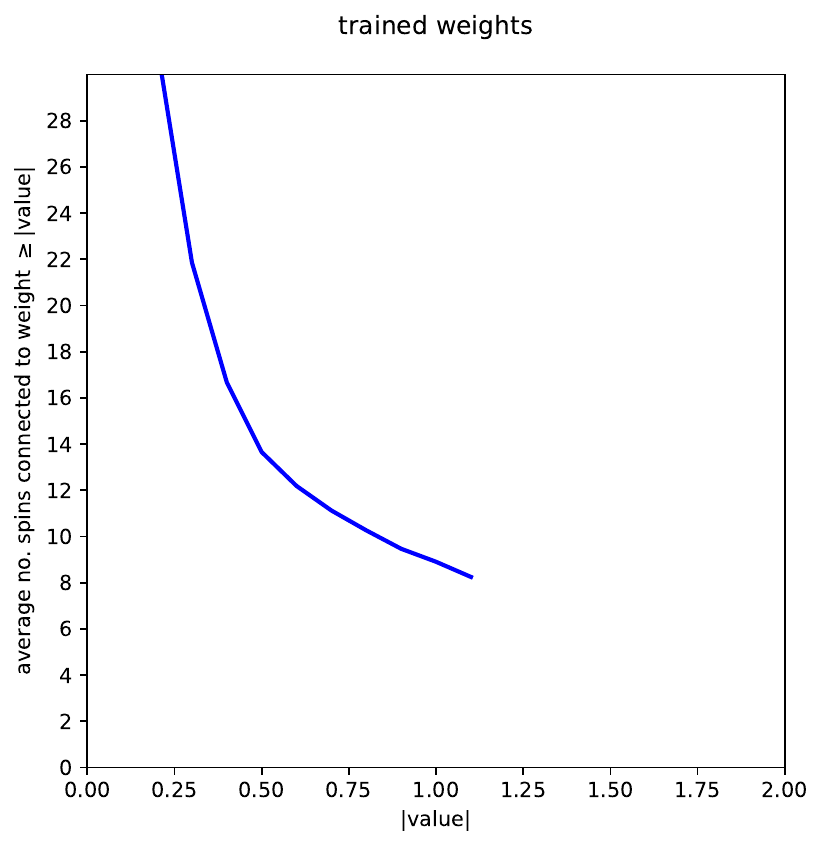}
         \caption{Layer 3 Weights}
     \end{subfigure}
     \hfill
         \begin{subfigure}{0.32\textwidth}
         \centering
         \includegraphics[width=\textwidth]{av_weights_3r__415.pdf}
         \caption{Layer 3 Receptive}
     \end{subfigure}
           \begin{subfigure}{0.32\textwidth}
         \centering
         \includegraphics[width=\textwidth]{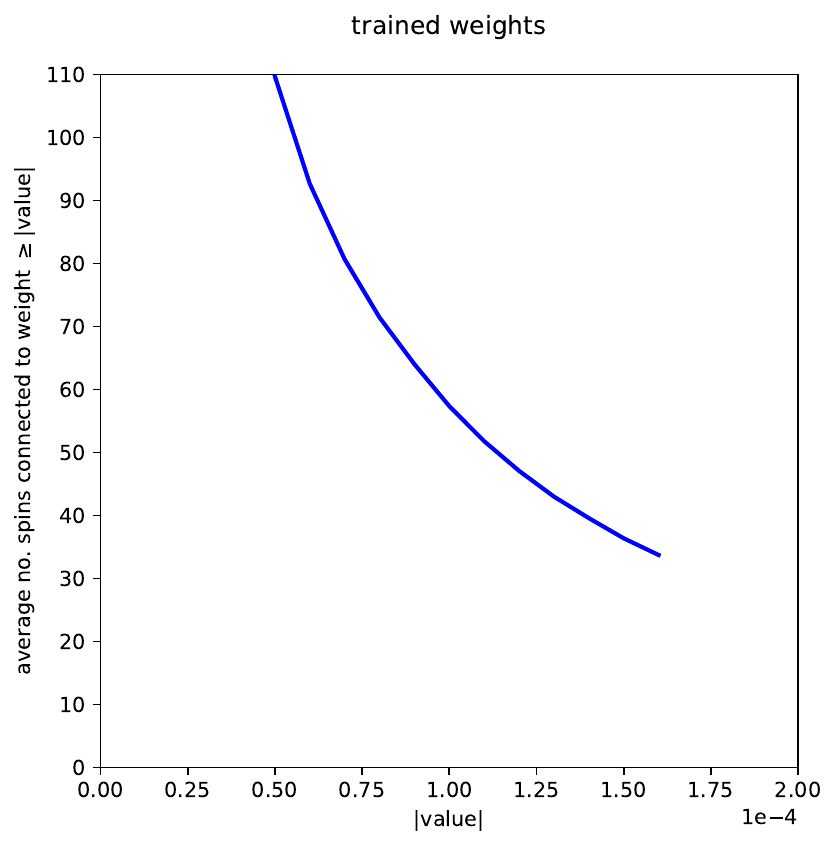}
         \caption{Layer 3 Receptive}
     \end{subfigure}
        \caption{Weight analysis plot for $\beta = .415.$}
        \label{fig:weight_analysis_.415}
\end{figure}

\begin{figure}
     \centering
     \begin{subfigure}{0.32\textwidth}
         \centering
         \includegraphics[width=\textwidth]{av_weights_1__42.pdf}
         \caption{Layer 1 Weights}
     \end{subfigure}
     \hfill
     \begin{subfigure}{0.32\textwidth}
         \centering
         \includegraphics[width=\textwidth]{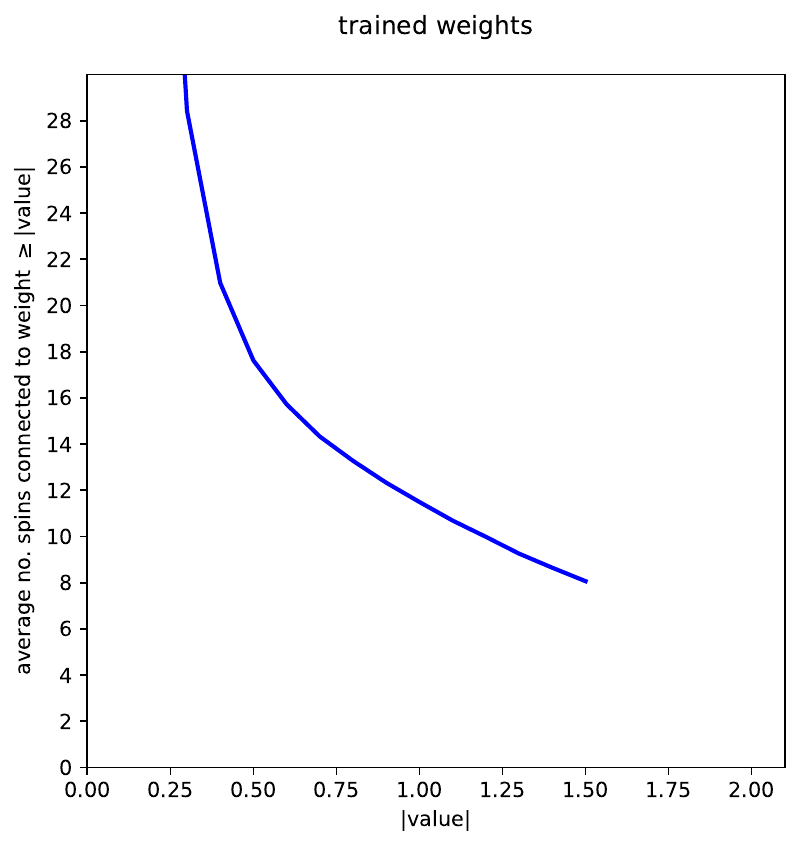}
         \caption{Layer 1 Weights}
     \end{subfigure}
     \hfill
     \begin{subfigure}{0.32\textwidth}
         \centering
         \includegraphics[width=\textwidth]{av_weights_2w__42.pdf}
         \caption{Layer 2 Weights}
     \end{subfigure}
     \hfill
     \begin{subfigure}{0.32\textwidth}
         \centering
         \includegraphics[width=\textwidth]{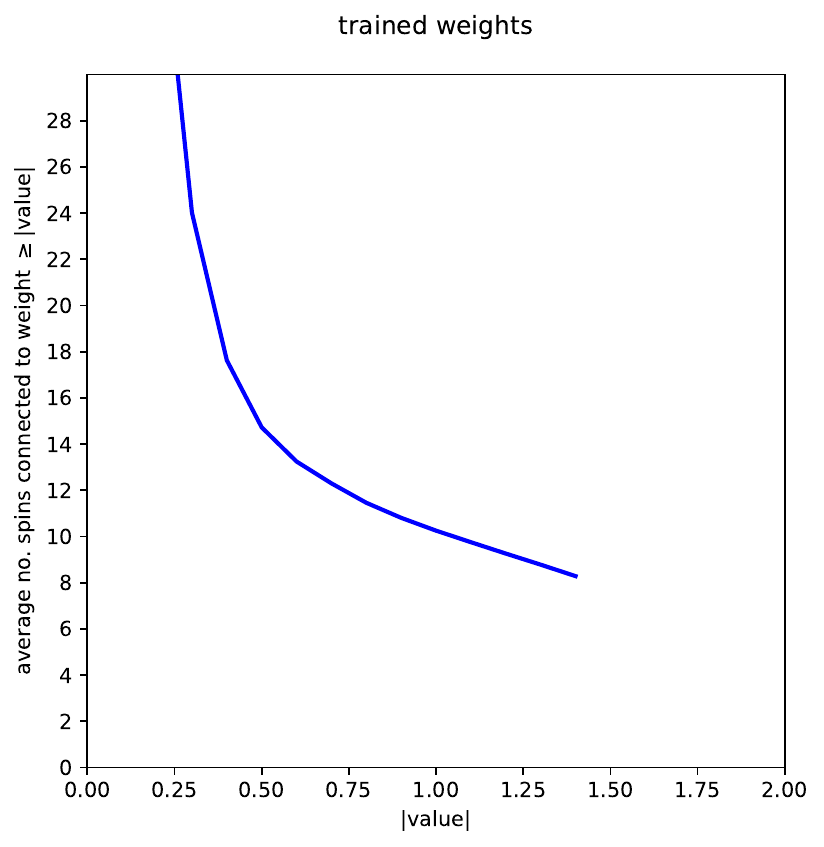}
         \caption{Layer 2 Weights}
     \end{subfigure}
          \begin{subfigure}{0.32\textwidth}
         \centering
         \includegraphics[width=\textwidth]{av_weights_2r__42.pdf}
         \caption{Layer 2 Receptive}
     \end{subfigure}
     \hfill
     \begin{subfigure}{0.32\textwidth}
         \centering
         \includegraphics[width=\textwidth]{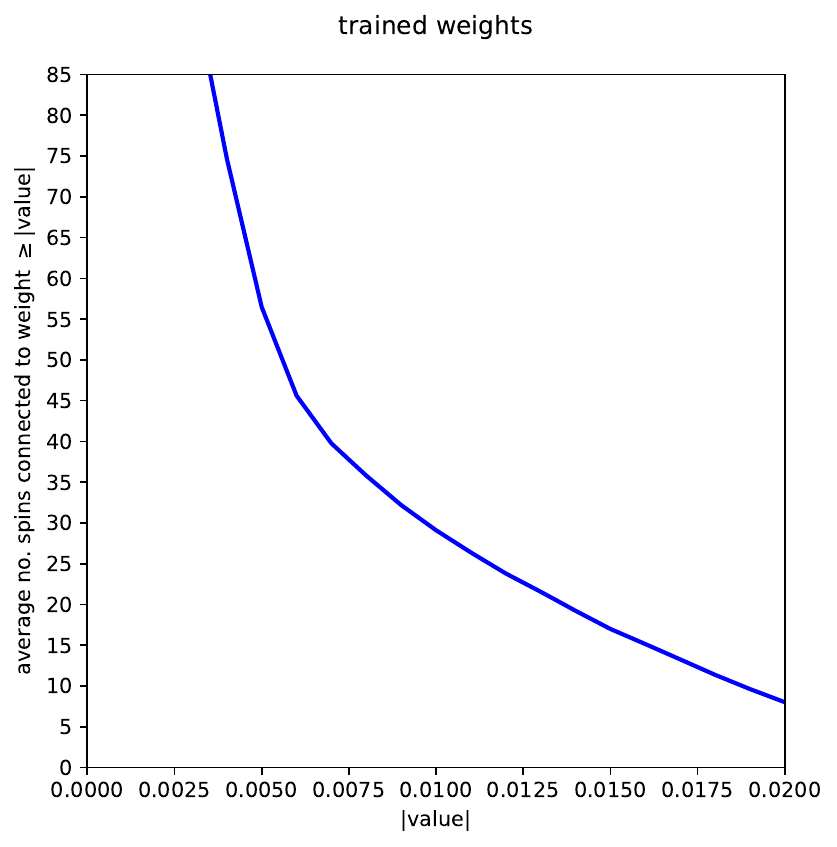}
         \caption{Layer 2 Receptive}
     \end{subfigure}
    \begin{subfigure}{0.32\textwidth}
         \centering
         \includegraphics[width=\textwidth]{av_weights_3w__42.pdf}
         \caption{Layer 3 Weights}
     \end{subfigure}
           \begin{subfigure}{0.32\textwidth}
         \centering
         \includegraphics[width=\textwidth]{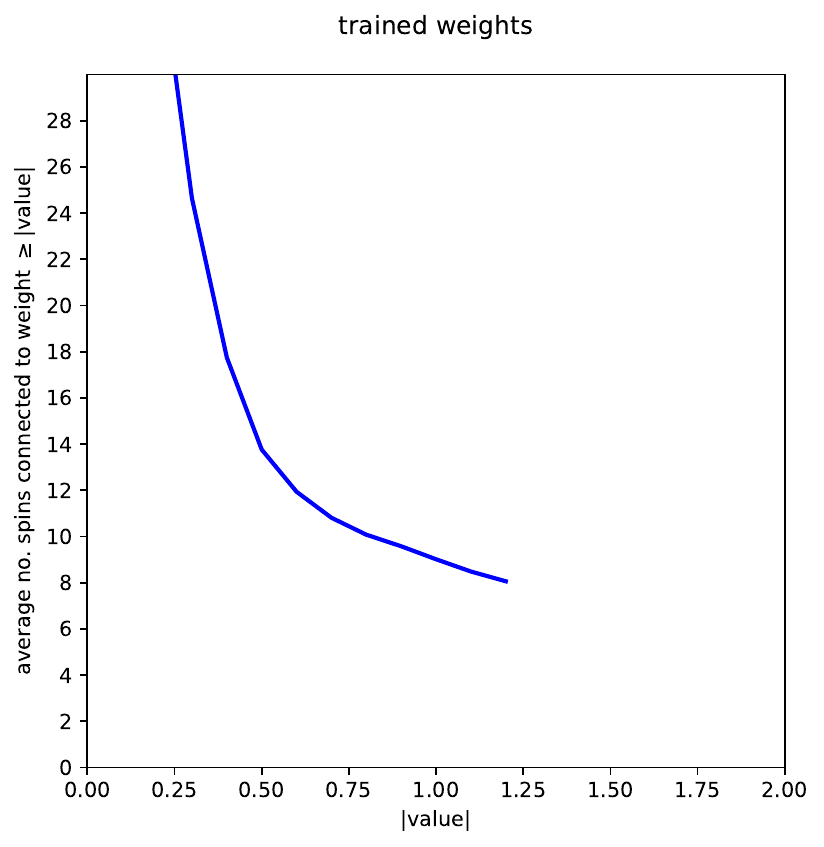}
         \caption{Layer 3 Weights}
     \end{subfigure}
     \hfill
         \begin{subfigure}{0.32\textwidth}
         \centering
         \includegraphics[width=\textwidth]{av_weights_3r__42.pdf}
         \caption{Layer 3 Receptive}
     \end{subfigure}
           \begin{subfigure}{0.32\textwidth}
         \centering
         \includegraphics[width=\textwidth]{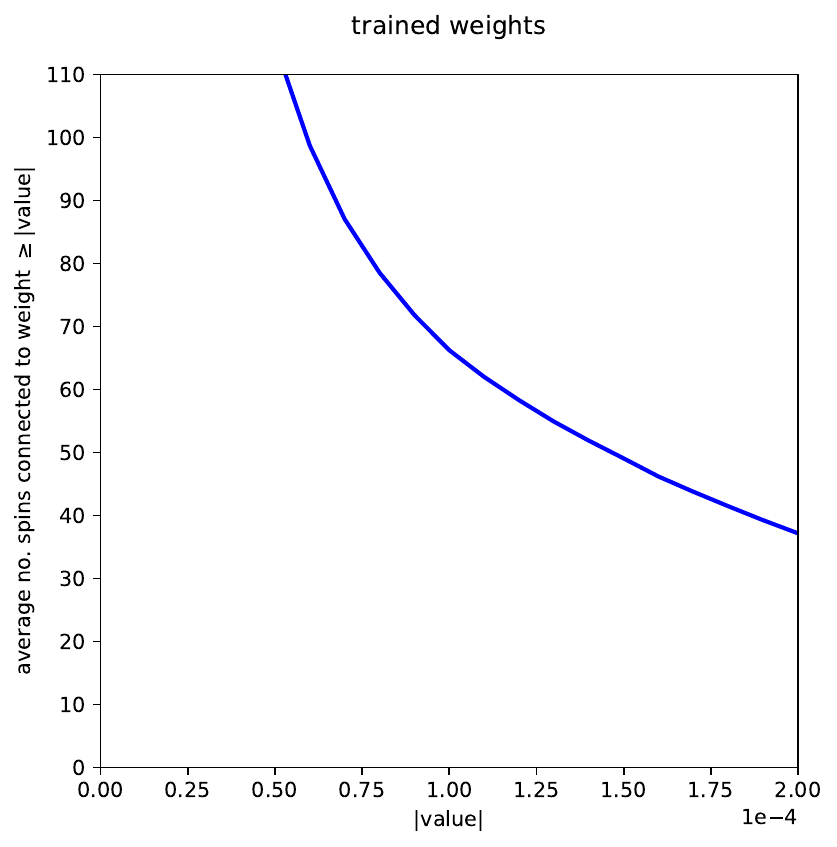}
         \caption{Layer 3 Receptive}
     \end{subfigure}
        \caption{Weight analysis plot for $\beta = .42.$}
        \label{fig:weight_analysis_.42}
\end{figure}

\begin{figure}[h!]
     \centering
     \begin{subfigure}{0.32\textwidth}
         \centering
         \includegraphics[width=\textwidth]{av_weights_1__425.pdf}
         \caption{Layer 1 Weights}
     \end{subfigure}
     \hfill
     \begin{subfigure}{0.32\textwidth}
         \centering
         \includegraphics[width=\textwidth]{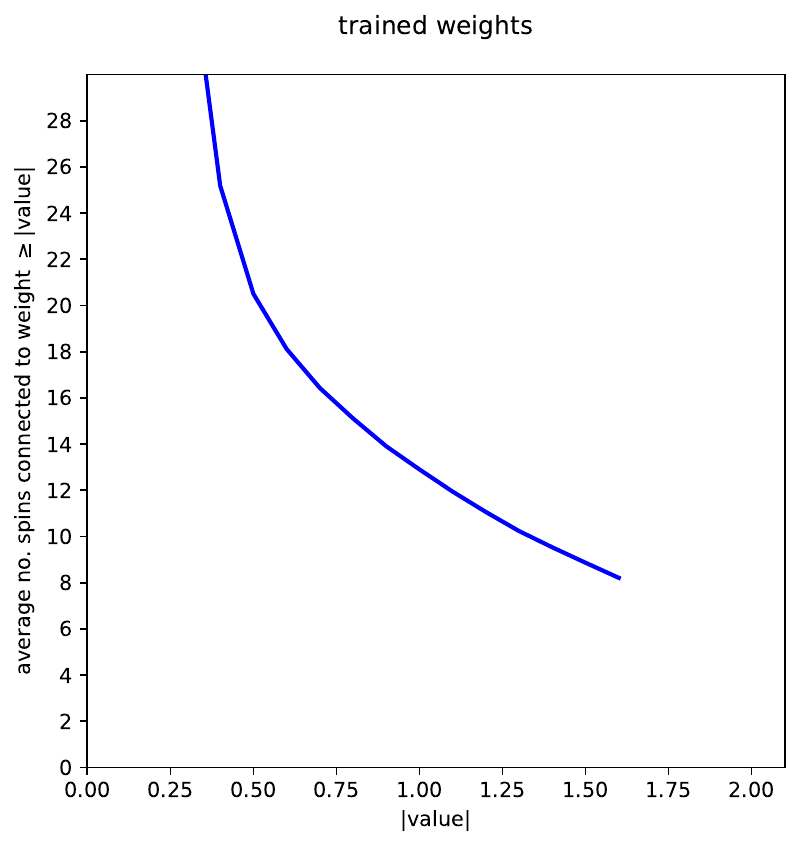}
         \caption{Layer 1 Weights}
     \end{subfigure}
     \hfill
     \begin{subfigure}{0.32\textwidth}
         \centering
         \includegraphics[width=\textwidth]{av_weights_2w__425.pdf}
         \caption{Layer 2 Weights}
     \end{subfigure}
     \hfill
     \begin{subfigure}{0.32\textwidth}
         \centering
         \includegraphics[width=\textwidth]{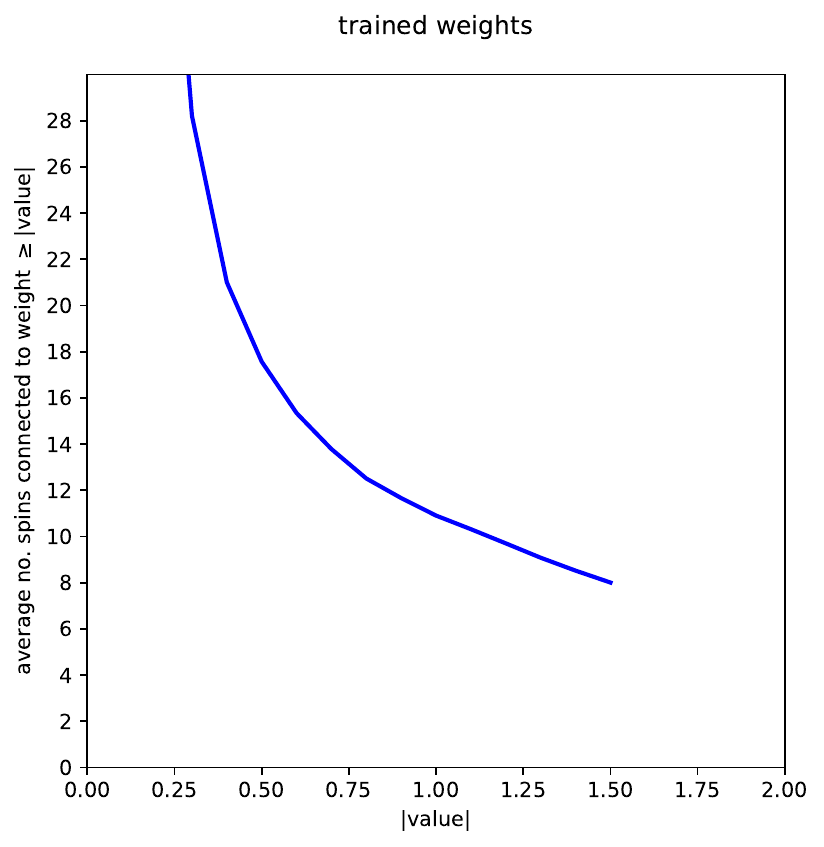}
         \caption{Layer 2 Weights}
     \end{subfigure}
          \begin{subfigure}{0.32\textwidth}
         \centering
         \includegraphics[width=\textwidth]{av_weights_2r__425.pdf}
         \caption{Layer 2 Receptive}
     \end{subfigure}
     \hfill
     \begin{subfigure}{0.32\textwidth}
         \centering
         \includegraphics[width=\textwidth]{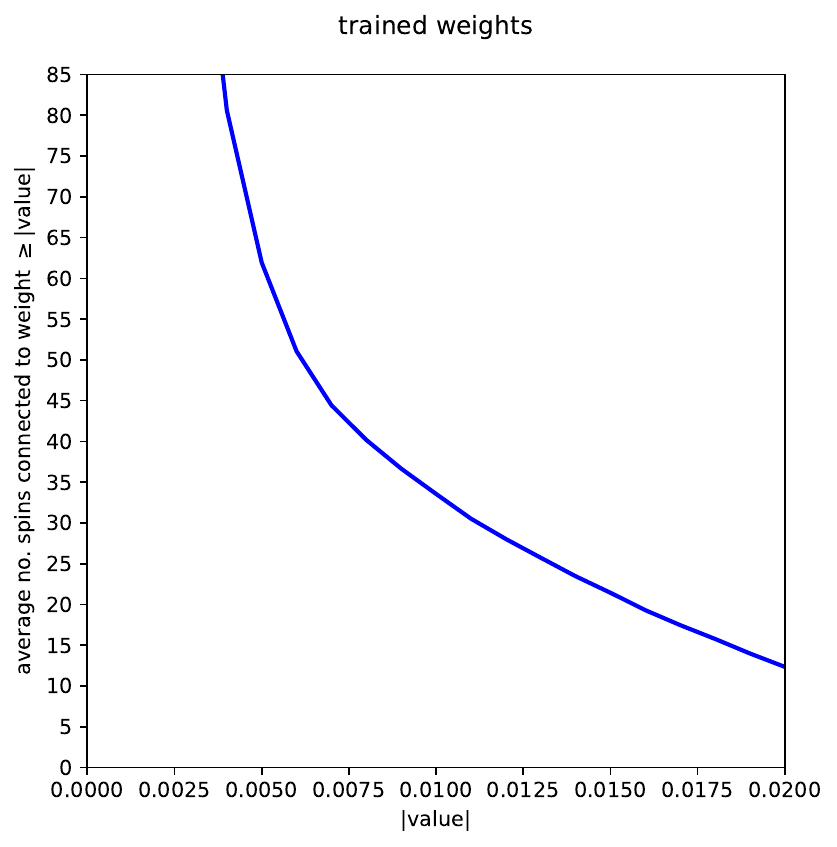}
         \caption{Layer 2 Receptive}
     \end{subfigure}
    \begin{subfigure}{0.32\textwidth}
         \centering
         \includegraphics[width=\textwidth]{av_weights_3w__425.pdf}
         \caption{Layer 3 Weights}
     \end{subfigure}
           \begin{subfigure}{0.32\textwidth}
         \centering
         \includegraphics[width=\textwidth]{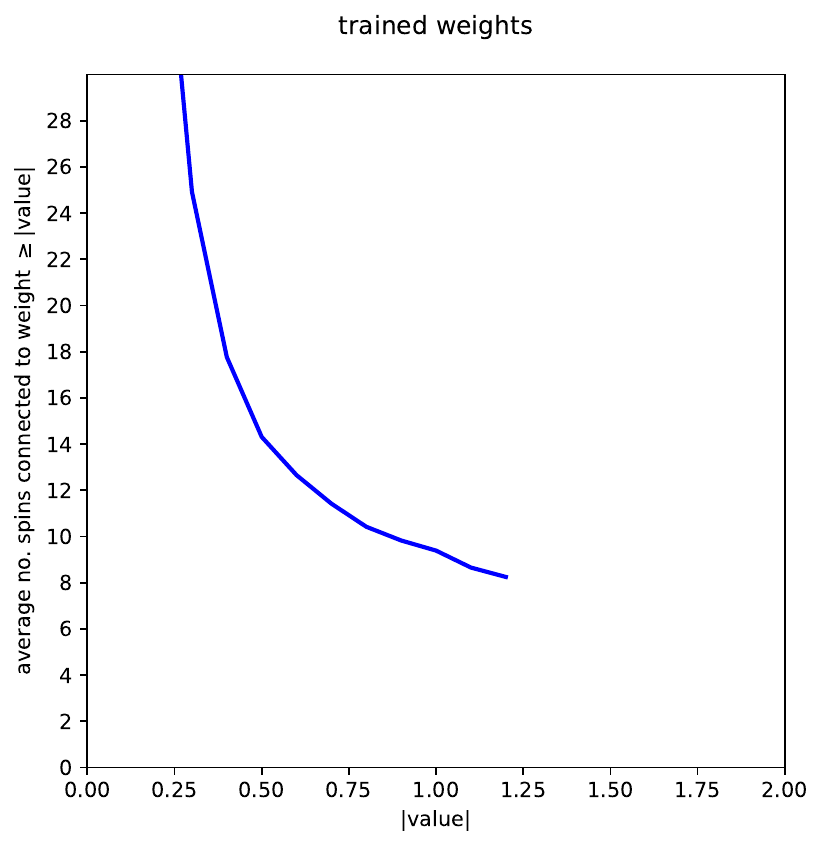}
         \caption{Layer 3 Weights}
     \end{subfigure}
     \hfill
         \begin{subfigure}{0.32\textwidth}
         \centering
         \includegraphics[width=\textwidth]{av_weights_3r__425.pdf}
         \caption{Layer 3 Receptive}
     \end{subfigure}
           \begin{subfigure}{0.32\textwidth}
         \centering
         \includegraphics[width=\textwidth]{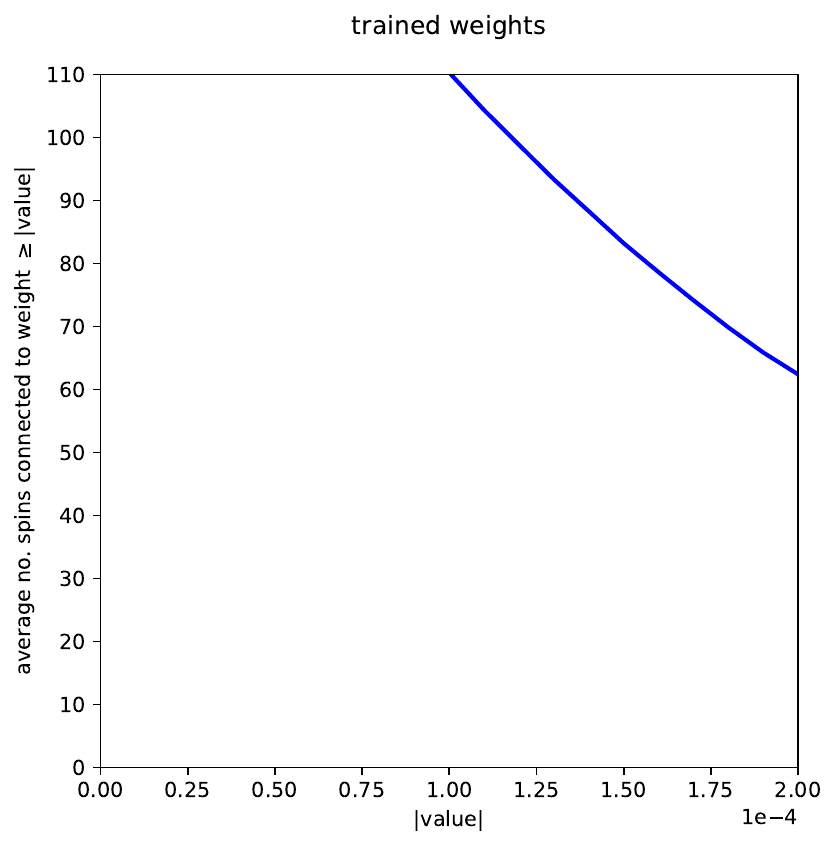}
         \caption{Layer 3 Receptive}
     \end{subfigure}
        \caption{Weight analysis plot for $\beta = .425.$}
        \label{fig:weight_analysis_.425}
\end{figure}

\begin{figure}[h!]
     \centering
     \begin{subfigure}{0.32\textwidth}
         \centering
         \includegraphics[width=\textwidth]{av_weights_1__43.pdf}
         \caption{Layer 1 Weights}
     \end{subfigure}
     \hfill
     \begin{subfigure}{0.32\textwidth}
         \centering
         \includegraphics[width=\textwidth]{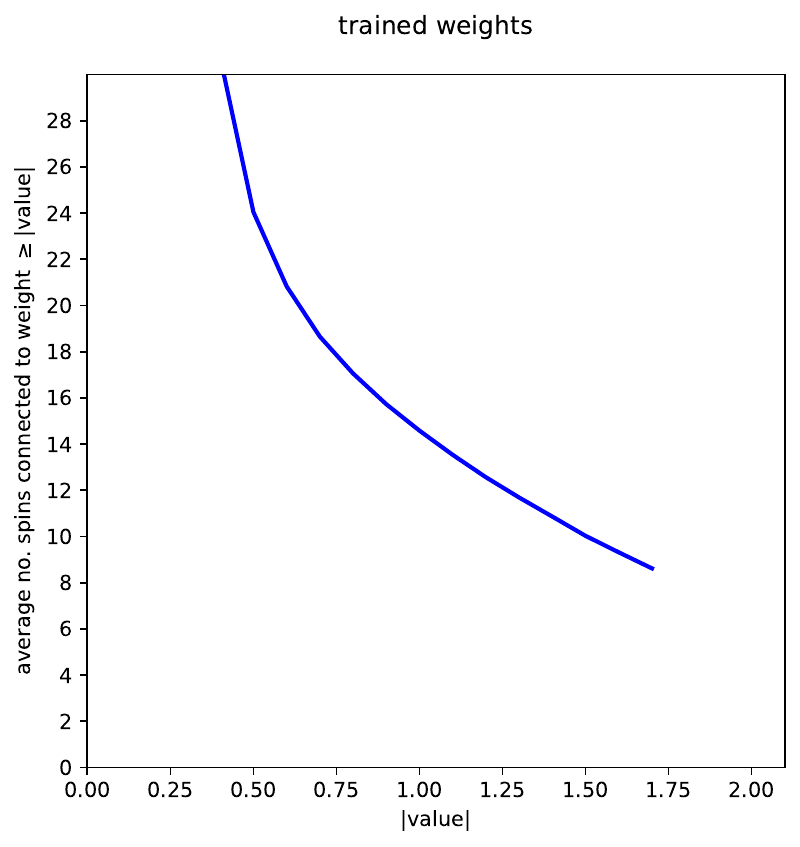}
         \caption{Layer 1 Weights}
     \end{subfigure}
     \hfill
     \begin{subfigure}{0.32\textwidth}
         \centering
         \includegraphics[width=\textwidth]{av_weights_2w__43.pdf}
         \caption{Layer 2 Weights}
     \end{subfigure}
     \hfill
     \begin{subfigure}{0.32\textwidth}
         \centering
         \includegraphics[width=\textwidth]{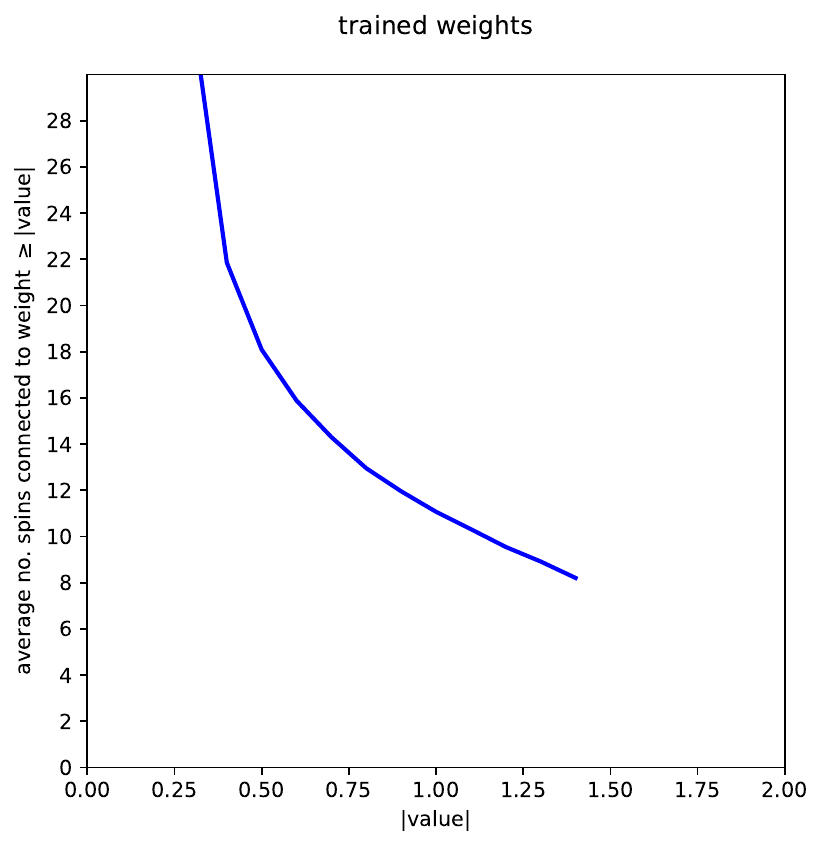}
         \caption{Layer 2 Weights}
     \end{subfigure}
          \begin{subfigure}{0.32\textwidth}
         \centering
         \includegraphics[width=\textwidth]{av_weights_2r__43.pdf}
         \caption{Layer 2 Receptive}
     \end{subfigure}
     \hfill
     \begin{subfigure}{0.32\textwidth}
         \centering
         \includegraphics[width=\textwidth]{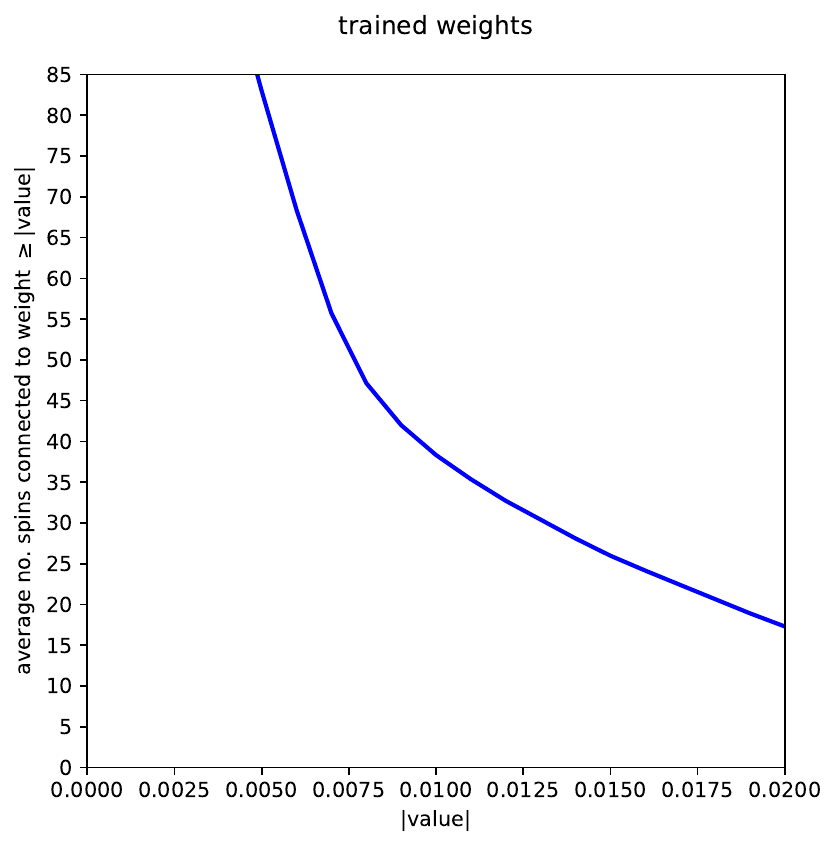}
         \caption{Layer 2 Receptive}
     \end{subfigure}
    \begin{subfigure}{0.32\textwidth}
         \centering
         \includegraphics[width=\textwidth]{av_weights_3w__43.pdf}
         \caption{Layer 3 Weights}
     \end{subfigure}
           \begin{subfigure}{0.32\textwidth}
         \centering
         \includegraphics[width=\textwidth]{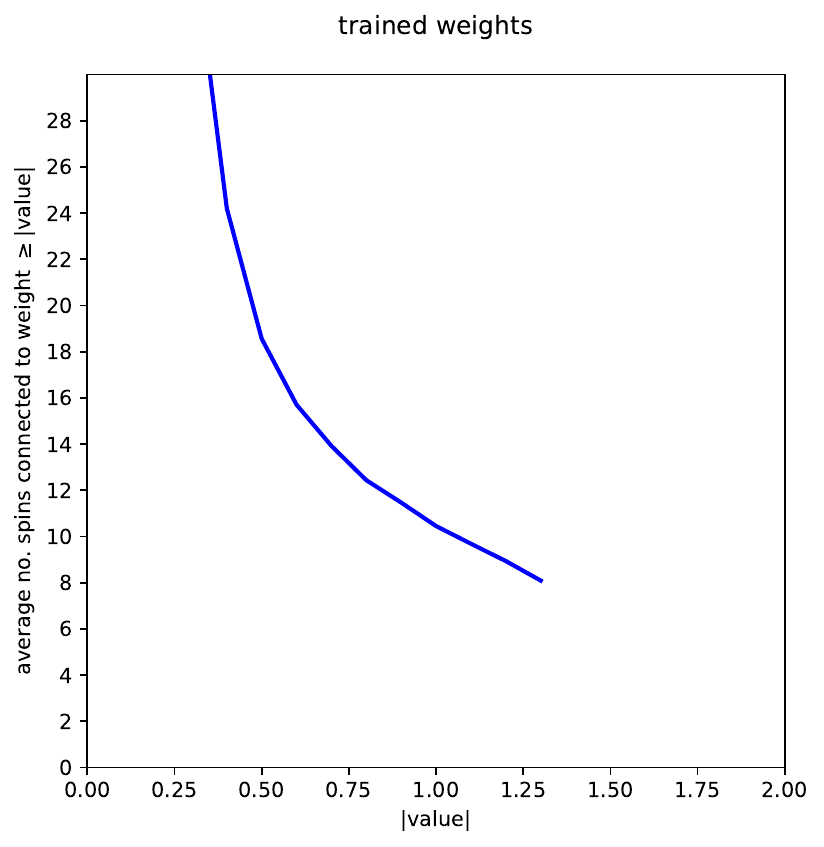}
         \caption{Layer 3 Weights}
     \end{subfigure}
     \hfill
         \begin{subfigure}{0.32\textwidth}
         \centering
         \includegraphics[width=\textwidth]{av_weights_3r__43.pdf}
         \caption{Layer 3 Receptive}
     \end{subfigure}
           \begin{subfigure}{0.32\textwidth}
         \centering
         \includegraphics[width=\textwidth]{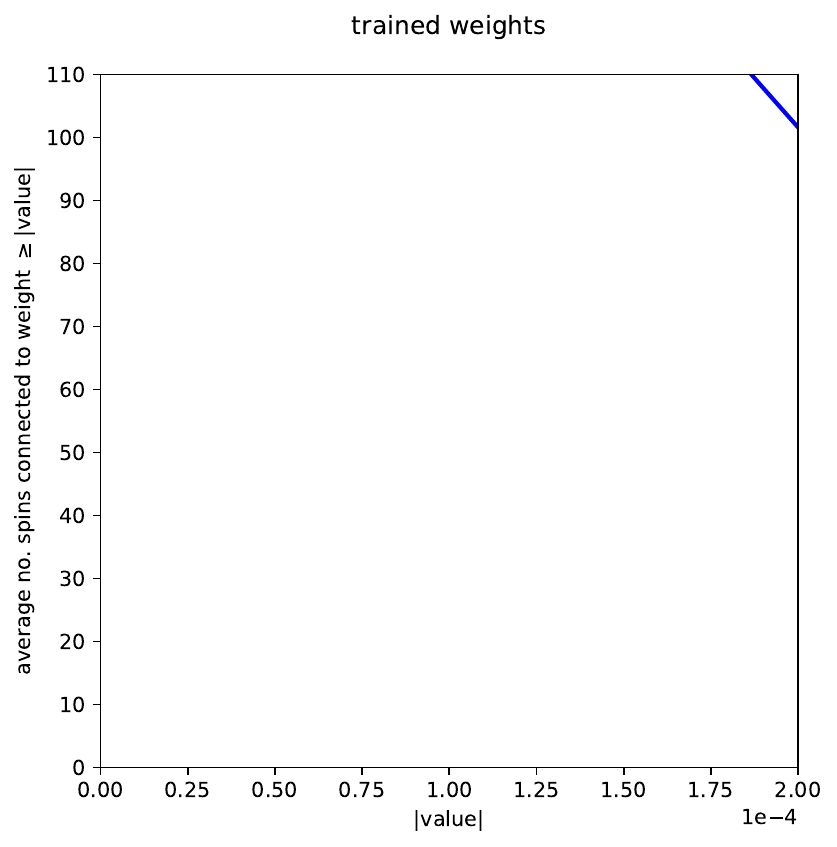}
         \caption{Layer 3 Receptive}
     \end{subfigure}
        \caption{Weight analysis plot for $\beta = .43.$}
        \label{fig:weight_analysis_.43}
\end{figure}

\clearpage

\section{Block Analysis Histograms }
\label{sec:block_plots}

The plots starting on the next page analyze the blocking structure for each value of $\beta$, analyzing the structure for both the weight tensors and receptive field tensors. Each tensor has three histograms. The first histogram is the number of spins per block averaged over all the lattices. The second is the maximum distance between spins per block averaged over all the lattices. The third is just all the distances between spins averaged over all the lattices. These plots are discussed in detail in Section \ref{sec:blocks}.

\clearpage

\begin{figure}[h!]
     \centering
     \begin{subfigure}{0.25\textwidth}
         \centering
         \includegraphics[width=\textwidth]{dist_spins_1__395.pdf}
         \caption{Layer 1}
     \end{subfigure}
     \hfill
     \begin{subfigure}{0.25\textwidth}
         \centering
         \includegraphics[width=\textwidth]{dist_1__395.pdf}
         \caption{Layer 1}
     \end{subfigure}
     \hfill
     \begin{subfigure}{0.25\textwidth}
         \centering
         \includegraphics[width=\textwidth]{dist_max_1__395.pdf}
         \caption{Layer 1}
     \end{subfigure}
     \hfill
     \begin{subfigure}{0.25\textwidth}
         \centering
         \includegraphics[width=\textwidth]{dist_spins_2w__395.pdf}
         \caption{Layer 2 Weights}
     \end{subfigure}
     \hfill
     \begin{subfigure}{0.25\textwidth}
         \centering
         \includegraphics[width=\textwidth]{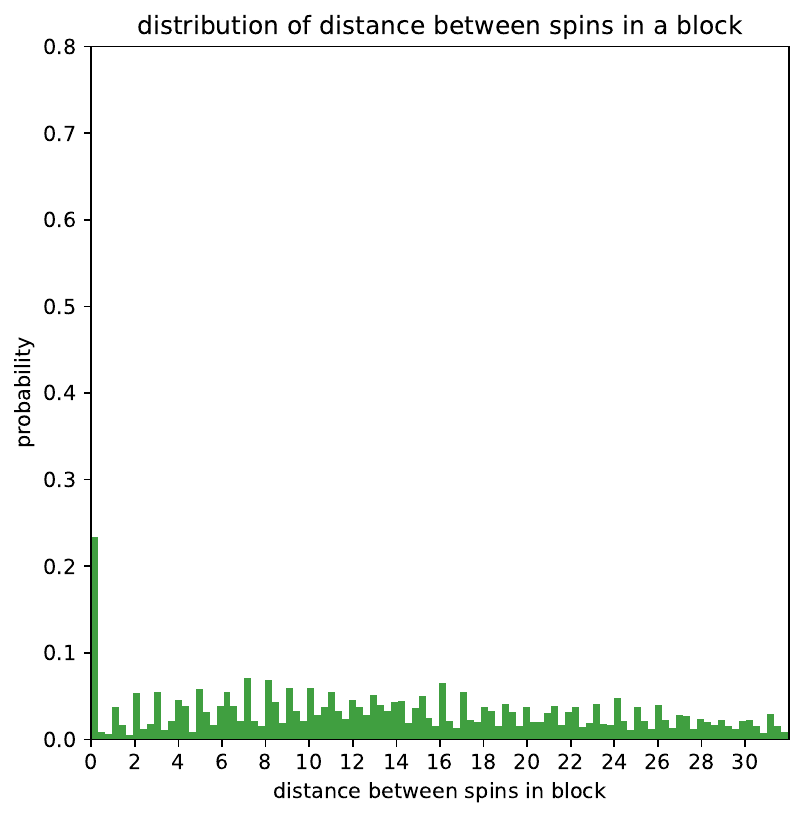}
         \caption{Layer 2 Weights}
     \end{subfigure}
          \hfill
     \begin{subfigure}{0.25\textwidth}
         \centering
         \includegraphics[width=\textwidth]{dist_max_2w__395.pdf}
         \caption{Layer 2 Weights}
     \end{subfigure}
     \hfill
     \begin{subfigure}{0.25\textwidth}
         \centering
         \includegraphics[width=\textwidth]{dist_spins_3w__395.pdf}
         \caption{Layer 3 Weights}
     \end{subfigure}
     \hfill
          \begin{subfigure}{0.25\textwidth}
         \centering
         \includegraphics[width=\textwidth]{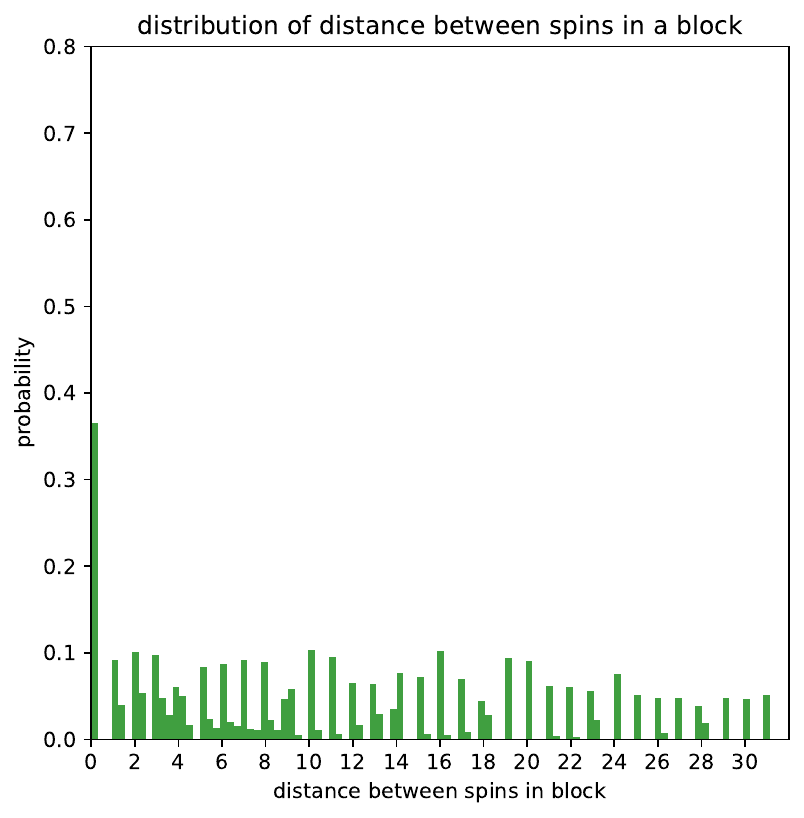}
         \caption{Layer 3 Weights}
     \end{subfigure}
     \hfill
     \begin{subfigure}{0.25\textwidth}
         \centering
         \includegraphics[width=\textwidth]{dist_max_3w__395.pdf}
         \caption{Layer 3 Weights}
     \end{subfigure}
          \hfill
          \begin{subfigure}{0.25\textwidth}
         \centering
         \includegraphics[width=\textwidth]{dist_spins_2r__395.pdf}
         \caption{Layer 2 Receptive}
     \end{subfigure}
     \hfill
     \begin{subfigure}{0.25\textwidth}
         \centering
         \includegraphics[width=\textwidth]{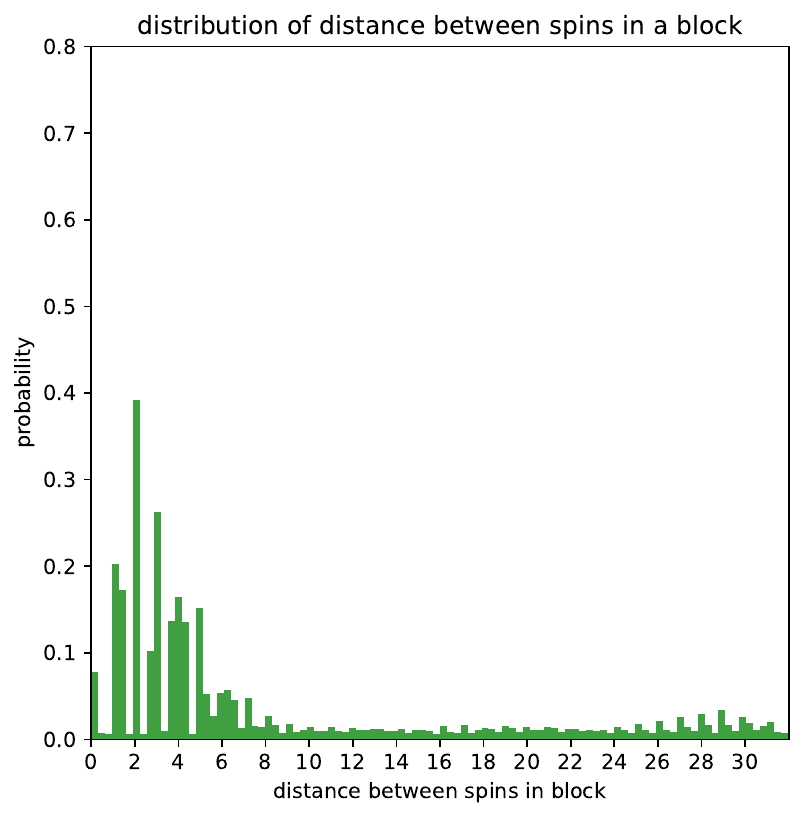}
         \caption{Layer 2 Receptive}
     \end{subfigure}
          \hfill
     \begin{subfigure}{0.25\textwidth}
         \centering
         \includegraphics[width=\textwidth]{dist_max_2r__395.pdf}
         \caption{Layer 2 Receptive}
     \end{subfigure}
     \hfill
     \begin{subfigure}{0.25\textwidth}
         \centering
         \includegraphics[width=\textwidth]{dist_spins_3r__395.pdf}
         \caption{Layer 3 Receptive}
     \end{subfigure}
     \hfill
          \begin{subfigure}{0.25\textwidth}
         \centering
         \includegraphics[width=\textwidth]{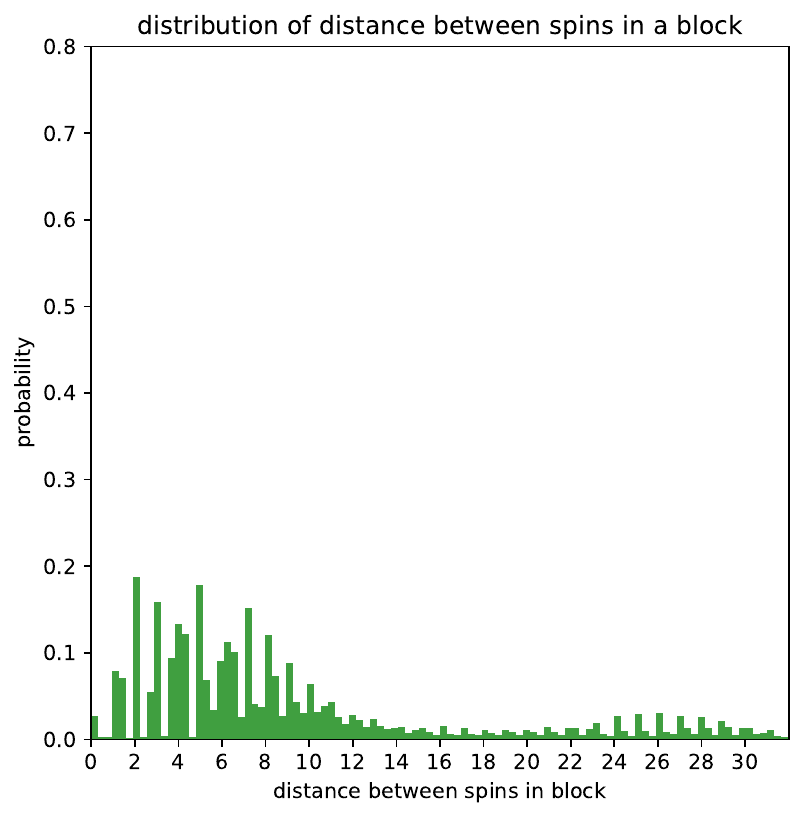}
         \caption{Layer 3 Receptive}
     \end{subfigure}
     \hfill
     \begin{subfigure}{0.25\textwidth}
         \centering
         \includegraphics[width=\textwidth]{dist_max_3r__395.pdf}
         \caption{Layer 3 Receptive}
     \end{subfigure}
     \hfill
        \caption{Block analysis plot for $\beta = .395.$}
        \label{fig:dist_analysis_.395}
\end{figure}

\clearpage

\begin{figure}[h!]
     \centering
     \begin{subfigure}{0.25\textwidth}
         \centering
         \includegraphics[width=\textwidth]{dist_spins_1__4.pdf}
         \caption{Layer 1}
     \end{subfigure}
     \hfill
     \begin{subfigure}{0.25\textwidth}
         \centering
         \includegraphics[width=\textwidth]{dist_1__4.pdf}
         \caption{Layer 1}
     \end{subfigure}
     \hfill
     \begin{subfigure}{0.25\textwidth}
         \centering
         \includegraphics[width=\textwidth]{dist_max_1__4.pdf}
         \caption{Layer 1}
     \end{subfigure}
     \hfill
     \begin{subfigure}{0.25\textwidth}
         \centering
         \includegraphics[width=\textwidth]{dist_spins_2w__4.pdf}
         \caption{Layer 2 Weights}
     \end{subfigure}
     \hfill
     \begin{subfigure}{0.25\textwidth}
         \centering
         \includegraphics[width=\textwidth]{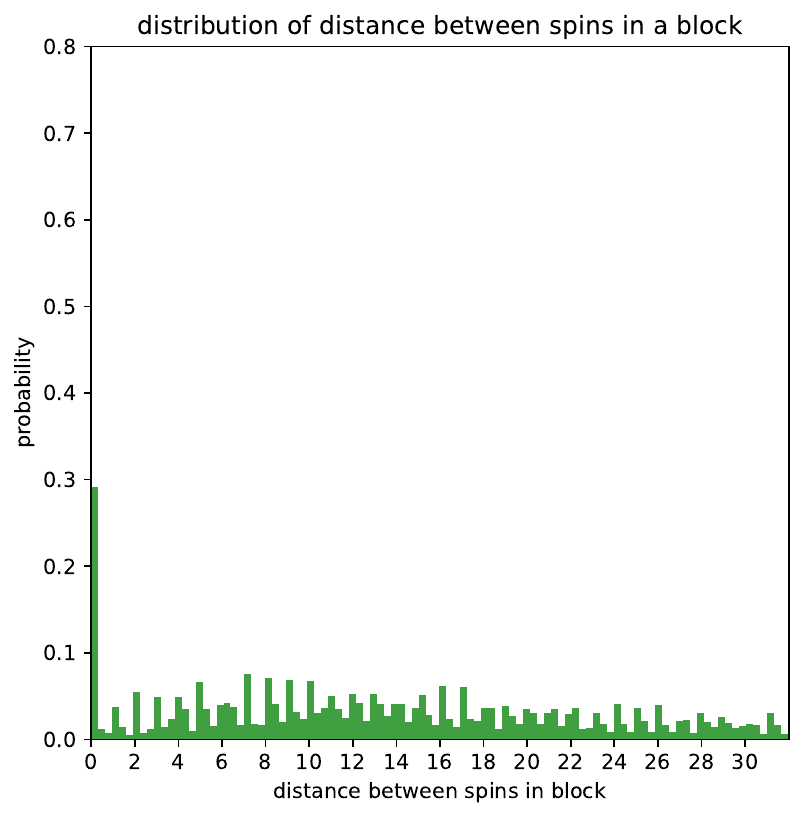}
         \caption{Layer 2 Weights}
     \end{subfigure}
          \hfill
     \begin{subfigure}{0.25\textwidth}
         \centering
         \includegraphics[width=\textwidth]{dist_max_2w__4.pdf}
         \caption{Layer 2 Weights}
     \end{subfigure}
     \hfill
     \begin{subfigure}{0.25\textwidth}
         \centering
         \includegraphics[width=\textwidth]{dist_spins_3w__4.pdf}
         \caption{Layer 3 Weights}
     \end{subfigure}
     \hfill
          \begin{subfigure}{0.25\textwidth}
         \centering
         \includegraphics[width=\textwidth]{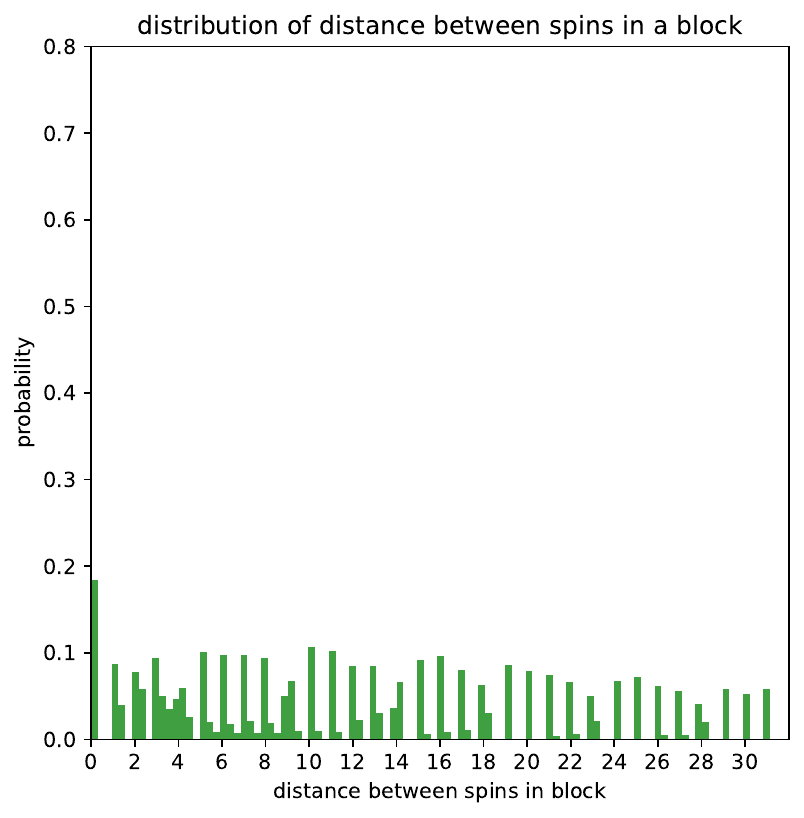}
         \caption{Layer 3 Weights}
     \end{subfigure}
     \hfill
     \begin{subfigure}{0.25\textwidth}
         \centering
         \includegraphics[width=\textwidth]{dist_max_3w__4.pdf}
         \caption{Layer 3 Weights}
     \end{subfigure}
          \hfill
          \begin{subfigure}{0.25\textwidth}
         \centering
         \includegraphics[width=\textwidth]{dist_spins_2r__4.pdf}
         \caption{Layer 2 Receptive}
     \end{subfigure}
     \hfill
     \begin{subfigure}{0.25\textwidth}
         \centering
         \includegraphics[width=\textwidth]{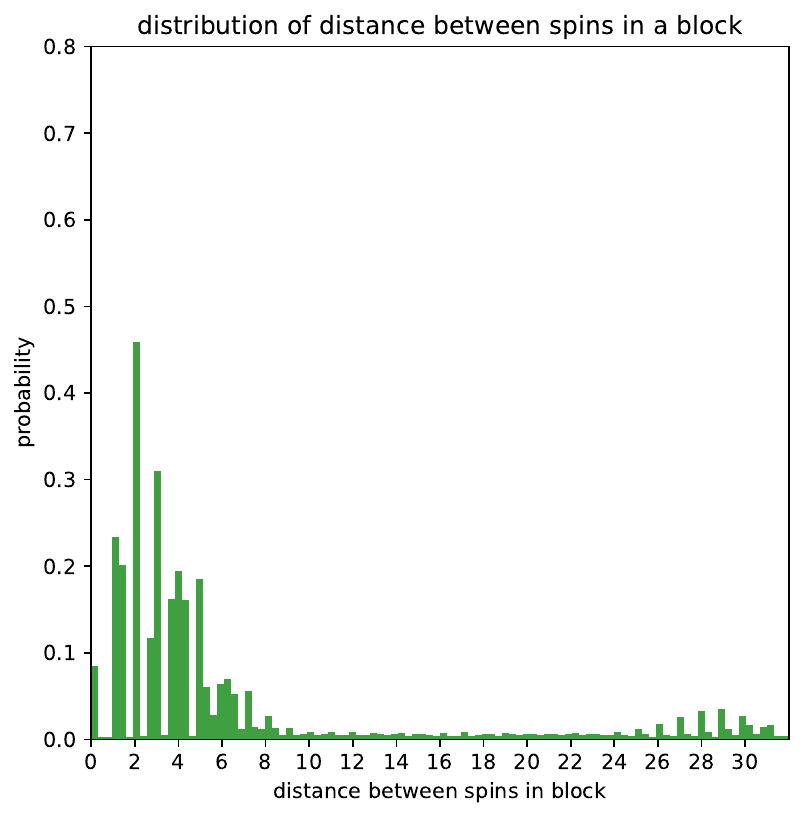}
         \caption{Layer 2 Receptive}
     \end{subfigure}
          \hfill
     \begin{subfigure}{0.25\textwidth}
         \centering
         \includegraphics[width=\textwidth]{dist_max_2r__4.pdf}
         \caption{Layer 2 Receptive}
     \end{subfigure}
     \hfill
     \begin{subfigure}{0.25\textwidth}
         \centering
         \includegraphics[width=\textwidth]{dist_spins_3r__4.pdf}
         \caption{Layer 3 Receptive}
     \end{subfigure}
     \hfill
          \begin{subfigure}{0.25\textwidth}
         \centering
         \includegraphics[width=\textwidth]{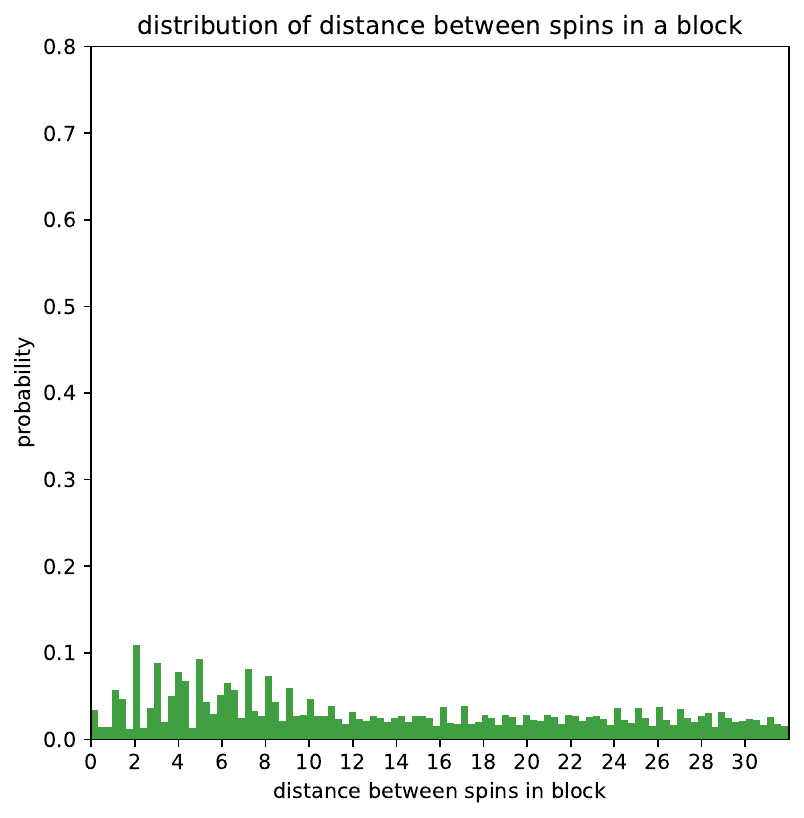}
         \caption{Layer 3 Receptive}
     \end{subfigure}
     \hfill
     \begin{subfigure}{0.25\textwidth}
         \centering
         \includegraphics[width=\textwidth]{dist_max_3r__4.pdf}
         \caption{Layer 3 Receptive}
     \end{subfigure}
     \hfill
        \caption{Block analysis plot for $\beta = .4.$}
        \label{fig:dist_analysis_.4}
\end{figure}

\clearpage

\begin{figure}[h!]
     \centering
     \begin{subfigure}{0.25\textwidth}
         \centering
         \includegraphics[width=\textwidth]{dist_spins_1__405.pdf}
         \caption{Layer 1}
     \end{subfigure}
     \hfill
     \begin{subfigure}{0.25\textwidth}
         \centering
         \includegraphics[width=\textwidth]{dist_1__405.pdf}
         \caption{Layer 1}
     \end{subfigure}
     \hfill
     \begin{subfigure}{0.25\textwidth}
         \centering
         \includegraphics[width=\textwidth]{dist_max_1__405.pdf}
         \caption{Layer 1}
     \end{subfigure}
     \hfill
     \begin{subfigure}{0.25\textwidth}
         \centering
         \includegraphics[width=\textwidth]{dist_spins_2w__405.pdf}
         \caption{Layer 2 Weights}
     \end{subfigure}
     \hfill
     \begin{subfigure}{0.25\textwidth}
         \centering
         \includegraphics[width=\textwidth]{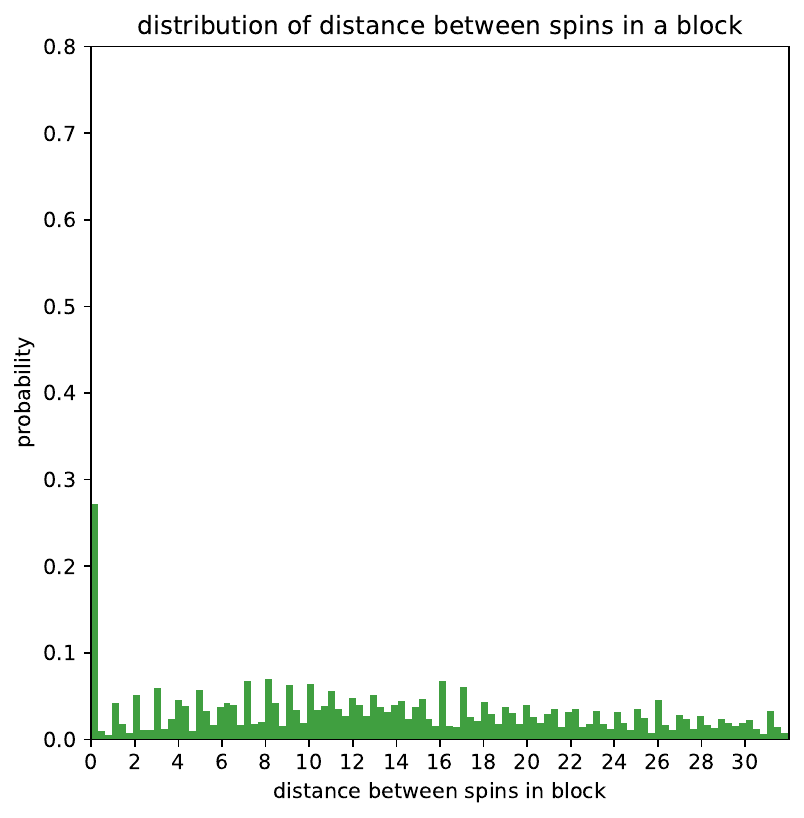}
         \caption{Layer 2 Weights}
     \end{subfigure}
          \hfill
     \begin{subfigure}{0.25\textwidth}
         \centering
         \includegraphics[width=\textwidth]{dist_max_2w__405.pdf}
         \caption{Layer 2 Weights}
     \end{subfigure}
     \hfill
     \begin{subfigure}{0.25\textwidth}
         \centering
         \includegraphics[width=\textwidth]{dist_spins_3w__405.pdf}
         \caption{Layer 3 Weights}
     \end{subfigure}
     \hfill
          \begin{subfigure}{0.25\textwidth}
         \centering
         \includegraphics[width=\textwidth]{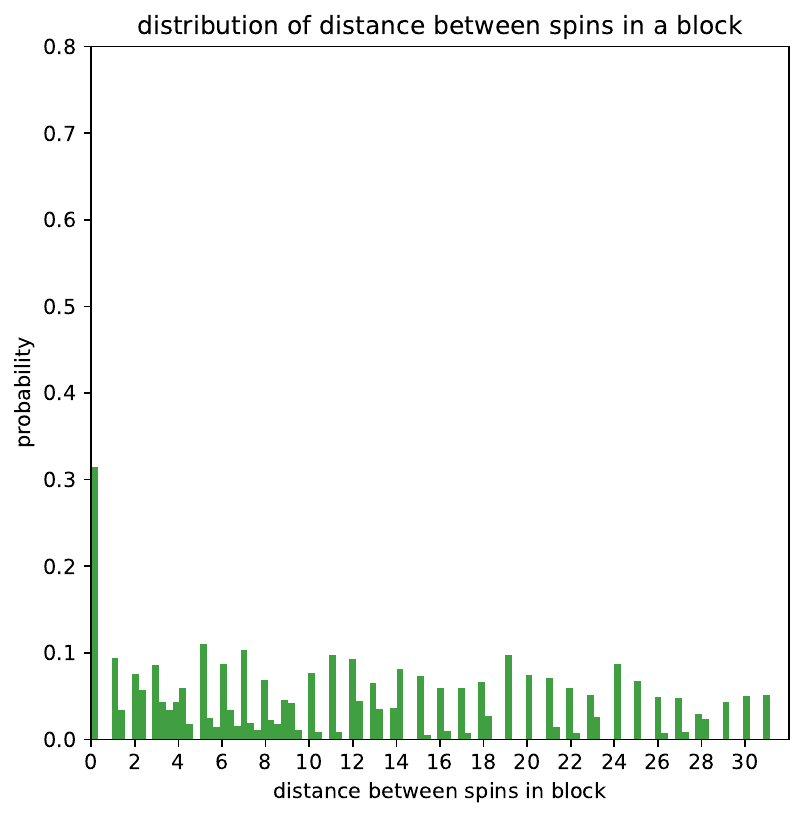}
         \caption{Layer 3 Weights}
     \end{subfigure}
     \hfill
     \begin{subfigure}{0.25\textwidth}
         \centering
         \includegraphics[width=\textwidth]{dist_max_3w__405.pdf}
         \caption{Layer 3 Weights}
     \end{subfigure}
          \hfill
          \begin{subfigure}{0.25\textwidth}
         \centering
         \includegraphics[width=\textwidth]{dist_spins_2r__405.pdf}
         \caption{Layer 2 Receptive}
     \end{subfigure}
     \hfill
     \begin{subfigure}{0.25\textwidth}
         \centering
         \includegraphics[width=\textwidth]{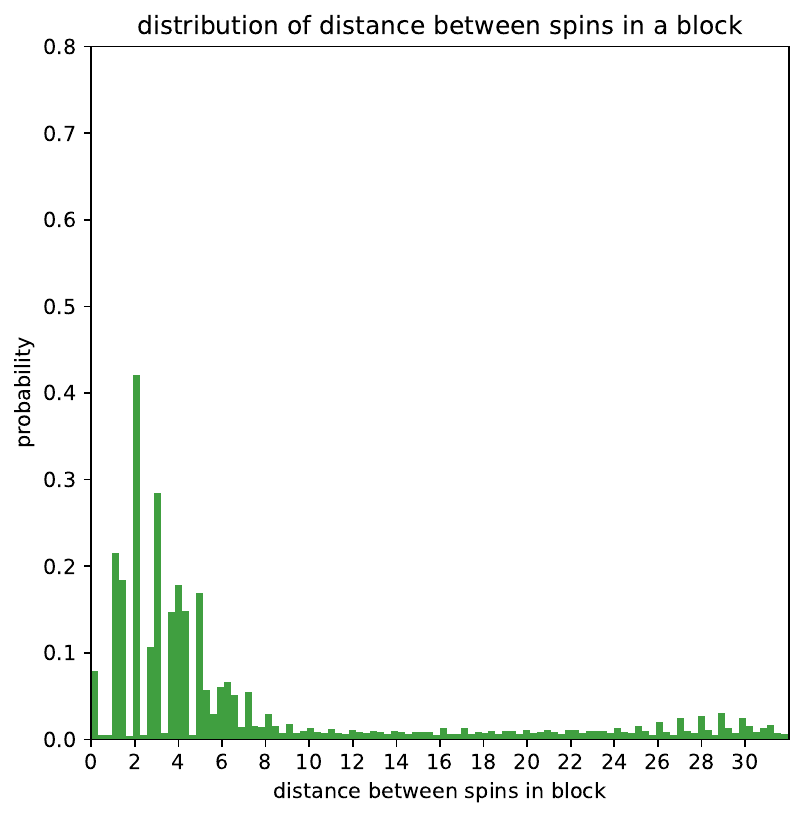}
         \caption{Layer 2 Receptive}
     \end{subfigure}
          \hfill
     \begin{subfigure}{0.25\textwidth}
         \centering
         \includegraphics[width=\textwidth]{dist_max_2r__405.pdf}
         \caption{Layer 2 Receptive}
     \end{subfigure}
     \hfill
     \begin{subfigure}{0.25\textwidth}
         \centering
         \includegraphics[width=\textwidth]{dist_spins_3r__405.pdf}
         \caption{Layer 3 Receptive}
     \end{subfigure}
     \hfill
          \begin{subfigure}{0.25\textwidth}
         \centering
         \includegraphics[width=\textwidth]{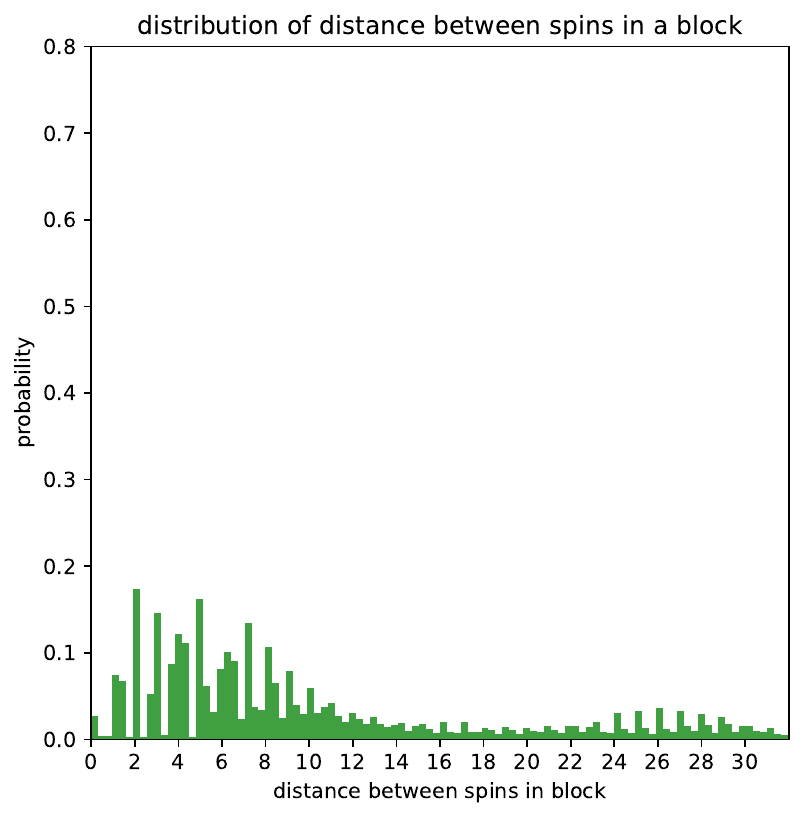}
         \caption{Layer 3 Receptive}
     \end{subfigure}
     \hfill
     \begin{subfigure}{0.25\textwidth}
         \centering
         \includegraphics[width=\textwidth]{dist_max_3r__405.pdf}
         \caption{Layer 3 Receptive}
     \end{subfigure}
     \hfill
        \caption{Block analysis plot for $\beta = .405.$}
        \label{fig:dist_analysis_.405}
\end{figure}

\clearpage

\begin{figure}[h!]
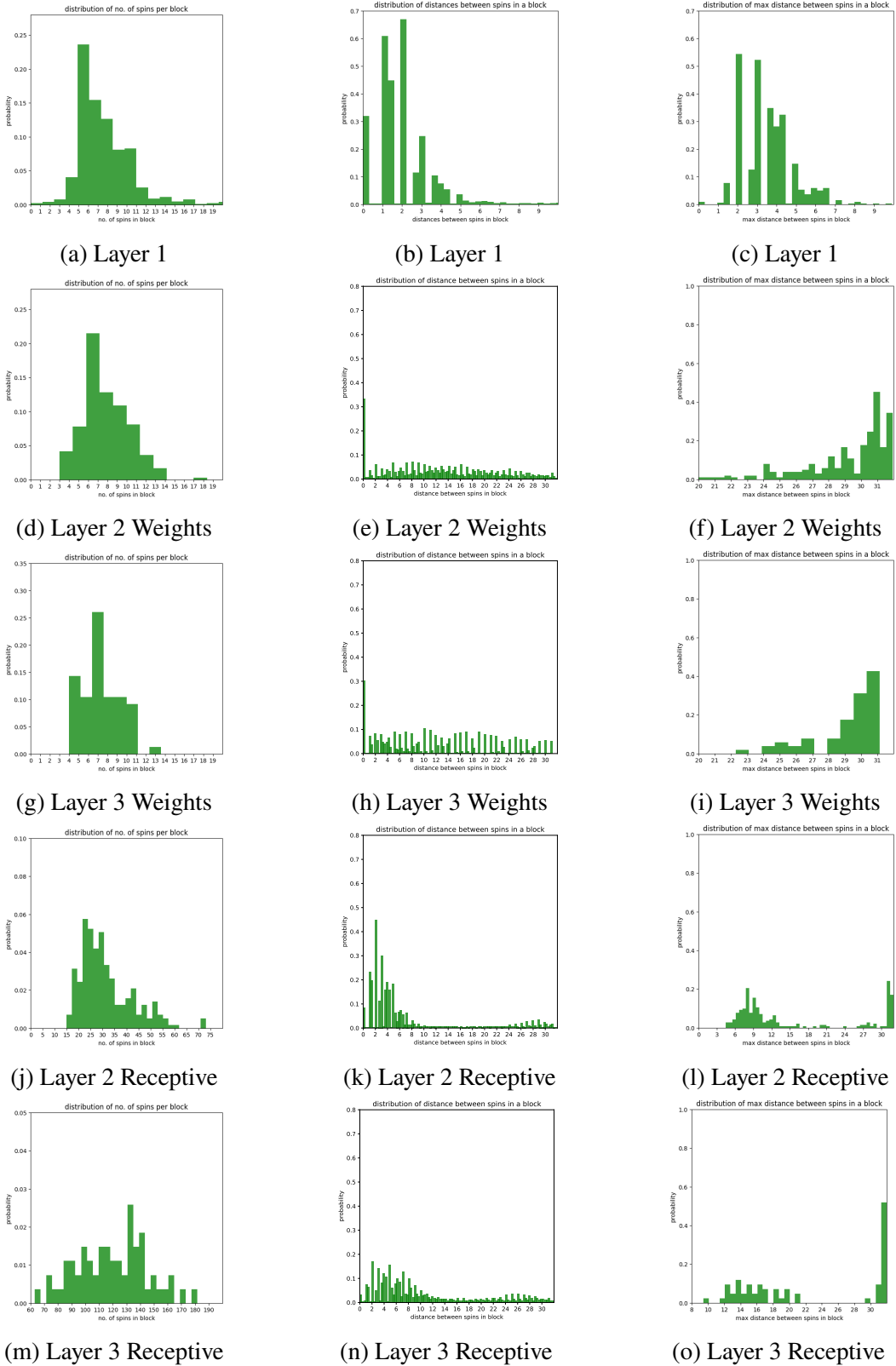

     \centering
     \begin{subfigure}{0.25\textwidth}
         \centering
         \includegraphics[width=\textwidth]{dist_spins_1__41.pdf}
         \caption{Layer 1}
     \end{subfigure}
     \hfill
     \begin{subfigure}{0.25\textwidth}
         \centering
         \includegraphics[width=\textwidth]{dist_1__41.pdf}
         \caption{Layer 1}
     \end{subfigure}
     \hfill
     \begin{subfigure}{0.25\textwidth}
         \centering
         \includegraphics[width=\textwidth]{dist_max_1__41.pdf}
         \caption{Layer 1}
     \end{subfigure}
     \hfill
     \begin{subfigure}{0.25\textwidth}
         \centering
         \includegraphics[width=\textwidth]{dist_spins_2w__41.pdf}
         \caption{Layer 2 Weights}
     \end{subfigure}
     \hfill
     \begin{subfigure}{0.25\textwidth}
         \centering
         \includegraphics[width=\textwidth]{dists_2w_0_41_.pdf}
         \caption{Layer 2 Weights}
     \end{subfigure}
          \hfill
     \begin{subfigure}{0.25\textwidth}
         \centering
         \includegraphics[width=\textwidth]{dist_max_2w__41.pdf}
         \caption{Layer 2 Weights}
     \end{subfigure}
     \hfill
     \begin{subfigure}{0.25\textwidth}
         \centering
         \includegraphics[width=\textwidth]{dist_spins_3w__41.pdf}
         \caption{Layer 3 Weights}
     \end{subfigure}
     \hfill
          \begin{subfigure}{0.25\textwidth}
         \centering
         \includegraphics[width=\textwidth]{dists_3w_0_41_.pdf}
         \caption{Layer 3 Weights}
     \end{subfigure}
     \hfill
     \begin{subfigure}{0.25\textwidth}
         \centering
         \includegraphics[width=\textwidth]{dist_max_3w__41.pdf}
         \caption{Layer 3 Weights}
     \end{subfigure}
          \hfill
          \begin{subfigure}{0.25\textwidth}
         \centering
         \includegraphics[width=\textwidth]{dist_spins_2r__41.pdf}
         \caption{Layer 2 Receptive}
     \end{subfigure}
     \hfill
     \begin{subfigure}{0.25\textwidth}
         \centering
         \includegraphics[width=\textwidth]{dists_2r_0_41_.pdf}
         \caption{Layer 2 Receptive}
     \end{subfigure}
          \hfill
     \begin{subfigure}{0.25\textwidth}
         \centering
         \includegraphics[width=\textwidth]{dist_max_2r__41.pdf}
         \caption{Layer 2 Receptive}
     \end{subfigure}
     \hfill
     \begin{subfigure}{0.25\textwidth}
         \centering
         \includegraphics[width=\textwidth]{dist_spins_3r__41.pdf}
         \caption{Layer 3 Receptive}
     \end{subfigure}
     \hfill
          \begin{subfigure}{0.25\textwidth}
         \centering
         \includegraphics[width=\textwidth]{dists_3r_0_41_.pdf}
         \caption{Layer 3 Receptive}
     \end{subfigure}
     \hfill
     \begin{subfigure}{0.25\textwidth}
         \centering
         \includegraphics[width=\textwidth]{dist_max_3r__41.pdf}
         \caption{Layer 3 Receptive}
     \end{subfigure}
     \hfill
        \caption{Block analysis plot for $\beta = .41.$}
        \label{fig:dist_analysis_.41}
\end{figure}

\clearpage

\begin{figure}[h!]
     \centering
     \begin{subfigure}{0.25\textwidth}
         \centering
         \includegraphics[width=\textwidth]{dist_spins_1__415.pdf}
         \caption{Layer 1}
     \end{subfigure}
     \hfill
     \begin{subfigure}{0.25\textwidth}
         \centering
         \includegraphics[width=\textwidth]{dist_1__415.pdf}
         \caption{Layer 1}
     \end{subfigure}
     \hfill
     \begin{subfigure}{0.25\textwidth}
         \centering
         \includegraphics[width=\textwidth]{dist_max_1__415.pdf}
         \caption{Layer 1}
     \end{subfigure}
     \hfill
     \begin{subfigure}{0.25\textwidth}
         \centering
         \includegraphics[width=\textwidth]{dist_spins_2w__415.pdf}
         \caption{Layer 2 Weights}
     \end{subfigure}
     \hfill
     \begin{subfigure}{0.25\textwidth}
         \centering
         \includegraphics[width=\textwidth]{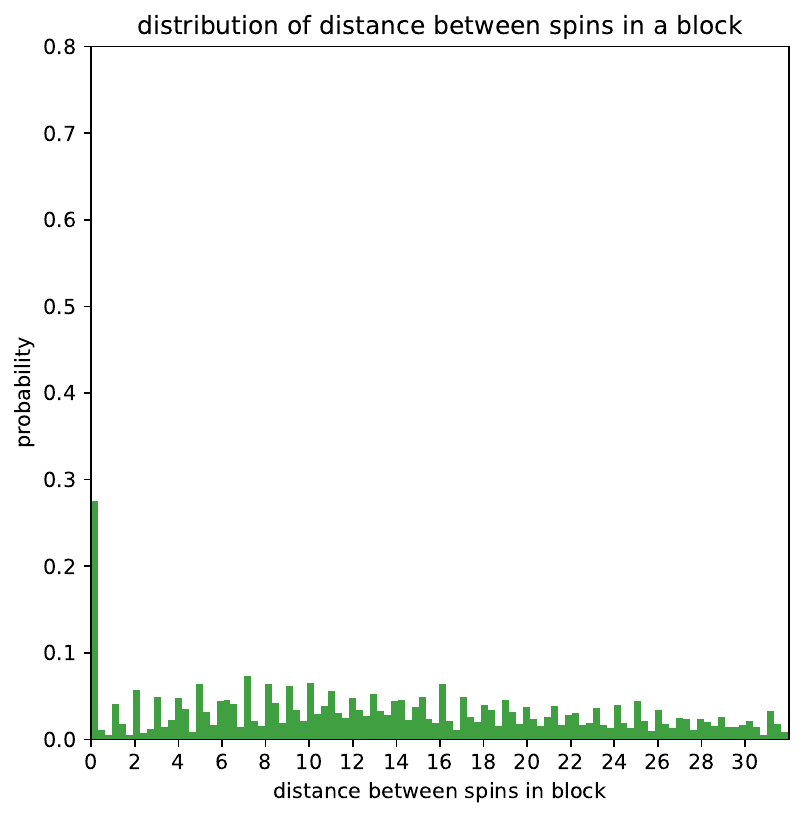}
         \caption{Layer 2 Weights}
     \end{subfigure}
          \hfill
     \begin{subfigure}{0.25\textwidth}
         \centering
         \includegraphics[width=\textwidth]{dist_max_2w__415.pdf}
         \caption{Layer 2 Weights}
     \end{subfigure}
     \hfill
     \begin{subfigure}{0.25\textwidth}
         \centering
         \includegraphics[width=\textwidth]{dist_spins_3w__415.pdf}
         \caption{Layer 3 Weights}
     \end{subfigure}
     \hfill
          \begin{subfigure}{0.25\textwidth}
         \centering
         \includegraphics[width=\textwidth]{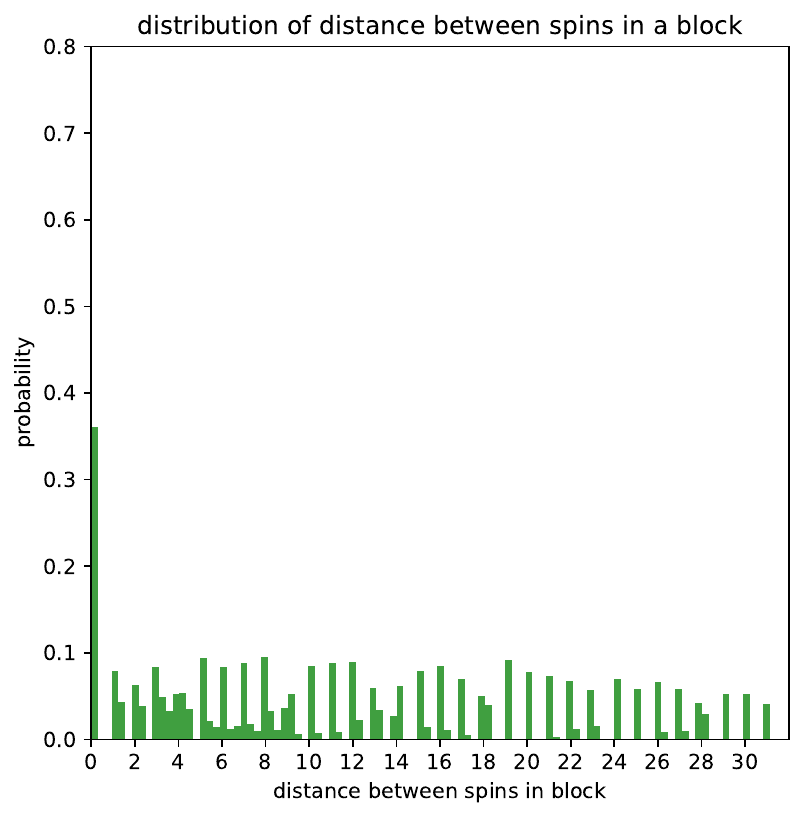}
         \caption{Layer 3 Weights}
     \end{subfigure}
     \hfill
     \begin{subfigure}{0.25\textwidth}
         \centering
         \includegraphics[width=\textwidth]{dist_max_3w__415.pdf}
         \caption{Layer 3 Weights}
     \end{subfigure}
          \hfill
          \begin{subfigure}{0.25\textwidth}
         \centering
         \includegraphics[width=\textwidth]{dist_spins_2r__415.pdf}
         \caption{Layer 2 Receptive}
     \end{subfigure}
     \hfill
     \begin{subfigure}{0.25\textwidth}
         \centering
         \includegraphics[width=\textwidth]{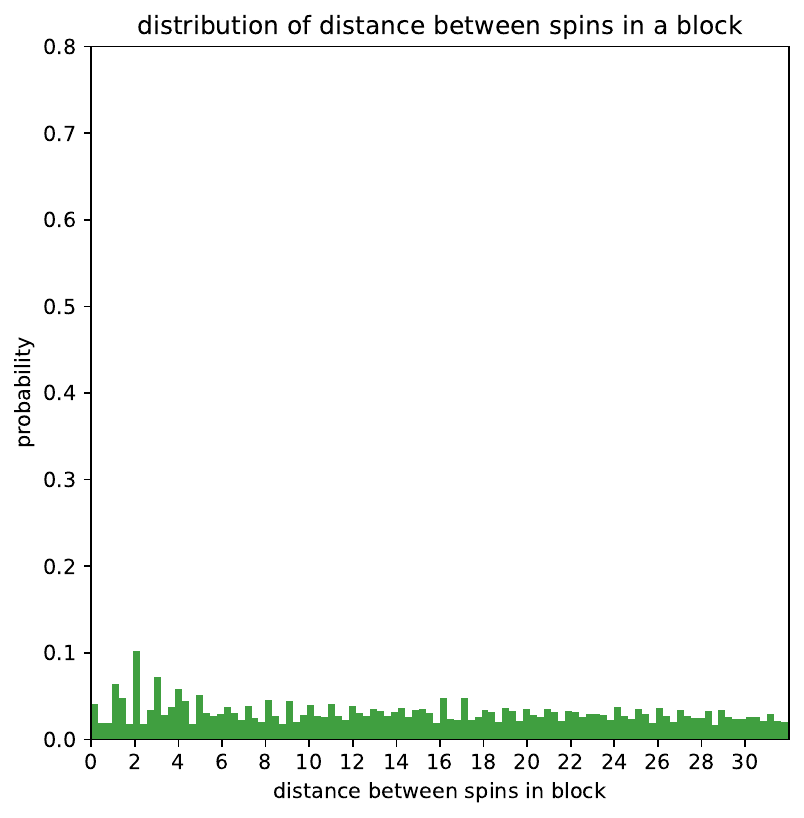}
         \caption{Layer 2 Receptive}
     \end{subfigure}
          \hfill
     \begin{subfigure}{0.25\textwidth}
         \centering
         \includegraphics[width=\textwidth]{dist_max_2r__415.pdf}
         \caption{Layer 2 Receptive}
     \end{subfigure}
     \hfill
     \begin{subfigure}{0.25\textwidth}
         \centering
         \includegraphics[width=\textwidth]{dist_spins_3r__415.pdf}
         \caption{Layer 3 Receptive}
     \end{subfigure}
     \hfill
          \begin{subfigure}{0.25\textwidth}
         \centering
         \includegraphics[width=\textwidth]{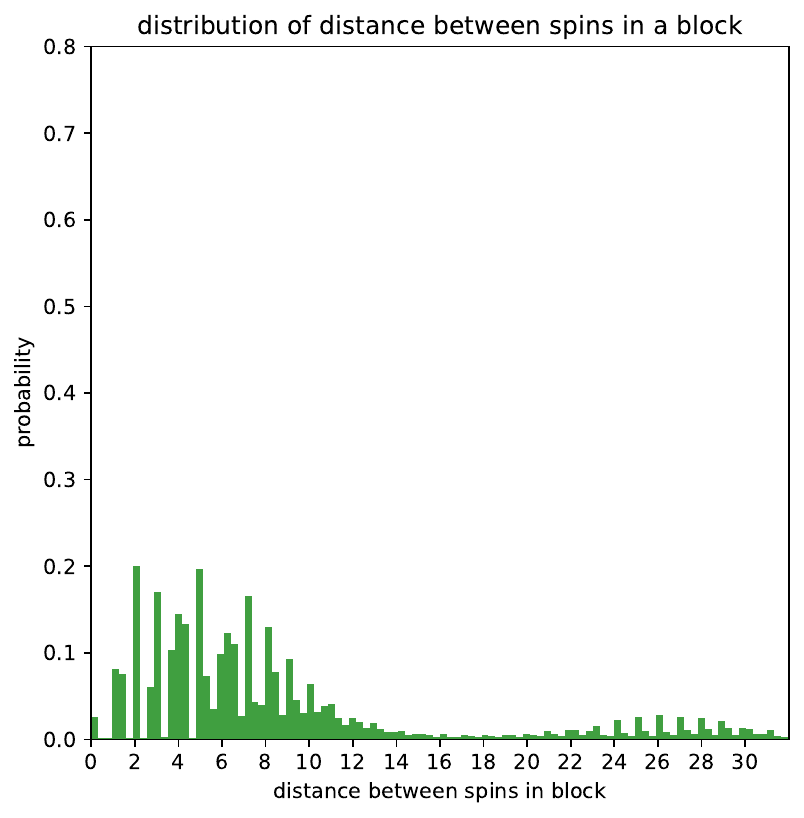}
         \caption{Layer 3 Receptive}
     \end{subfigure}
     \hfill
     \begin{subfigure}{0.25\textwidth}
         \centering
         \includegraphics[width=\textwidth]{dist_max_3r__415.pdf}
         \caption{Layer 3 Receptive}
     \end{subfigure}
     \hfill
        \caption{Block analysis plot for $\beta = .415.$}
        \label{fig:dist_analysis_.415}
\end{figure}

\clearpage

\begin{figure}[h!]
     \centering
     \begin{subfigure}{0.25\textwidth}
         \centering
         \includegraphics[width=\textwidth]{dist_spins_1__42.pdf}
         \caption{Layer 1}
     \end{subfigure}
     \hfill
     \begin{subfigure}{0.25\textwidth}
         \centering
         \includegraphics[width=\textwidth]{dist_1__42.pdf}
         \caption{Layer 1}
     \end{subfigure}
     \hfill
     \begin{subfigure}{0.25\textwidth}
         \centering
         \includegraphics[width=\textwidth]{dist_max_1__42.pdf}
         \caption{Layer 1}
     \end{subfigure}
     \hfill
     \begin{subfigure}{0.25\textwidth}
         \centering
         \includegraphics[width=\textwidth]{dist_spins_2w__42.pdf}
         \caption{Layer 2 Weights}
     \end{subfigure}
     \hfill
     \begin{subfigure}{0.25\textwidth}
         \centering
         \includegraphics[width=\textwidth]{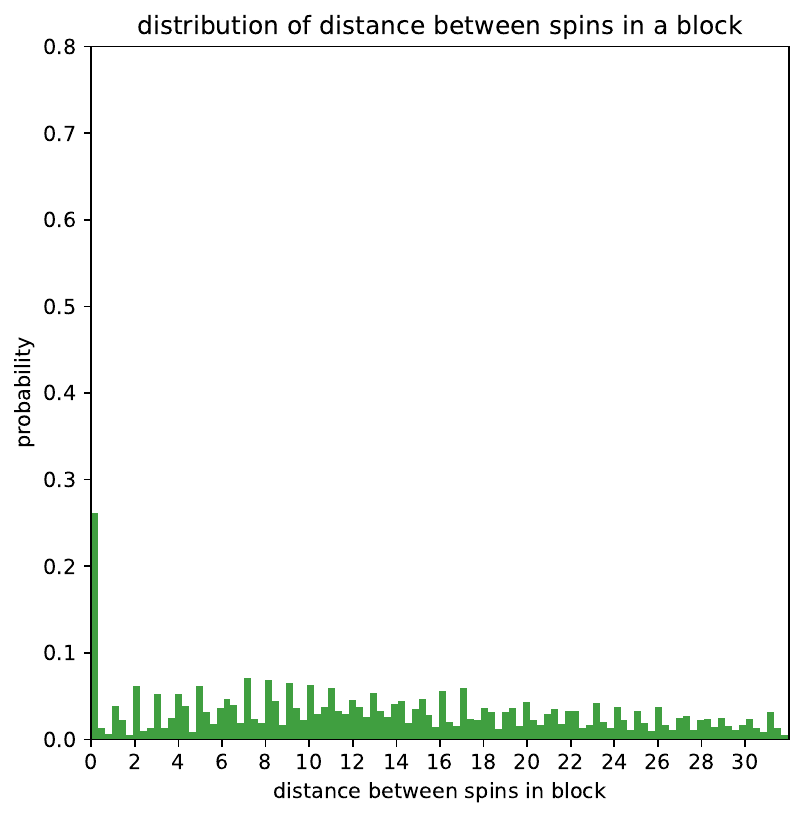}
         \caption{Layer 2 Weights}
     \end{subfigure}
          \hfill
     \begin{subfigure}{0.25\textwidth}
         \centering
         \includegraphics[width=\textwidth]{dist_max_2w__42.pdf}
         \caption{Layer 2 Weights}
     \end{subfigure}
     \hfill
     \begin{subfigure}{0.25\textwidth}
         \centering
         \includegraphics[width=\textwidth]{dist_spins_3w__42.pdf}
         \caption{Layer 3 Weights}
     \end{subfigure}
     \hfill
          \begin{subfigure}{0.25\textwidth}
         \centering
         \includegraphics[width=\textwidth]{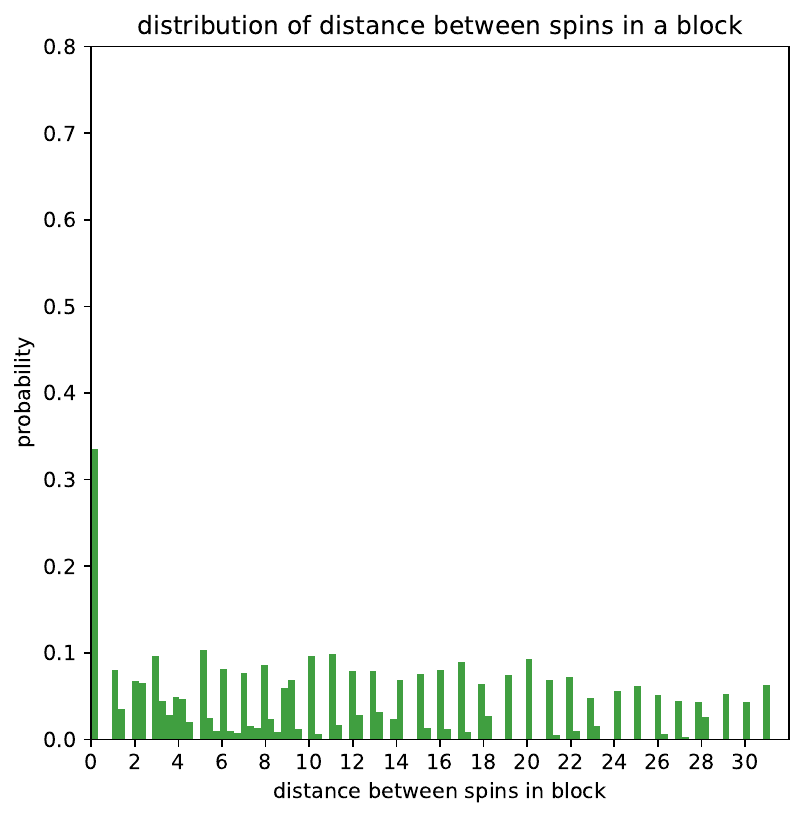}
         \caption{Layer 3 Weights}
     \end{subfigure}
     \hfill
     \begin{subfigure}{0.25\textwidth}
         \centering
         \includegraphics[width=\textwidth]{dist_max_3w__42.pdf}
         \caption{Layer 3 Weights}
     \end{subfigure}
          \hfill
          \begin{subfigure}{0.25\textwidth}
         \centering
         \includegraphics[width=\textwidth]{dist_spins_2r__42.pdf}
         \caption{Layer 2 Receptive}
     \end{subfigure}
     \hfill
     \begin{subfigure}{0.25\textwidth}
         \centering
         \includegraphics[width=\textwidth]{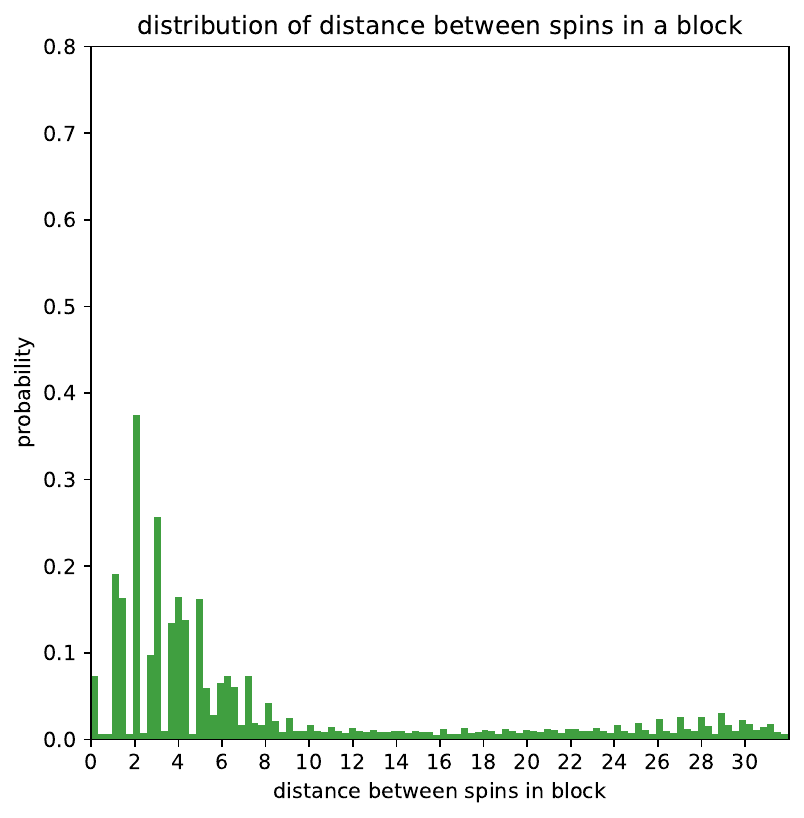}
         \caption{Layer 2 Receptive}
     \end{subfigure}
          \hfill
     \begin{subfigure}{0.25\textwidth}
         \centering
         \includegraphics[width=\textwidth]{dist_max_2r__42.pdf}
         \caption{Layer 2 Receptive}
     \end{subfigure}
     \hfill
     \begin{subfigure}{0.25\textwidth}
         \centering
         \includegraphics[width=\textwidth]{dist_spins_3r__42.pdf}
         \caption{Layer 3 Receptive}
     \end{subfigure}
     \hfill
          \begin{subfigure}{0.25\textwidth}
         \centering
         \includegraphics[width=\textwidth]{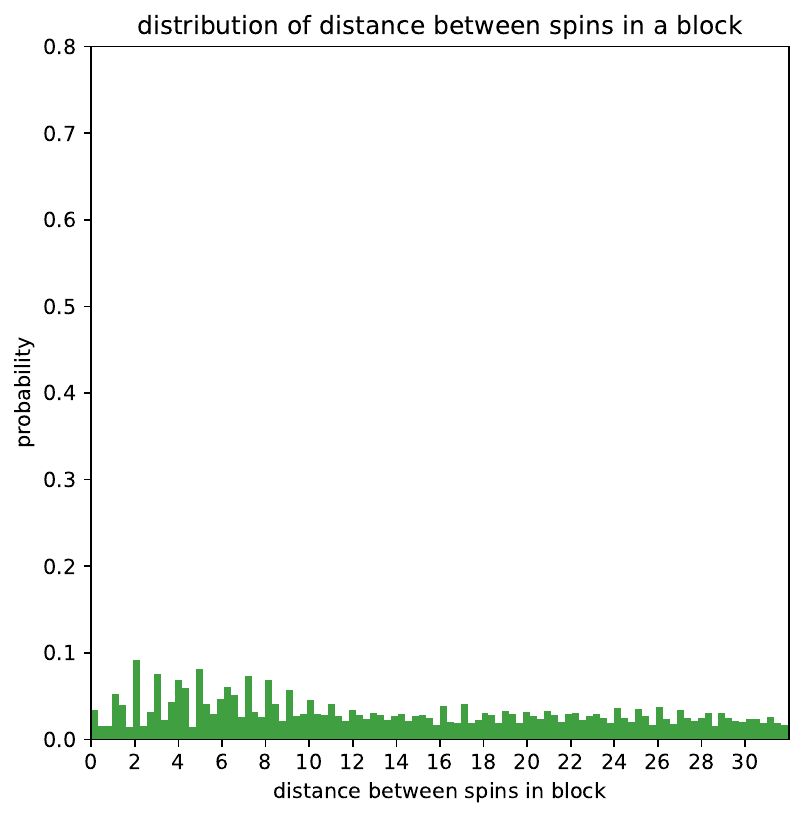}
         \caption{Layer 3 Receptive}
     \end{subfigure}
     \hfill
     \begin{subfigure}{0.25\textwidth}
         \centering
         \includegraphics[width=\textwidth]{dist_max_3r__42.pdf}
         \caption{Layer 3 Receptive}
     \end{subfigure}
     \hfill
        \caption{Block analysis plot for $\beta = .42.$}
        \label{fig:dist_analysis_.42}
\end{figure}

\clearpage

\begin{figure}[h!]
     \centering
     \begin{subfigure}{0.25\textwidth}
         \centering
         \includegraphics[width=\textwidth]{dist_spins_1__425.pdf}
         \caption{Layer 1}
     \end{subfigure}
     \hfill
     \begin{subfigure}{0.25\textwidth}
         \centering
         \includegraphics[width=\textwidth]{dist_1__425.pdf}
         \caption{Layer 1}
     \end{subfigure}
     \hfill
     \begin{subfigure}{0.25\textwidth}
         \centering
         \includegraphics[width=\textwidth]{dist_max_1__425.pdf}
         \caption{Layer 1}
     \end{subfigure}
     \hfill
     \begin{subfigure}{0.25\textwidth}
         \centering
         \includegraphics[width=\textwidth]{dist_spins_2w__425.pdf}
         \caption{Layer 2 Weights}
     \end{subfigure}
     \hfill
     \begin{subfigure}{0.25\textwidth}
         \centering
         \includegraphics[width=\textwidth]{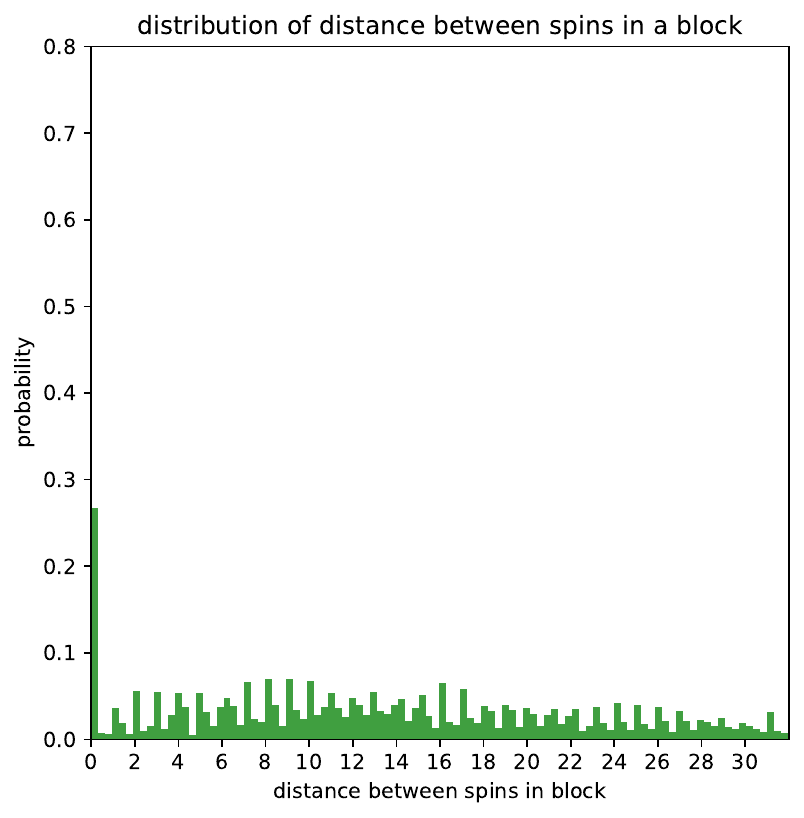}
         \caption{Layer 2 Weights}
     \end{subfigure}
          \hfill
     \begin{subfigure}{0.25\textwidth}
         \centering
         \includegraphics[width=\textwidth]{dist_max_2w__425.pdf}
         \caption{Layer 2 Weights}
     \end{subfigure}
     \hfill
     \begin{subfigure}{0.25\textwidth}
         \centering
         \includegraphics[width=\textwidth]{dist_spins_3w__425.pdf}
         \caption{Layer 3 Weights}
     \end{subfigure}
     \hfill
          \begin{subfigure}{0.25\textwidth}
         \centering
         \includegraphics[width=\textwidth]{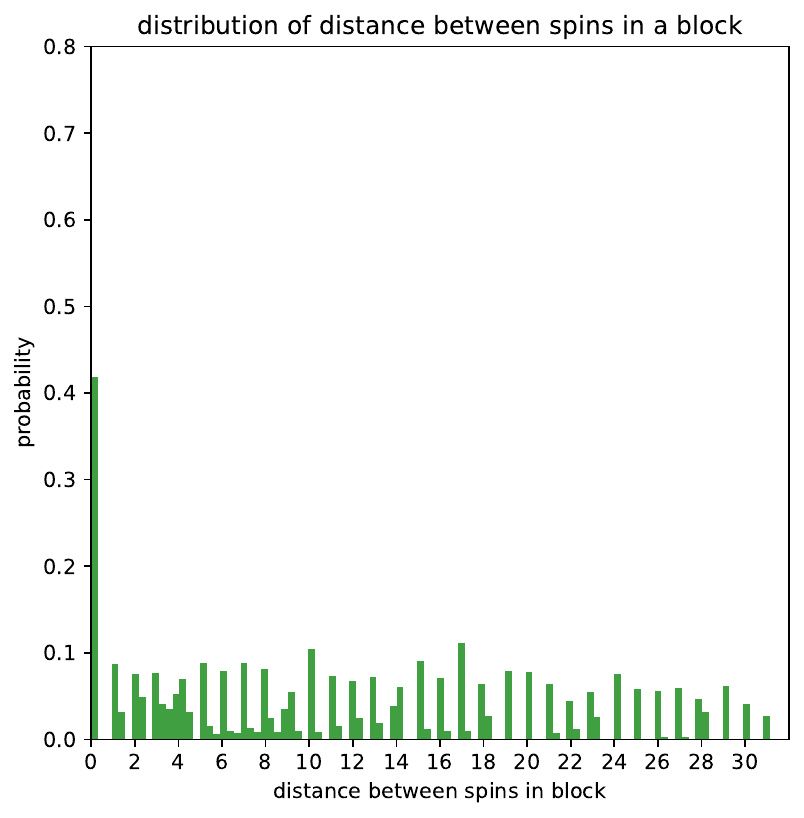}
         \caption{Layer 3 Weights}
     \end{subfigure}
     \hfill
     \begin{subfigure}{0.25\textwidth}
         \centering
         \includegraphics[width=\textwidth]{dist_max_3w__425.pdf}
         \caption{Layer 3 Weights}
     \end{subfigure}
          \hfill
          \begin{subfigure}{0.25\textwidth}
         \centering
         \includegraphics[width=\textwidth]{dist_spins_2r__425.pdf}
         \caption{Layer 2 Receptive}
     \end{subfigure}
     \hfill
     \begin{subfigure}{0.25\textwidth}
         \centering
         \includegraphics[width=\textwidth]{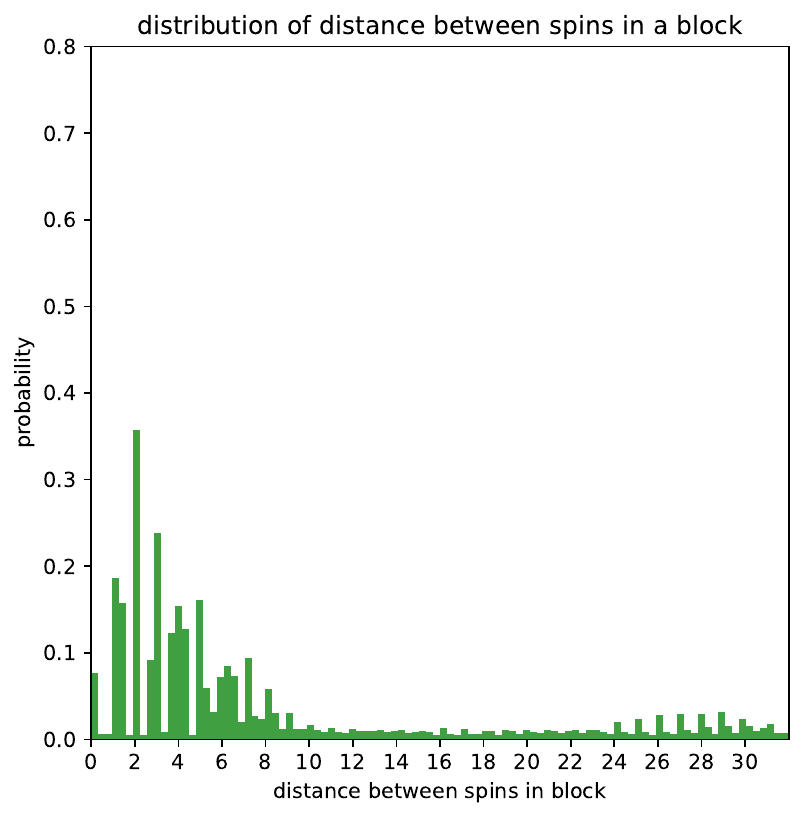}
         \caption{Layer 2 Receptive}
     \end{subfigure}
          \hfill
     \begin{subfigure}{0.25\textwidth}
         \centering
         \includegraphics[width=\textwidth]{dist_max_2r__425.pdf}
         \caption{Layer 2 Receptive}
     \end{subfigure}
     \hfill
     \begin{subfigure}{0.25\textwidth}
         \centering
         \includegraphics[width=\textwidth]{dist_spins_3r__425.pdf}
         \caption{Layer 3 Receptive}
     \end{subfigure}
     \hfill
          \begin{subfigure}{0.25\textwidth}
         \centering
         \includegraphics[width=\textwidth]{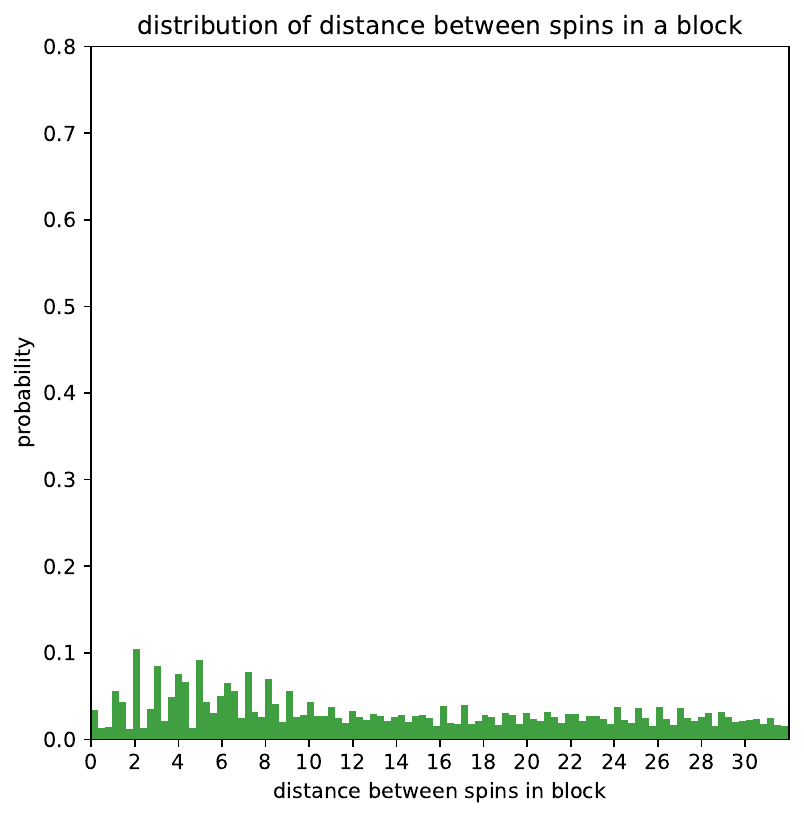}
         \caption{Layer 3 Receptive}
     \end{subfigure}
     \hfill
     \begin{subfigure}{0.25\textwidth}
         \centering
         \includegraphics[width=\textwidth]{dist_max_3r__425.pdf}
         \caption{Layer 3 Receptive}
     \end{subfigure}
     \hfill
        \caption{Block analysis plot for $\beta = .425.$}
        \label{fig:dist_analysis_.425}
\end{figure}

\clearpage

\begin{figure}[h!]
     \centering
     \begin{subfigure}{0.25\textwidth}
         \centering
         \includegraphics[width=\textwidth]{dist_spins_1__43.pdf}
         \caption{Layer 1}
     \end{subfigure}
     \hfill
     \begin{subfigure}{0.25\textwidth}
         \centering
         \includegraphics[width=\textwidth]{dist_1__43.pdf}
         \caption{Layer 1}
     \end{subfigure}
     \hfill
     \begin{subfigure}{0.25\textwidth}
         \centering
         \includegraphics[width=\textwidth]{dist_max_1__43.pdf}
         \caption{Layer 1}
     \end{subfigure}
     \hfill
     \begin{subfigure}{0.25\textwidth}
         \centering
         \includegraphics[width=\textwidth]{dist_spins_2w__43.pdf}
         \caption{Layer 2 Weights}
     \end{subfigure}
     \hfill
     \begin{subfigure}{0.25\textwidth}
         \centering
         \includegraphics[width=\textwidth]{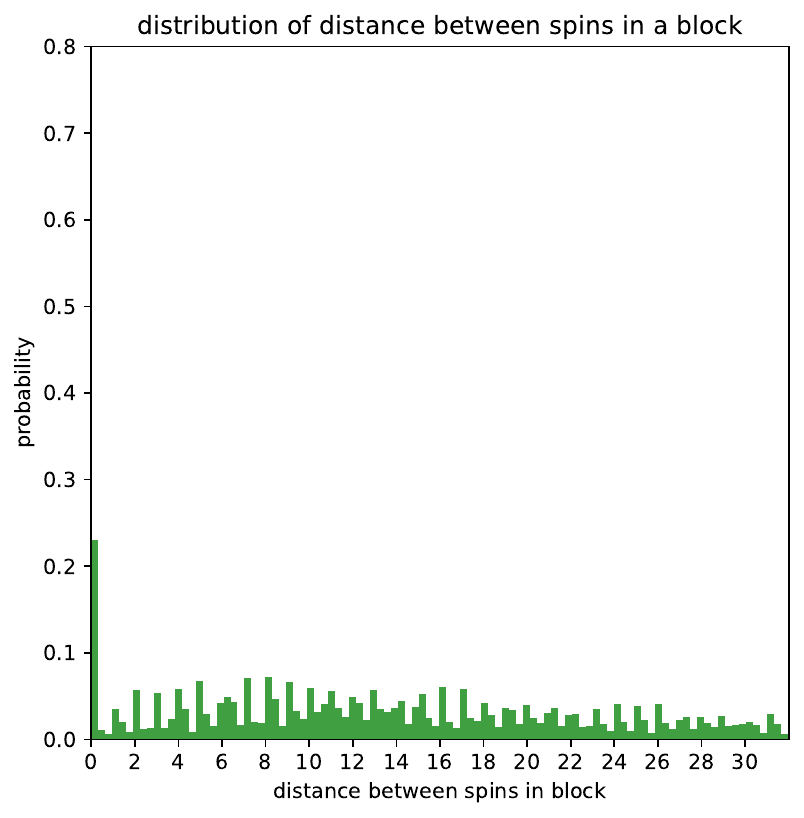}
         \caption{Layer 2 Weights}
     \end{subfigure}
          \hfill
     \begin{subfigure}{0.25\textwidth}
         \centering
         \includegraphics[width=\textwidth]{dist_max_2w__43.pdf}
         \caption{Layer 2 Weights}
     \end{subfigure}
     \hfill
     \begin{subfigure}{0.25\textwidth}
         \centering
         \includegraphics[width=\textwidth]{dist_spins_3w__43.pdf}
         \caption{Layer 3 Weights}
     \end{subfigure}
     \hfill
          \begin{subfigure}{0.25\textwidth}
         \centering
         \includegraphics[width=\textwidth]{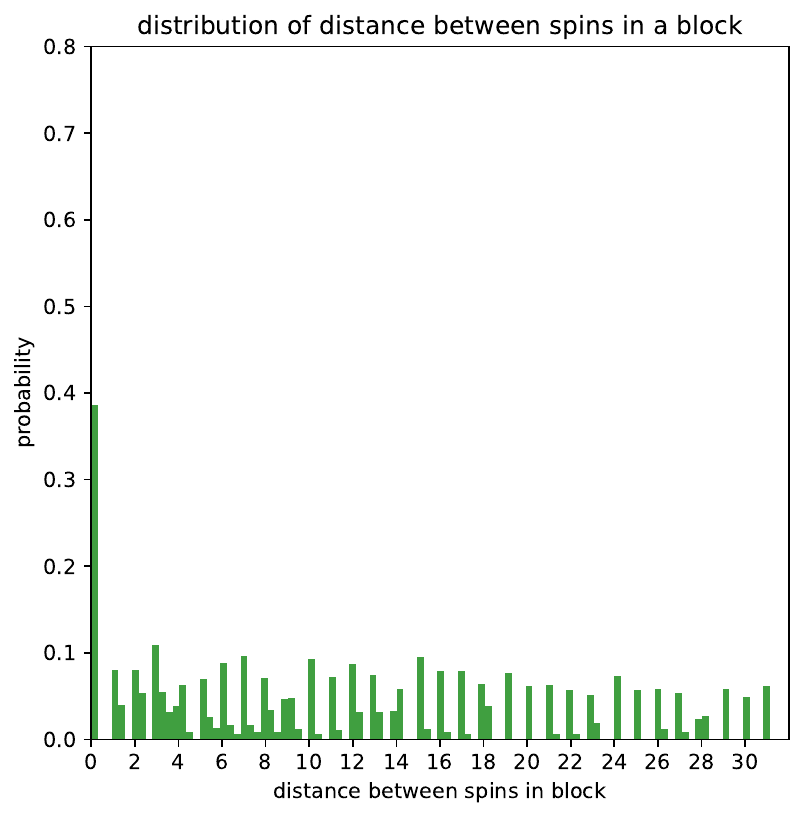}
         \caption{Layer 3 Weights}
     \end{subfigure}
     \hfill
     \begin{subfigure}{0.25\textwidth}
         \centering
         \includegraphics[width=\textwidth]{dist_max_3w__43.pdf}
         \caption{Layer 3 Receptive}
     \end{subfigure}
          \hfill
          \begin{subfigure}{0.25\textwidth}
         \centering
         \includegraphics[width=\textwidth]{dist_spins_2r__43.pdf}
         \caption{Layer 2 Receptive}
     \end{subfigure}
     \hfill
     \begin{subfigure}{0.25\textwidth}
         \centering
         \includegraphics[width=\textwidth]{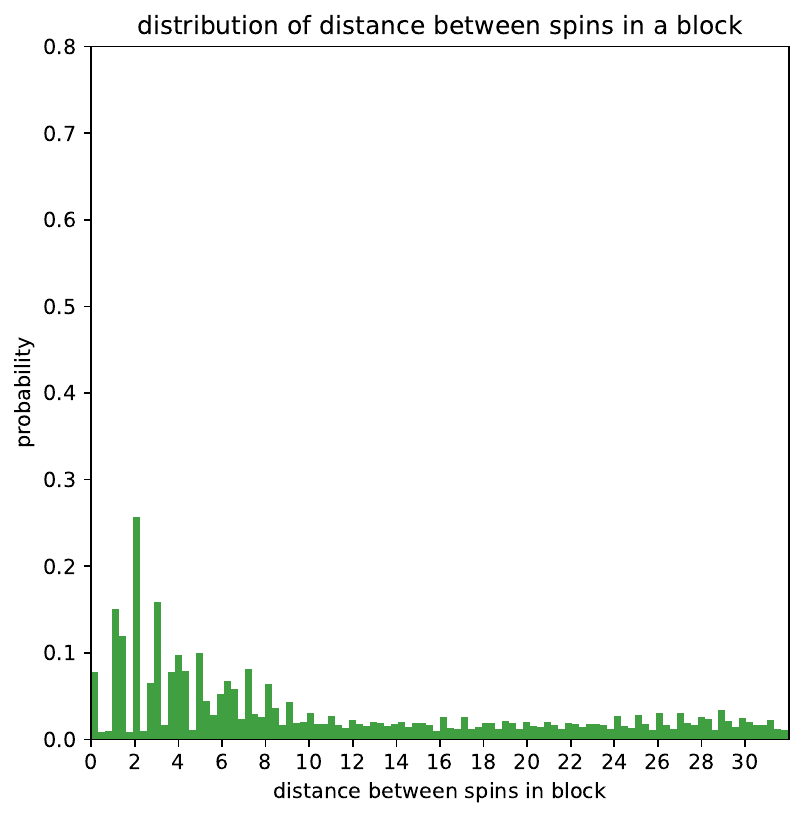}
         \caption{Layer 2 Receptive}
     \end{subfigure}
          \hfill
     \begin{subfigure}{0.25\textwidth}
         \centering
         \includegraphics[width=\textwidth]{dist_max_2r__43.pdf}
         \caption{Layer 2 Receptive}
     \end{subfigure}
     \hfill
     \begin{subfigure}{0.25\textwidth}
         \centering
         \includegraphics[width=\textwidth]{dist_spins_3r__43.pdf}
         \caption{Layer 3 Receptive}
     \end{subfigure}
     \hfill
          \begin{subfigure}{0.25\textwidth}
         \centering
         \includegraphics[width=\textwidth]{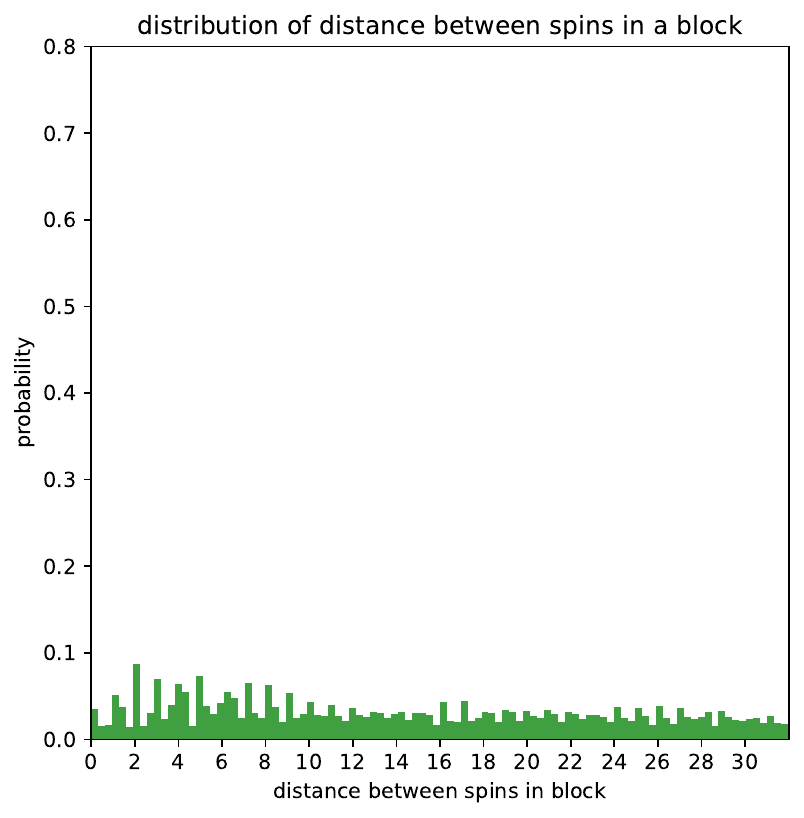}
         \caption{Layer 3 Receptive}
     \end{subfigure}
     \hfill
     \begin{subfigure}{0.25\textwidth}
         \centering
         \includegraphics[width=\textwidth]{dist_max_3r__43.pdf}
         \caption{Layer 3 Receptive}
     \end{subfigure}
     \hfill
        \caption{Block analysis plot for $\beta = .43.$}
        \label{fig:dist_analysis_.43}
\end{figure}

\clearpage

\section{Reconstruction Plots}
\label{sec:reco_plots}

This section contains plots made by using the deep learning network as an autoencoder to reconstruct the original Ising model. For various values of $\beta$ In each plot, the first row consists of the original Ising input, rows 2-4 consist of the reconstructed Ising models using 1, 2, and 3 layers respectively, and rows 5-7 do the same thing but only using the largest two receptive field values to calculate the reconstruction. These results are discussed in detail in Section \ref{sec:reconstruction}.

\begin{figure}[h!]
    \centering
    \includegraphics[width=\textwidth]{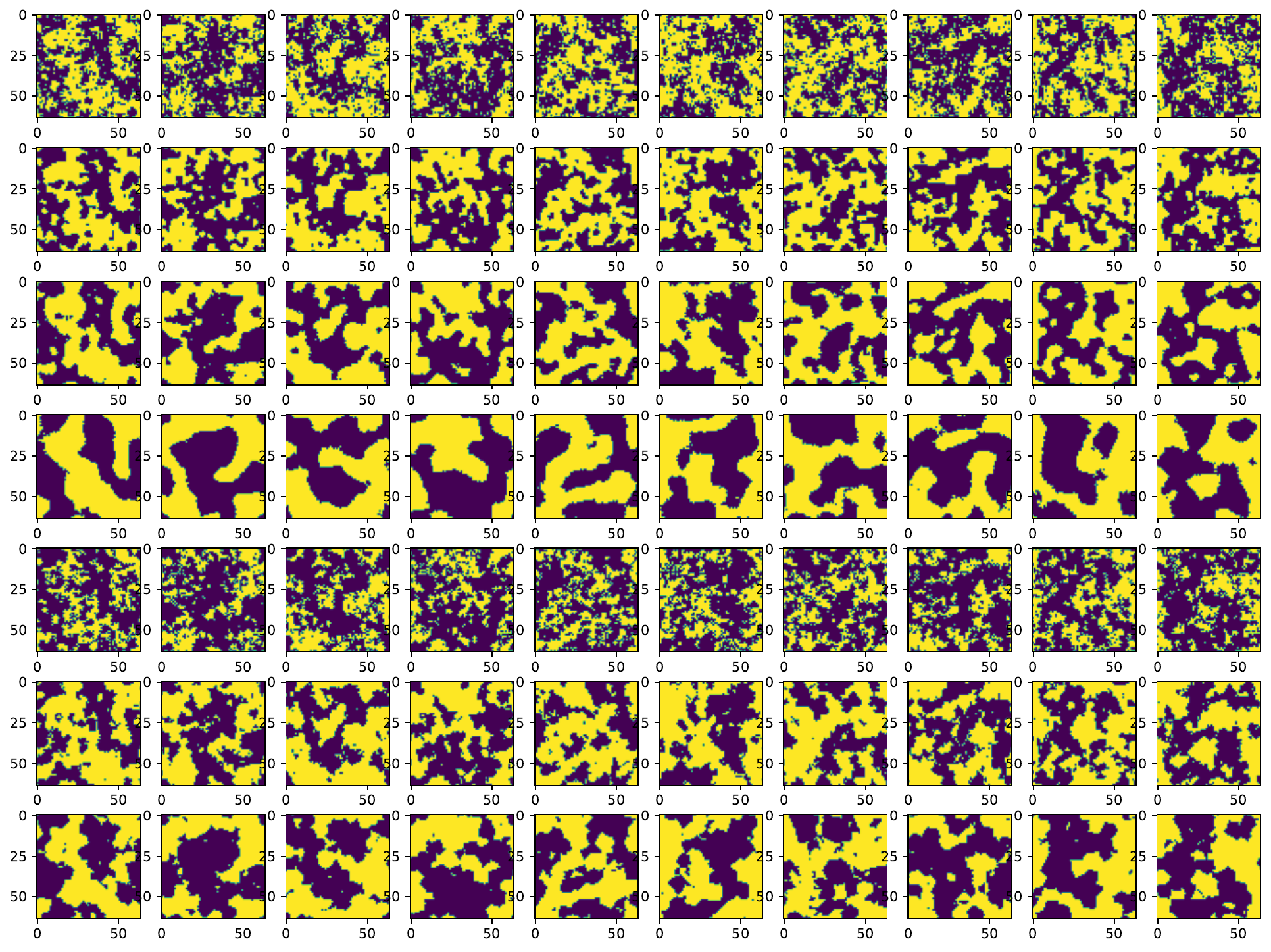}
    \caption{Reconstructed model for $\beta = .395$.}
    \label{fig:reco_.395}
\end{figure}

\begin{figure}[h!]
    \centering
    \includegraphics[width=\textwidth]{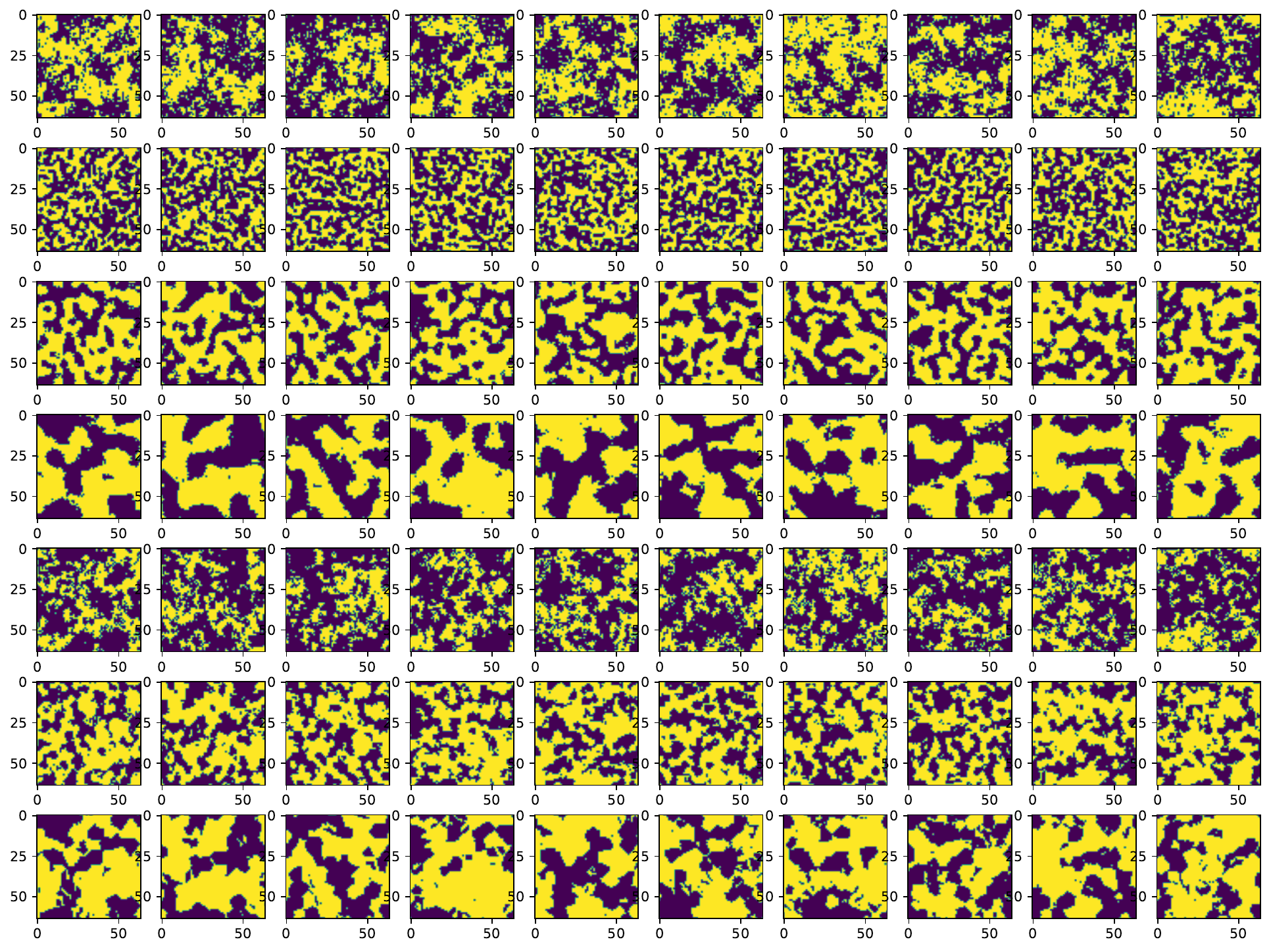}
    \caption{Reconstructed model for $\beta =. 4$.}
    \label{fig:reco_.4}
\end{figure}

\begin{figure}[h!]
    \centering
    \includegraphics[width=\textwidth]{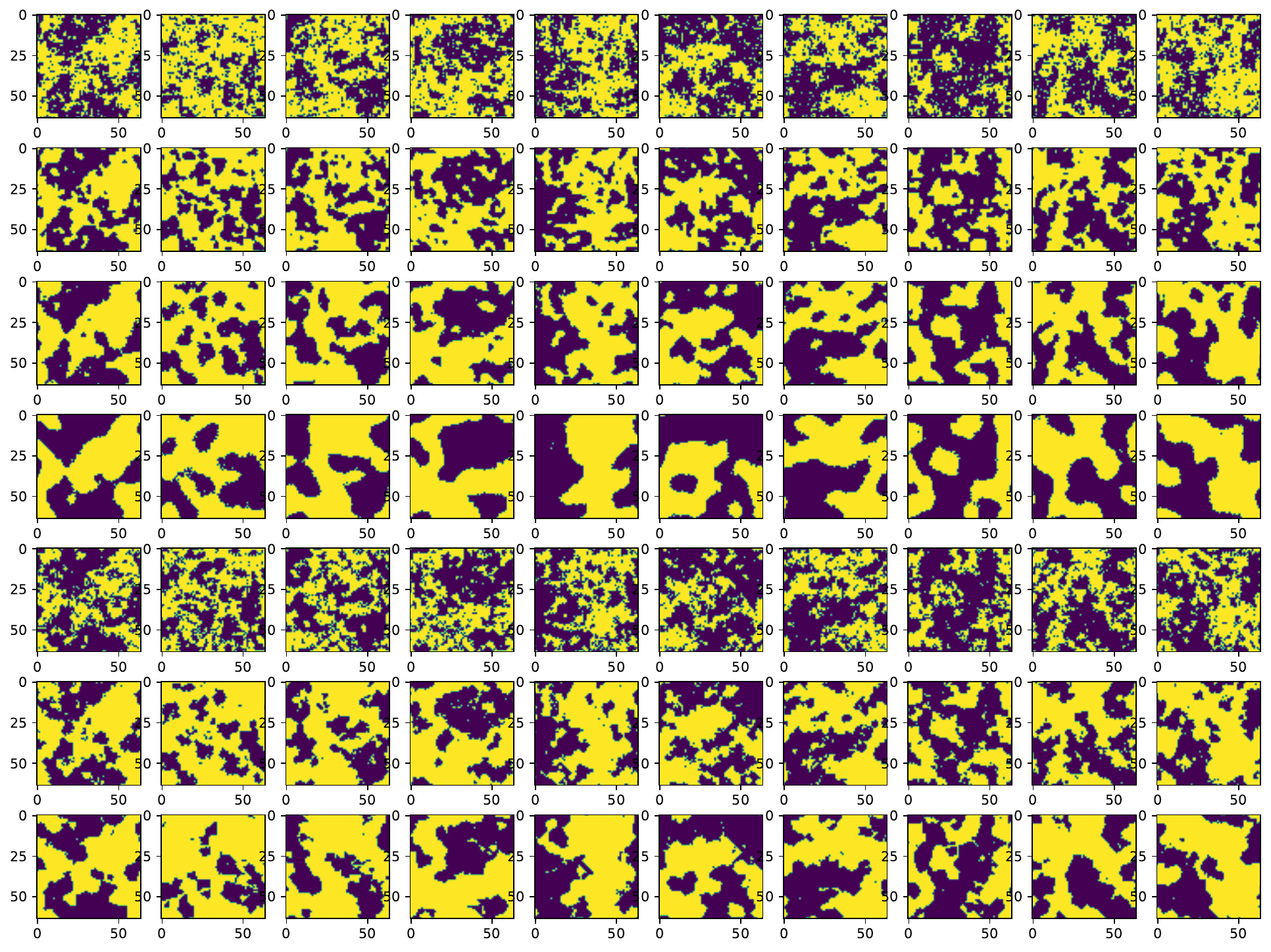}
    \caption{Reconstructed model for $\beta = .405$.}
    \label{fig:reco_.405}
\end{figure}

\begin{figure}[h!]
    \centering
    \includegraphics[width=\textwidth]{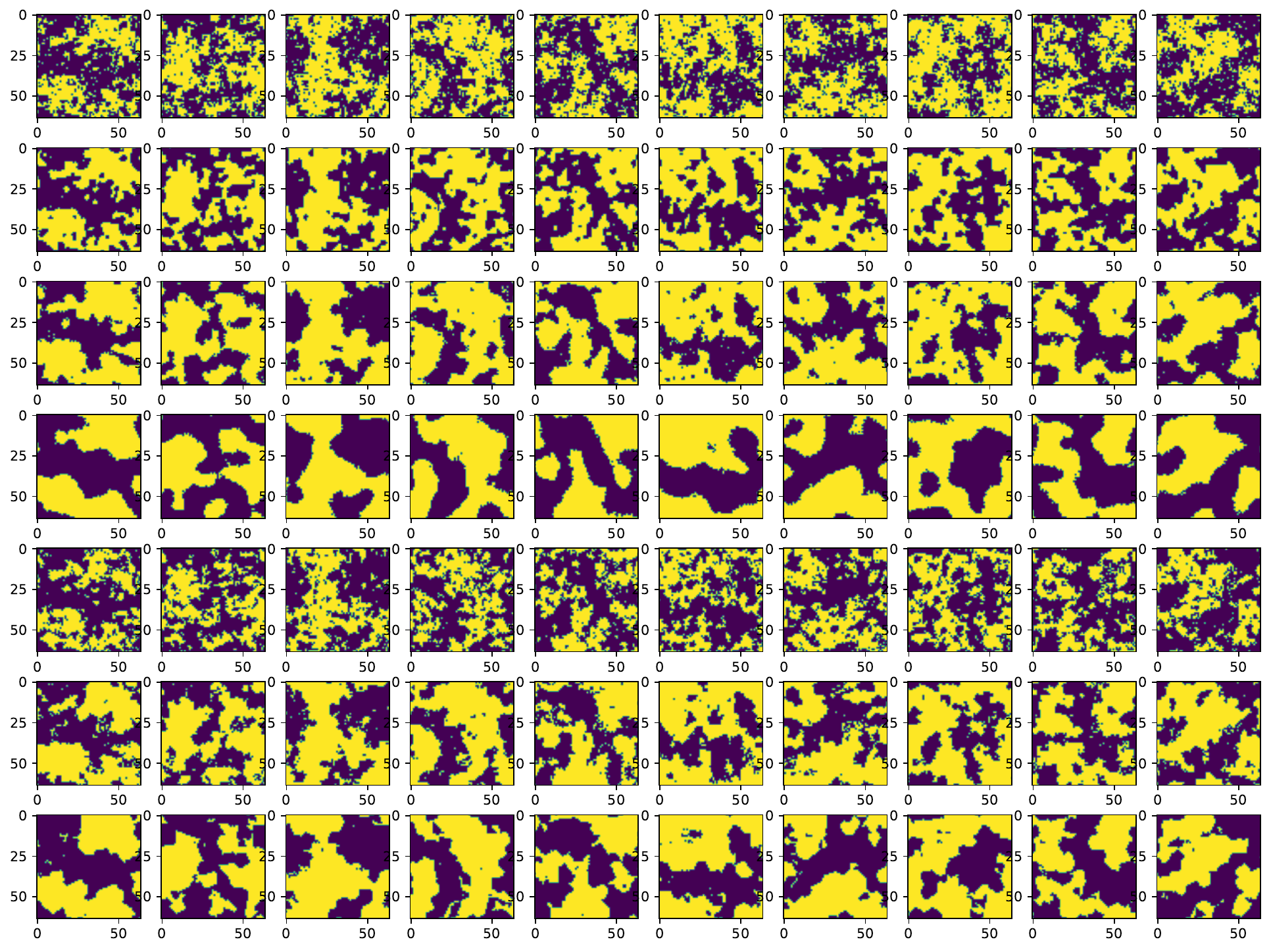}
    \caption{Reconstructed model for $\beta = .41$.}
    \label{fig:reco_.41}
\end{figure}

\begin{figure}[h!]
    \centering
    \includegraphics[width=\textwidth]{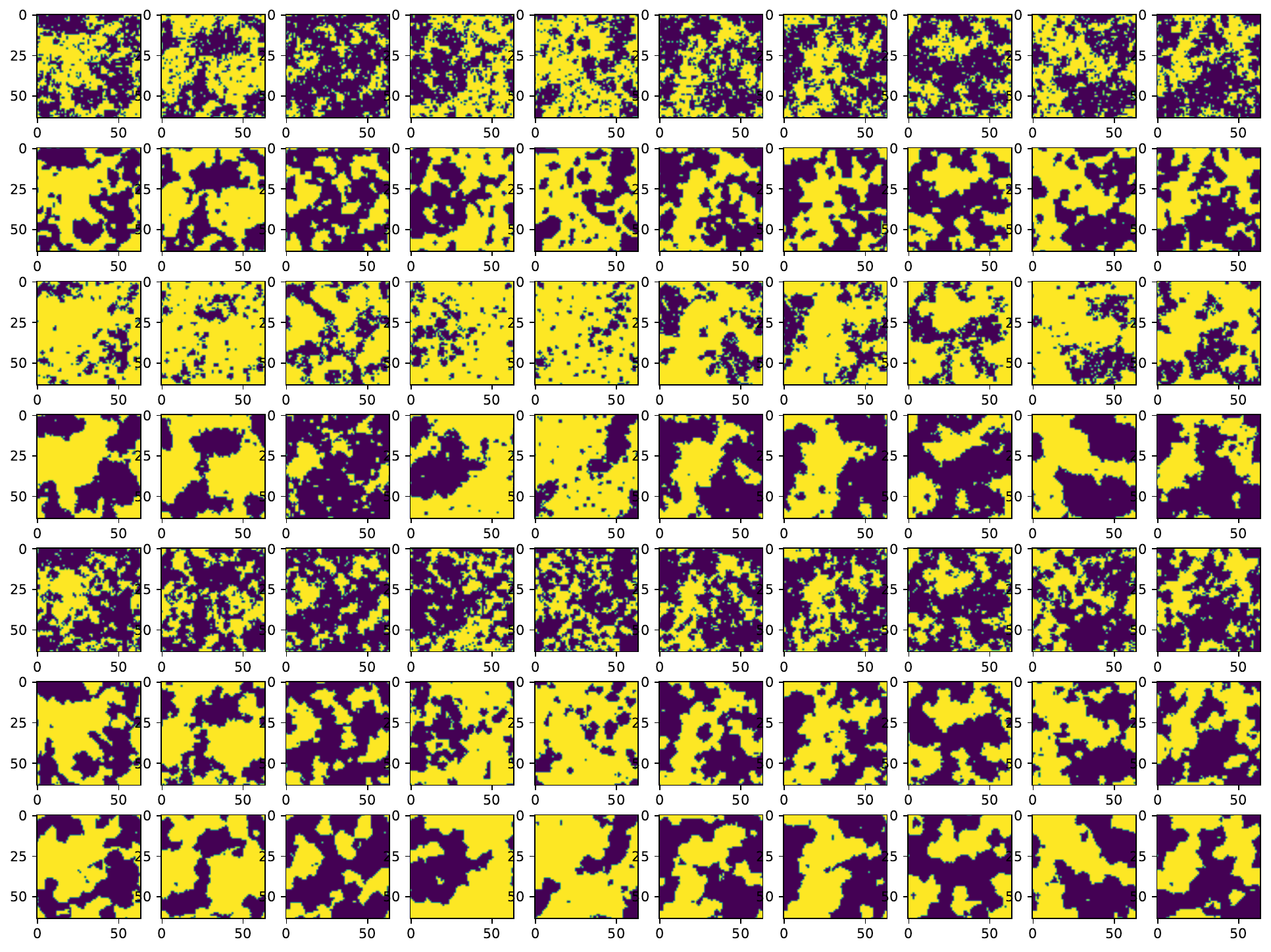}
    \caption{Reconstructed model for $\beta = .415$.}
    \label{fig:reco_.415}
\end{figure}

\begin{figure}[h!]
    \centering
    \includegraphics[width=\textwidth]{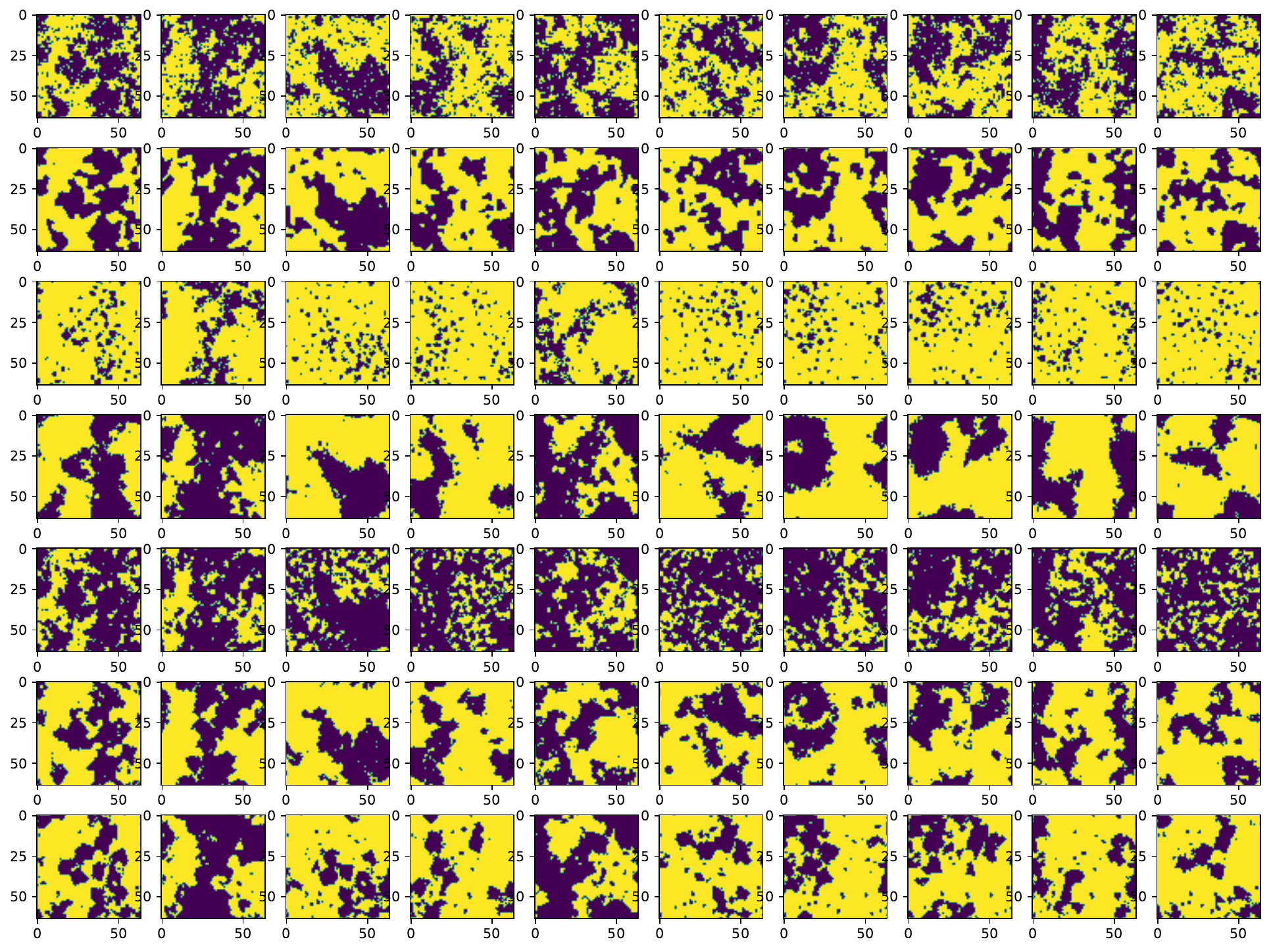}
    \caption{Reconstructed model for $\beta = .42$.}
    \label{fig:reco_.42}
\end{figure}

\begin{figure}[h!]
    \centering
    \includegraphics[width=\textwidth]{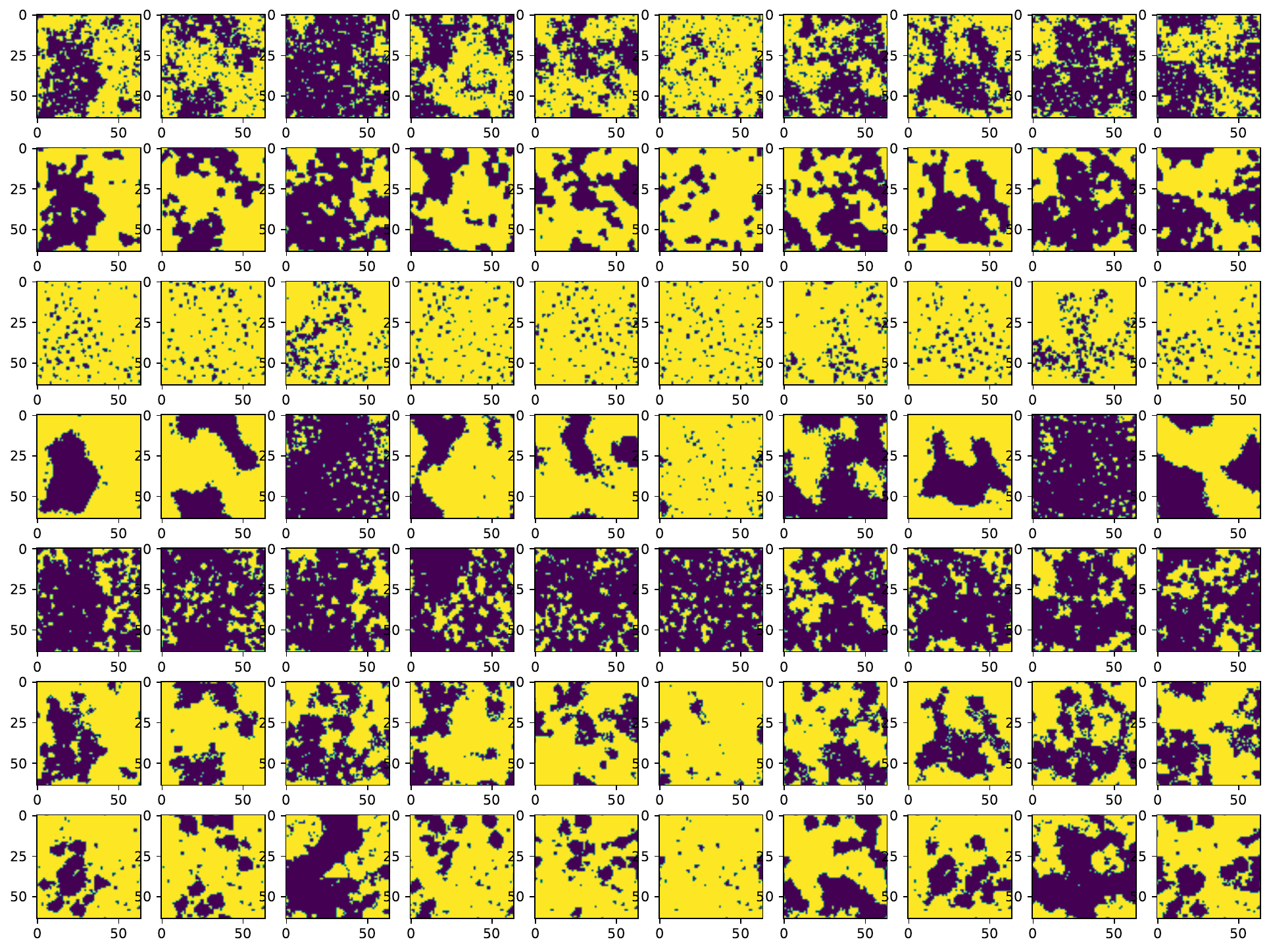}
    \caption{Reconstructed model for $\beta = .425$.}
    \label{fig:reco_.425}
\end{figure}

\begin{figure}[h!]
    \centering
    \includegraphics[width=\textwidth]{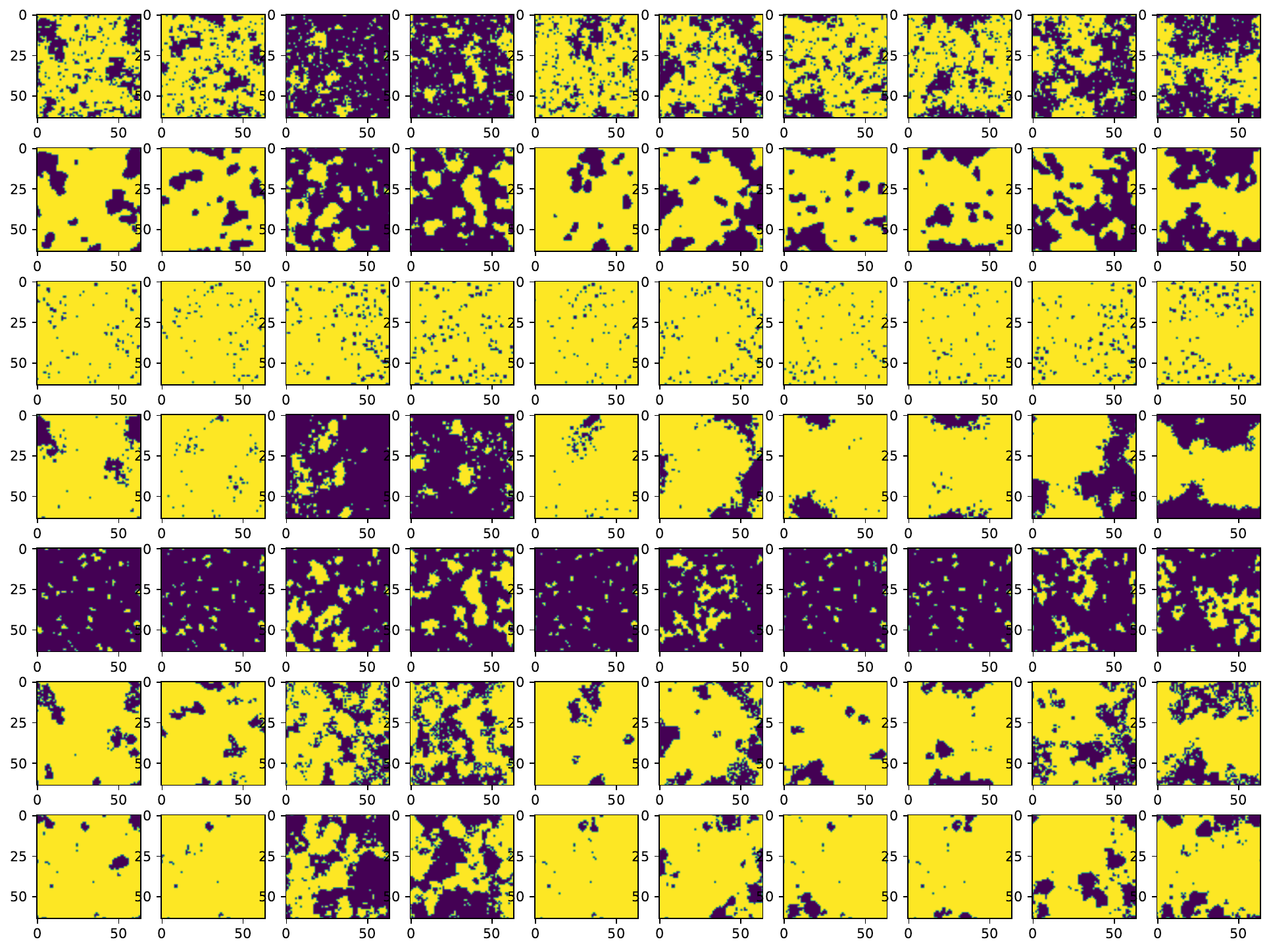}
    \caption{Reconstructed model for $\beta = .43$.}
    \label{fig:reco_.43}
\end{figure}

\chapter{Renormalization Techniques and Deep Learning Plots}
\label{chap:connection_plots}

\section{Generative Model Plots}
\label{sec:gen_plots}

This section contains plots showing the generative models as discussed in Section \ref{sec:gen_models}. The first set of plots contains plots of the models themselves, with the first row containing the original 8 by 8 lattice models and the second row containing the 64 by 64 generated model. The second set of plots consists of fits to the correlation function for each model, in order to derive the correlation lengths.

\begin{figure}[h!]
    \centering
    \includegraphics[width=\textwidth]{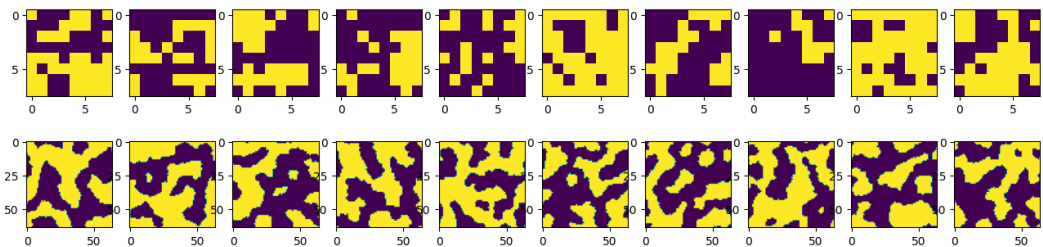}
    \caption{Generative model for $\beta = .395$.}
    \label{fig:gen_.395}
\end{figure}

\begin{figure}[h!]
    \centering
    \includegraphics[width=\textwidth]{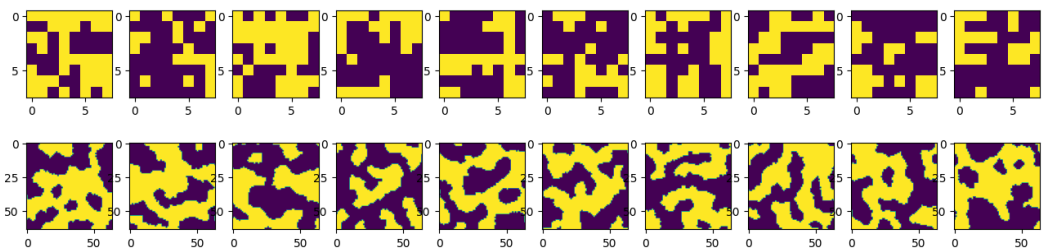}
    \caption{Generative model for $\beta = .4$.}
    \label{fig:gen_.4}
\end{figure}

\begin{figure}[h!]
    \centering
    \includegraphics[width=\textwidth]{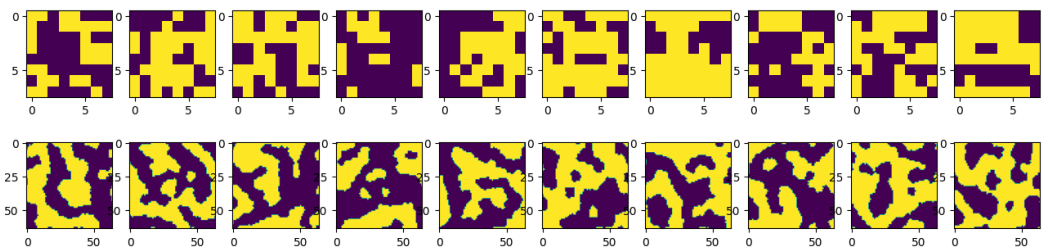}
    \caption{Generative model for $\beta = .405$.}
    \label{fig:gen_.405}
\end{figure}

\begin{figure}[h!]
    \centering
    \includegraphics[width=\textwidth]{gen_model__41.pdf}
    \caption{Generative model for $\beta = .41$.}
    \label{fig:gen_.41}
\end{figure}

\begin{figure}[h!]
    \centering
    \includegraphics[width=\textwidth]{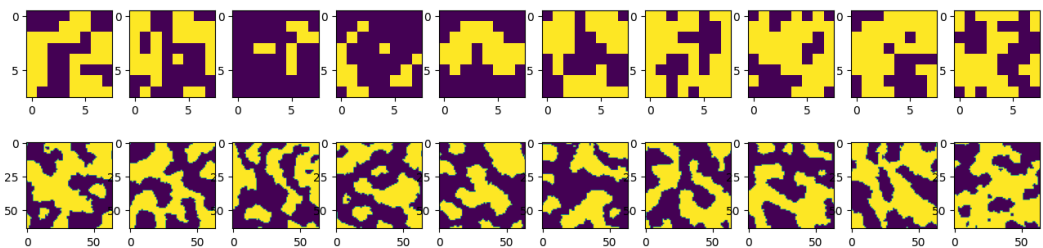}
    \caption{Generative model for $\beta = .415$.}
    \label{fig:gen_.415}
\end{figure}

\begin{figure}[h!]
    \centering
    \includegraphics[width=\textwidth]{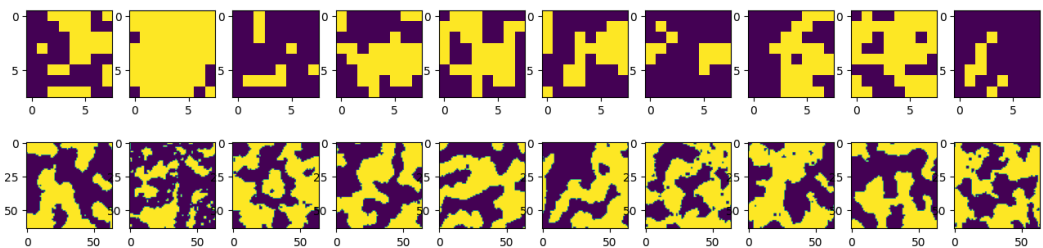}
    \caption{Generative model for $\beta = .42$.}
    \label{fig:gen_.42}
\end{figure}

\begin{figure}[h!]
    \centering
    \includegraphics[width=\textwidth]{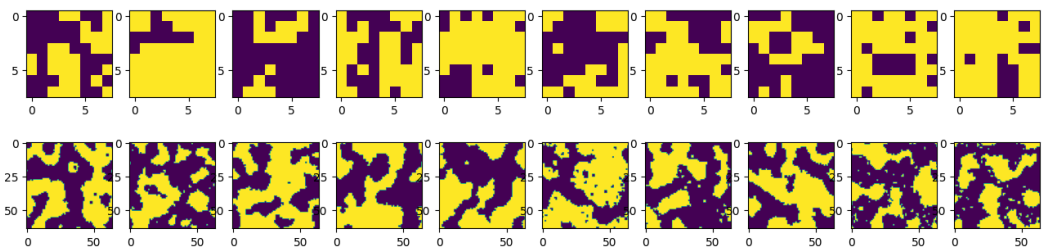}
    \caption{Generative model for $\beta = .425$.}
    \label{fig:gen_.425}
\end{figure}

\begin{figure}[h!]
    \centering
    \includegraphics[width=\textwidth]{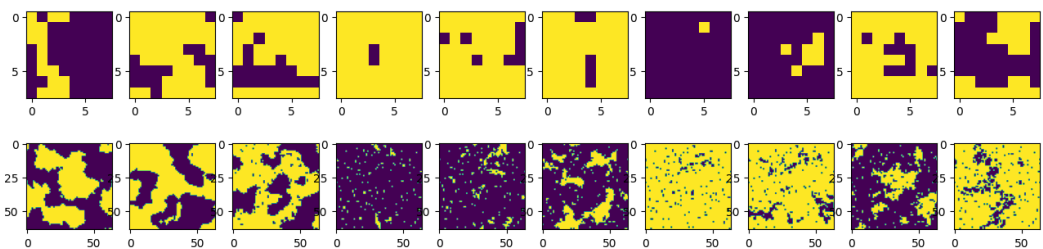}
    \caption{Generative model for $\beta = .43$.}
    \label{fig:gen_.43}
\end{figure}

\begin{figure}[h!]
     \centering
     \begin{subfigure}{0.49\textwidth}
         \centering
         \includegraphics[width=\textwidth]{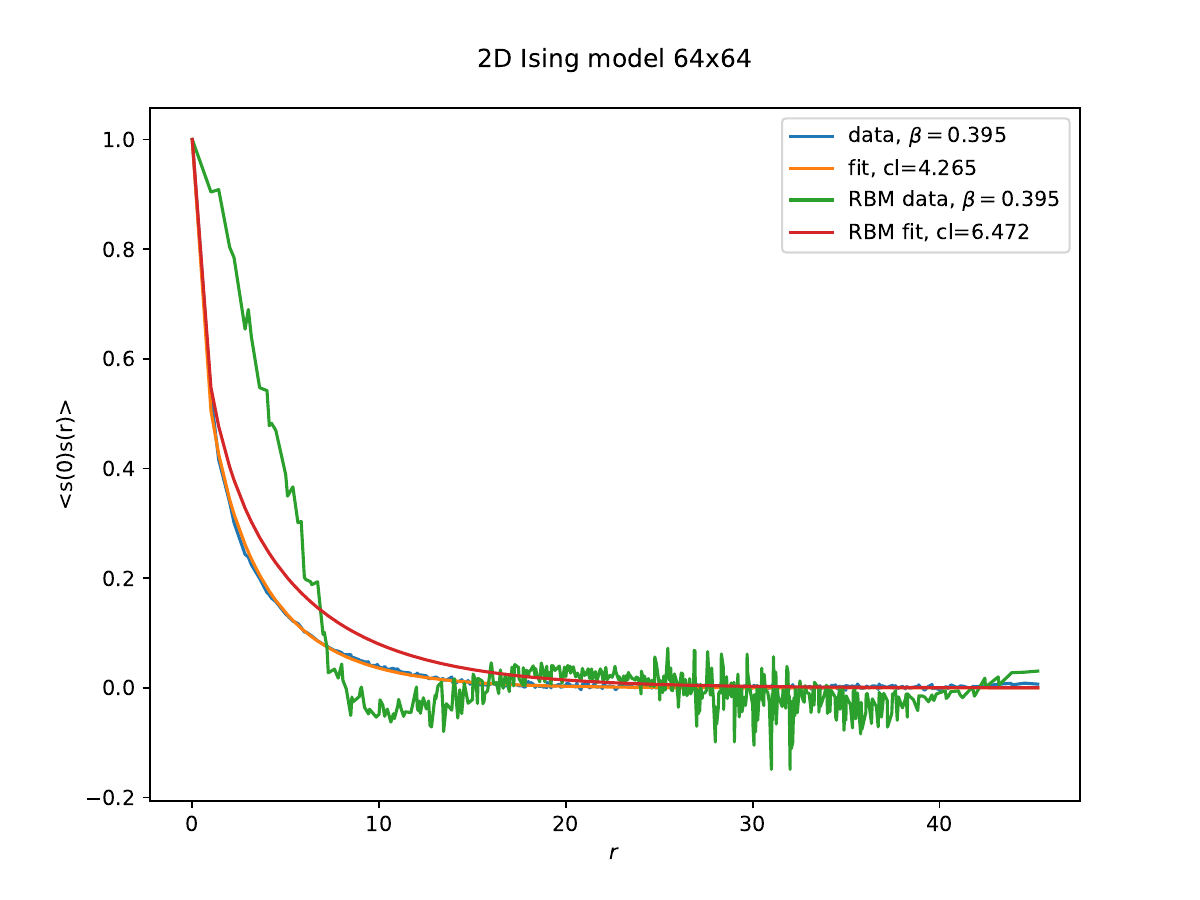}
         \caption{$\beta = .395.$}
     \end{subfigure}
     \hfill
     \begin{subfigure}{0.49\textwidth}
         \centering
         \includegraphics[width=\textwidth]{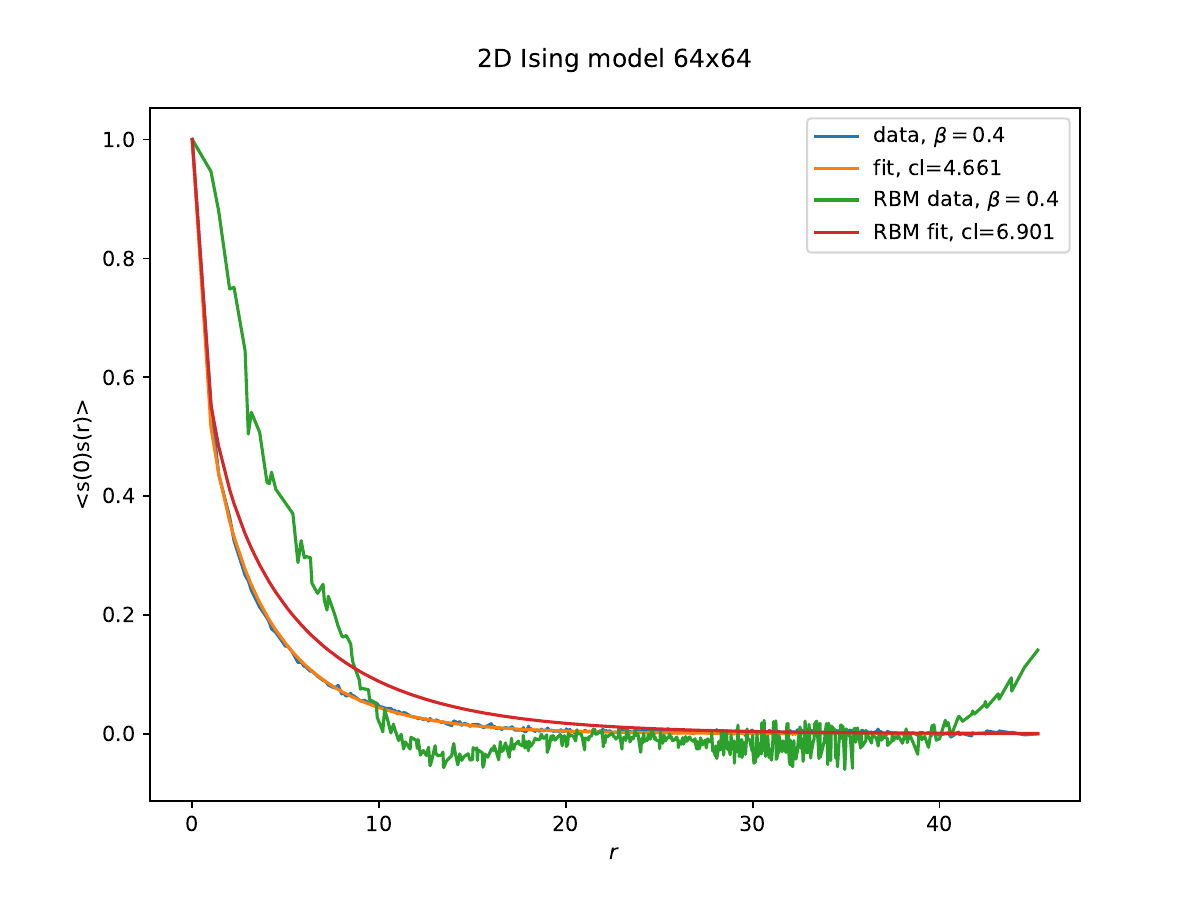}
         \caption{$\beta = .4.$}
     \end{subfigure}
     \hfill
     \begin{subfigure}{0.49\textwidth}
         \centering
         \includegraphics[width=\textwidth]{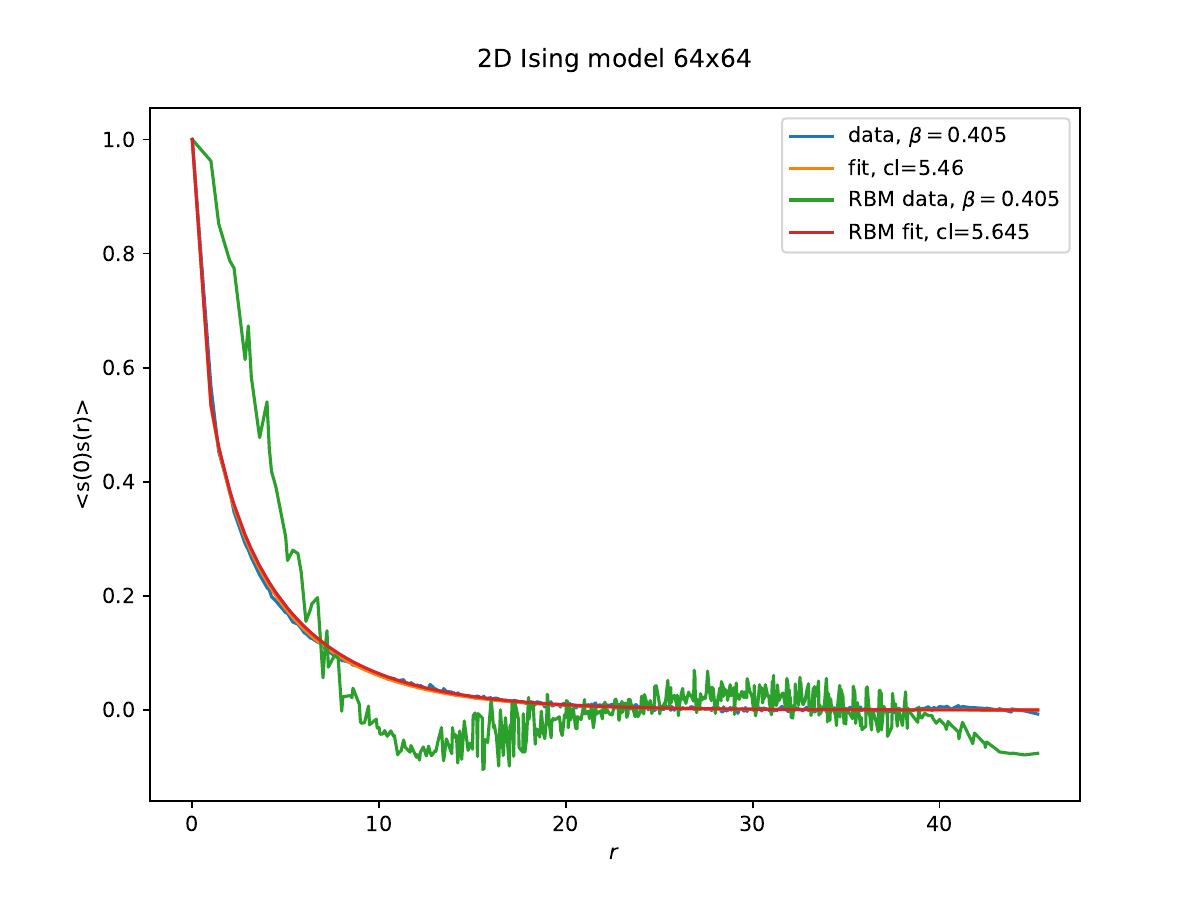}
         \caption{$\beta = .405.$}
     \end{subfigure}
     \hfill
     \begin{subfigure}{0.49\textwidth}
         \centering
         \includegraphics[width=\textwidth]{generative_correlation_length_0_41.pdf}
         \caption{$\beta = .41.$}
     \end{subfigure}
     \hfill
          \begin{subfigure}{0.49\textwidth}
         \centering
         \includegraphics[width=\textwidth]{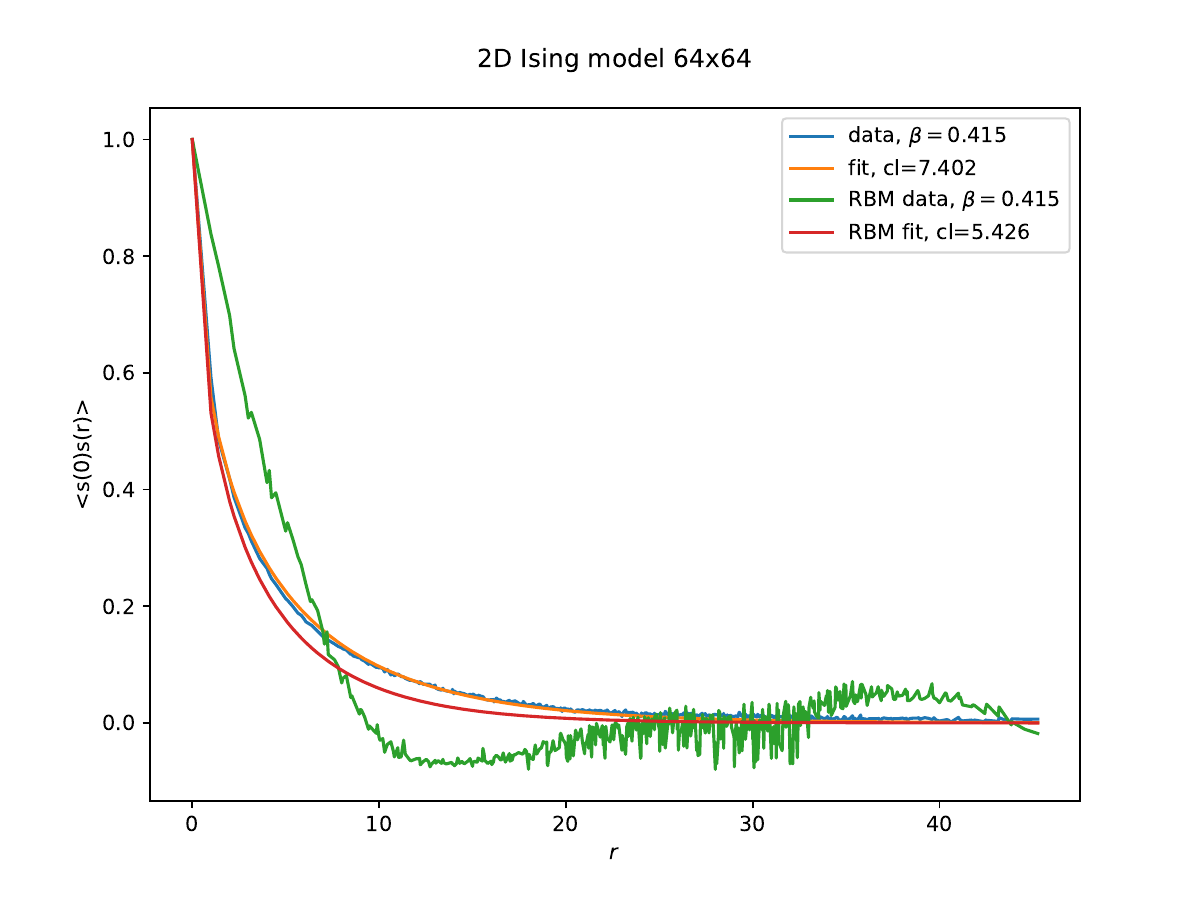}
         \caption{$\beta = .415.$}
     \end{subfigure}
     \hfill
     \begin{subfigure}{0.49\textwidth}
         \centering
         \includegraphics[width=\textwidth]{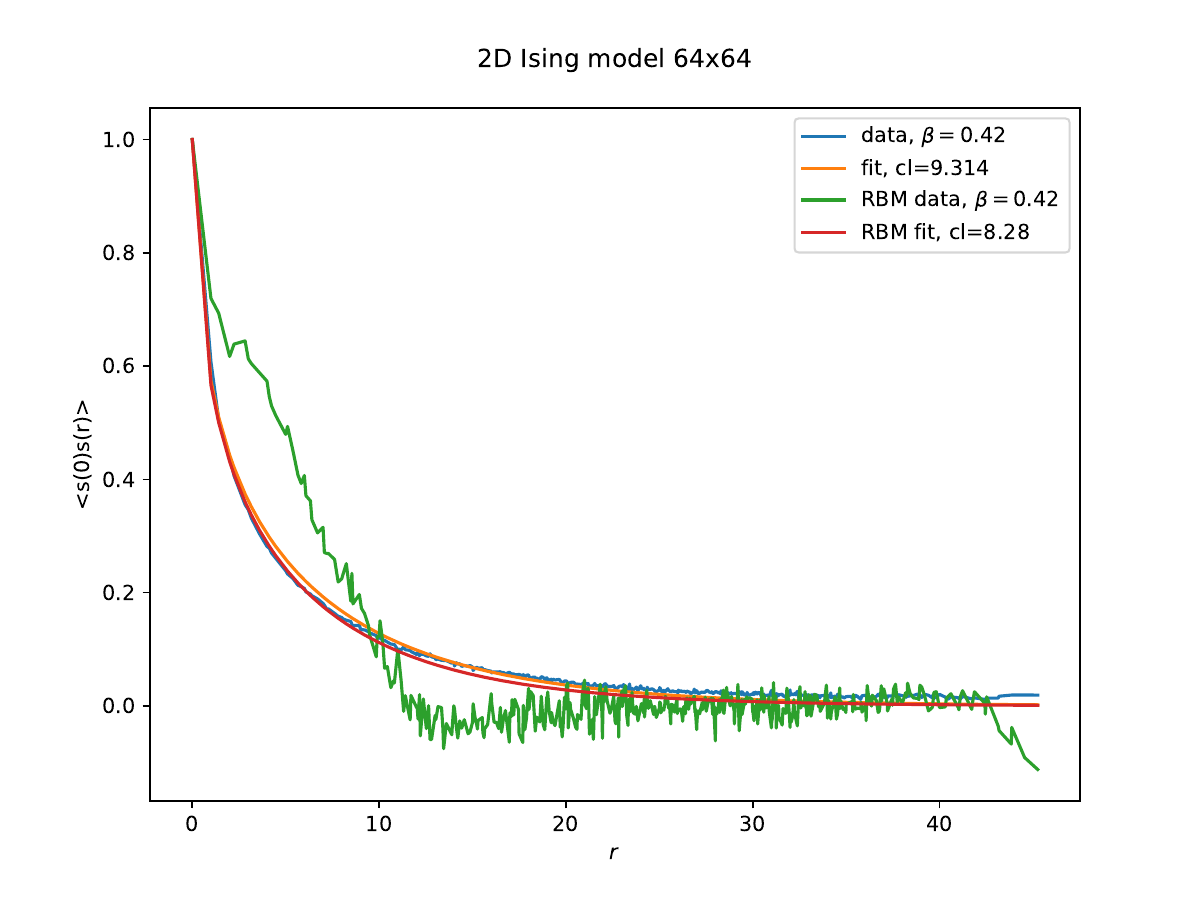}
         \caption{$\beta = .42.$}
     \end{subfigure}
     \hfill
    \begin{subfigure}{0.49\textwidth}
         \centering
         \includegraphics[width=\textwidth]{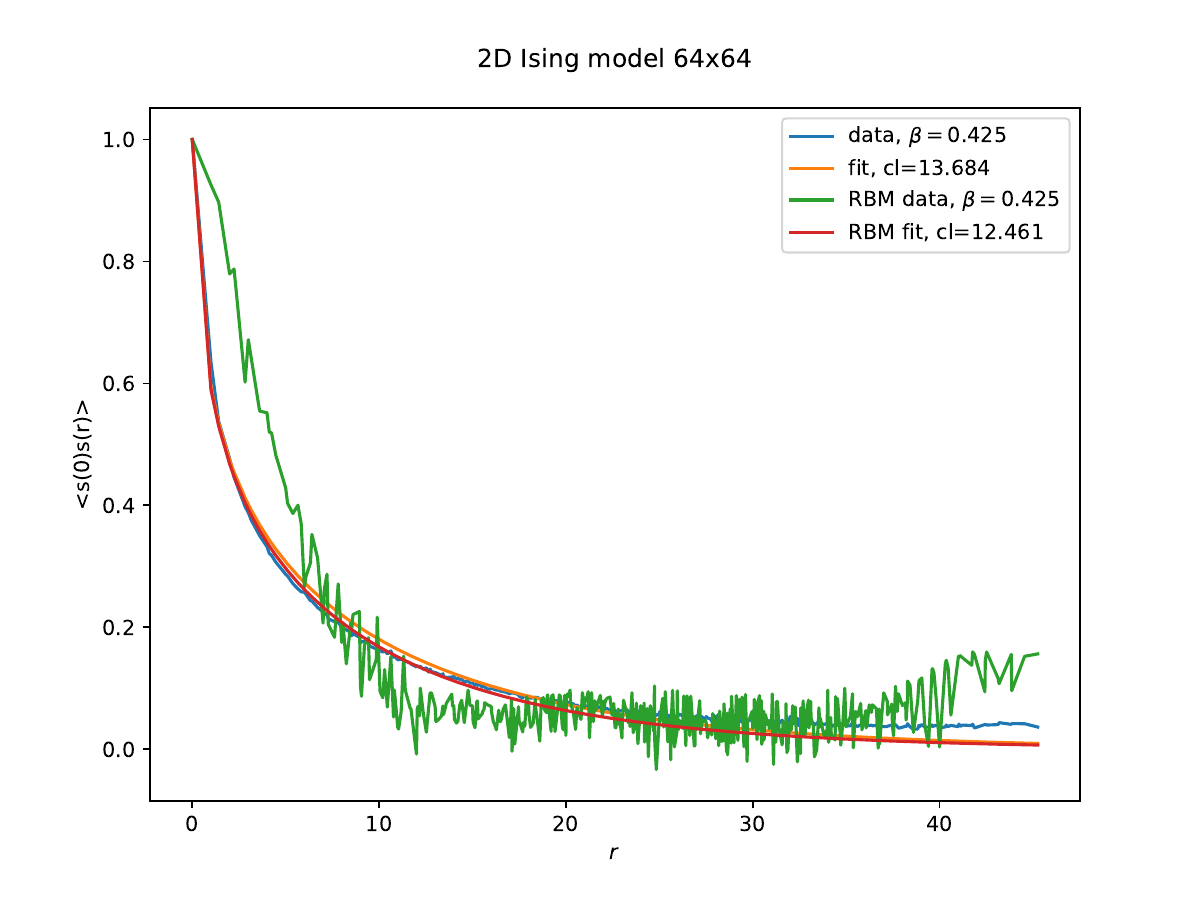}
         \caption{$\beta = .425.$}
     \end{subfigure}
     \hfill
           \begin{subfigure}{0.49\textwidth}
         \centering
         \includegraphics[width=\textwidth]{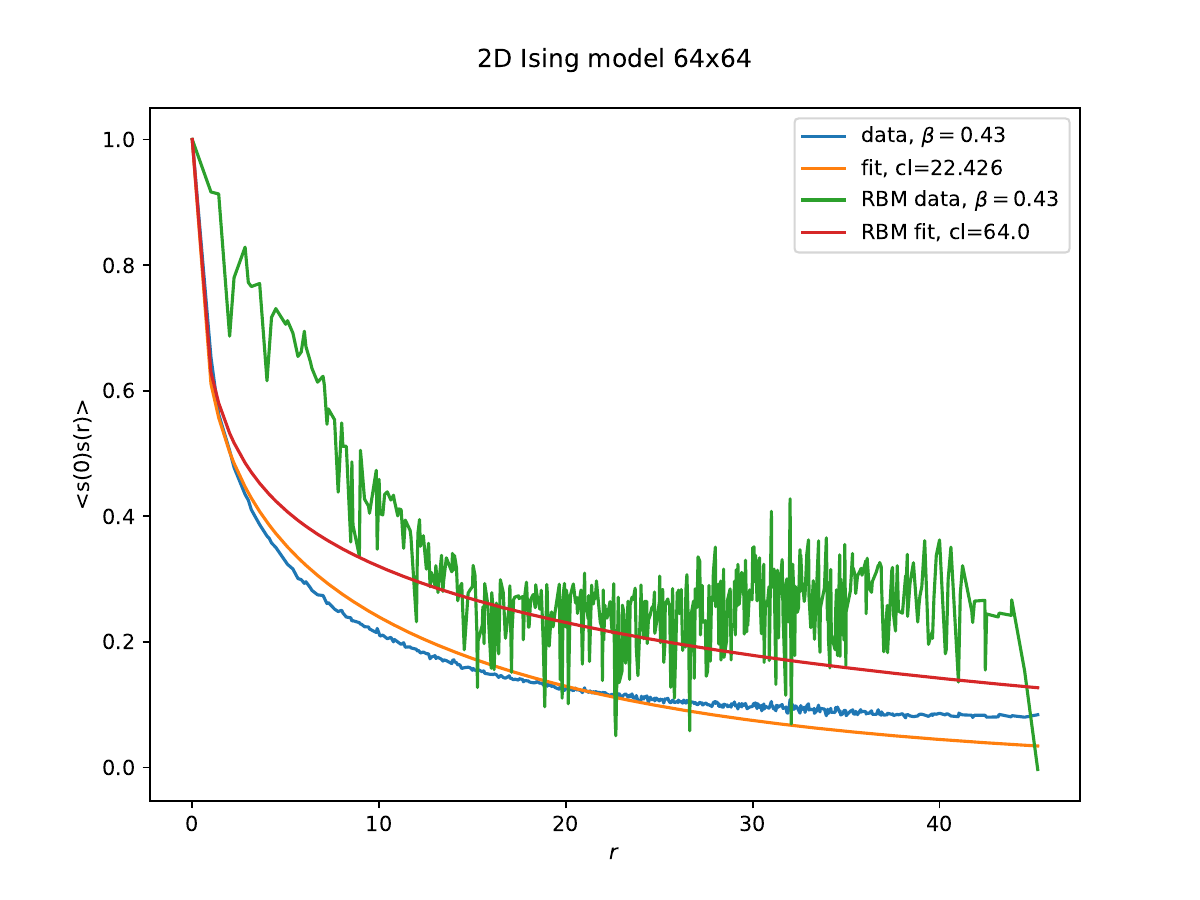}
         \caption{$\beta = .43.$}
     \end{subfigure}
        \caption{Correlation length comparisons for generative model.}
        \label{fig:corr_layer_gen}
\end{figure}

\clearpage

\section{Ising Spin Plots}
\label{sec:spin_plots}

This section contains plots showing the RBM spin models as discussed in Section \ref{sec:weights_and_spins}. The first set of plots contains the models themselves, with the first row containing the original Wolff generated model, and the next three rows consisting of the RBM coarse-grained modes. The second set of plots consists of fits to the correlation function for each model, in order to derive the correlation lengths.

\begin{figure}[h!]
    \centering
    \includegraphics[width=\textwidth]{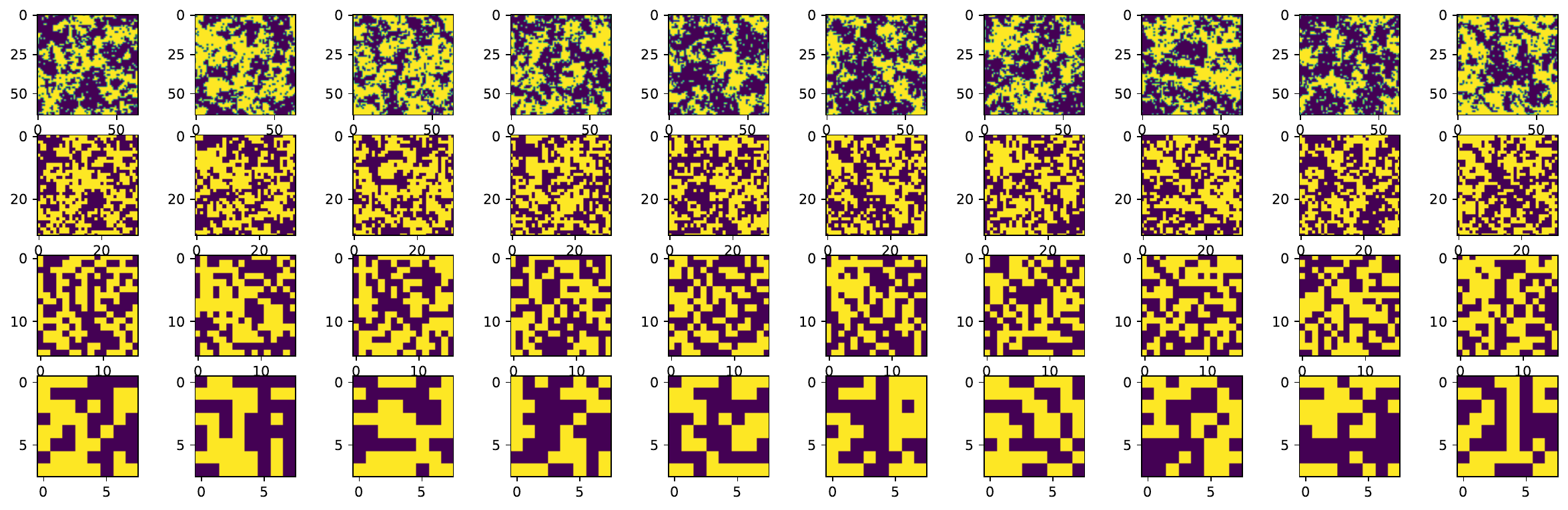}
    \caption{RBM coarse graining for $\beta = .395$.}
    \label{fig:.395_grain}
\end{figure}

\begin{figure}[h!]
    \centering
    \includegraphics[width=\textwidth]{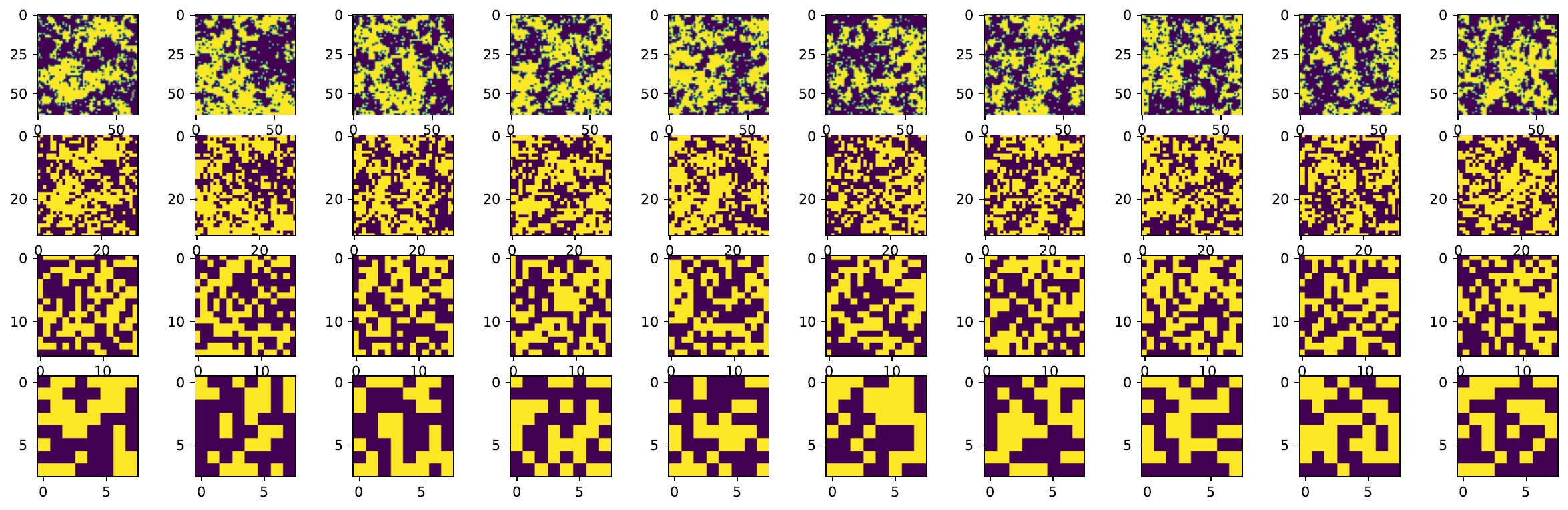}
    \caption{RBM coarse graining for $\beta = .4$.}
    \label{fig:.4_grain}
\end{figure}

\begin{figure}[h!]
    \centering
    \includegraphics[width=\textwidth]{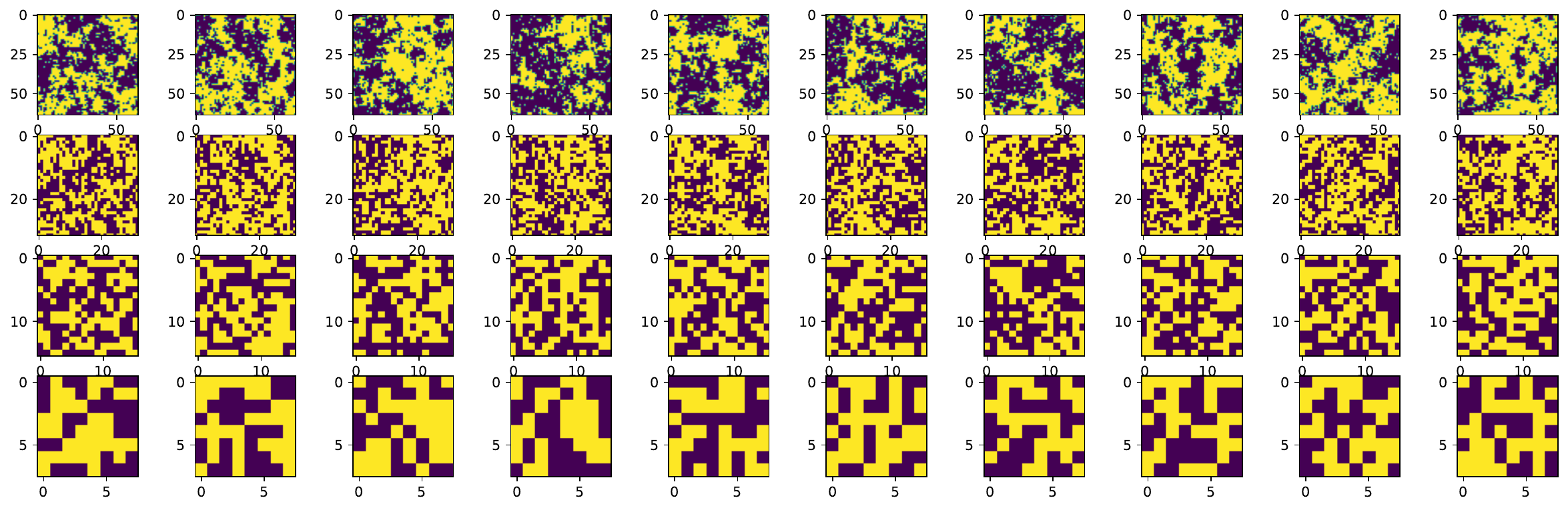}
    \caption{RBM coarse graining for $\beta = .405$.}
    \label{fig:.405_grain}
\end{figure}

\begin{figure}[h!]
    \centering
    \includegraphics[width=\textwidth]{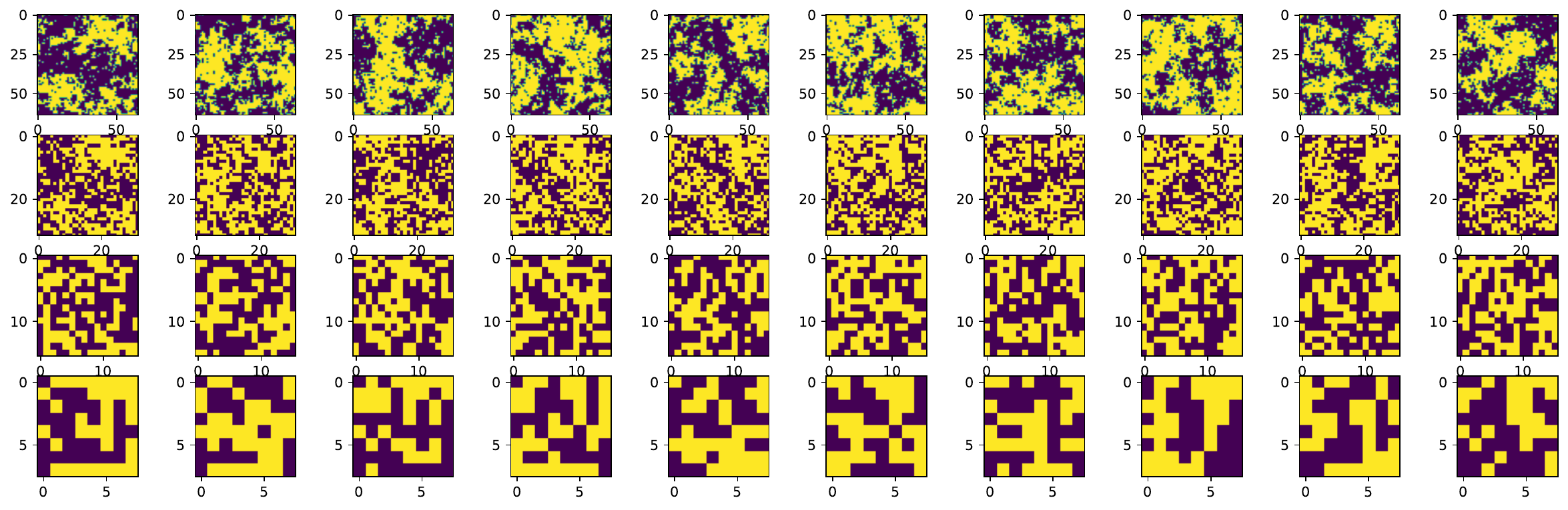}
    \caption{RBM coarse graining for $\beta = .41$.}
    \label{fig:.41_grain}
\end{figure}

\begin{figure}[h!]
    \centering
    \includegraphics[width=\textwidth]{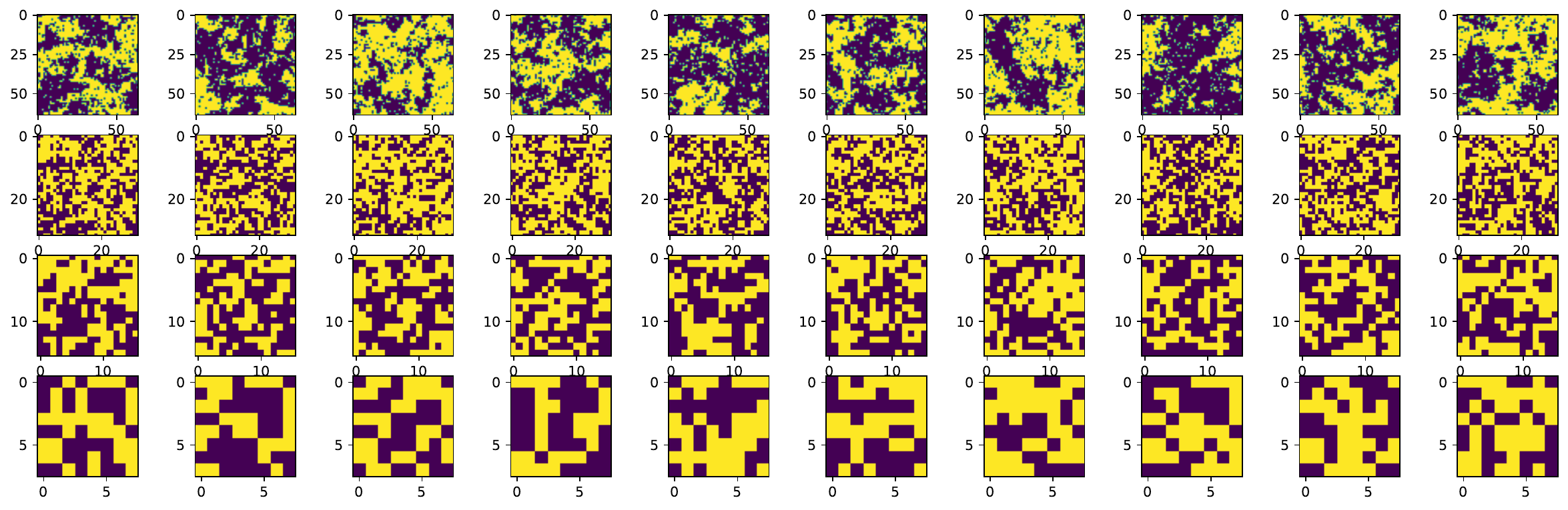}
    \caption{RBM coarse graining for $\beta = .415$.}
    \label{fig:.415_grain}
\end{figure}

\begin{figure}[h!]
    \centering
    \includegraphics[width=\textwidth]{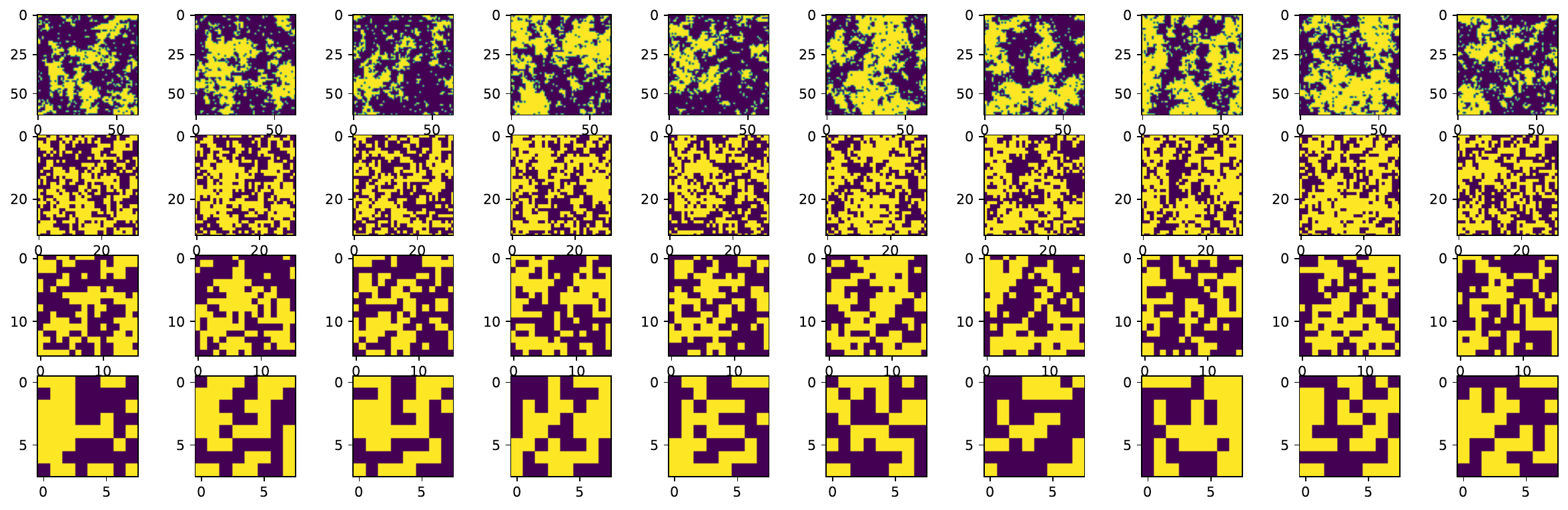}
    \caption{RBM coarse graining for $\beta = .42$.}
    \label{fig:.42_grain}
\end{figure}

\begin{figure}[h!]
    \centering
    \includegraphics[width=\textwidth]{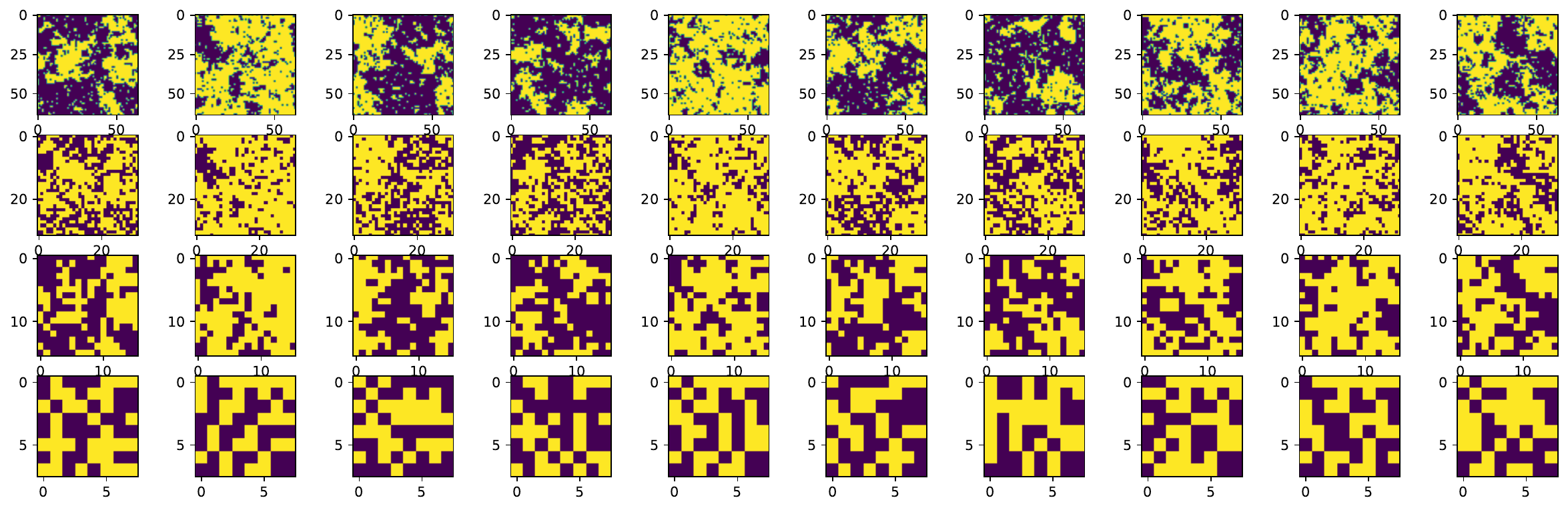}
    \caption{RBM coarse graining for $\beta = .425$.}
    \label{fig:.425_grain}
\end{figure}

\begin{figure}[h!]
    \centering
    \includegraphics[width=\textwidth]{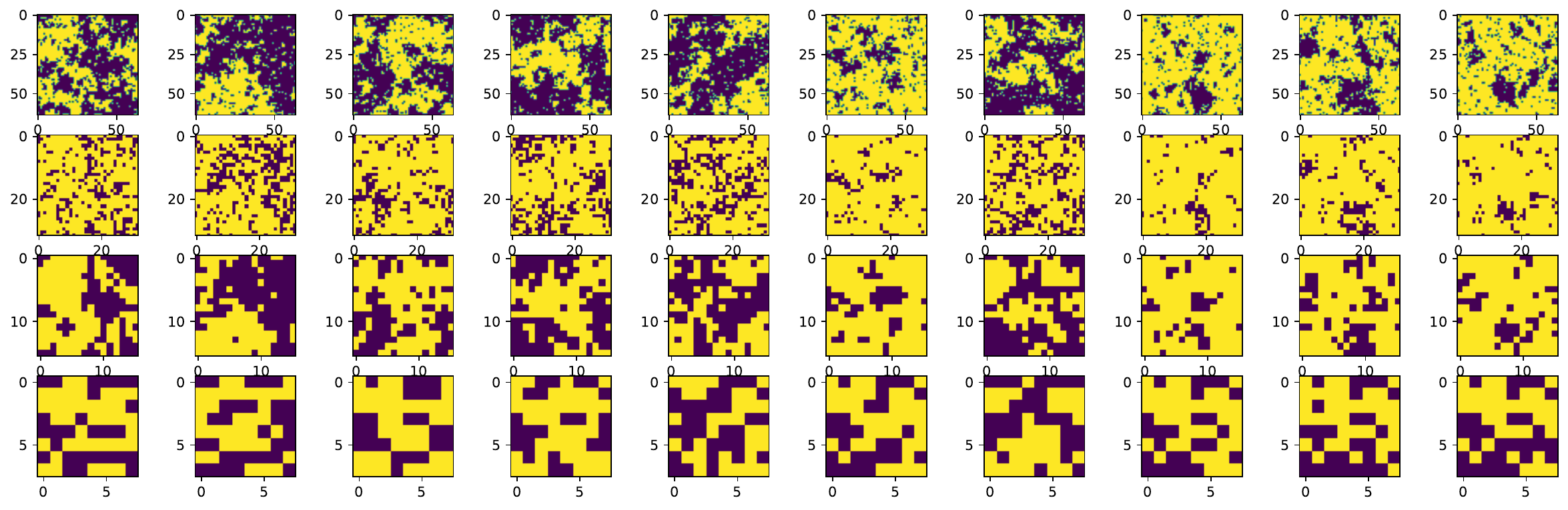}
    \caption{RBM coarse graining for $\beta = .43$.}
    \label{fig:.43_grain}
\end{figure}

\begin{figure}[h!]
     \centering
     \begin{subfigure}{0.49\textwidth}
         \centering
         \includegraphics[width=\textwidth]{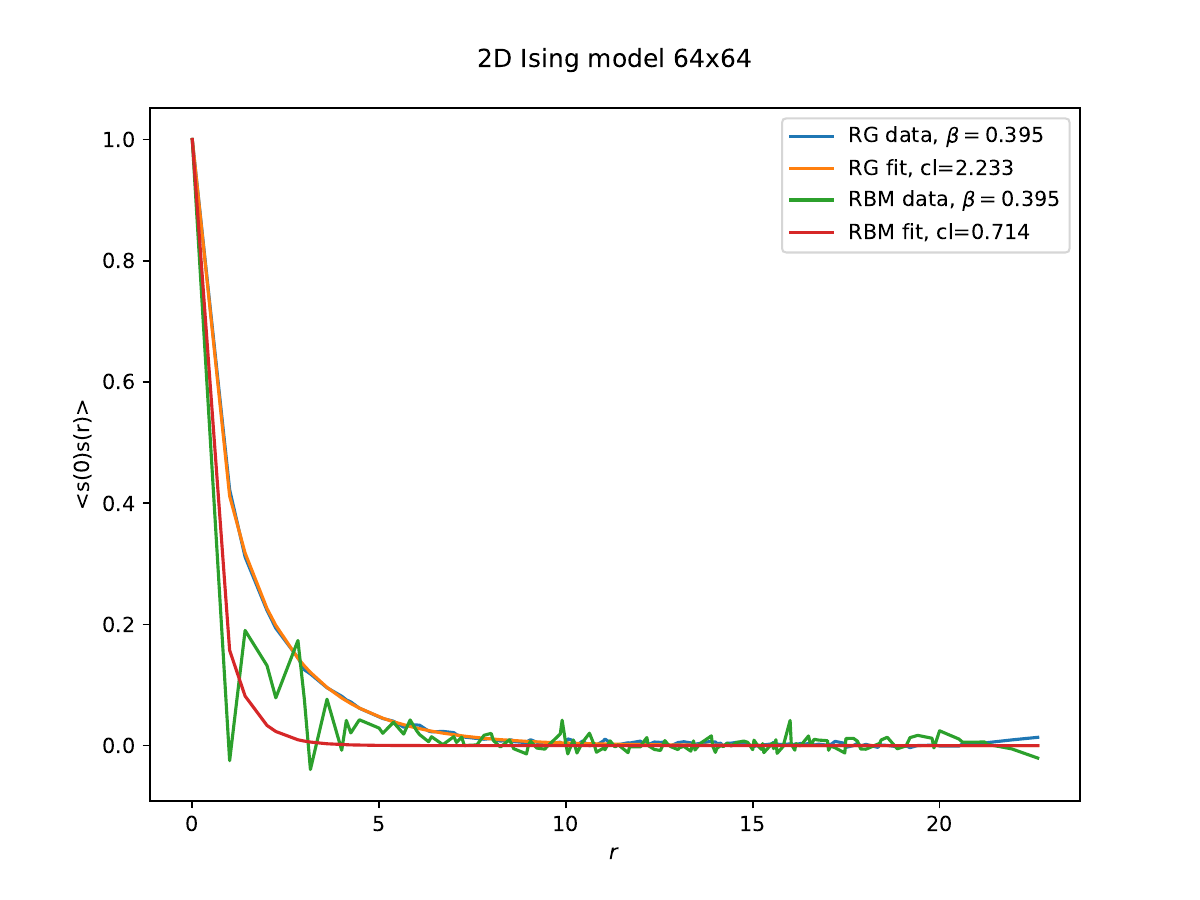}
         \caption{$\beta = .395.$}
     \end{subfigure}
     \hfill
     \begin{subfigure}{0.49\textwidth}
         \centering
         \includegraphics[width=\textwidth]{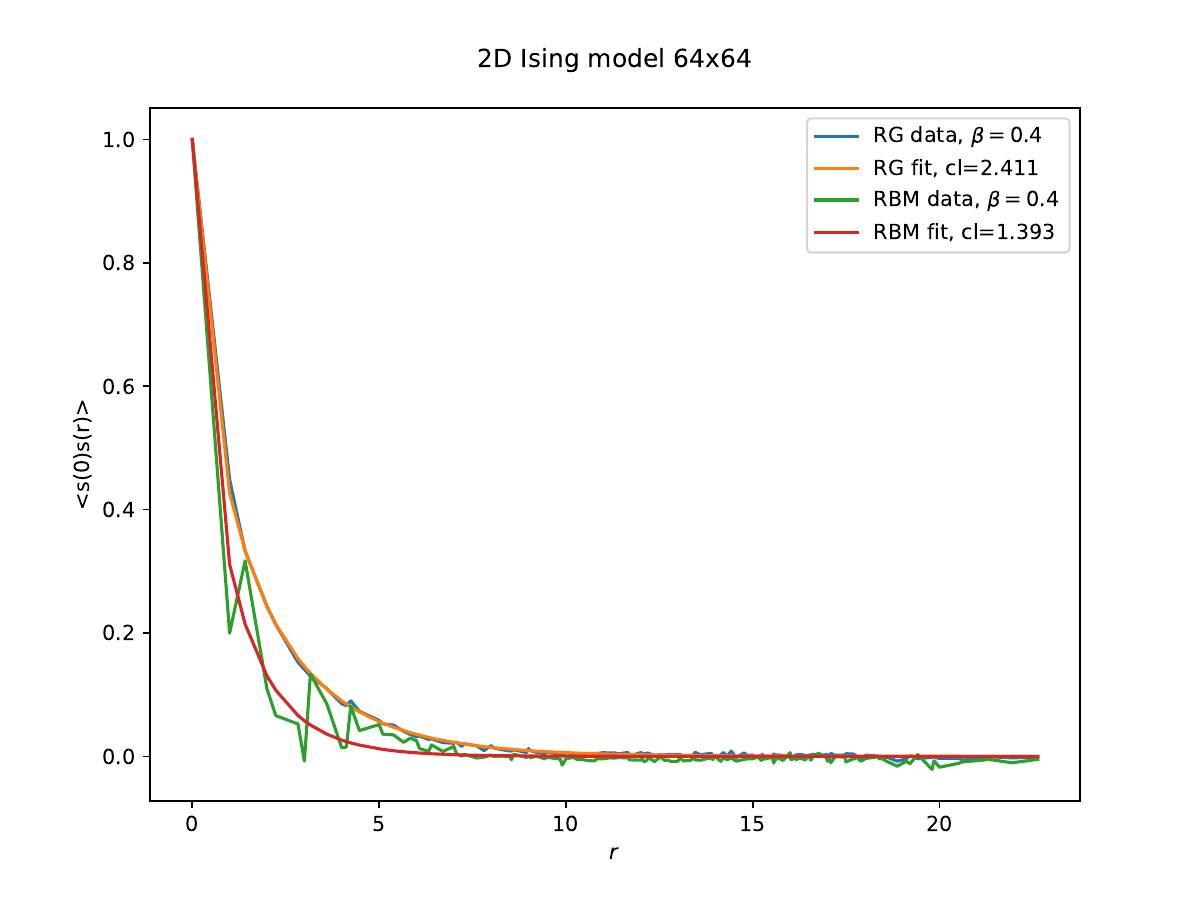}
         \caption{$\beta = .4.$}
     \end{subfigure}
     \hfill
     \begin{subfigure}{0.49\textwidth}
         \centering
         \includegraphics[width=\textwidth]{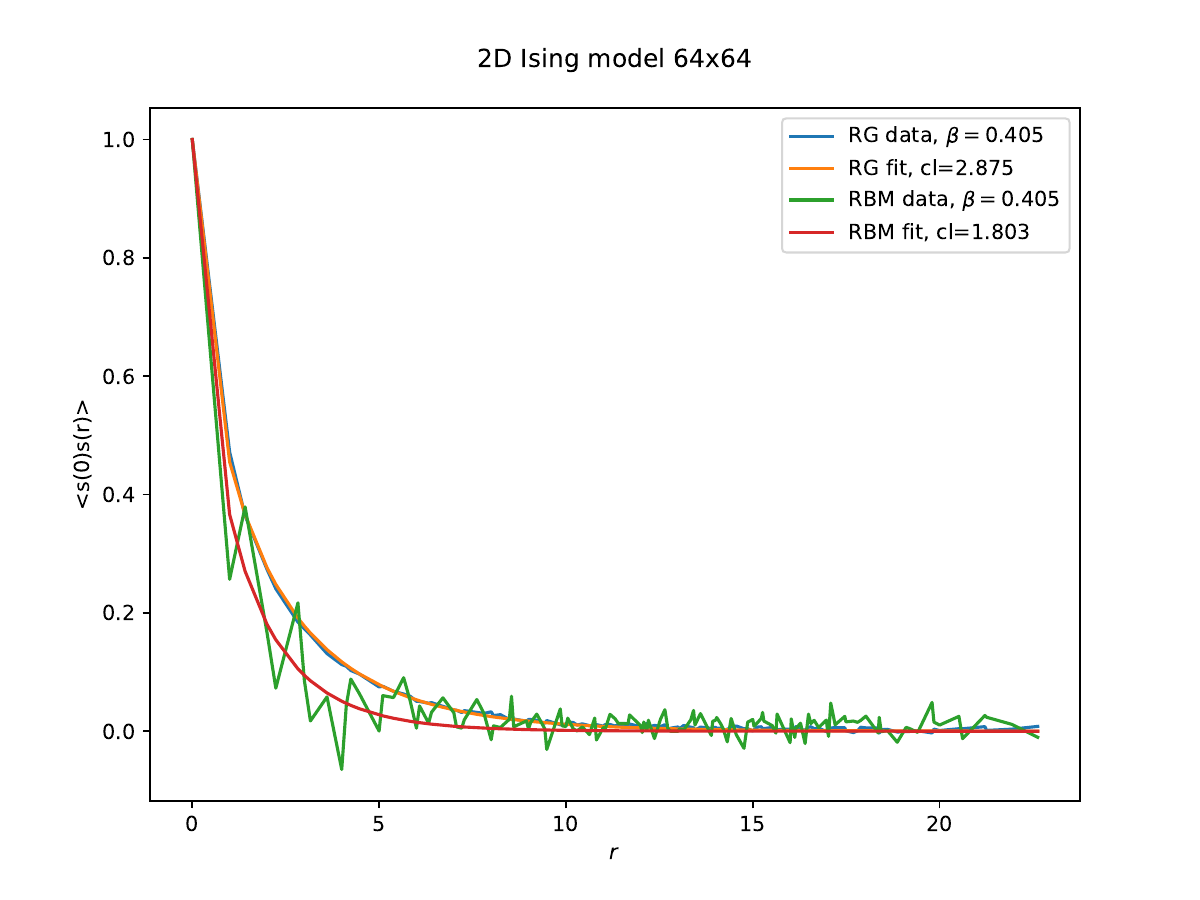}
         \caption{$\beta = .405.$}
     \end{subfigure}
     \hfill
     \begin{subfigure}{0.49\textwidth}
         \centering
         \includegraphics[width=\textwidth]{layer_1_model_0_41.pdf}
         \caption{$\beta = .41.$}
     \end{subfigure}
     \hfill
          \begin{subfigure}{0.49\textwidth}
         \centering
         \includegraphics[width=\textwidth]{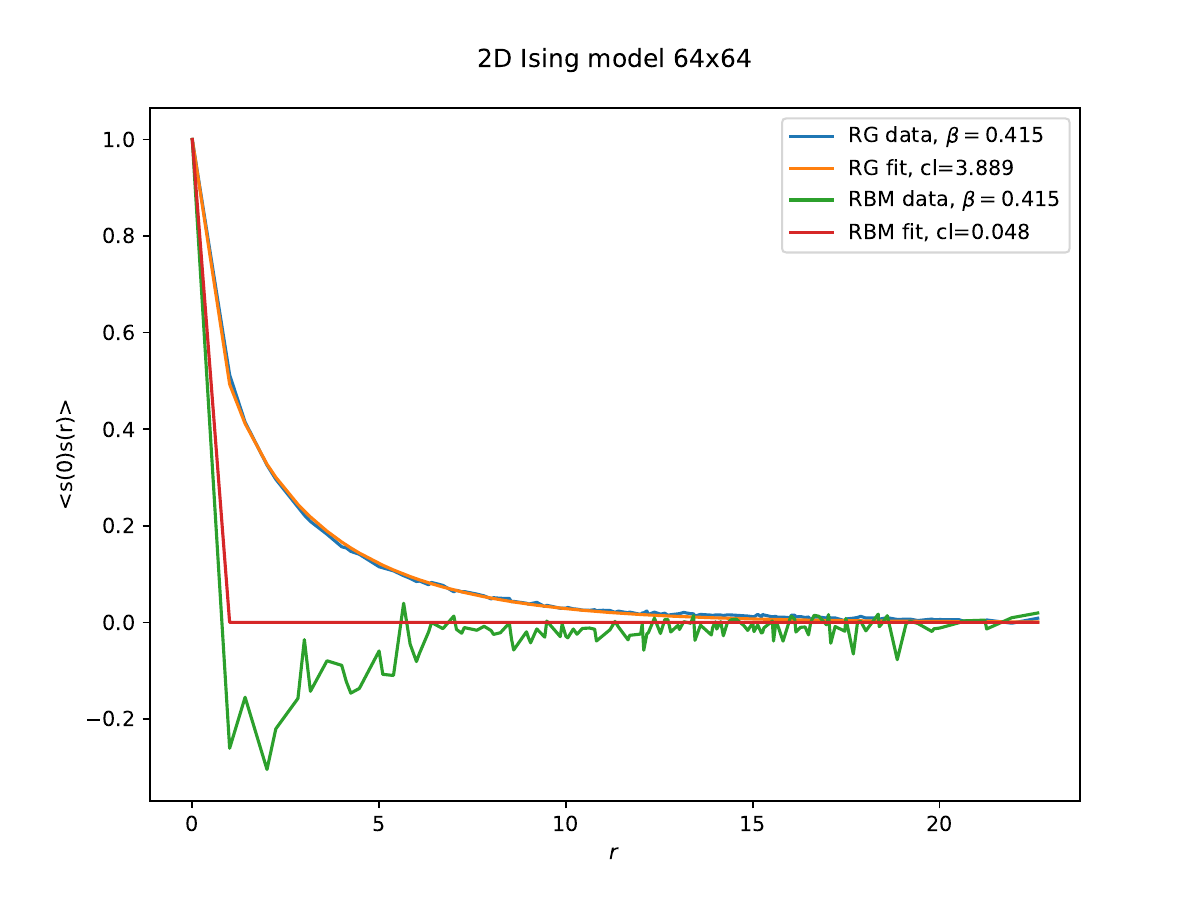}
         \caption{$\beta = .415.$}
     \end{subfigure}
     \hfill
     \begin{subfigure}{0.49\textwidth}
         \centering
         \includegraphics[width=\textwidth]{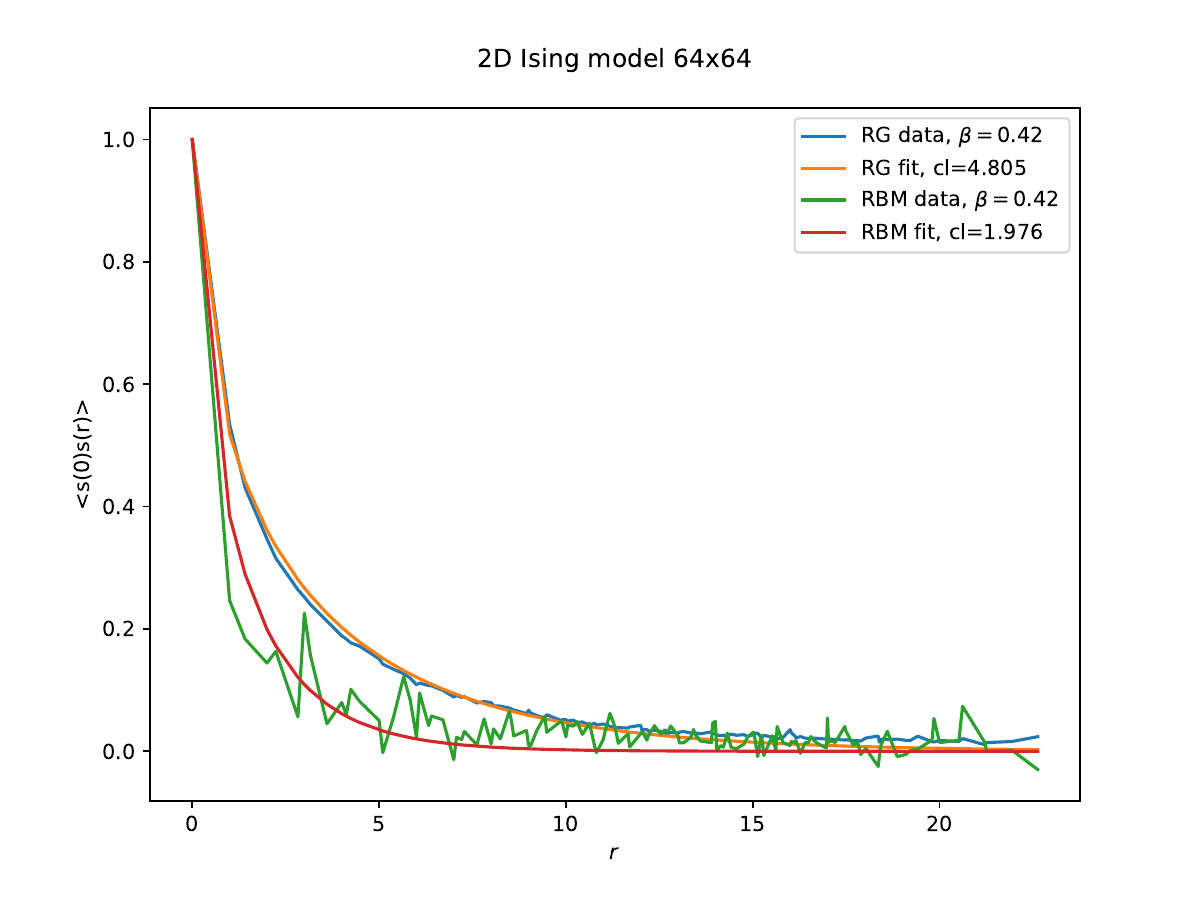}
         \caption{$\beta = .42.$}
     \end{subfigure}
     \hfill
    \begin{subfigure}{0.49\textwidth}
         \centering
         \includegraphics[width=\textwidth]{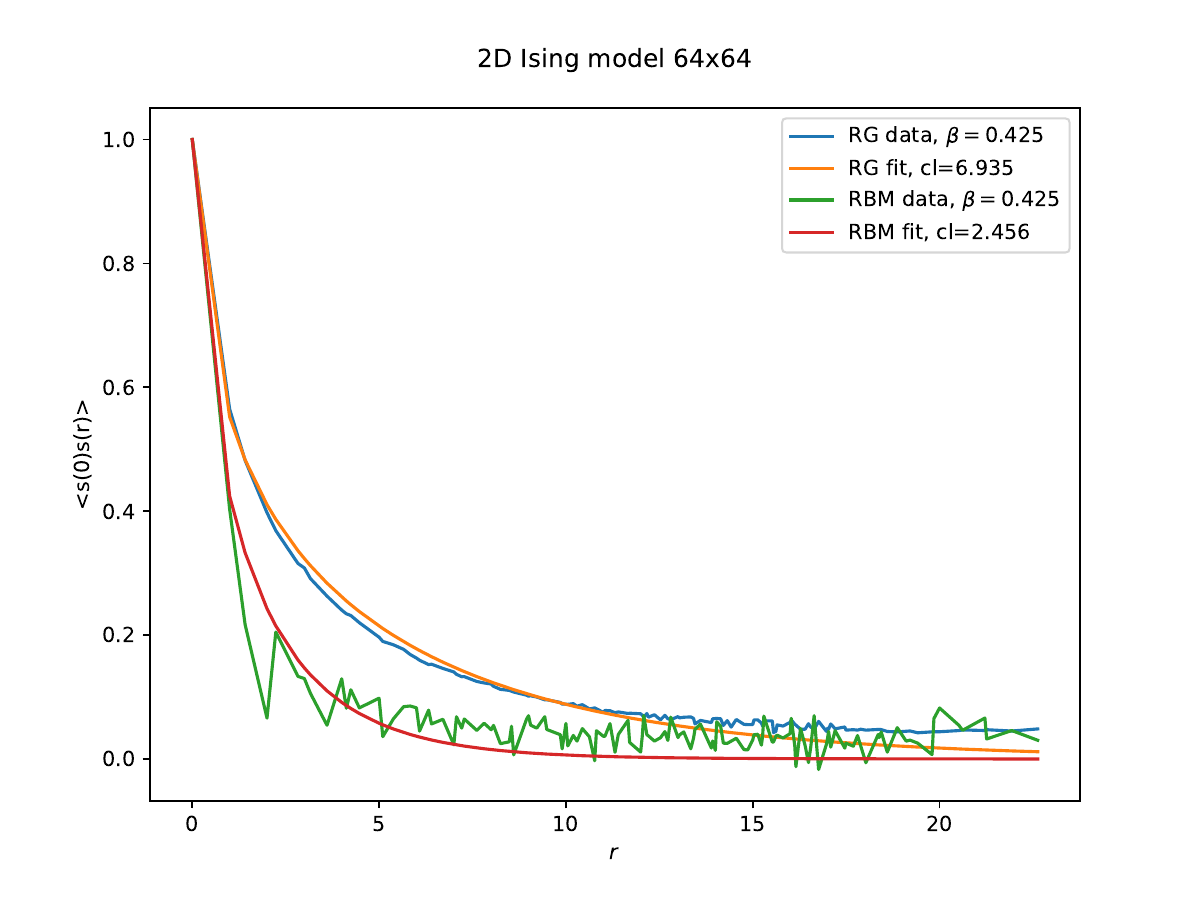}
         \caption{$\beta = .425.$}
     \end{subfigure}
     \hfill
           \begin{subfigure}{0.49\textwidth}
         \centering
         \includegraphics[width=\textwidth]{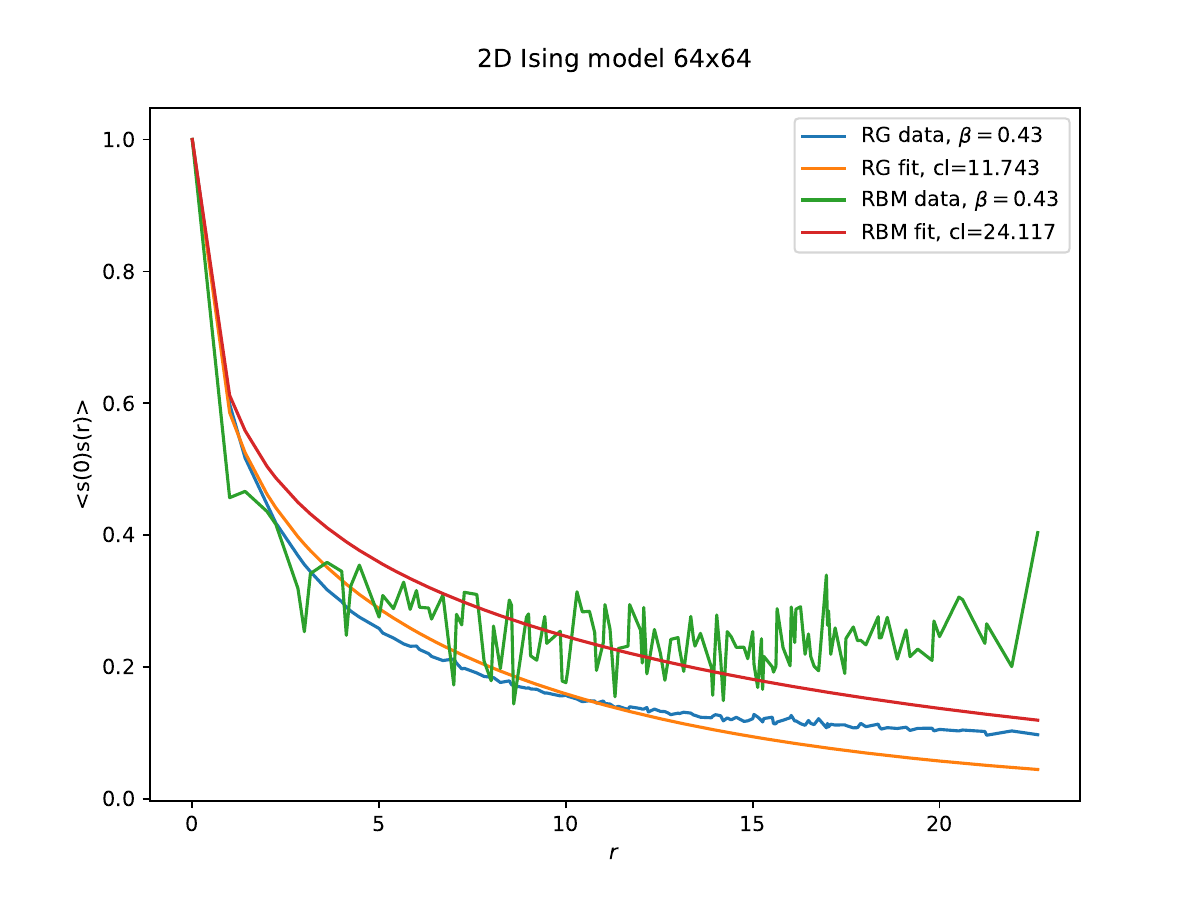}
         \caption{$\beta = .43.$}
     \end{subfigure}
        \caption{Correlation length comparisons for Layer 1.}
        \label{fig:corr_layer_1}
\end{figure}

\begin{figure}[h!]
     \centering
     \begin{subfigure}{0.49\textwidth}
         \centering
         \includegraphics[width=\textwidth]{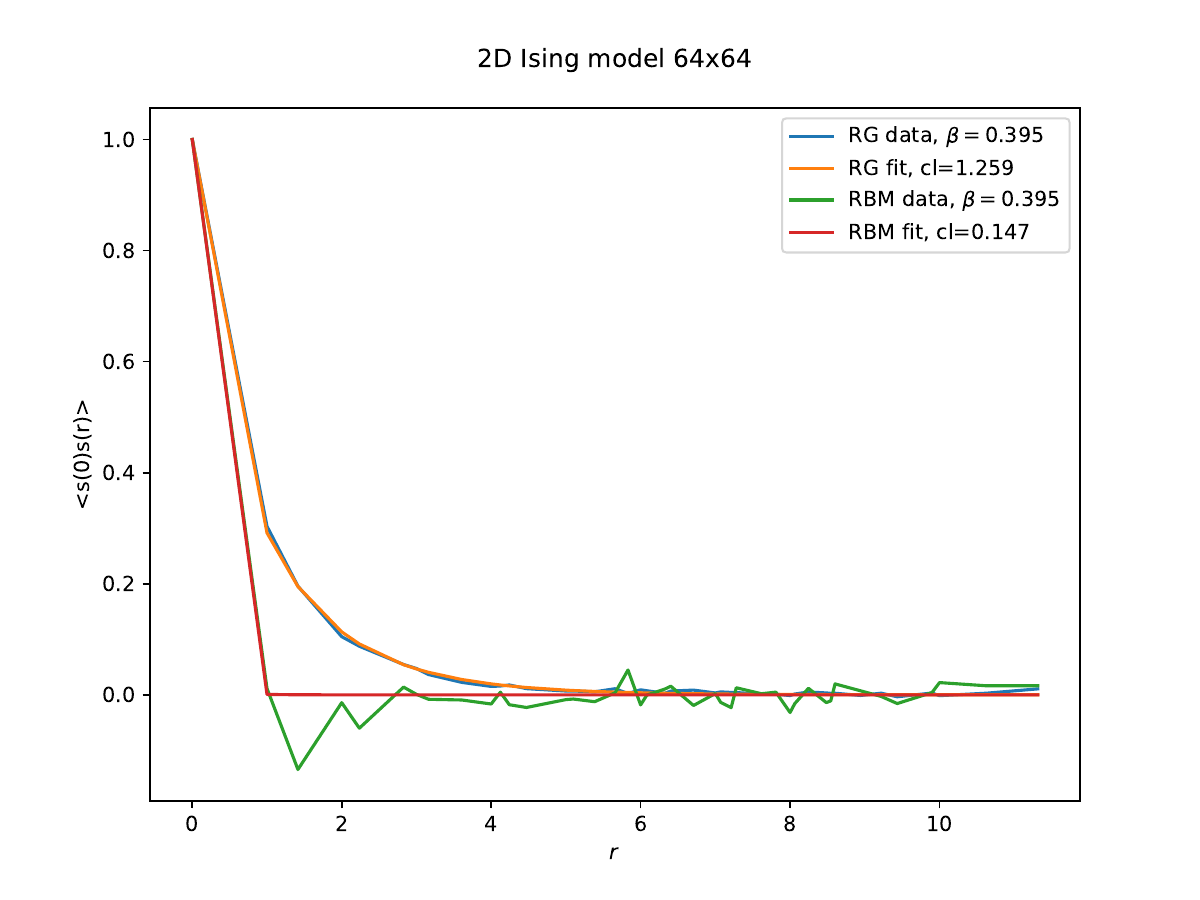}
         \caption{$\beta = .395.$}
     \end{subfigure}
     \hfill
     \begin{subfigure}{0.49\textwidth}
         \centering
         \includegraphics[width=\textwidth]{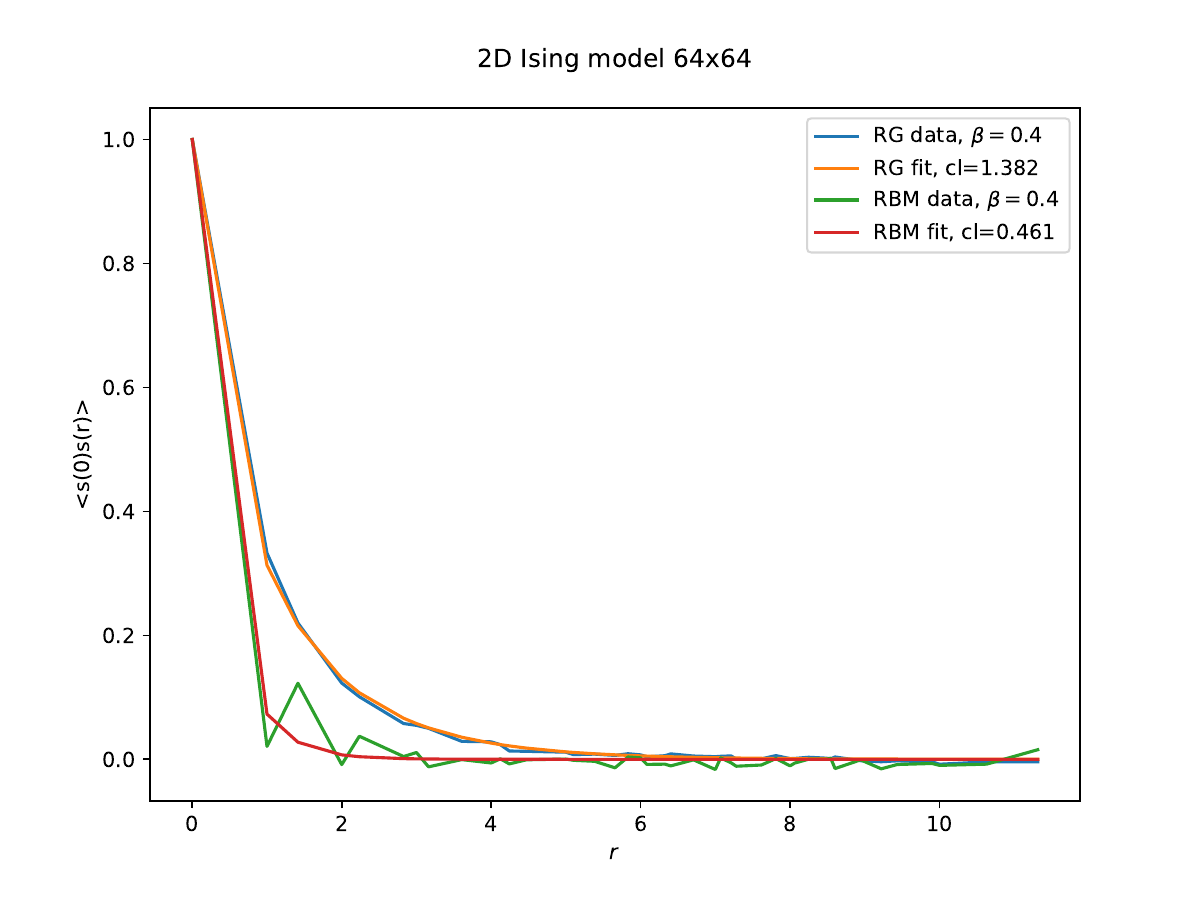}
         \caption{$\beta = .4.$}
     \end{subfigure}
     \hfill
     \begin{subfigure}{0.49\textwidth}
         \centering
         \includegraphics[width=\textwidth]{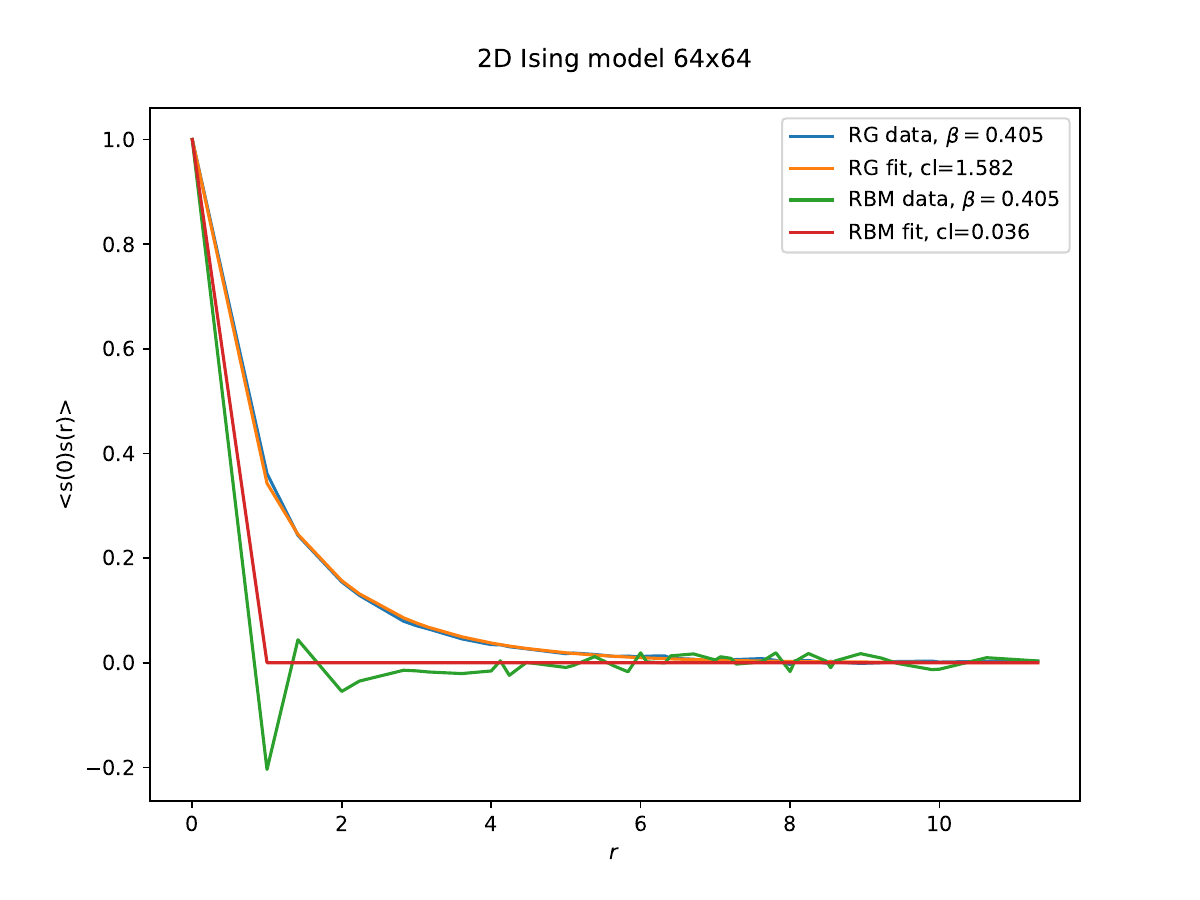}
         \caption{$\beta = .405.$}
     \end{subfigure}
     \hfill
     \begin{subfigure}{0.49\textwidth}
         \centering
         \includegraphics[width=\textwidth]{layer_2_model_0_41.pdf}
         \caption{$\beta = .41.$}
     \end{subfigure}
     \hfill
          \begin{subfigure}{0.49\textwidth}
         \centering
         \includegraphics[width=\textwidth]{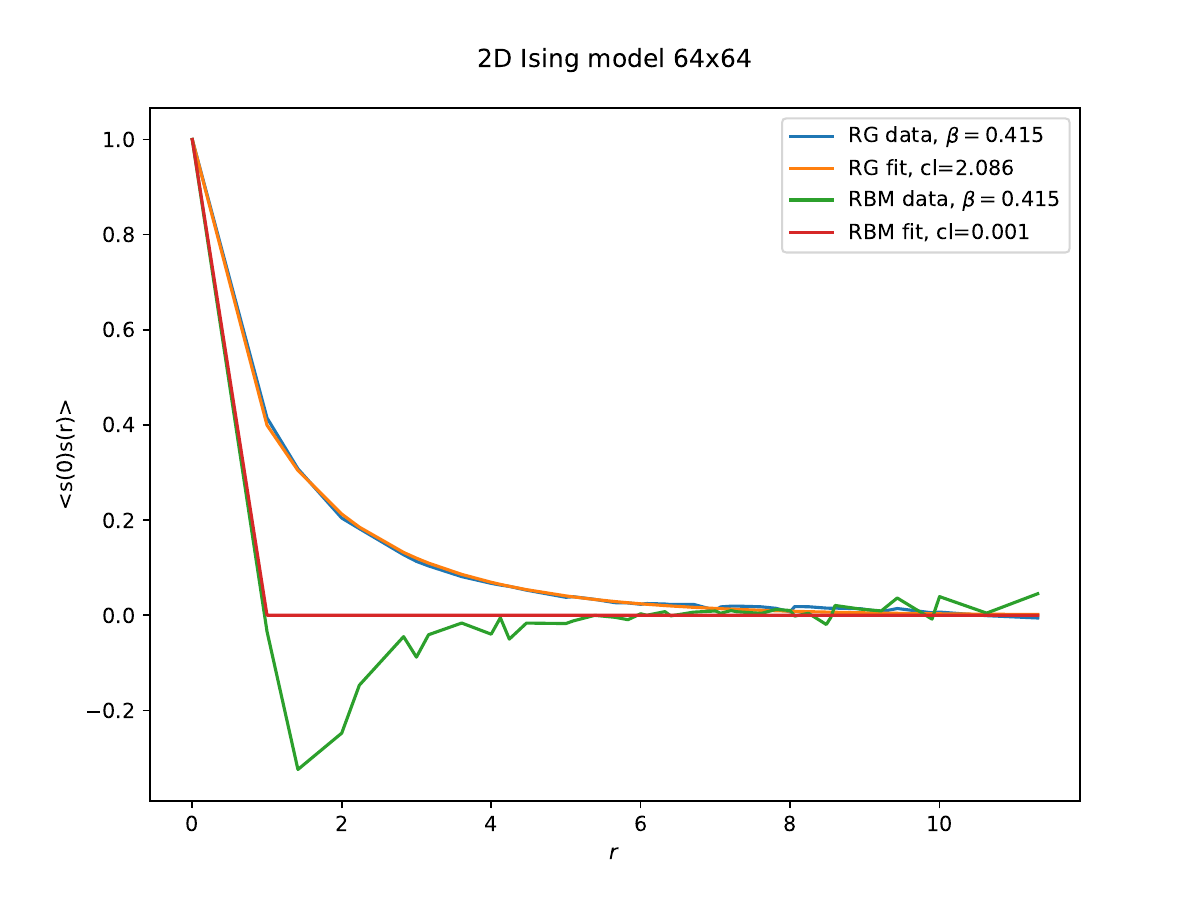}
         \caption{$\beta = .415.$}
     \end{subfigure}
     \hfill
     \begin{subfigure}{0.49\textwidth}
         \centering
         \includegraphics[width=\textwidth]{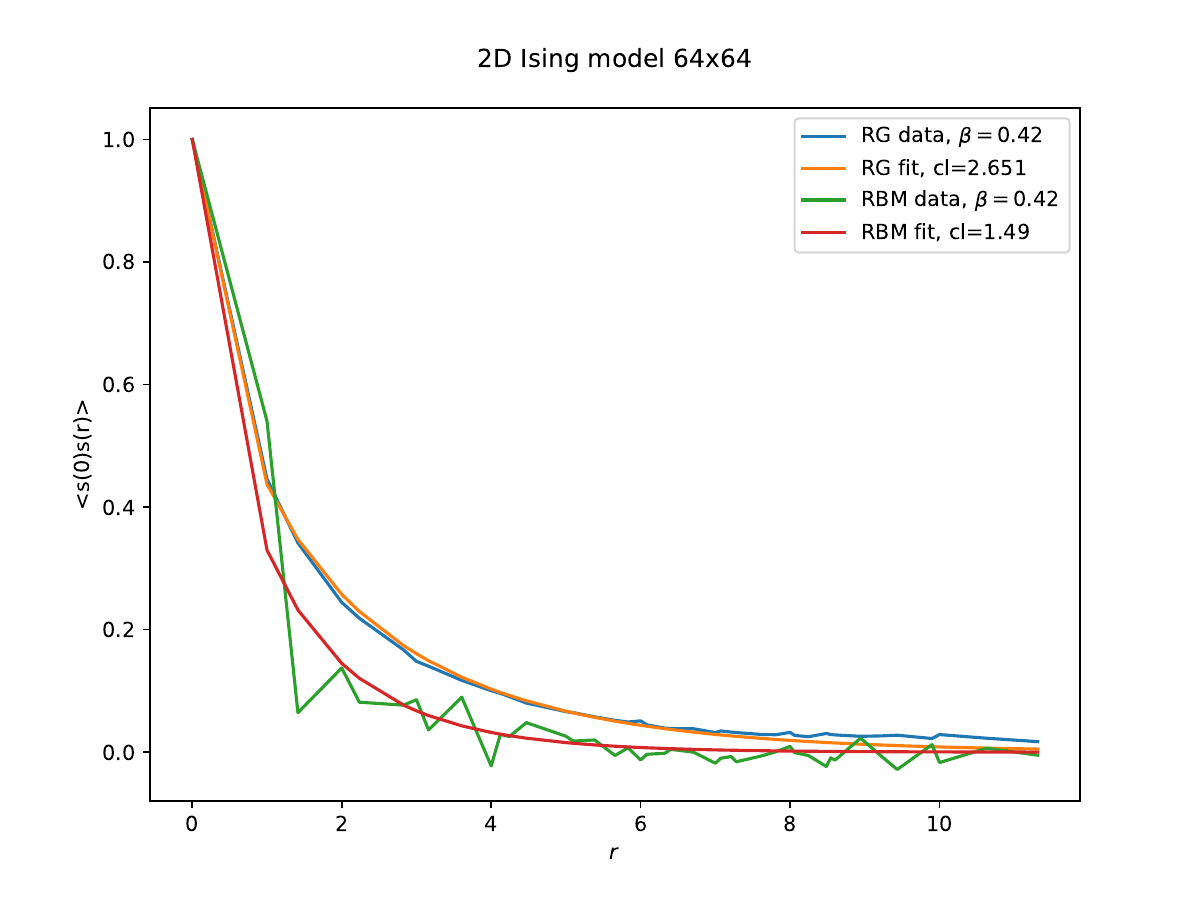}
         \caption{$\beta = .42.$}
     \end{subfigure}
     \hfill
    \begin{subfigure}{0.49\textwidth}
         \centering
         \includegraphics[width=\textwidth]{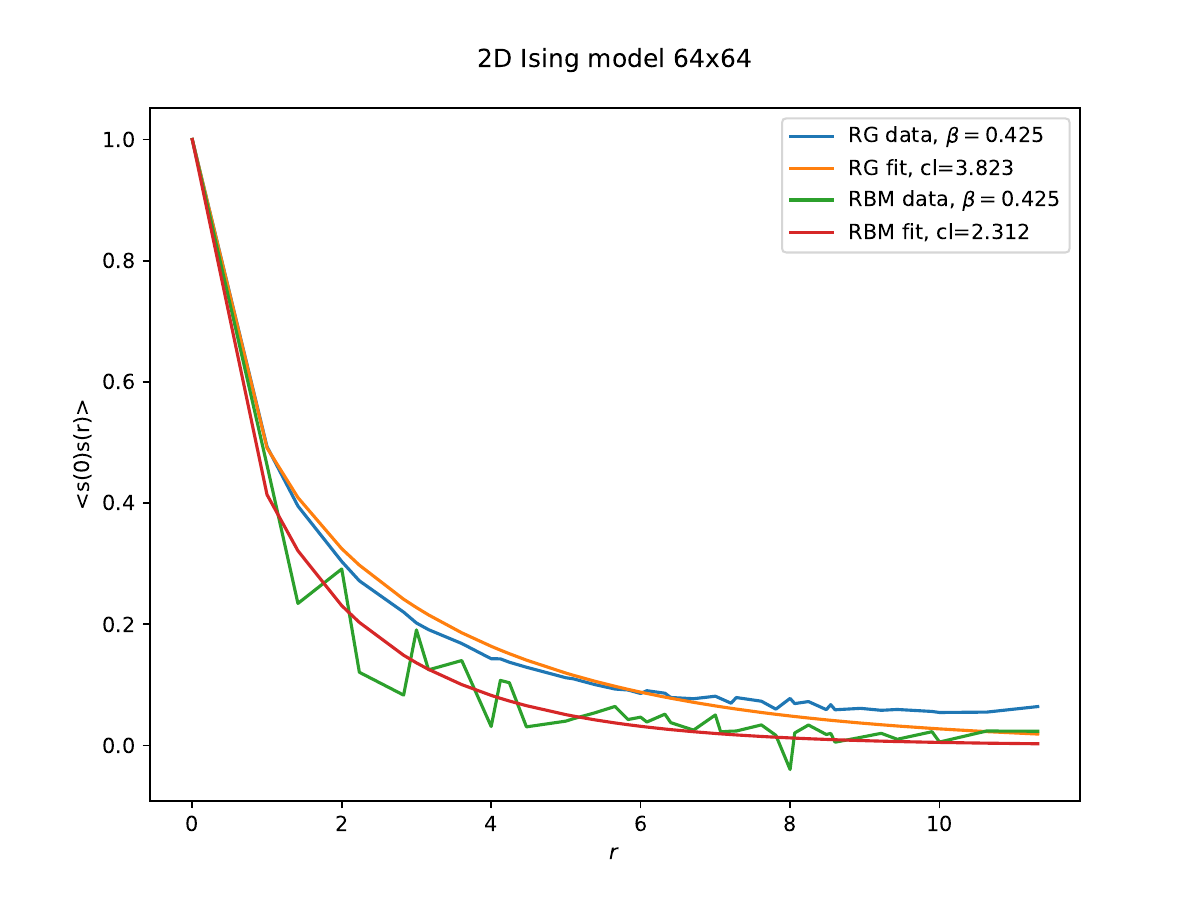}
         \caption{$\beta = .425.$}
     \end{subfigure}
     \hfill
           \begin{subfigure}{0.49\textwidth}
         \centering
         \includegraphics[width=\textwidth]{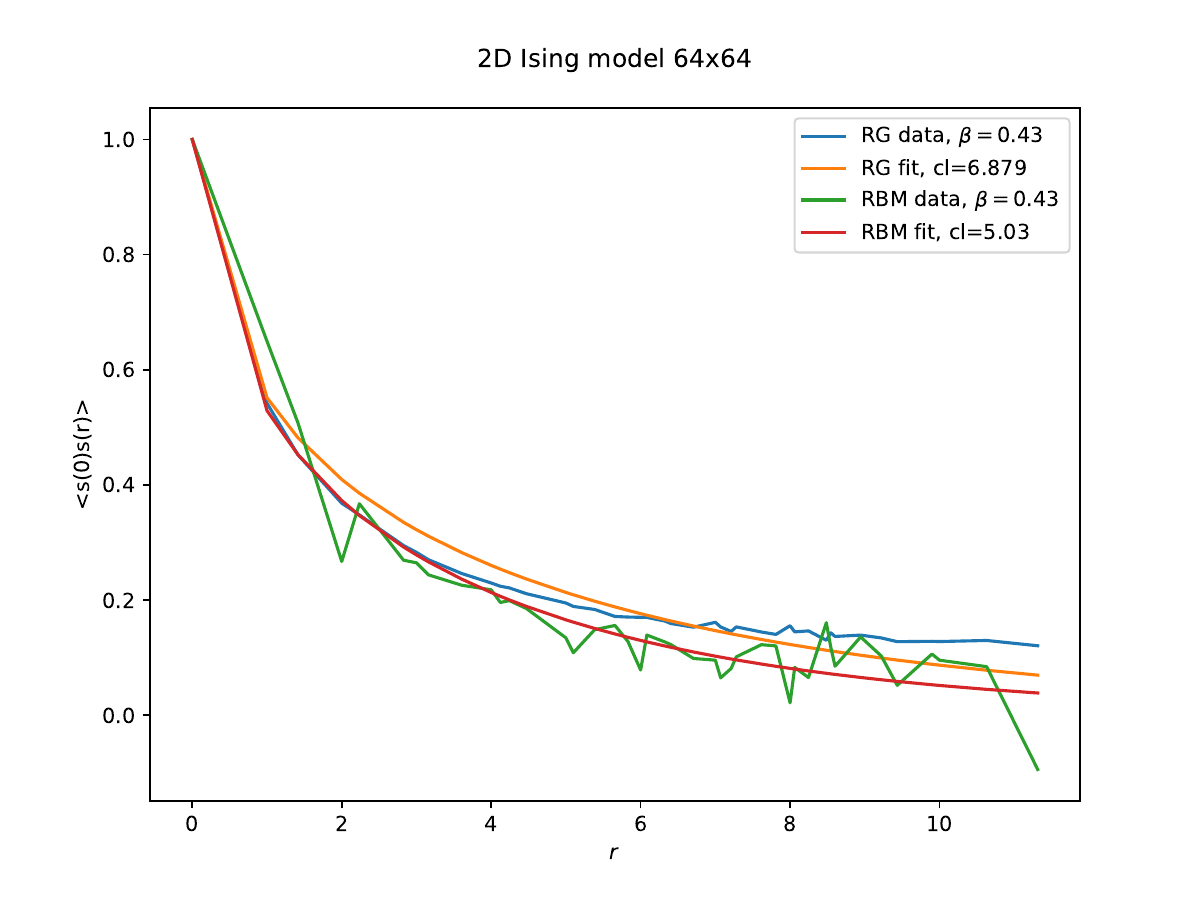}
         \caption{$\beta = .43.$}
     \end{subfigure}
        \caption{Correlation length comparisons for Layer 2.}
        \label{fig:corr_layer_2}
\end{figure}

\begin{figure}[h!]
     \centering
     \begin{subfigure}{0.49\textwidth}
         \centering
         \includegraphics[width=\textwidth]{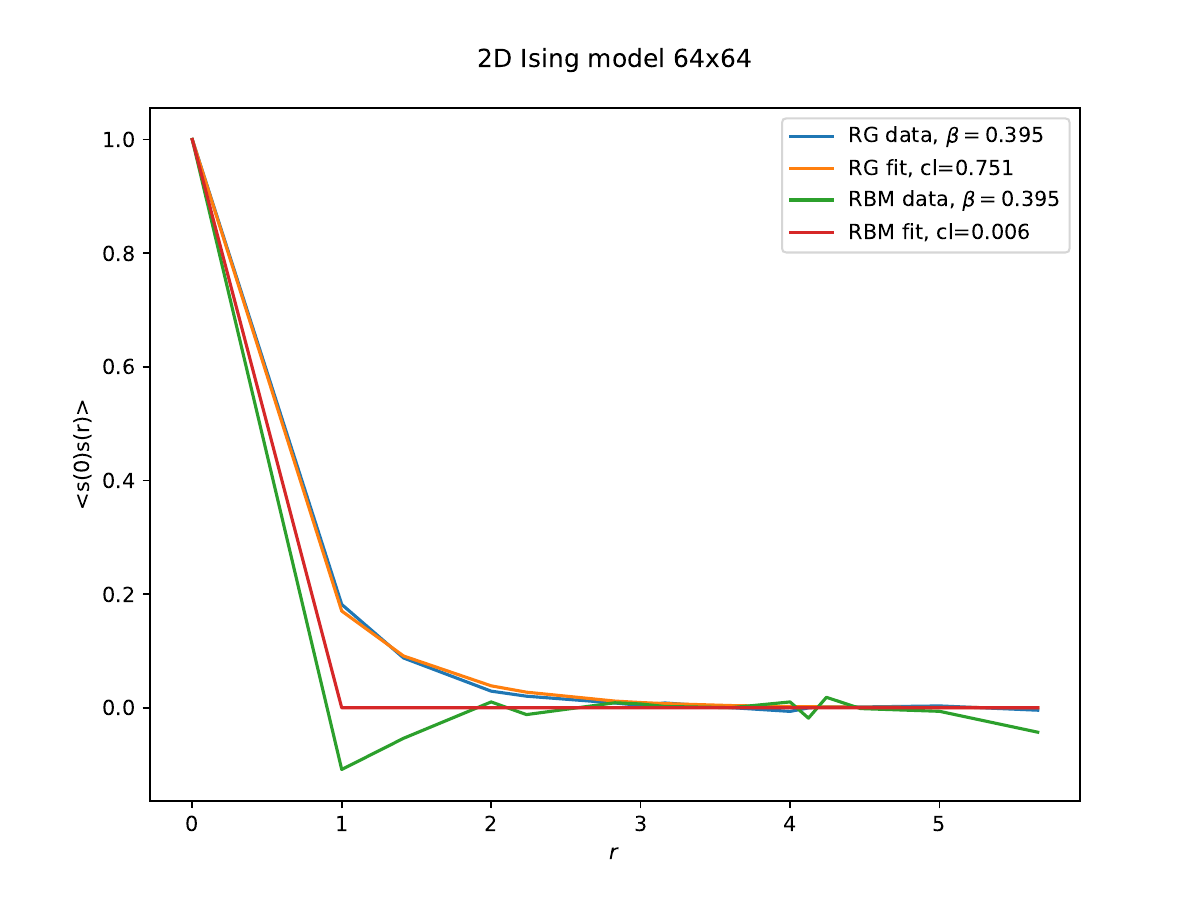}
         \caption{$\beta = .395.$}
     \end{subfigure}
     \hfill
     \begin{subfigure}{0.49\textwidth}
         \centering
         \includegraphics[width=\textwidth]{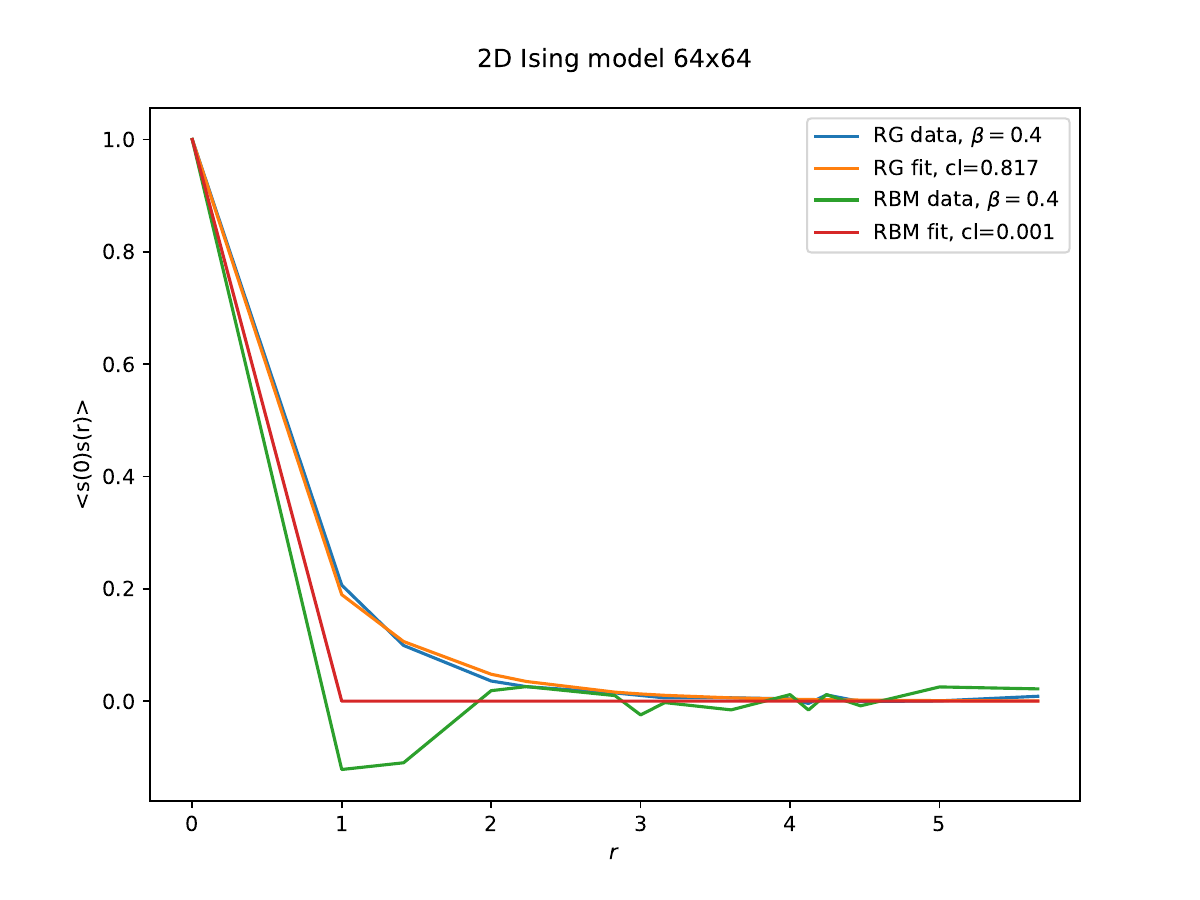}
         \caption{$\beta = .4.$}
     \end{subfigure}
     \hfill
     \begin{subfigure}{0.49\textwidth}
         \centering
         \includegraphics[width=\textwidth]{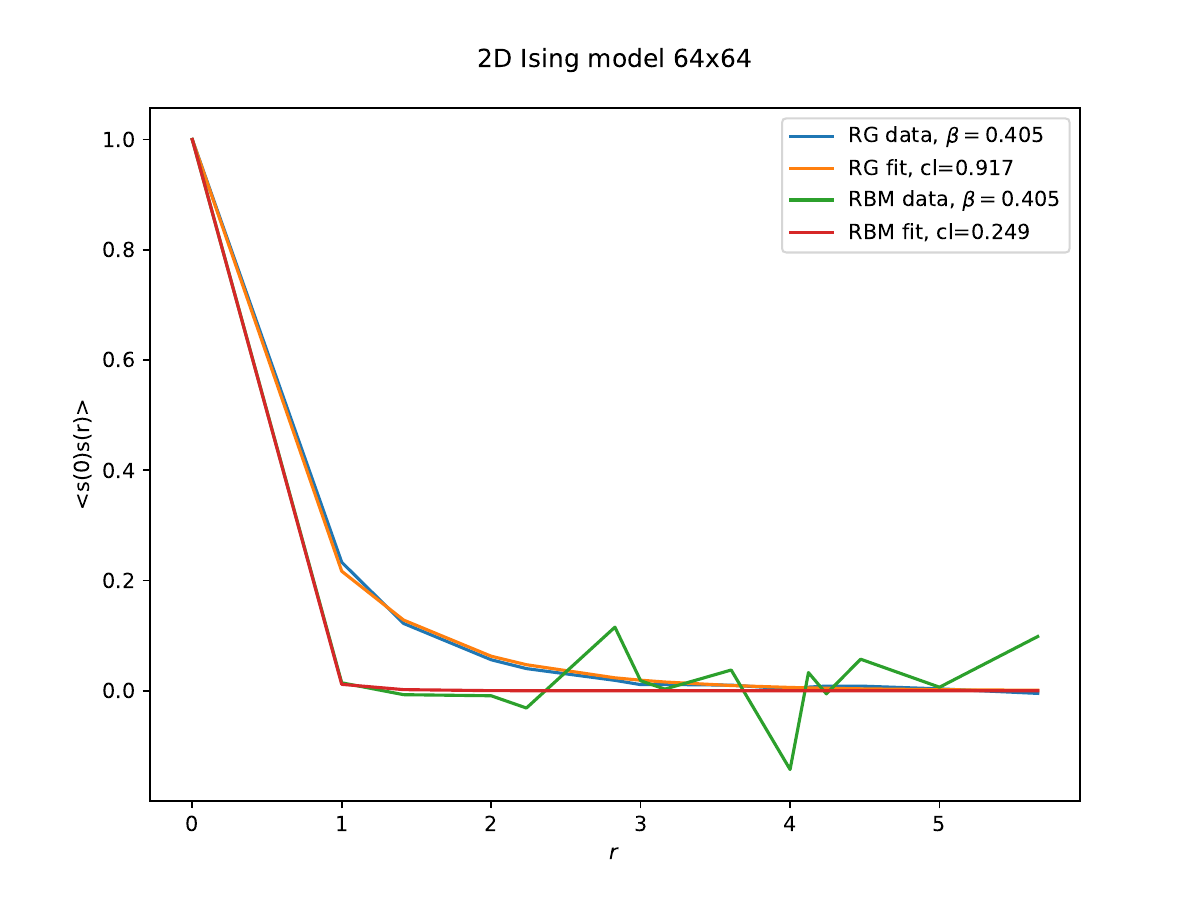}
         \caption{$\beta = .405.$}
     \end{subfigure}
     \hfill
     \begin{subfigure}{0.49\textwidth}
         \centering
         \includegraphics[width=\textwidth]{layer_3_model_0_41.pdf}
         \caption{$\beta = .41.$}
     \end{subfigure}
     \hfill
          \begin{subfigure}{0.49\textwidth}
         \centering
         \includegraphics[width=\textwidth]{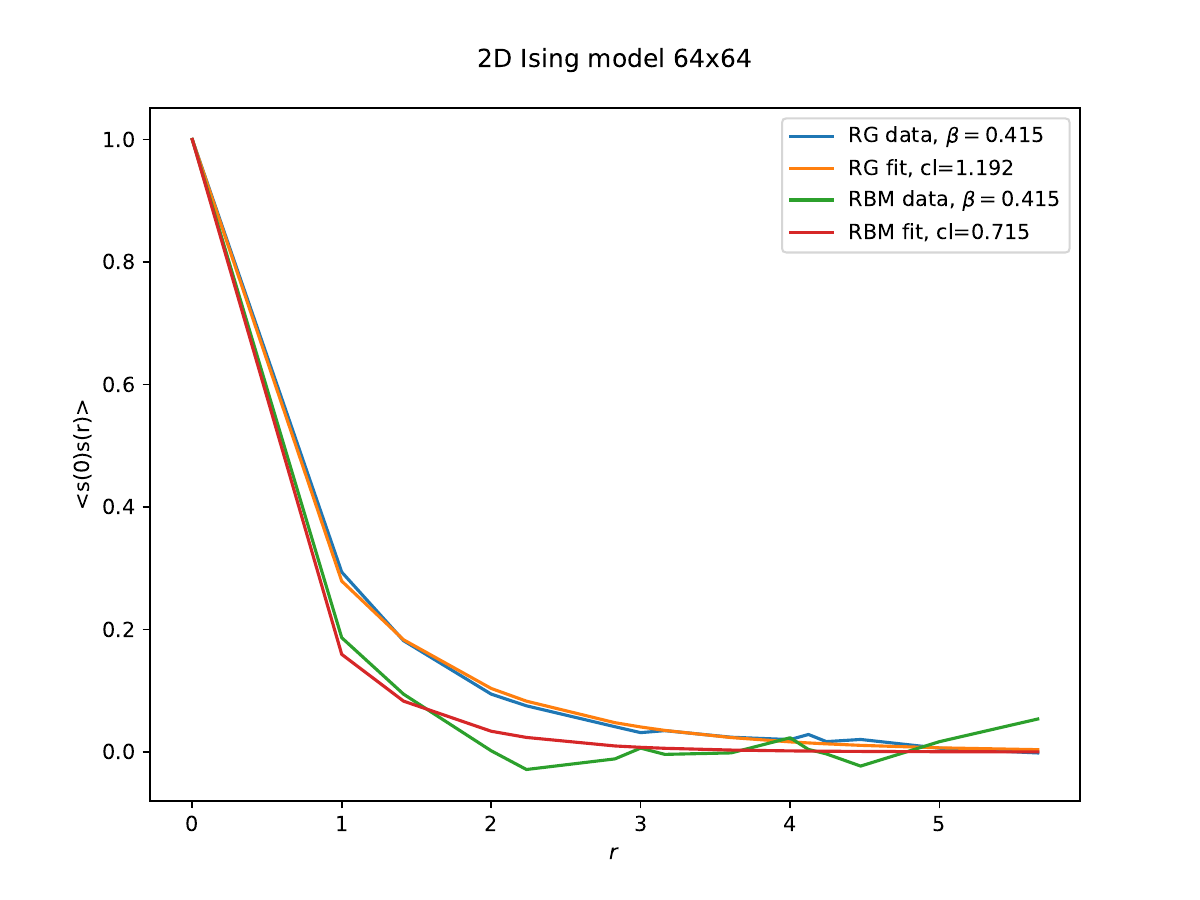}
         \caption{$\beta = .415.$}
     \end{subfigure}
     \hfill
     \begin{subfigure}{0.49\textwidth}
         \centering
         \includegraphics[width=\textwidth]{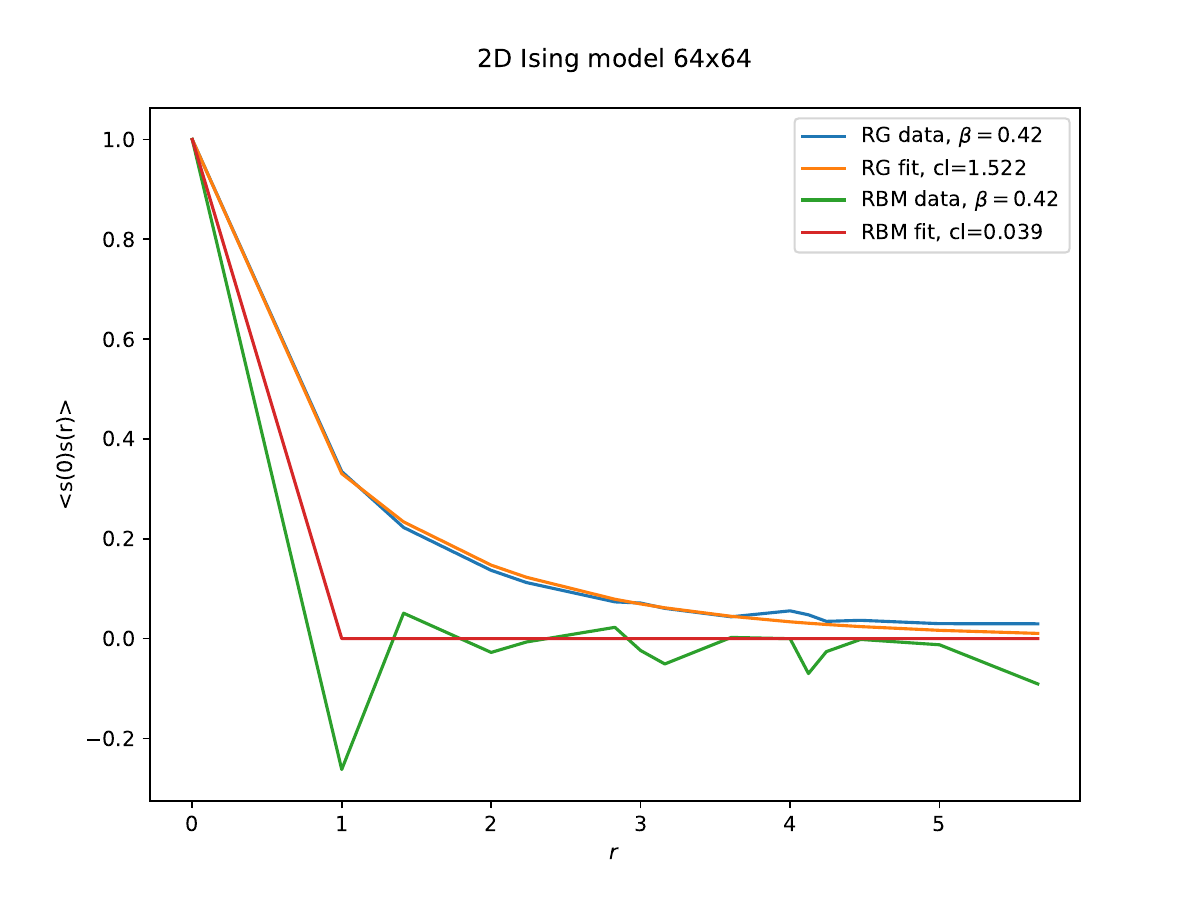}
         \caption{$\beta = .42.$}
     \end{subfigure}
     \hfill
    \begin{subfigure}{0.49\textwidth}
         \centering
         \includegraphics[width=\textwidth]{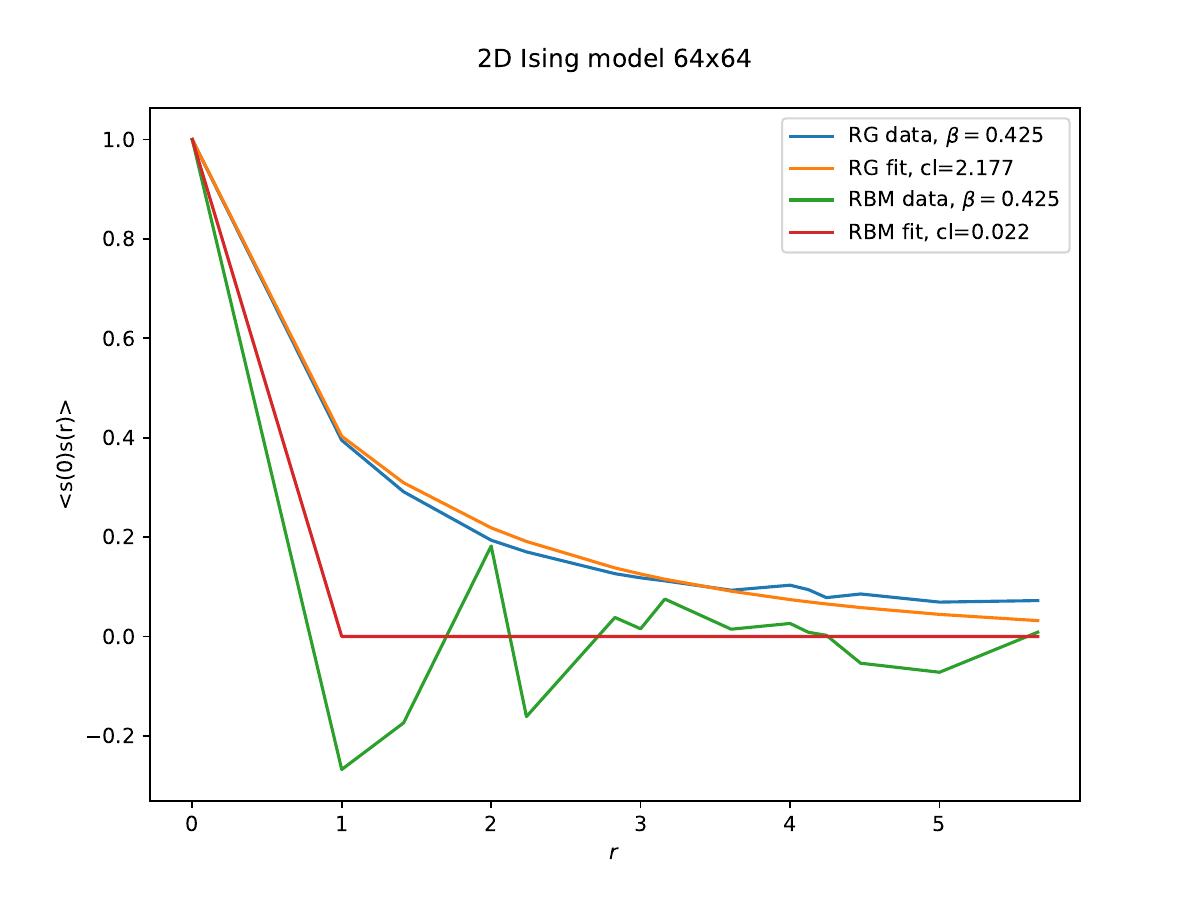}
         \caption{$\beta = .425.$}
     \end{subfigure}
     \hfill
           \begin{subfigure}{0.49\textwidth}
         \centering
         \includegraphics[width=\textwidth]{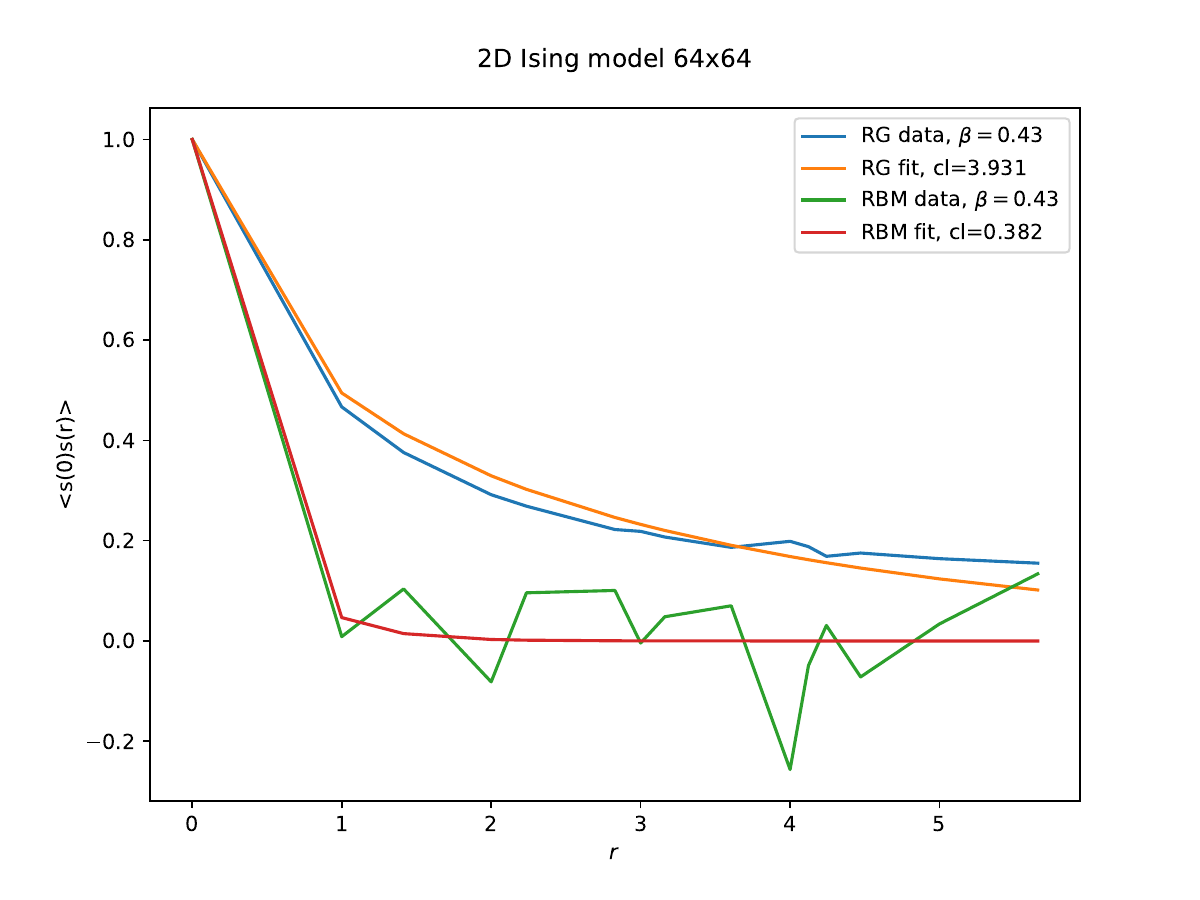}
         \caption{$\beta = .43.$}
     \end{subfigure}
        \caption{Correlation length comparisons for Layer 3.}
        \label{fig:corr_layer_3}
\end{figure}
\printindex

%Nomenclature section!
\nomenclature{Renormalization Group (RG) Flow}{The process by which multi-scaled phenomena in physics are analyzed across scales. The process changes a model's parameters based on the scale of the model}

\nomenclature{Restricted Boltzmann Machine (RBM)}{Simple, energy-based learning models that model data via introducing a set of "hidden" spins that couple to the original spins via a weight tensor. When stacked, RBMs can form DNNs}

\nomenclature{Adam Optimization}{A common optimization algorithm that uses both an adaptive learning rate and a moving average of previous gradients to optimize a function through stochastic gradient descent}

\nomenclature{Visible Spin}{The input data into an RBM}

\nomenclature{Hidden Spin}{The output data of an RBM. It connects to visible spins through a weight tensor}

\nomenclature{Ising Model}{Simple classical model that describes ferromagnetism in materials as a lattice of binary spins}

\nomenclature{Deep Neural Network (DNN)}{Graphical statistical machine learning models in which multiple layers of nodes stack onto one another, and each layer receives input from the layer before it}

\nomenclature{Universality Class}{The class of all models with the same critical behavior and renormalization parameters, and thus the same macroscopic behavior}

\nomenclature{Contrastive Divergence}{The function that we choose to minimize when optimizing for the RBMs}

\nomenclature{Kadanoff Block Spinning}{Process by which the Ising model is renormalized through taking blocks of spins and coarse graining them into single spins}

\nomenclature{Receptive Field Tensor}{In a set of stacked RBMs, this tensor relates the output of the entire stack of RBMs to the input of the entire stack of RBMs}

\nomenclature{Weight Tensor}{The tensor that connects the hidden spins to the visible spins within a given RBM}

\nomenclature{Wolff Algorithm}{Monte Carlo cluster algorithm designed for simulating the Ising model near criticality}

\nomenclature{Stochastic Gradient Descent (SGD) with Momentum}{A common optimization algorithm that uses a random subset of a dataset to optimize a function (SGD), and averages over previous gradients to speed this process up (momentum)}

\end{document}